\documentclass[a4paper, 12pt]{book}

\usepackage[utf8]{inputenc}
\usepackage[english]{babel}
\usepackage{graphicx}
\usepackage{amsmath}
\usepackage{amssymb}
\usepackage{bbold}
\usepackage[abs]{overpic}

\usepackage{cancel}

\newcommand{\Di}{\mathcal{D}}

\newcommand{\Ai}{\mathcal{A}}
\newcommand{\eS}{\mathcal{S}}

\newcommand{\Dsla}{\cancel{\mathcal{D}}}
\newcommand{\Asla}{\cancel{\mathcal{A}}}
\newcommand{\sigsla}{\cancel{\sigma}}

\newcommand{\trace}{\text{Tr}}

\newcommand{\Id}{{\mathbb 1}}
\newcommand{\Etra}{{\mathbb E}}

\author{Pietro Silvi}
\date{\today}
\title{XXX}

\begin{document}

\begin{titlepage}
  \begin{center}
  \hspace{-35pt}
    \vbox to0pt{%
    \vbox to\textheight{\vfil
      \vspace{-6.5cm}
    
    \includegraphics[width=15cm]{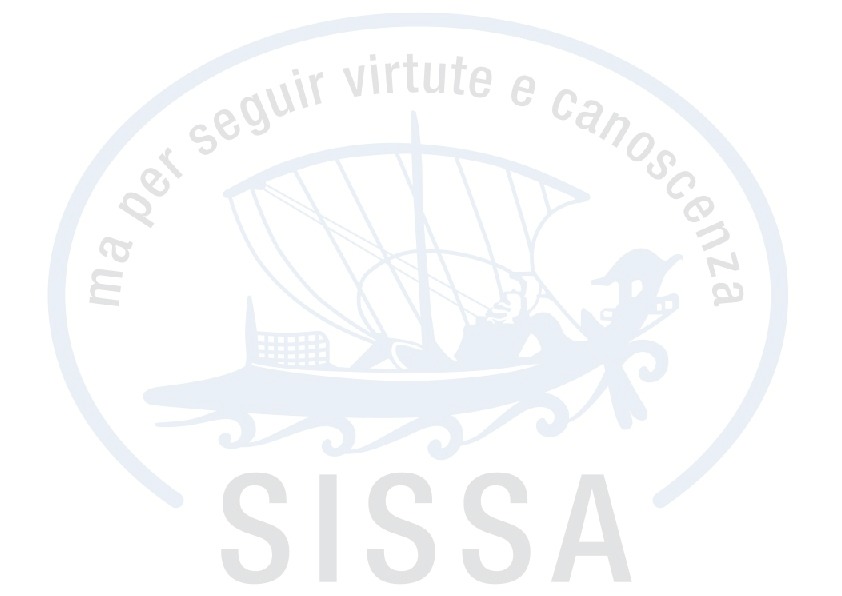}%
    \vfil}\vss}
  \end{center}
\vspace{-4cm}
\begin{center}
\hspace{-2.25cm} 
\begin{minipage}{.23\textwidth}%
 %\centering%
  \includegraphics[height=2.25cm]{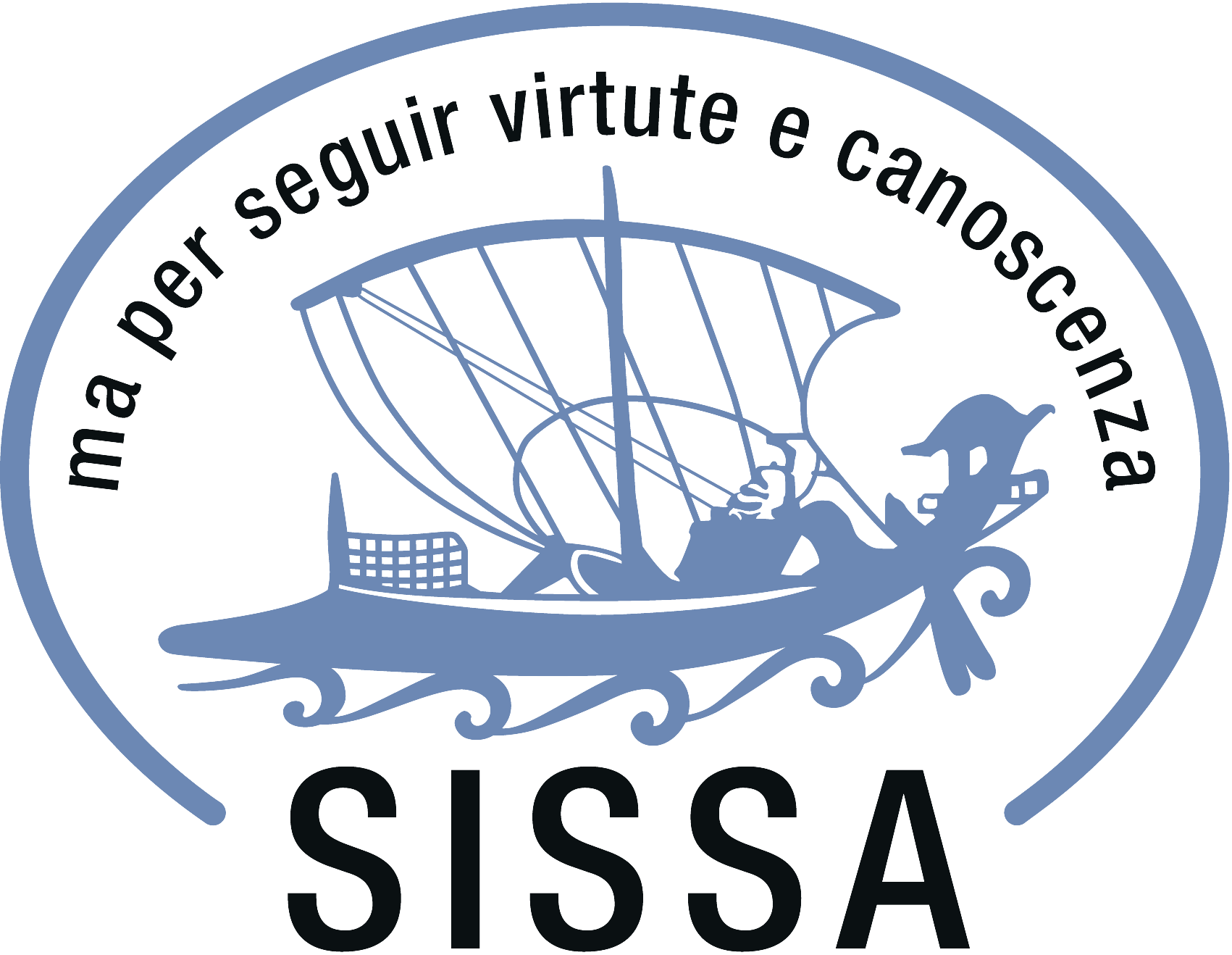}%Ricambiare a 2.5cm per 10pt e 11pt
\end{minipage}%
\begin{minipage}{.77\textwidth}%
%\begin{table}[H]
	\begin{tabular}{l}
\scshape{{INTERNATIONAL SCHOOL FOR ADVANCED STUDIES}}\\%Ricambiare a \large per 10pt e 11pt
\hline
\scshape{
\large{Condensed Matter Physics Sector}}\\ 
\end{tabular}%
%\end{table}%
\end{minipage}
\end{center}

\begin{center}
\vspace{1cm}
\upshape{\Large{Academic Year 2010/2011}}\\

\vspace{3.3cm}
\linespread{1}
{\Huge{\underline{\textbf{Tensor Networks:}}}} \\ \vspace{0.2cm}
{\huge{\textbf{ a quantum-information \\ \vspace{0.2cm}
  perspective on numerical \\ \vspace{0.45cm}
 renormalization groups}}}\\
\linespread{1}
\vspace{4.5cm}
\Large{thesis submitted for the degree of}\\
\Large{\textit{Doctor Philosophiae}}
\end{center}

\vspace{1cm}

\begin{tabbing}
\=\large{Advisors:} \hspace{7.7cm} \=\large{Candidate:}\\%Ricambiare a 5cm per 10pt e 11pt
\vspace{0.5cm}

\>\large{\bfseries{Prof. Giuseppe E. Santoro}} \>\large{\bfseries{Pietro Silvi}}\\
\>\large{\bfseries{Prof. Rosario Fazio}}\\
\>\large{\bfseries{Prof. Vittorio Giovannetti}}\\
\end{tabbing}

\begin{center}
\large{October 28, 2011}
\end{center}

\end{titlepage}
\clearpage
\thispagestyle{empty}
\phantom{a}

\tableofcontents

\chapter[Introduction]{Introduction}

Quantum many-body problems in condensed matter physics are a context of everlasting interest and
relentless investigation in physical research. The macroscopic amount of interacting degrees of
freedom is such that even the simplest models become extremely hard problems, in exact description as well as
in perturbative regime. The analytical and computational complexity of many-body physics is deeply
rooted in the foundations of quantum mechanics themselves: the hardness of these problems undergoes
a scaling-law with the number of elementary constituents (size) of the system $L$, which
is typically much more abrupt than extensive behavior; it usually grows exponentially with $L$.
In physical literature a variety of models which have proven to be
particularly suitable for analytical study was developed, e.g. due to some peculiar local structure or to some
wide symmetry-group invariance; however, the large majority of known non-perturbative Hamiltonians
manifest no attitude towards analytical simplification and must be faced head-on with numerical techniques.
Efficient simulation methods for condensed matter settings are many, often capable of integrating any bit of theoretical
knowledge, then crossing the remaining gap with computation.
At the same time, when no previous hint from theory is available,
the exact problem turns hard again, and computational costs scale fast with system specifics, so
that only very small sizes $L$ are manageable for practical purposes.

The family of \emph{variational} algorithms has always been regarded as one of the most natural and
promising paths in order to address many-body problems at zero temperature:
as ground states of Hamiltonians are minima of the spectrum, searching them through variational procedures
seems most appealing. Yet the crucial point of any variational paradigm is
the capability of reducing the whole amount of degrees of freedom into a small number of effective,
important ones: these must embed all the relevant physics of the target state, while requiring a limited number
of numerical resources. Such primary descriptors, or variational parameters (also known as
reaction coordinates in some contexts) need to be identified and discriminated from the non-influential
ones. This is clearly a delicate issue, especially if no knowledge on the model is available.
In other words, the selection a priori of some appropriate basis of variational
wavefunctions is the fundamental step to undertake, determining the faithfulness and efficiency
(and thus overall success) of any variational algorithm we might want to develop.

In this thesis, we will focus on a very general family of variational wavefunctions,
whose main peculiarity is that their descriptors/parameters are tailored according to simple linear algebraic
relations. The computational power and success of these tools descends from arguments
that were born within quantum information framework: entanglement \cite{Nielsen}. Quantum entanglement
is indeed a resource, but it is also a measure of internal correlations in multipartite systems.
Once we characterized general entanglement properties of many-body ground states,
then by controlling entanglement of a variational trial wavefunction we can exclusively address physical
states, and disregard non-physical states, even before the simulation takes place.
This is the central concept which Tensor Network architectures are based upon.

Historically, the realization and profound understanding of this class of states, was  possible
only after the introduction of Density Matrix Renormalization Group (DMRG), curiously, an algorithm
which is not formulated in variational terms at all.

\section[The age of DMRG]{The age of Density Matrix\\Renormalization Group}

The idea of adapting a Renormalization Group algorithm to a lattice Density Matrix was proposed
by Steven R. White, considered undoubtedly one of the founders of the DMRG methods. In his
first approaches \cite{White92} to the technique, he was inspired by a paper of K.W. Wilson \cite{Wilson75}
where a numerical renormalization group paradigm is applied to the Hamiltonian of a Kondo problem.

The simple, yet brilliant, idea behind White's formulation of DMRG was to replace the traditional
procedure of renormalization group, which acted extensively in the real space, and thus actually
performing a coarse-graining transformation upon the system, with a scheme that applied extensively
in the Hilbert space dimension itself, leading to a site-by-site renormalization scheme.
In practice, assume that we are describing the state of
a given portion of the system (in terms of a density matrix) with a fixed amount of computational resources $D$.
Now when another single constituent (a site) is added to the subsystem, the resulting dimension
grows linearly with the local dimension $d$ of the new component, $\sim d D$.
Then renormalization is performed, allowing us to represent the new subsystem with the same initial amount of
resources $D$, cutting the least relevant density matrix-eigenstates out of the description.
Obviously, such operations still manifest a group structure, and they are summoned every time the density matrix
dimension (and not the real-space size) increases of a given factor $d$: thus DMRG.

The great amount of credit and interest gathered by DMRG is surely due to its outstanding
successfulness for low-dimensionality quantum systems. In particular, for one-dimensional (open-boundary)
systems, DMRG achieved variational precisions (compared to experiment, and theory whenever possible)
that challenged other simulation approaches. At the end of the '90s, it was considered
probably the most powerful numerical method to address 1D problems, with practically no
knowledge on the model required a priori. In literature, DMRG picture has been exploited in several
settings of both physics and quantum chemistry, and numerous variants of its original formulation
were proposed \cite{WhiteOneSite, SchollDMRG};
in the end the basic idea was proven to be winning, even though within its dimensionality limits.

\section[The advent of MPS]{The advent of Matrix Product States}

The DMRG concept was quite renown, but it was in 2004 that the in-depth reason of its success was
fully understood: when F. Verstraete, D. Porras and J. I. Cirac started to investigate quantum
states built via a DMRG algorithm under a quantum information perspective \cite{PrimoMPS}. In fact, they realized
that DMRG states had a strict equivalence relation with finitely correlated states, i.e. lattice states
whose entanglement is upper-bounded by an arbitrary finite value, which does not scale with the system size.

Moreover, ground states of short-ranged (non-critical) Hamiltonians, have been known for quite some time to satisfy the
so-called \emph{area-law} of entanglement \cite{Arealaws, Arealaws2, Eisarea}. This general rule
was developed in quantum information contexts, but carries important physical prescriptions.
It claims that the partition entanglement of a non-critical ground state does \emph{not} scale
as the volume of the parted spatial region $\sim L^{\#N}$
(which is the typical entanglement scaling for random states), but rather with the parting surface
$\sim L^{\#N-1}$, where $\#N$ is the number of spatial dimensions.

It is clear that for 1D systems, the correct area law is given by a non-scaling constant $\sim L^{0}$:
indeed entanglement of 1D non-critical ground states typically saturates to a finite bound.
This means, in turn, that ground states of 1D non-critical Hamiltonians are finitely-correlated
states, and therefore, that DMRG procedure can approximate them with arbitrary precision, and their entanglement as well.

Moreover, ref.~\cite{PrimoMPS} has shown that finitely-correlated states allow a direct, simple,
immediate algebraic representation in terms of a product of matrices, each of these
matrices storing all the information related to a single renormalization process.
This provides a one-to-one local correspondence between DMRG, and these Matrix Product State (MPS)
\cite{MPSreview, MPSVidal, MPScontinuous, TuttoMPS}
which, as they allow an explicit analytic expression, actually form a
class of tailored variational wavefunctions.

Although not-homogeneously perceived by the condensed-matter physics community, this discovery
was definitely a breakthrough, for several reasons.
First, a variational formulation of DMRG opened the possibility for new numerical strategies
based on finitely-correlated states, so that several minimum-search algorithms could
be applied, but still using just the right amount of necessary computational resources.
Secondly, the algebraic MPS expression provided faster ways to access physical information,
and, at the same time, it
allowed innovative problem-solving options even from the analytical point of view \cite{MPSground}.
Finally, it is easy to generalize the MPS concept to suit physical settings other than 1D, and
still taking care of appropriate area-laws.
This argument lead, for instance, to the design of Product of Entangled Pairs States (PEPS) \cite{PEPSgo, PowaPEPS}.

\section{Entanglement Renormalization}

The understanding of the relationship between DMRG, MPS, and finitely-correlated states provided
an unquestionable paradigm for dealing numerically
with non-critical 1D system in a fully-contextualized theoretical framework.
Despite the fair success of adapting these algorithms to critical problems as well
(although in a hand-waving and unnatural way), it was guessed that due to the presence of scale-invariance
symmetry, an old-fashioned real-space renormalization group would be more appropriate to simulate strongly
correlated systems.
Indeed, a state which is locally stationary under the action of a coarse-graining transformation would
definitely be scale invariant. At the same time Wilson's numerical RG had the unpractical feature of suffering
loss of short-range detailed structure, identified by translational instability of entanglement.

An intriguing proposal to work around this trouble was introduced by G. Vidal in 2007, who fist spoke
about Entanglement Renormalization \cite{MERAzero}. The idea is intuitive, yet very effective:
we are still performing a real-space renormalization group, but prior to the renormalization process
itself, we apply a quasi-local unitary transformation, whose purpose is to decrease the correlation
among regions who are to be renormalized separately. Since the goal of these
unitary gates is to absorb, and thus store, entanglement out of the pre-RG state, they are commonly
called \emph{disentanglers}. The disentangling operation is then scheduled before every real-space RG operation
takes place; in the end, such entanglement renormalization acts at every lenghtscale, since every RG step performs
actually a scale transformation.

Similarly to DMRG and MPS, these entanglement-RG states have a variational counterpart as well.
Precisely, it is possible to define a class of tailored variational wavefunctions, whose descriptors
are tied together by linear relations, reproducing exactly the entire set of those states.
Such states are thus called \emph{Multiscale Entanglement Renormalization Ansatz} (MERA)
\cite{MERAalgo, MERAevo, MERAgauge, MERAtopo},
and manifest a natural attitude towards describing strong correlation and criticality \cite{QuMERAchan}.

\vspace{1em}
MPS, MERA, PEPS, are different classes of variational states sharing some important attributes:
they are able to capture interesting physics, yet they require a small, manageable number of parameters,
with simple algebraic rules and direct access to relevant physical information.
Physicists started to regard them as belonging together to a larger,
comprehensive family of tailored variational states, whose entanglement can be directly controlled through
the selection of a related graph geometry. This is the concept of Tensor Network states
\cite{TNevenbly, TNfujii, TNdenny, TNchan}.

\section{Outline}

The thesis is organized as follows:

\begin{itemize}
 \item In chapter \ref{chap:OBCMPS} we will present an in-depth review on Matrix
 Product States, in 1D open-boundary conditions settings.
 We will show how the MPS analytical expression is derived by the DMRG algorithm, show its entanglement bounds
 and sketch their algebraic manipulation features. We will explain how to achieve physical information on these states
 in a computationally-fast scheme, and present some protocols to simulate ground states.

 \item In chapter \ref{chap:PBCMPS} we will generalize the concept of MPS
 to periodic boundary conditions systems, and discuss how
 translational homogeneity of the representation allows us to well-define the thermodynamical limit for MPS.
 This will be the proper setting to show that MPS manifest natural non-criticality, whose signature
 is an exponential decay of two-point correlations.
 We will then present possible generalizations to the PBC case of MPS minimization algorithms, focusing on
 some tricks of the trade useful to speed-up and stabilize the procedure.

 \item In chapter \ref{chap:TN} we will explain how the MPS architecture can be generalized to more complex Tensor
 Network geometries. We will investigate the entanglement properties of Tensor Network states, and some
 of their common algebraic features, like efficient contraction schemes or the adaptability to fermionic
 contexts. We will also present some remarkable subclasses of Tensor Networks, like PEPS and CPS,
 and discuss on how they relate to each other.

 \item In chapter \ref{chap:TTNMERA} we will put our attention on Trees and MERA, two classes of Tensor Networks that
 share most of their main features. We will show their natural attitude to describe critical states in 1D,
 identified both by a logarithmic violation of the area law, and more importantly by manifesting power-law
 decaying correlations. Critical exponents, as well as the TTN/MERA state properties in the thermodynamical limit,
 are completely characterizable by adopting a completely positive trace-preserving map formalism.
 We will investigate other general properties of such TN-architectures, e.g. the possibility to construct a parent Hamiltonian.
\end{itemize}

\subsection{Original content}

Here I will list the original contribution I developed personally, either as brand-new material
or as a reinterpretation of previous knowledge, during my Philosophiae Doctorateship.

\begin{itemize}
 \item Section \ref{sec:Slater}: analytical MPS representation of Slater Determinants, many-body basis change,
 and configuration interaction states.

 \item Section \ref{sec:MPJastrow}: matrix product representation for correlator product states (Jastrow factors).

 \item Most analytical results of chapter \ref{chap:TTNMERA}: scaling properties for TTN, parent Hamiltonians,
 fluctuations, boundaries, hybrid geometries; and respective generalizations to MERA.
\end{itemize}
Part of this research was published in \cite{IoTree, BERA, MatteoMera}.

\chapter{Matrix Product States} \label{chap:OBCMPS}

It was in 2004 that computational physicists started to consider Density Matrix Renormalization Group
according to a Quantum Information perspective \cite{PrimoMPS};
they realized that it is possible to understand DMRG
in a variational sense, in which the role played by entanglement and quantum correlation is clear.
Indeed, as a quantum many-body state achieved by DMRG procedure is uniquely defined
by the renormalization transformations (intended as endomorphisms upon the density matrices space)
one can regard such transformations as variational elements, and every single choice of those
elements defines a state within a set of tailored variational wavefunctions.
Nicely enough, it was discovered that such states allow a simple and immediate analytical description,
where their many-body wavefunction, wrote upon a product basis of one-body levels, appears just
as a product of variational matrices, thus leading to the name of Matrix Product States (MPS).
As a matter of fact, such transparent description allowed research to further investigate the
properties of these states, leading to a deeper understanding of DMRG as well, and in the
end the new knowledge served well the purpose of gaining more computational power in simulations,
through a wider range of algebraic manipulations and the adaptability of variational-based algorithms.

In the end, it was the Matrix Product State picture that helped to understand the deep physical
reason of DMRG successfulness in 1D. Indeed, MPS have proven to be in tight relation with
1D finitely correlated states \cite{FannesNacht, AKLT},
and in turn this set is known to include ground states of short-range
interacting non-critical Hamiltonians. Such argument holds not only for finite, isolated
systems, but extends naturally to open and/or thermodynamical limit systems (as long as a zero temperature
can be defined), allowing DMRG/MPS to succeed even in these cases.

It is important to point out that Matrix Product State methods can be successfully
adopted for dealing with fermionic
systems, and they naturally avoid the sign problem which, in stead, is a major issue on other
variational fermionic algorithms like Montecarlo.
Finally, MPS are not merely a numerical tool, they have proven in various contexts to be
fundamental to address analytically several condensed matter models \cite{Karimipour, TuttoMPS}.

\section[MPS construction from DMRG]{Matrix Product State construction\\from DMRG} \label{sec:dmrgtomps}

Following the formalism of \cite{White92, MPSreview}, we start with a one-dimensional lattice,
$L$ (length) being the total number of sites, $d$ (local Hilbert dimension) the number of levels per site,
and where open boundary conditions (OBC) are chosen for simplicity.
Let us assume that we are describing a given quantum state of this system obtained via a DMRG algorithm:
$D \ll d^L$ is the maximal number of states allowed for the each renormalization.
Now let $\ell < L$ be the site where the last density matrix renormalization was
applied in the algorithm (while moving, say, to the right), this means that we know
the reduced density matrix of the state $\rho_{\ell}^{L}$, involving all the sites from $\ell$ to the leftmost,
but we only have access to its renormalized form: namely in stead of keeping in memory
all its $d^{\ell}$ eigenvectors $|L_j\rangle_{\ell}^{L}$
and their relative probability $p_j$ ($\sum_j^{d^{\ell}} p_j = 1$, in decreasing order $p_j \geq p_{j+1}$),
only the $D < d^{\ell}$ of such vectors are kept, of course those with highest probability:
\begin{equation} \label{eq:Piribango}
 \tilde{\rho}_{\ell}^{L} = \sum_{j=1}^{D} \tilde{p}_j \;|L_j\rangle_{\ell}^{L} \langle L_j|,
 \quad \mbox{where} \quad \tilde{p}_j = \frac{p_j}{\sum_{k = 1}^D p_k}
\end{equation}
ensures that the new statistic $\tilde{p}_j$ is properly renormalized.
The $D$ vectors $|L_j\rangle_{\ell}^{L}$ appearing in \eqref{eq:Piribango} are orthogonal by construction,
and are normalized on their space of definition (the left part of the system, i.e. sites to the left of $\ell$);
these shall be the only relevant vectors in the left part of the system which will contribute to the
full analytical expression of the global state.
Now, the trick of the trade, is considering that $\tilde{\rho}_{\ell}^{L}$ was obtained from the
reduced density matrix of the previous DMRG step
$\tilde{\rho}_{\ell-1}^{L} =\sum_{j}^{D} \tilde{q}_j \;|L_j\rangle_{\ell-1}^{L} \langle L_j|$,
which of course was renormalized $\sum_{j}^{D} \tilde{q}_j = 1$.
This means that, the $D$ states $|L_j\rangle_{\ell-1}^{L}$ joint together with the local levels $|s\rangle$
at site $\ell$, are enough to generate the set of $|L_j\rangle_{\ell}^{L}$:
\begin{equation} \label{eq:wata}
 |L_j\rangle_{\ell}^{L} = \sum_{k = 1}^D \sum_{s = 1}^d 
 A_{k,j}^{[\ell] s} \; |L_k\rangle_{\ell-1}^{L} \otimes |s\rangle_{\ell}.
\end{equation}
Here $A_{k,j}^{[\ell] s}$ represents the decomposition over the product basis; it can be either understood as
a three-indices tensor (indices being $s$, $k$, and $j$), or, since typically $D \gg d$, a
$d$-long array (through $s$) of square $D \times D$ matrices (from $j$ to $k$).
Preservation of orthonormality among $|L_j\rangle^L$ states determines a condition upon
$A$; indeed, assuming that the local basis $|s\rangle$ is orthonormal by definition,
one finds that the transformation must satisfy the equation
\begin{equation} \label{eq:phoneL1}
 \sum_{k = 1}^D \sum_{s = 1}^d {A^{\star}}_{ki}^{[\ell] s} \; A_{kj}^{[\ell] s} = \delta_{i,j},
\end{equation}
where the superscript $\vphantom{A}^{\star}$ stands for complex conjugation.
Eq.~\eqref{eq:phoneL1} can be rewritten in an even clearer form once the $A$ are intended as $D$ by $D$ matrices:
\begin{equation} \label{eq:phoneL2}
 \sum_{s = 1}^d {A^{\dagger}}_{s}^{[\ell]} \cdot A_{s}^{[\ell]} = \Id,
\end{equation}
with $\cdot$ being the standard rows-by-columns matrix product.
Equation \eqref{eq:phoneL1} follows directly from the fact that
\begin{equation}
 \delta_{i,j} = \langle L_i|L_j\rangle_{\ell}^{L} =
 \sum_{k, m}^D \sum_{s, t}^d {A^{\star}}_{ik}^{[\ell] s} \; A_{mj}^{[\ell] t}
 \langle L_k|L_m\rangle_{\ell-1}^{L} \langle s|t \rangle_{\ell}.
\end{equation}
but $\langle s|t \rangle_{\ell} = \delta_{s,t}$ by assumption, and
$\langle L_k|L_m\rangle_{\ell-1}^{L} = \delta_{k,m}$ is the inductive hypothesis, thus \eqref{eq:phoneL1}.

Moreover, looking at \eqref{eq:phoneL2} under a quantum information perspective,
we clearly understand that the $A^{[\ell]}_{s}$ actually
form a set of Kraus operators for a completely positive trace preserving (CPT) map \cite{Nielsen};
CPT maps are the most generic transformations mapping density matrices into density matrices,
they represent the action of a quantum channel on an open system
(for details, see appendix \ref{app:cptchap}).
On our case, the set of $A_{s}^{[\ell]}$ define exactly
the CPT map $\mathcal{M}_{\text{CPT}}$ performing the inverse DMRG transformation
$\tilde{\rho}_{\ell}^{L} \longrightarrow \tilde{\rho}_{\ell-1}^{L}$ as follows
\begin{equation} \label{eq:mappuzL}
 \tilde{\rho}_{\ell-1}^{L} = \mathcal{M}_{\text{CPT}}\left[ \tilde{\rho}_{\ell}^{L} \right] \equiv
 \sum_{s = 1}^d A_{s}^{[\ell]} \cdot \tilde{\rho}_{\ell}^{L} \cdot {A^{\dagger}}_{s}^{[\ell]}.
\end{equation}

Let us now go back at \eqref{eq:wata}; as DMRG procedure is recursive, one can apply the same
argument several times, for instance, until reaching the first site. This leads to
\begin{equation} \label{eq:MPopen}
 |L_j\rangle_{\ell}^{L} = \sum_{s_1 \ldots s_{\ell} = 1}^d 
 \left( A_{s_1}^{[1]} \cdot A_{s_2}^{[2]} \cdot \ldots \cdot A_{s_{\ell}}^{[\ell]} \right)
 |s_1\rangle_{1} \otimes |s_2\rangle_{2} \otimes \ldots \otimes |s_{\ell}\rangle_{\ell}
\end{equation}
the component of $|L_j\rangle_{\ell}^{L}$ over the local homogeneous product basis made of
$|s_1 \ldots s_{\ell}\rangle$ is now expressed in terms of a product of matrices.
These matrices have a number of rows and columns always bounded by $D$, although in general
it is impossible to require for all of them to be $D \times D$
square matrices and satisfy \eqref{eq:phoneL1}
at the same time; typically, as the left boundary grows near, their size shrink, up to the
first site, whose $A^{[1]}_{s_1}$ are all one-row matrices. Similarly, $A^{[L]}_{s_L}$ are
one-column matrices.

In fact, we can associate a \emph{correlation space dimension} $D_{\ell'}$
to every site to the left of $\ell$ (more appropriately: to every \emph{bond} $\ell'$),\
and state that the matrices $A^{[\ell']}_{s_{\ell'}}$ have common size $D_{\ell'} \times D_{\ell'+1}$.
In order for \eqref{eq:phoneL1} to hold, the following inequality is a necessary condition:
\begin{equation} \label{eq:ConsistenL}
 D_{\ell'} \leq d \cdot D_{\ell'-1}
\end{equation}
for $\ell' \leq \ell$, where it is intended that $D_0 = 1$.
As \eqref{eq:ConsistenL} provides an upper bound to the correlation dimension, so does the
parametric renormalization dimension $D$, typically acting as a simple cutoff.
Indeed, in standard DMRG algorithms it is a natural choice to adopt $D_{\ell'} = \min\{d^{\ell'}, D\}$.

\subsection[Single center site DMRG]{Completing the picture:\\single center site DMRG}

So far we understood how to represent in a clear, simple analytical way of representing the left block
states $|L_j\rangle_{\ell}^{L}$ of our DMRG. In order to complete the picture to include the whole
system we first need to identify which specific architecture of DMRG (of those proposed in literature)
is being used. For simplicity we start with the case where the DMRG optimization is performed by
considering at every step a single center site (system) and the left and right blocks (environment),
as in ref.~\cite{WhiteOneSite}.

According to such description, the $D$ left environment renormalized states $|L_j\rangle_{\ell-1}^{L}$,
joint with the right environment renormalized states $|R_j\rangle_{\ell}^{R}$, and the site levels
$|s\rangle_\ell$, generate the DMRG state of the whole system:
\begin{equation} \label{eq:DMfull}
 |\Psi_{\text{DMRG}}\rangle = \sum_{j,k = 1}^{D} \sum_{s = 1}^{d} C^{[\ell] s}_{j,k} \;
 |L_j\rangle_{\ell-1}^{L} \otimes |s\rangle \otimes |R_k\rangle_{\ell}^{R},
\end{equation}
the components tensor $C^{[\ell] s}_{j,k}$ defines uniquely the precise state within the DMRG space.

Now, the very argument we used previously to prove that for $|L_j\rangle_{\ell-1}^{L}$
eq.~\eqref{eq:MPopen} holds, can be applied in a similar fashion to right environment vectors as well.
Precisely, if
$\tilde{\rho}_{\ell}^{R} =\sum_{j}^{D} \tilde{p}_j \;|R_j\rangle_{\ell}^{R} \langle R_j|$ is a
density matrix obtained by recursive renormalizations starting from the right boundary (site $L$),
we have that
\begin{equation} \label{eq:MPopen2}
 |R_j\rangle_{\ell}^{R} = \sum_{s_{\ell+1} \ldots s_{L} = 1}^d 
 \left( B_{s_{\ell+1}}^{[\ell+1]} \cdot \ldots \cdot B_{s_{L-1}}^{[L-1]} \cdot B_{s_{L}}^{[L]} \right)
 |s_{\ell+1}\rangle_{\ell+1} \otimes \ldots \otimes |s_{L}\rangle_{L}.
\end{equation}
Like before, we encounter a product of matrices $B_{s_{\ell'}}^{[\ell']}$ ($\ell' > \ell$);
but notice that this time the preservation of orthonormality property for the $|R_j\rangle_{\ell}^{R}$
states goes from the right boundary towards the center site, i.e. propagating toward the left.
This means that the matrices $B_{s_{\ell'}}^{[\ell']}$ should satisfy a relation which is different
from \eqref{eq:phoneL2}, namely:
\begin{equation} \label{eq:phoneR2}
 \sum_{s = 1}^d {B}_{s}^{[\ell']} \cdot {B^{\dagger}}_{s}^{[\ell']} = \Id.
\end{equation}
To satisfy the present equation the following constraint on matrices dimensions (${B}_{s}^{[\ell']}$ being
a $D_{\ell'-1} \times D_{\ell'}$ complex matrix) is due:
\begin{equation}
 D_{\ell'} \leq d \cdot D_{\ell'+1}, \quad \forall \ell'>\ell; \quad D_L = 1.
\end{equation}
Consistently, the CPT mapping associated to ${B}_{s}^{[\ell']}$ performs the inverse RG transformation,
i.e. towards the right
\begin{equation} \label{eq:mappuzR}
 \tilde{\rho}_{\ell'+1}^{R} = \mathcal{M}_{\ell'}^{R} \left[ \tilde{\rho}_{\ell'}^{R} \right] \equiv
 \sum_{s = 1}^d {B^{\dagger}}_{s}^{[\ell']} \cdot \tilde{\rho}_{\ell'}^{R} \cdot {B}_{s}^{[\ell']}.
\end{equation}

After all these considerations, we can put \eqref{eq:MPopen} and \eqref{eq:MPopen2} into \eqref{eq:DMfull},
the state $|\Psi_{\text{DMRG}}\rangle$ appears automatically expanded in the natural separable basis:
\begin{equation} \label{eq:PrehistoMPS}
 |\Psi\rangle = \sum_{s_1 \ldots s_L = 1}^d 
 \left( A_{s_{1}}^{[1]} \cdot \ldots \cdot A_{s_{\ell-1}}^{[\ell-1]} \cdot C_{s_{\ell}}^{[\ell]}
 \cdot B_{s_{\ell+1}}^{[\ell+1]} \cdot \ldots \cdot B_{s_{L}}^{[L]} \right)
 | s_1 \ldots s_L \rangle,
\end{equation}
where $C_{s_{\ell}}^{[\ell]}$ has been written as an array of matrices as well. Notice that the term within the
parentheses is a scalar due to the fact that the $A^{[1]}_{s_1}$ matrices are actually row vectors
(one-row matrix) and the $B_{s_{L}}^{[L]}$ are column vectors (one-column matrix).
Equation \eqref{eq:PrehistoMPS} tells us that all the components of $|\Psi\rangle$
over the canonical basis are the product of $L$ matrices, which are local objects and depend only
on the state of the site they are associated with. This is the definition of Matrix Product State \cite{PrimoMPS}.

Once we require that matrices $A_{s}^{[\ell']}$ and $B_{s}^{[\ell']}$
respectively satisfy \eqref{eq:phoneL2} and \eqref{eq:phoneR2} in order to preserve orthonormality
of environment states, the proper normalization of the state $|\Psi\rangle$ becomes an equation
involving the element $C_{s_{\ell}}^{[\ell]}$ alone.
Indeed we can explicitly calculate the norm of $|\Psi\rangle$ by exploiting its MPS representation
\eqref{eq:PrehistoMPS}, as
\begin{multline}
 \langle \Psi | \Psi \rangle = \sum_{s_1 \ldots s_L = 1}^{d}
 \left( A_{s_1}^{[1]} \otimes {A^{\star}}_{s_1}^{[1]} \right) \ldots
 \left( A_{s_{\ell-1}}^{[\ell-1]} \otimes {A^{\star}}_{s_{\ell-1}}^{[\ell-1]} \right)
 \times \\ \times	
 \left( C_{s_{\ell}}^{[\ell]} \otimes {C^{\star}}_{s_{\ell}}^{[\ell]} \right)
 \left( B_{s_{\ell+1}}^{[\ell+1]} \otimes {B^{\star}}_{s_{\ell+1}}^{[\ell+1]} \right) \ldots
 \left( B_{s_L}^{[L]} \otimes {B^{\star}}_{s_L}^{[L]} \right),
\end{multline}
which, after some algebraic manipulation, reads
\begin{multline} \label{eq:MPSmanip}
 \langle \Psi | \Psi \rangle = \sum_{s_1 \ldots s_L = 1}^{d}
 \trace \left[ A_{s_1}^{[1]} A_{s_2}^{[2]} \ldots
 A_{s_{\ell-1}}^{[\ell-1]} C_{s_{\ell}}^{[\ell]} B_{s_{\ell+1}}^{[\ell+1]}
 \ldots \right. \\ \left.
 \ldots B_{s_{L-1}}^{[L-1]} B_{s_L}^{[L]} {B^{\dagger}}_{s_L}^{[L]} {B^{\dagger}}_{s_L-1}^{[L-1]}
 \ldots {B^{\dagger}}_{s_{\ell+1}}^{[\ell+1]} {C^{\dagger}}_{s_{\ell}}^{[\ell]}
 A_{s_{\ell-1}}^{[\ell-1]} \ldots {A^{\dagger}}_{s_2}^{[2]} {A^{\dagger}}_{s_1}^{[1]} \right].
\end{multline}
And, by exploiting \eqref{eq:phoneL2}, \eqref{eq:phoneR2} and cyclicity of the trace, all the
$A_{s}^{[\ell']}$ and $B_{s}^{[\ell']}$ matrices disappear from the equation. In the end,
we are left with
\begin{equation}
 1 = \langle \Psi | \Psi \rangle =
 \sum_{s_{\ell} = 1}^{d} \trace\left[ C_{s_{\ell}}^{[\ell]} {C^{\dagger}}_{s_{\ell}}^{[\ell]} \right]
 = \sum_{j,k = 1}^{D} \sum_{s = 1}^{d} C_{jk}^{[\ell]s} {C^{\star}}_{jk}^{[\ell]s},
\end{equation}
the desired normalization condition.
We would like to remark that manipulations performed in order to derive
\eqref{eq:MPSmanip} are identically suitable if we were to calculate the one-site reduced density
matrix $\rho^{1}$ of $|\Psi\rangle$ at site $\ell$, namely
\begin{equation}
 \rho^{1}_{\ell} = \trace_{\ell^\text{c}} \left[ | \Psi \rangle| \langle \Psi | \right] =
 \sum_{s, t = 1}^{d} \trace\left[ C_{s}^{[\ell]} {C^{\dagger}}_{t}^{[\ell]} \right]
 |s\rangle \langle t|,
\end{equation}
where, clearly, the partial trace spans $\ell^{\text{c}}$, the complementary of $\ell$.
Similarly, the reduced density matrices $\tilde{\rho}_{\ell}^{L}$ and $\tilde{\rho}_{\ell}^{R}$
are easily accessible one the representation \eqref{eq:PrehistoMPS} is at our disposal.
In fact, those read
\begin{equation}
 \tilde{\rho}_{\ell}^{L} = \sum_{j,k,m = 1}^{D} \sum_{s = 1}^{d}
 C_{jm}^{[\ell]s} {C^{\star}}_{km}^{[\ell]s} |j\rangle\langle k|
 , \qquad
 \tilde{\rho}_{\ell}^{R} = \sum_{j,k,m = 1}^{D} \sum_{s = 1}^{d}
 C_{mj}^{[\ell]s} {C^{\star}}_{mk}^{[\ell]s} |j\rangle\langle k|,
\end{equation}
and all the other reduced density matrices achieved through the original DMRG algorithm can
be generated starting from the previous expressions via \eqref{eq:mappuzL} and \eqref{eq:mappuzR}.
In practice, the MPS representation \eqref{eq:PrehistoMPS} provides us a quick access
to the whole information of the DMRG algorithm, and at the same time is more immediate and
flexible than DMRG itself, proving a useful computational tool as we will see later on.

\subsection{Double center site DMRG}

The original DMRG protocol proposed by White \cite{White92}, and most of the DMRG architectures still
in use nowadays adopt a slightly different picture than the one we presented in \eqref{eq:DMfull}.
The basic idea is to consider the active system block on which to perform the minimization as it
were composed by two adjacent sites in stead of just one, coupling to the left and right environments
as before. Of course, at fixed renormalization dimension parameter $D$, this procedure is more
expensive from a computational point of view, but provides a big gain in algorithm precision  and
rapid convergence; moreover, it allows to manipulate symmetries in a more
natural and flexible fashion, thus improving algorithm stability.
\begin{equation} \label{eq:DM2full}
 |\Psi_{\text{DMRG}}\rangle = \sum_{j,k = 1}^{D} \sum_{s,t = 1}^{d} T^{s,t}_{j,k} \;
 |L_j\rangle_{\ell-1}^{L} \otimes |s\rangle_{\ell} \otimes
 |t\rangle_{\ell+1} \otimes |R_k\rangle_{\ell+1}^{R},
\end{equation}
In order to recover a complete analytical expression of the form \eqref{eq:PrehistoMPS},
some manipulation on the components tensor $T^{s,t}_{j,k}$ has to be made.
The simplest path to take, is to consider two composite indexes $\alpha$ and $\beta$:
$\alpha$ representing the pair $\{j,s\}$, while $\beta$ representing $\{k,t\}$. This
allows us to write $T_{\alpha \beta}$
as a matrix from index $\alpha$ to $\beta$, of dimension $dD \times dD$.
At this point we perform a Singular Value Decomposition (SVD) upon $T$:
\begin{equation} \label{eq:SVDone}
 T_{\alpha \beta} = A^{[\ell]}_{\alpha \gamma} \;\lambda_{\gamma}\; B^{[\ell+1]}_{\gamma \beta},
\end{equation}
where $A^{[\ell]}$ and $B^{[\ell+1]}$ are unitary matrices, and the diagonal matrix of singular
values is positive semidefinite, i.e. $\lambda_{\gamma} \geq 0$ $\forall \gamma$.
If we write again $A^{[\ell]}$ and $B^{[\ell+1]}$ in the original $j$, $s$ and $k$, $t$ indices
it is clear that they satisfy the proper orthonormalization propagation requirements,
respectively \eqref{eq:phoneL2} and \eqref{eq:phoneR2}.
This tells us that \eqref{eq:DM2full} can also be interpreted as follows
\begin{equation} \label{eq:Shmidtdeco}
 |\Psi_{\text{DMRG}}\rangle = \sum_{\gamma = 1}^{dD}
 \lambda_{\gamma} \; |L_{\gamma}\rangle_{\ell}^{L} \otimes |R_{\gamma} \rangle_{\ell}^{R},
\end{equation}
where, following the formalism of \eqref{eq:wata} we substituted
\begin{equation}
 \begin{aligned}
 |L_{\gamma} \rangle_{\ell}^{L} &= \sum_{j = 1}^D \sum_{s = 1}^d 
 A_{j,{\gamma}}^{[\ell] s} \; |L_j\rangle_{\ell-1}^{L} \otimes |s\rangle_{\ell}
 \qquad \mbox{and} \\
 |R_{\gamma} \rangle_{\ell}^{R} &= \sum_{k = 1}^D \sum_{t = 1}^d 
 B_{{\gamma},k}^{[\ell+1] t} \; |R_k\rangle_{\ell+1}^{R} \otimes |t\rangle_{\ell+1}.
 \end{aligned}
\end{equation}
Indeed, as $|L_{\gamma}\rangle_{\ell}^{L}$ (resp. $|R_{\gamma} \rangle_{\ell}^{R}$)
form a set of orthonormal vectors for the left (right) partition of the system,
equation \eqref{eq:Shmidtdeco} actually represents the Schmidt decomposition of $|\Psi_{\text{DMRG}}\rangle$,
cut at site $\ell$.
The Schmidt coefficients $\lambda_{\gamma}$ must satisfy the normalization condition
$1 = \langle \Psi | \Psi \rangle = \sum_{\gamma} \lambda_{\gamma}^2$; they
are the positive square roots of the probabilities $\tilde{p}$
in the ($dD$-renormalized) reduced density matrices of either partition of the system. The latter read
\begin{equation}
 \tilde{\rho}_{\ell}^{L} = \sum_{\gamma = 1}^{dD}
 \lambda_{\gamma}^{2} |L_{\gamma} \rangle_{\ell}^{L} \langle L_{\gamma}|
 , \qquad
 \tilde{\rho}_{\ell}^{R} = \sum_{\gamma = 1}^{dD}
 \lambda_{\gamma}^{2} |R_{\gamma} \rangle_{\ell}^{R} \langle R_{\gamma}|.
\end{equation}
In conclusion, the SVD decomposition \eqref{eq:SVDone} allows us to recover a matrix product expression, substantially
identical to \eqref{eq:PrehistoMPS}. Precisely
\begin{equation} \label{eq:SchmidtMPS}
 |\Psi\rangle = \sum_{s_1 \ldots s_L = 1}^d
 \left( A_{s_{1}}^{[1]} \cdot \ldots \cdot A_{s_{\ell}}^{[\ell]} \cdot \lambda^{[\ell]}
 \cdot B_{s_{\ell+1}}^{[\ell+1]} \cdot \ldots \cdot B_{s_{L}}^{[L]} \right)
 | s_1 \ldots s_L \rangle,
\end{equation}
where $\lambda^{[\ell]}$ is intended as the diagonal matrix with elements $\lambda^{[\ell]}_{\gamma}$.
Normally, in order to press further on with DMRG algorithm, the left (or right) density matrix
should be properly renormalized to be $D \times D$ dimensioned; but this is straightforward, by
just cutting off the smallest singular values $\lambda_{\gamma}$ until only the $D$ largest of them remain,
and renormalize as follows
\begin{equation} 
 {\tilde{\lambda}}_{\gamma} = \frac{\lambda_{\gamma}}{\sqrt{\sum_{\eta}^{D} \lambda^2_{\eta}}}
 \quad \gamma \in \{1..D \}
 \qquad \longrightarrow \qquad \sum_{\gamma}^{D} {\tilde{\lambda}}_{\gamma}^2 = 1,
\end{equation}
so that state normalization $\langle \Psi | \Psi \rangle = 1$ is preserved.
Even after this cutoff, $A_{s_{\ell}}^{[\ell]}$ will still satisfy the condition
\eqref{eq:phoneL2}: this descends automatically from the fact that if any number of columns are cut
out of a unitary matrix, a left-isometric rectangular matrix ($A^{\dagger} A = \Id$, but
$A A^{\dagger} = P = P^2 = P^{\dagger} \neq \Id$) is obtained, thus \eqref{eq:phoneL2}.
Similarly, the $B_{s_{\ell+1}}^{[\ell+1]}$ resulting from the cutoff will still satisfy \eqref{eq:phoneR2}.

It is trivial to make \eqref{eq:SchmidtMPS} formally match \eqref{eq:PrehistoMPS},
we can either identify $C_{s}^{[\ell]} = A_{s}^{[\ell]} \lambda^{[\ell]}$, or alternatively
$C_{s}^{[\ell+1]} = \lambda^{[\ell]} B_{s}^{[\ell+1]}$ and recover the previous formalism.
Similarly, we can manipulate \eqref{eq:PrehistoMPS} to appear in the latter form,
such operation will be clearer once we introduced a
state-invariant transformation (gauge) of the MPS representation, which we are going to review in section \ref{sec:MPSGauge}.

\section[Valence bond picture]{Valence bond picture and\\MPS entanglement} \label{sec:valencebond}

Interestingly enough, it is possible to interpret Matrix Product States in a way 
\cite{PrimoMPS, MPSreview} that clarifies many of their quantum correlation properties,
often referred to as \emph{valence bond picture}.
The basic idea is to start by considering an auxiliary system space that is actually larger
than the proper Hilbert $\mathcal{H} = \mathbb{C}^{d^L}$ of our 1D (open boundary, so far) system;
then we project onto the original system by means of a local transformation.
Let us associate to any site of our quantum chain a \emph{pair} of $D$-dimensional spins, one per bond formed by that site.
the starting state is prepared so that every pair of virtual spins corresponding
to the same bond, is initially in a maximally entangled state
$|\Phi^{+}\rangle = D^{-\frac{1}{2}}\sum_{\alpha}^{D} | \alpha \alpha \rangle$,
known in literature as \emph{entangled bond}.
Then apply a local on-site map
\begin{equation} \label{eq:VBmap}
 \mathcal{A}^{[\ell]} = \sum_{s = 1}^{d} \sum_{j,k = 1}^{D} A^{[\ell]s}_{j,k} |s\rangle_{\ell}
 \,\langle j,k |^{\text{aux}}_{\ell}
\end{equation}
to every site $\ell \in \{1 .. L\}$, where $|s\rangle_{\ell}$ is a canonical state in the local physical
space at site $\ell$ while $| j,k \rangle^{\text{aux}}_{\ell}$ is a vector of the (double spin)
respective auxiliary space. Equation \eqref{eq:VBmap} applied on the initial valence bond state
$\bigotimes_{\ell} \mathcal{A}^{[\ell]} (\bigotimes_{\ell'} |\Phi^{+}\rangle^{\text{aux}}_{\ell', \ell'+1})$
leads to an expression where auxiliary indexes of neighboring $A^{[\ell]s}_{j,k}$ are contracted.
Then, by writing any tensor $A^{[\ell]s}_{j,k}$ as a set of $d$ complex $D \times D$ matrices,
the state we are describing is naturally expressed in the matrix product form
\begin{equation} \label{eq:NaturMPS}
 |\Psi\rangle = \sum_{s_1 \ldots s_L = 1}^d 
 \left( A_{s_{1}}^{[1]} \cdot A_{s_{2}}^{[2]} \cdot
 \ldots \cdot A_{s_{L-1}}^{[L-1]} \cdot A_{s_{L}}^{[L]} \right)
 | s_1 \ldots s_L \rangle.
\end{equation}
In general, not only the $\mathcal{A}^{[\ell]}$ operations, but even the auxiliary dimension $D$ of the
entangled pair $|\Phi^{+}\rangle$ can be site dependent; this way the $A_{s}^{[L]}$ matrices
are $D_{\ell-1} \times D_{\ell}$ dimensioned (where $D_0 = D_L = 1$ to ensure that the complete Matrix
Product expression is a scalar quantity).

It is important to focus on the fact that, since the $\mathcal{A}^{[\ell]}$ are basically a
LOCC transformation (i.e. achievable by means of Local Operations and Classical Communication)
its action can only degrade entanglement, thus
the entanglement of the resulting state $|\Psi\rangle$ is bound by that of the initial state,
which is known and straightforward to calculate.
Precisely, consider the entanglement entropy related to a left-right partition of the state
$|\Psi\rangle$, say at bond $\{\ell, \ell+1\}$. This is by definition the Von Neumann entropy of the reduced
density matrix to the left (or right) part of the system, and it is bounded by the entanglement
of the original pair across the bond:
\begin{equation}
 \mathcal{S}_{\text{VN}}\left( {\rho}_{\ell}^{L} \right) \equiv
 -\trace \left[ {\rho}_{\ell}^{L} \log {\rho}_{\ell}^{L}\right] \leq \log D_{\ell};
\end{equation}
where $\log D$ is the entanglement of a maximally entangled spin pair of dimension $D$, like $|\Phi^{+}\rangle$
(to check this, just consider that
${\bar{\rho}}^B_D \equiv \trace[|\Phi^{+}\rangle \langle \Phi^{+} |] = \frac{1}{D} \,\Id$, thus
$\mathcal{S}_{\text{VN}} ({\bar{\rho}}^B_D) = \log D$).

In conclusion, a Matrix Product State, i.e. quantum state on a 1D lattice allowing an analytic representation
as in eq.~\eqref{eq:NaturMPS}, has well-defined upper bounds on its entanglement.
The entropy related to a left-right partition of the system is bounded by the logarithm of $D_{\ell}$,
with $D_{\ell}$ being the dimension of the Matrix Product \emph{bondlink} $\ell$ we are breaking.

\section[Completeness of MPS representation]{Completeness of Matrix Product\\State representation}
\label{sec:MPScompleteness}

The previous observation involving entanglement in MPS becomes even more meaningful once we will provide
a theorem of completeness of MPS representations. Indeed, we are going to prove that, as long as we are NOT
imposing a finite bound to the maximal MPS bondlink dimension $D$, \emph{any} 1D finite lattice state
can be expressed exactly as an MPS.

The argument behind this claim is quite simple indeed. Let us choose a site $\ell$ within the
(open boundary) lattice, $1 < \ell < L$. Let $|\Psi\rangle$ be the global quantum state,
and let us consider the Schmidt decomposition of $|\Psi\rangle$ where the first subsystem
is made by sites $\{1..\ell\}$ and the second by $\{\ell+1 .. L\}$:
\begin{equation} \label{eq:Schm230}
 |\Psi\rangle = \sum_{\alpha}^{D_\ell} \lambda^{[\ell]}_{\alpha} |L_{\alpha} \rangle^{L}_{\ell}
 \otimes |R_{\alpha} \rangle^{R}_{\ell}.
\end{equation}
Following the formalism of previous sections, $|L_{\alpha} \rangle^{L}_{\ell}$ are left block
Schmidt vectors and $|R_{\alpha} \rangle^{R}_{\ell}$ the right block ones.
$\lambda^{[\ell]}_{\alpha}$ are the Schmidt coefficients ($\sum_{\alpha} {\lambda^{[\ell]}_{\alpha}}^2 = 1$),
but now the number $D_{\ell}$ of values the index $\alpha$ can assume is not anymore defined \emph{a priori};
instead it depends of the specifics of the $|\Psi\rangle$, precisely on its partition entanglement
across the bond $\{\ell,\ell+1\}$.
Similarly, we could adopt the same argument when partitioning the system between sites $\ell-1$ and $\ell$,
namely
\begin{equation}
 |\Psi\rangle = \sum_{\alpha}^{D_{\ell - 1}} \lambda^{[\ell-1]}_{\alpha} |L_{\alpha} \rangle^{L}_{\ell-1}
 \otimes |R_{\alpha} \rangle^{R}_{\ell-1}.
\end{equation}
Now, since both descriptions are exact, and the fact that the block $\{1 .. \ell \}$ is actually
the composition of block $\{1 .. \ell-1 \}$ with the site $\ell$ alone, we must conclude that
the set of product states of the form
$|L_{\alpha} \rangle^{L}_{\ell-1} \otimes |s\rangle_\ell$ generate every
$|L_{\alpha} \rangle^{L}_{\ell}$ state (completeness argument).
In fact, we may define the decomposition tensor $A^{[\ell]}$ as follows
\begin{equation} \label{eq:MconstrL}
 A_{\alpha, \beta}^{[\ell] s} =
 \left( \vphantom{\sum} \langle L_{\alpha} |_{\ell-1}^{L} \otimes
 \langle s|_{\ell} \right) |L_{\beta} \rangle_{\ell}^{L}.
\end{equation}
so that we can expand $|L_{\alpha} \rangle^{L}_{\ell}$ in equation \eqref{eq:Schm230} in the
new product basis
\begin{equation}
 |\Psi\rangle = \sum_{\alpha = 1}^{D_{\ell - 1}} \sum_{\beta = 1}^{D_{\ell}} \sum_{s = 1}^{d}
 \left( A_{\alpha, \beta}^{[\ell] s} \; \lambda^{[\ell]}_{\beta} \right)
 |L_{\alpha} \rangle^{L}_{\ell-1} \otimes |s\rangle_\ell \otimes |R_{\alpha} \rangle^{R}_{\ell}.
\end{equation}
Of course, the completeness argument we used poses a relevant constraint upon dimensions
of Schmidt decompositions; in particular as $|L_{\alpha} \rangle^{L}_{\ell}$ are orthogonal,
they are linearly independent, and since the $|L_{\alpha} \rangle^{L}_{\ell-1} \otimes |s\rangle_\ell$
basis can generate them, it must be that $D_{\ell} \leq d \cdot D_{\ell-1}$. Then, by construction,
if we write $A_{\alpha, \beta}^{[\ell] s}$ as a set of matrices (from $\beta$ to $\alpha$) then it holds
\begin{equation} \label{eq:phoneL3}
 \begin{aligned}
  \sum_{s = 1}^d {A^{\dagger}}_{s}^{[\ell]} \cdot A_{s}^{[\ell]} &= \Id \\
  \sum_{s = 1}^d {A}_{s}^{[\ell]} \cdot \Lambda^{[\ell]} \cdot {A^{\dagger}}_{s}^{[\ell]} &= \Lambda^{[\ell-1]},
 \end{aligned}
\end{equation}
where the positive diagonal matrices $\Lambda^{[\ell]}$ are given by
$\Lambda^{[\ell]}_{\alpha, \beta} = \delta_{\alpha, \beta} (\lambda^{[\ell]}_{\beta})^2$,
and correspond to the Schmidt-basis reduced density matrices of the partition,
i.e. $\Lambda^{[\ell]} = \rho_{\ell}^{L} = \rho_{\ell}^{R}$.
The previous equations resume together the orthonormalization preservation relation \eqref{eq:phoneL2},
and the CPT mapping propagation of reduced density matrices \eqref{eq:mappuzL}.

Similarly to \eqref{eq:MconstrL}, one can perform the formal expansion into site $\ell$
and reduced environment for the right block of the partition, where we can define
\begin{equation} \label{eq:MconstrR}
 B_{\alpha, \beta}^{[\ell] s} =
 \left( \vphantom{\sum} \langle R_{\beta} |_{\ell}^{R}
 \otimes \langle s|_{\ell} \right) |R_{\alpha}\rangle_{\ell-1}^{R},
\end{equation}
which allows us to write, provided the completeness constraint upon Schmidt dimensions
$D_{\ell-1} \leq d \cdot D_{\ell}$ holds,
\begin{equation}
 |\Psi\rangle = \sum_{\alpha = 1}^{D_{\ell - 1}} \sum_{\beta = 1}^{D_{\ell}} \sum_{s = 1}^{d}
 \left( \lambda^{[\ell-1]}_{\alpha} \; B_{\alpha, \beta}^{[\ell] s} \right)
 |L_{\alpha} \rangle^{L}_{\ell-1} \otimes |s\rangle_\ell \otimes |R_{\alpha} \rangle^{R}_{\ell};
\end{equation}
and of course, complete positivity relations read
\begin{equation} \label{eq:phoneR3}
 \begin{aligned}
  \sum_{s = 1}^d B_{s}^{[\ell]} \cdot {B^{\dagger}}_{s}^{[\ell]} &= \Id \\
  \sum_{s = 1}^d {B^{\dagger}}_{s}^{[\ell]} \cdot \Lambda^{[\ell-1]} \cdot {B}_{s}^{[\ell]} &= \Lambda^{[\ell]}.
 \end{aligned}
\end{equation}
In the end, by applying recursively either the left-block or right-block argument presented
in this section, we are allowed to build the analytical MPS representation of the original state.

In fact, for any given state $|\Psi\rangle$ and any choice of $\ell$ ($1 \leq \ell < L$), one can
formally express it as
\begin{equation}
 |\Psi\rangle = \sum_{s_1 \ldots s_L = 1}^d
 \left( A_{s_{1}}^{[1]} \cdot \ldots \cdot A_{s_{\ell}}^{[\ell]} \cdot \lambda^{[\ell]}
 \cdot B_{s_{\ell+1}}^{[\ell+1]} \cdot \ldots \cdot B_{s_{L}}^{[L]} \right)
 | s_1 \ldots s_L \rangle,
\end{equation}
Where the matrices $A_{s_{\ell'}}^{[\ell']}$ and $B_{s_{\ell'}}^{[\ell']}$ are respectively
given by \eqref{eq:MconstrL} and \eqref{eq:MconstrR}; they are $D_{\ell'-1} \times D_{\ell'}$ dimensioned,
and are well defined since the Schmidt decomposition exists for any partition of the system.
Since the choice for the site $\ell$
to start from, is completely arbitrary, the constraint on Schmidt dimensions holds both left-ways and right-ways
for every site in the lattice, namely
\begin{equation} \label{eq:dimconstr}
 D_{\ell'-1} \leq d\:D_{\ell'} \qquad \mbox{and} \qquad D_{\ell'} \leq d\:D_{\ell'-1} \qquad
 \forall \ell',\;0 < \ell' \leq L,
\end{equation}
where, of course, $D_0 = D_L = 1$. This concludes the proof.

It is now important to point out a major fact concerning the completeness of MPS representation;
since the dimension constraint \eqref{eq:dimconstr} are quite weak, if the state we are dealing with
has no limitations on its entanglement properties (which is the typical case for, say,
a random state in the many-body Hilbert) such MPS representation is poorly efficient.
Indeed, \eqref{eq:dimconstr} tells us that the largest correlation dimensions
$D_{\text{max}} = \max\{D_{\ell}\}$ are typically
reached next to the middle of the 1D chain: precisely we have
\begin{equation}
D_{\ell} \leq \min\{ d^\ell, d^{L - \ell} \}
 \qquad \longrightarrow \quad D_{\text{max}} \leq d^{L/2}.
\end{equation}
Therefore, in general, the typical dimension (number of rows and columns of $A_{s_{1}}^{[1]}$)
of the MPS representation \emph{does} scale with the full size $L$ of the system,
and in the worst case scenario it grows exponentially.

This is the main reason why, in literature, when speaking of Matrix Product States
most of the time one actually refers to the manifold of quantum states allowing an MPS representation
for which the maximal bondlink dimension $D$ is finite, does not scale with the system size $L$,
and is typically small. By putting together the valence bond picture
(introduced in section \ref{sec:valencebond}) and the completeness argument, one can conclude that
a Matrix Product State representation of bondlink $D$ can describe exactly any state whose
partition entanglement is bound by $\log D$. Equivalently, \emph{every finitely-correlated state is a MPS}.

\section{Area Law and successfulness of 1D MPS}

After such preliminary considerations, interpreting matrix product states as variational tools
turns straightforward. The typical problem we want to address is finding the ground
states of a given, typically short-ranged, Hamiltonian upon an OBC system with $L$ sites.
Similarly to the DMRG procedure, we choose arbitrarily a maximal bondlink dimension $D$ allowed
for the simulation that should lead us to the ground state itself, and regard the elements $A_{s}^{[\ell]}$
in MPS representation as variational tensors/matrices. Then, we adjust variational parameters
according to some algorithm (see section \ref{sec:minimizOBC}) in order to minimize
the energy.

Due to the completeness theorem of MPS representations, we know that, for any global system size $L$
it exists a finite $D$ for which the \emph{exact} ground state is representable by a $D$-bondlinked MPS,
and such $D$ is related to the estimated entanglement $\varepsilon$ of the state itself,
like $\varepsilon \sim \log D$. But now we can exploit some theoretical knowledge involving ground states
of many-body systems, known in literature
as the area-law of entanglement \cite{Arealaws, Arealaws2, Eisarea}: The partition entanglement in
a ground state of a non-critical local Hamiltonian scales with the surface of the partition itself,
and not with the parted volume.
For 1D non-critical systems, this means that $\varepsilon$ does not scale with the size of the system $L$, but
rather saturates to a finite value. This also suggests that the bond dimension $D$ required to
achieve good precisions in representing the ground state does \emph{not} scale with $L$. In practice,
for many tested models, the $D$ necessary to get an outstanding approximation to the GS is surprisingly small,
regardless to system size \cite{MPSground}. This very argument allows us to address even problems with
a large number of sites, and yet deal with them in a quasi-exact fashion. Of course this also explains
the great success of DMRG for 1D non-critical systems, a reason which was not yet fully understood in the '90s.

Indeed, the area law argument also suggests that finite-$D$ MPS should also be capable to characterize
a 1D problem \emph{directly in the thermodynamical limit} as $\varepsilon$ converges to a finite value
(we will discuss this approach in section \ref{sec:MPSTD}).
A special interest within this framework is raised by critical 1D systems \cite{QptMps}. They are known for violating
the area law of entanglement by a logarithmic (with $L$) correction to the partition entropy, with
the proportionality constant given by the central charge $C$ of the model
\cite{CFTHolzhey, CFTCalab}:
\begin{equation}
 \varepsilon (\Psi) \equiv \mathcal{S}_{\text{VN}}(\rho_{1..L/2}) \sim
 \frac{C}{6} \log L + C'
\end{equation}
and therefore the appropriate $D$ to represent the ground state faithfully, does scale in the end with
the system size, according a power-law like behavior $D \propto L^{C/6}$ where the exponent is $C/6$ itself.
Now since the large majority of the famous 1D critical models have typically small central charges
(e.g. crit. Ising, crit. XXZ, Heisenberg, have $C \leq 1$), even though $D$ scales with $L$, the scaling function
is so concave that even in that case we can address efficiently quite large system sizes with good precision.

Nevertheless, it is important to remember that for critical 1D systems, their efficient MPS 
representability depends directly on the central charge, while for non-critical systems it
is natural, an automatic consequence of the area law of entanglement. In chapter \ref{chap:TTNMERA}
we will introduce families of variational states more suitable to address criticality than mere MPS

\section[Gauge group of MPS]{Gauge group of\\Matrix Product State representation} \label{sec:MPSGauge}

By now, it should be clear that, given a quantum state on an OBC chain,
its exact  Matrix Product State representation is in general not unique.
The issue is simple: the state components expanded in the canonical basis are composite products of
matrices, and the same product can be matrix-factorized in many ways.
We will now define and explain the usage of a group of transformations that
manipulate the set of matrices in the representation, under which the physical state is invariant:
by definition this is the \emph{gauge group} of MPS representation.

Let us start again from the state $|\Psi\rangle$, whose MPS representation has bondlink
dimension $D$, and is given by
\begin{equation} \label{eq:MPSgau}
 |\Psi\rangle = \sum_{s_1 \ldots s_L = 1}^d
 \left( A_{s_{1}}^{[1]} \cdot A_{s_{2}}^{[2]}
 \cdot \ldots \cdot A_{s_{L-1}}^{[L-1]} \cdot A_{s_{L}}^{[L]} \right)
 | s_1 \ldots s_L \rangle,
\end{equation}
where matrices $A_{s_{\ell}}^{[\ell]}$ are $D_{\ell - 1} \times D_{\ell}$ dimensioned
($D_{\ell} \leq D$, $\forall \ell$).
For every $\ell < L$, we now define an \emph{invertible} square matrix $X_{\ell}$, of dimension
$D_{\ell} \times D_{\ell}$. The expression within parentheses in eq.~\eqref{eq:MPSgau} is left invariant by
\begin{multline} \label{eq:mulgau}
  A_{s_{1}}^{[1]} \cdot A_{s_{2}}^{[2]}
 \cdot \ldots \cdot A_{s_{L-1}}^{[L-1]} \cdot A_{s_{L}}^{[L]}
 =\\= 
 \left( A_{s_{1}}^{[1]} \cdot X_1 \right) \left( X_1^{-1} \cdot A_{s_{2}}^{[2]} \cdot X_2 \right)
 \ldots \left( X_{L-1}^{-1} \cdot A_{s_{L-1}}^{[L-1]} \cdot
 X_{L-1} \right) \left( X_{L-1}^{-1} \cdot A_{s_{L}}^{[L]} \right).
\end{multline}
But now, any term of the form $( X_{\ell-1}^{-1} \cdot A_{s}^{[\ell]} \cdot X_{\ell} )$ is again a
$D_{\ell - 1} \times D_{\ell}$ matrix, and as above we have $d$ of them per site, indexed by $s$.
In conclusion, the latter expression in eq.~\eqref{eq:mulgau} is again a Matrix Product, where
the bondlink dimensions $D_{\ell}$ are preserved site-by-site, and the original Matrices of the
representation underwent the (gauge) transformation
\begin{equation} \label{eq:Gaugetransform}
 A_{s}^{[\ell]} \longrightarrow B_{s}^{[\ell]} \equiv X_{\ell-1}^{-1} \cdot A_{s}^{[\ell]} \cdot X_{\ell}
 \qquad \forall s \in \{1..d\},
\end{equation}
while the state is left invariant, i.e.
\begin{equation}
 |\Psi\rangle = \sum_{s_1 \ldots s_L = 1}^d
 \left( B_{s_{1}}^{[1]} \cdot B_{s_{2}}^{[2]}
 \cdot \ldots \cdot B_{s_{L-1}}^{[L-1]} \cdot B_{s_{L}}^{[L]} \right)
 | s_1 \ldots s_L \rangle.
\end{equation}
For any nearest-neighboring bond, $X_{\ell}$ defines an allowed transformation as long as its inverse is defined.
Therefore, the gauge group of Matrix Product States is equivalent to the direct sum of the groups of
Isomorphisms of $D_{\ell}$ dimensioned complex vector spaces
\begin{equation}
 \mathcal{G}_{\text{MPS}} \equiv \bigoplus_{\ell = 1}^{L-1} \text{Iso}\left( \mathbb{C}^{D_\ell} \right).
\end{equation}

In order to define properly $\mathcal{G}_{\text{MPS}}$ we did not need to summon the Hilbert structure:
the invertibility condition is a rank dimension requirement, \emph{not} a metric constraint.
This remark is definitely sensible, since
the correlation spaces are fictitious, virtual, and therefore there is no reason for a gauge group
to mingle with the physical-space metric properties.

Finally, notice that the gauge group we built $\mathcal{G}_{\text{MPS}}$
is identified by the initial choice of site-dependent bondlink dimensions $D_{\ell}$, which we required
to be left unaltered from the transformation. Actually, in \eqref{eq:mulgau} we could
have used any rectangular $D_{\ell} \times D'$ matrix $X_{\ell}$ (with $D' < D_{\ell}$)
which is right-invertible, i.e.
\begin{equation}
 X_{\ell} \cdot X_{\ell}^{-1} = \Id, \quad \mbox{but} \quad
 X_{\ell}^{-1} \cdot X_{\ell} = P = P^2 \neq \Id,
\end{equation}
and adopt such $X_{\ell}$ in \eqref{eq:Gaugetransform}; this, of course, leaves the matrix product
invariant, but the bondlinks of the representations are altered, their dimensions increased
($B_{s}^{[\ell]}$ now are $D_{\ell -1} \times D'$, and
$B_{s}^{[\ell+1]}$ are $D' \times D_{\ell +1}$). However, the state contains the same amount of
entanglement as before, but we are spending more resources to describe it:
we are working in a non-optimal numerical framework.
Moreover, this extension $\mathcal{G}'$
to the previously defined $\mathcal{G}_{\text{MPS}}$ is clearly a group lacking
an inverse-element property.
For these reasons, in most cases it is interesting to limit the study of
MPS gauge features on $\mathcal{G}_{\text{MPS}}$ itself, under which the MPS representation space,
given by the $\{D_{\ell}\}_{\ell}$, is stable.

\section{The canonical form} \label{sec:MPScanonical}

The presence of a gauge group for MPS provides an
computational advantage, since freedom and manipulability of our description tools are increased.
At the same time, the capability of quickly recognizing state properties, or comparison between states
is reduced, as even MPS representations of two identical states may look very different, when
their gauges are incompatible. The simplest way to avoid such difficulty is to break the gauge invariance
by hand, i.e. by characterizing a representative in the class of equivalence for MPS, which
is easy to achieve, recognize, and completely general.
This concept realizes in the definition of a \emph{canonical form}
for MPS representations.

We say that a Matrix Product State, of bond dimension $D$
\begin{equation} \label{eq:MPS21}
 |\Psi\rangle = \sum_{s_1 \ldots s_L = 1}^d
 \left( A_{s_{1}}^{[1]} \cdot A_{s_{2}}^{[2]}
 \cdot \ldots \cdot A_{s_{L-1}}^{[L-1]} \cdot A_{s_{L}}^{[L]} \right)
 | s_1 \ldots s_L \rangle,
\end{equation}
where $A_{s_{\ell}}^{[\ell]}$ are $D_{\ell - 1} \times D_{\ell}$ dimensioned matrices,
with open boundary conditions ($D_0 = D_L = 1$), is in the (right-) canonical form if it holds
\begin{equation} \label{eq:phoneRX}
 \begin{aligned}
  &\sum_{s = 1}^d A_{s}^{[\ell]} \cdot {A^{\dagger}}_{s}^{[\ell]} = \Id
  \qquad \forall \ell, 1 \leq \ell \leq L \\
  &\sum_{s = 1}^d {A^{\dagger}}_{s}^{[\ell]} \cdot \Lambda^{[\ell-1]} \cdot {A}_{s}^{[\ell]} = \Lambda^{[\ell]}
  \qquad \forall \ell, 1 \leq \ell \leq L \\
  &\Lambda^{[0]} = \Lambda^{[L]} = 1, \mbox{ and every $\Lambda^{[\ell]}$ is positive and full rank.}
 \end{aligned}
\end{equation}
We recall that the first equation is the CPT-condition to preserve orthonormality
among Schmidt vectors, when propagating from the right: for this reason, we will, from now on, refer to
this gauge-invariance breaking, \eqref{eq:phoneR2} \eqref{eq:phoneR3}, as \emph{right gauge}, for brevity.

Given any MPS, it is \emph{always} possible to write a canonical matrix product representation
for the same state; the bondlink dimensions are equal or smaller than the original ones.
An operational proof of this statement is explained in detail in ref.~\cite{MPSreview}, we are now going
to sketch the fundamentals, as many of the involved manipulations will be useful later on,
and this is the perfect context to introduce them.

\subsection{Proof of canonical form generality}

Let us take an OBC-MPS representation given by $B_{s}^{[\ell]}$ matrices
\begin{equation}
 |\Psi\rangle = \sum_{s_1 \ldots s_L = 1}^d
 \left( B_{s_{1}}^{[1]} \cdot B_{s_{2}}^{[2]}
 \cdot \ldots \cdot B_{s_{L-1}}^{[L-1]} \cdot B_{s_{L}}^{[L]} \right)
 | s_1 \ldots s_L \rangle,
\end{equation}
we are going to define explicitly a set of rectangular matrices
$Y_{\ell}$ and $Z_{\ell}$, with $Y_{\ell} Z_{\ell} = \Id$, such that by applying
\begin{equation}
 \begin{aligned}
 A_{s}^{[1]} = B_{s}^{[1]} Z_{1}, \qquad A_{s}^{[L]} = Y_{L-1} B_{s}^{[L]},\\
 A_{s}^{[\ell]} = Y_{\ell-1} B_{s}^{[\ell]} Z_{\ell} \quad \text{for } 1 < \ell < L,
 \end{aligned}
\end{equation}
the resulting matrices $A_{s}^{[\ell]}$ satisfy \eqref{eq:phoneRX}, and
they represent again $|\Psi\rangle$ faithfully via \eqref{eq:MPS21}. Moreover,
the resulting bondlink dimensions $D_\ell$ will be equal or smaller than those of $B_{s}^{[\ell]}$ representation.
Precisely, the full rank condition (\ref{eq:phoneRX}.c)
for $\Lambda^{[\ell]}$ tells us that the resulting
$A_{s}^{[\ell]}$ representation uses the \emph{minimal} bondlink dimension, for every bond,
necessary to describe $|\Psi\rangle$.

The $Y_{\ell}$ and $Z_{\ell}$ represent a gauge transformation followed by a correlation space truncation;
constructing them is quite simple.
We start from the right of the 1D chain,
by performing a Singular Value Decomposition (SVD) of the matrix $B_{j,s_{L}}^{[L]}$ read
as if $s_{L}$ were an incoming index, and $j$ an outcoming one:
\begin{equation}
 B_{j,s_{L}}^{[L]} = \sum_{\beta} U_{j, \beta}^{[L]} \Delta_{\beta}^{[L]} A_{\beta,s_{L}}^{[L]},
\end{equation}
where $U^{[L]}$ and $A^{[L]}$ are respectively left and right isometric matrices, i.e.
$U^{\dagger} U = \Id$ and $A A^{\dagger} = \Id$, and the diagonal matrix $\Delta^{[L]}$ is
positive. Not only, but we can make $\Delta^{[L]}$ strictly positive, by just cutting
$\beta$ values for which $\Delta_{\beta}^{[L]} = 0$ out of the sum. If we do,
$U^{[L]}$ and $A^{[L]}$ continue to be isometries, as any subset of columns of $U^{[L]}$
(resp. of rows of $A^{[L]}$) is still an orthonormal set.
Then we just fix $Y_{L-1}$ and $Z_{L-1}$ matrices as
\begin{equation}
 Y_{L-1} =  {\Delta^{[L]}}^{-1} {U^{[L]}}^{\dagger} \qquad
 Z_{L-1} =  {U^{[L]}} {\Delta^{[L]}}.
\end{equation}
By construction $Y_{L-1} B_{s}^{[L]} = A_{s}^{[L]}$, which is an isometry and thus in the right gauge.
Similarly we define $C^{L-1}_{s} = B_{s}^{[L-1]} Z_{L-1}$, and of course
\begin{equation}
 |\Psi\rangle = \sum_{s_1 \ldots s_L = 1}^d
 \left( B_{s_{1}}^{[1]} \ldots
 B_{s_{L-2}}^{[L-2]} \, C_{s_{L-1}}^{[L-1]} \, A_{s_{L}}^{[L]} \right)
 | s_1 \ldots s_L \rangle,
\end{equation}
is still a faithful representation of the original state $|\Psi\rangle$: the
rightmost bondlink $D_{L-1}$ might be decreased after the transformation, but due to the SVD
argument we know we disregarded only zero components. In other words we could
say that $Z_{L-1} Y_{L-1} = P = P^2 = P^{\dagger}$
is the projector over the actual support of the bondlink space (in $B_{s}^{[\ell]}$ representation).

Now we proceed recursively: we consider the composite matrix
$C_{j, \alpha}^{[\ell]}$ (with $\ell$ starting from $L-1$ and moving left)
whose incoming index $\alpha$ is the pair of indices $\{k, s_{\ell}\}$, and
again we perform a SVD
\begin{equation}
 C_{j,\alpha}^{[\ell]} = \sum_{\beta} U_{j, \beta}^{[\ell]} \Delta_{\beta}^{[\ell]} A_{\beta, \alpha}^{[\ell]}.
\end{equation}
By construction $A^{[\ell]}$ satisfies the right gauge condition, since
\begin{equation}
 \sum_{\alpha} A_{\beta, \alpha}^{[\ell]} {A^{\star}}_{\gamma, \alpha}^{[\ell]} = 
 \delta_{\beta, \gamma}
 \quad \longrightarrow \quad
 \sum_{s_{\ell}} A_{s_{\ell}}^{[\ell]} {A^{\dagger}}_{s_{\ell}}^{[\ell]} = \Id.
\end{equation}
Then, bondlink $\ell$ space is truncated to the support of $\Delta$, and the (pseudo-) gauge
transformation given by
\begin{equation}
 Y_{\ell-1} =  {\Delta^{[\ell]}}^{-1} {U^{[\ell]}}^{\dagger} \qquad
 Z_{\ell-1} =  {U^{[\ell]}} {\Delta^{[\ell]}},
\end{equation}
which lead to $Y_{\ell-1} C_{s}^{[\ell]} = Y_{\ell-1} B_{s}^{[\ell]} Z_{\ell} = A_{s}^{[L]}$, and
redefine $B_{s}^{[\ell-1]} Z_{\ell-1} = C_{s}^{[\ell-1]}$ so that we can apply the procedure
again on site $\ell -1$. Once we arrive at the left-end of the chain, $C_{s_{1}}^{[1]}$ is
already in the right gauge by assumption of initial state normalization
\begin{equation}
 \sum_{s_1} C_{s_{1}}^{[1]} {C^{\dagger}}_{s_{1}}^{[1]} = \langle \Psi | \Psi \rangle = 1
\end{equation}
in conclusion $A_{s_{1}}^{[1]} = C_{s_{1}}^{[1]} = B_{s}^{[\ell]} Z_{\ell} = A_{s}^{[L]}$.
This proves the first statement of \eqref{eq:phoneRX}: converting a complete MPS representation
in order to be fully in the right gauge is operatively possible by a recursive application of
Singular Value Decompositions.

The second statement of \eqref{eq:phoneRX} is a direct consequence of the CPT mapping argument
we presented in previous sections, in particular it corresponds to equation \eqref{eq:mappuzR}.
Finally, the full rank condition follows from the fact that after every SVD steps we truncated
the bondlink space (dimension $D_\ell$) to the support of the corresponding reduced density matrix
$\Lambda^{[\ell]}$. This issue is argumented in details in \cite{MPSreview}.
\vspace{1em}

As a concluding remark to the present section, we would like to point out that the canonical form
we just presented is the right-directed one, i.e. is made so that every MPS block is in the right gauge
(and the correlation space used is minimal). Of course, we could similarly define a left-canonical form,
where MPS is completely in the left gauge, i.e. \eqref{eq:phoneL2} \eqref{eq:phoneL3},
and other statements still hold:
\begin{equation} \label{eq:phoneLX}
 \begin{aligned}
  &\sum_{s = 1}^d {A^{\dagger}}_{s}^{[\ell]} \cdot A_{s}^{[\ell]} = \Id
  \qquad \forall \ell, 1 \leq \ell \leq L \\
  &\sum_{s = 1}^d {A}_{s}^{[\ell]} \cdot \tilde{\Lambda}^{[\ell]} \cdot {A^{\dagger}}_{s}^{[\ell]} =
  \tilde{\Lambda}^{[\ell-1]}
  \qquad \forall \ell, 1 \leq \ell \leq L \\
  &\tilde{\Lambda}^{[0]} = \tilde{\Lambda}^{[L]} = 1,
\mbox{ and every $\tilde{\Lambda}^{[\ell]}$ is positive and full rank,}
 \end{aligned}
\end{equation}
such is the left-canonical form for MPS representations.
The demonstration adopted in this section to
achieve the canonical form is operational in the sense that is
exactly the algorithm we apply in numerical settings: the computational
advantage of using canonical MPS, apart from immediate estimation of entanglement, will be clear as soon as we
explain how to achieve expectation values onto an MPS state.

\section{MPS and Observables} \label{sec:OBCMpsObs}

By now, we understood that MPS are outstanding candidates as tools for simulating condensed
matter one-dimensional many-body systems. Then, it is fundamental that we realize how to achieve expectation values
of observables, in a clear and efficient way. If we are to adopt, say, the global energy as a simulation benchmark,
so that our goal becomes achieving the absolute energy minimum, we need first to
calculate the expectation value $\langle \Psi | H | \Psi \rangle$ of the Hamiltonian $H$: and operator which is
nonlocal, but it is explicitly written as a sum of local (separable) terms. For simplicity we can assume it
couples only nearest neighboring sites
\begin{equation} \label{eq:NNhamilt}
 H = \sum_{\ell = 1}^{L} \sum_{q} g^{\ell}_{q} \;\Theta^{[\ell]}_{q} \;+\;
 \sum_{\ell = 2}^{L} \sum_{p} h^{\ell}_{p} \;{\Theta'}^{[\ell-1]}_{p} \otimes {\Theta''}^{[\ell]}_{p}.
\end{equation}

Let us start from getting the expectation value over a MPS of a separable observable
$O = \bigotimes_{\ell} \Theta^{[\ell]}$, where every operator $\Theta^{[\ell]}$ can depend on the
site $\ell$ on which it acts. We have
\begin{multline}
 \langle \Psi | H | \Psi \rangle = \sum_{s_1 \ldots s_n} \sum_{r_1 \ldots r_n}
 \left( A_{s_{1}}^{[1]} \cdot \ldots \cdot A_{s_{L}}^{[L]} \right)
 \times \\ \times
 \left( {A^{\star}}_{r_{1}}^{[1]} \cdot \ldots \cdot {A^{\star}}_{r_{L}}^{[L]} \right)
 \langle r_1 \ldots r_L | \bigotimes_{\ell} \Theta^{[\ell]} | s_1 \ldots s_L \rangle.
\end{multline}
Now we define the so-called \emph{transfer matrices}, as follows
\begin{equation}
 \Etra_{X}^{[\ell]} \equiv \sum_{s,r = 1}^d \langle r | X | s \rangle
 \left( A_{s}^{[\ell]} \otimes {A^{\star}}_{r}^{[\ell]} \right),
\end{equation}
where $X$ is a one-site operator, acting on site $\ell$;
transfer matrices $\Etra_{X}^{[\ell]}$ are $D_{\ell-1}^2 \times D_{\ell}^2$ dimensioned.
We now calculate the $\Etra_{\Theta_\ell}^{[\ell]}$ for every $\ell$,
and the expectation value becomes simply a multiplication of the whole string of transfer matrices
\begin{equation} \label{eq:Transmatstring}
  \langle \Psi | H | \Psi \rangle = \Etra_{\Theta_1}^{[1]} \cdot
  \Etra_{\Theta_2}^{[2]} \cdot \ldots \cdot \Etra_{\Theta_L}^{[L]}.
\end{equation}
The computational cost for acquiring this expectation value scales only \emph{linearly} with the
system size (recall that now we treat $L$ and $D$ as independent parameters of the MPS variational ansatz);
unfortunately, there is still a harsh dependence on the chosen bondlink dimension $D$.
Indeed multiplying two $D^2 \times D^2$ matrices costs $\sim D^6$ elementary operations, yet by adopting some
technical tricks we can further improve this scaling law:
\begin{itemize}
 \item We should start performing the multiplication from the right (or left) boundary, as $| Q_{L-1} ) = \Etra_{\Theta}^{[L]}$ is
 a one-column matrix, i.e. a column vector. And multiplying a $D^2$ dimensioned vector $| Q_{\bullet} )$
 for a $D^2 \times D^2$ matrix has an overall $D^4$ cost.
 \item instead of multiplying directly $\Etra_{\Theta}^{[\ell]} | Q_{\ell}) = |Q_{\ell-1})$ one can first calculate
 $|\Gamma^{[\ell]}_{s}) = [ A_{s}^{[\ell]} \otimes \Id ] |Q_{\ell})$, followed by 
 $|\Pi^{[\ell]}_{r}) = \sum_{s} \langle r | \Theta^{[\ell]} | s \rangle |\Gamma^{[\ell]}_{s})$ and finally
 $|Q_{\ell-1}) = \sum_r [ \Id \otimes {A^{\star}}_{r}^{[\ell]} ] |\Pi^{[\ell]}_{r})$. These operations requires respectively
 a number of elementary operations equal to: $dD^3$, $d^2 D^2$, and $dD^3$.
\end{itemize}
In the end, the total computational cost to achieve the MPS-expectation value of a separable observable $O$
scales, with size $L$ and bond dimension $D$, as
\begin{equation} \label{eq:MPScost}
 \# \mbox{cost} \sim L \left(2 d D^3 + d^2 D^2 \right).
\end{equation}
Usually, the first term in the parentheses is the leading one (and the other is negligible), since the typical bond
dimensions chosen in simulations are sensibly larger than local space dimensions $D \gg d$.

The result we got holds in a quite general scenario (provided that $O$ is acts locally); we will now see that
if the involved operator has a small support, we can considerably improve this limit by exploiting the gauge group of MPS.

\subsection{Local support Observables} \label{sec:Localsupport}

Let us assume that the observable $O$ we are interested with does not involve all the sites within the 1D chain, but
only a small connected subset of those, say lattice sites between $\ell_1$ and $\ell_2$ ($1 < \ell_1 \leq \ell_2 < L$).
Recalling the previous argument involving transfer matrices, i.e. eq.~\eqref{eq:Transmatstring} we can write
\begin{equation} \label{eq:Transfmatlocal}
  \langle \Psi | O | \Psi \rangle = \left( \Etra_{\Id}^{[1]} \cdot \ldots \cdot \Etra_{\Id}^{[\ell_1-1]} \right)
  \left( \Etra_{\Theta_{\ell_1}}^{[\ell_1]} \cdot \ldots \cdot \Etra_{\Theta_{\ell_1}}^{[\ell_1]} \right)
  \left( \Etra_{\Id}^{[\ell_2+1]} \cdot \ldots \cdot  \Etra_{\Id}^{[L]} \right).
\end{equation}
Notice that on the sites outside $\{\ell_1..\ell_2\}$ the observable $O$ acts trivially, so we are considering
the transfer matrix of the local identity operator $\Id$ there.
Now, as the expectation value in \eqref{eq:Transfmatlocal} is a physical quantity, i.e. it depends on the properties
of the quantum state and \emph{not} on its specific MPS representation: it is a quantity invariant under the action
of the MPS gauge group. At the same time the transfer matrices are not gauge invariant, so it is advisable to
choose a gauge that reduces the computational cost of \eqref{eq:Transfmatlocal}.

Precisely, we choose a gauge that turns our MPS to look like this
\begin{multline} \label{eq:CompaMPS}
 |\Psi\rangle = \sum_{s_1 \ldots s_L = 1}^d 
 \left( A_{s_{1}}^{[1]} \cdot \ldots \cdot A_{s_{\ell_1-1}}^{[\ell_1-1]} \cdot
 C_{s_{\ell_1}}^{[\ell_1]} \cdot \ldots \cdot C_{s_{\ell_2}}^{[\ell_2]}
 \cdot \phantom{A} \right. \\ \left. \phantom{A} \cdot
 B_{s_{\ell_2+1}}^{[\ell_2+1]} \cdot \ldots \cdot B_{s_{L}}^{[L]} \right)
 | s_1 \ldots s_L \rangle,
\end{multline}
where MPS tensors to the left of the $\{\ell_1..\ell_2\}$ support are in the left gauge
($\sum_{s} A_s^{\dagger} A_s = \Id$), those to the right of the support are in the right gauge
($\sum_{s} B_s B_s^{\dagger} = \Id$), and those in the middle can be in any gauge chosen by the user, with
the only constraint that they must satisfy the global state normalization condition.
As before, the form \eqref{eq:CompaMPS} can be achieved by means of recursive Singular Value Decompositions
that define appropriate gauge transformations
(from site 1 to $\ell_1$ to fix the $A$, from site $L$ to $\ell_2$ to fix the $B$),
exactly like we did in the section \ref{sec:MPScanonical}.

Now we focus on the transfer matrices of the outside zone $\Etra_{\Id}^{[\ell]}$.
Consider site $L$, due to the assumptions we made, it holds
\begin{equation} \label{eq:phoneRTrans}
 |Q_{L-1}) = \Etra_{\Id}^{[L]} = \sum_{j,k} \left( \sum_{s,r}^{d} B^{[L]s}_j \,\delta_{s,r}\, {B^{\star}}_k^{[L]r} \right) |jk)
  = | \Phi^{+} )
\end{equation}
where $| \Phi^{+} )$ is the (unnormalized) maximally entangled canonical vector
\begin{equation}
 | \Phi^{+} ) \equiv \sum_j^{D_{L-1}} |jj) = \sum_{j,k}^{D_{L-1}} \delta_{j,k} |jk).
\end{equation}
It is east to see that eq.~\eqref{eq:phoneRTrans} holds recursively for all $| Q_{\ell} )$, $\ell > \ell_2$, since
\begin{multline}
 | Q_{\ell-1} ) = \Etra_{\Id}^{[\ell]} | \Phi^{+} ) = 
 \sum_{j,k}^{D_{\ell-1}} \left( \sum_{s,r}^{d} \sum_{\alpha, \beta}^{D_\ell} B^{[\ell]s}_{j,\alpha} \,\delta_{s,r}\,
 \delta_{\alpha, \beta}\, {B^{\star}}_{k, \beta}^{[\ell]r} \right) |jk) = \\
 \sum_{j,k}^{D_{\ell-1}} \delta_{j,k}\, |jk) = | \Phi^{+} ),
\end{multline}
where we used both the fact that the operator acts like identity on site $\ell$ (thus $\delta_{s,r}$), and
the recursive hypothesis $| Q_{\ell} ) = | \Phi^{+} )$. An identical argument can be applied to transfer matrices
to the left of the support of $O$, where left gauge condition can be exploited to see that
$(Q_{\ell}| = (\Phi^{+}|$ for any $\ell \leq \ell_1$.
The conclusion simply follows:
\begin{equation} \label{eq:Transfmatcompact}
  \langle \Psi | O | \Psi \rangle = (\Phi^{+}|
  \Etra_{\Theta_{\ell_1}}^{[\ell_1]} \cdot \ldots \cdot \Etra_{\Theta_{\ell_1}}^{[\ell_1]} | \Phi^{+} ),
\end{equation}
which means that the number ($\#$cost) of elementary operations we have to perform does \emph{not} even scale with
the full size of the system $L$, but merely with the size of the operator support:
\begin{equation} \label{eq:MPScostcompact}
 \# \mbox{cost} \sim (\ell_2 - \ell_1) \left(2 d D^3 + d^2 D^2 \right).
\end{equation}
Honestly, we traded the modest effort of performing the SVD, needed to convert the MPS in the proper gauge,
to obtain a faster (and non-scaling) computational speed in acquiring finite-range MPS physics.

We will see that this result can be partially exploited even when we are to compute expectation values of
observables which are not local and not even separable, but allow a natural decomposition into local terms,
such are the Hamiltonians of typical short-range interacting models.

\subsection{Hamiltonian-like Observables}

We are now interested defining an operational algorithm that, exploiting MPS properties,
computes efficiently the expectation values $\langle \Psi | H | \Psi \rangle$ of an operator
$H$ which is formally written as a nearest-neighboring Hamiltonian of the system, i.e. like \eqref{eq:NNhamilt}.
For algebraic reasons which shall be clear soon, we rewrite it as $H = \tilde{H}^{\rightarrow}_{0}$, where
\begin{equation}
 \tilde{H}^{\rightarrow}_{\ell'} = \sum_{\ell = \ell'+1}^{L} \sum_{q} g^{[\ell]}_{q} \;\Theta^{[\ell]}_{q} +
 \sum_{\ell = \ell'+2}^{L} \sum_{p} h^{[\ell]}_{p} \;{\Theta'}^{[\ell-1]}_{p} \otimes {\Theta''}^{[\ell]}_{p},
\end{equation}
As before, it is important that we focus on the computational cost of this data acquisition.
We learned that working in the proper MPS gauge is instrumental for economy of calculus, thus we already start
from a canonical MPS representation (say the right one)
\begin{equation}
 |\Psi\rangle = \sum_{s_1 \ldots s_L = 1}^d
 \left( B_{s_{1}}^{[1]} \, B_{s_{2}}^{[2]} \ldots B_{s_{L}}^{[L]} \right)
 | s_1 \ldots s_L \rangle,
\end{equation}
where all the $B_{s}^{[\ell]}$ are right-gauged, requirement which also guarantees proper state normalization
$\langle \Psi | \Psi \rangle = 1$.

Again, our scheme to acquire $\langle \Psi | H | \Psi \rangle$ has a recursive formulation:
we need to propagate the contraction of our MPS structure and at the same time include every term of $H$.
Since this is expensive by definition, we will try to regroup and sum the partially contracted
tensors every time we can.
Then the starting point, at the right boundary, is defined as follows:
\begin{equation}
 \begin{aligned}
 |\xi^{[L-1]}) &= \sum_{j,k}^{D_{L-1}}
 \left( \sum_q g^{[L]}_q  \sum_{r,s}^{d} \langle r | \Theta^{[L]}_q | s \rangle \right)
 B_{j}^{[L]s} {B^{\star}}_{k}^{[L]r} |jk)\\
 |\chi^{[L-1]}_p) &= \sum_{j,k}^{D_{L-1}}
 \left( \sum_{r,s}^{d} h_p^{[L]} \langle r | {\Theta''}^{[L]}_p | s \rangle \right)
 B_{j}^{[L]s} {B^{\star}}_{k}^{[L]r} |jk).
 \end{aligned}
\end{equation}
Now we propagate towards the left. The idea is that $|\xi^{[\ell]})$ shall contain all the elements of the Hamiltonian
who have support in $\{ \ell+1 .. L \}$, while $|\chi^{[\ell]}_p)$ will take care of nonlocal neighboring terms
across the bond $\{\ell-1, \ell\}$. Of course, we can exploit the gauge conditions, telling us that
$|Q_\ell) = | \Phi^{+} )$. This allows us to calculate every $|\chi^{[\ell]}_p)$ directly
\begin{equation} \label{eq:chi_media}
 |\chi^{[\ell-1]}_p) = \sum_{j,k}^{D_{L-1}} \sum_{\alpha}^{D_{\ell}}
 \left( \sum_{r,s}^{d} h_p^{[\ell]} \langle r | {\Theta''}^{[\ell]}_p | s \rangle \right)
 B_{j,\alpha}^{[\ell]s} {B^{\star}}_{k,\alpha}^{[\ell]r} |jk).
\end{equation}
Instead $|\xi^{[\ell-1]})$ is obtained via the recursive relation
\begin{multline} \label{eq:xi_cumbersome}
 |\xi^{[\ell-1]}) = \sum_{j,k}^{D_{\ell-1}} \sum_{\alpha}^{D_\ell}
 \left( \sum_q g^{[\ell]}_q  \sum_{r,s}^{d} \langle r | \Theta^{[\ell]}_q | s \rangle \right)
 B_{j,\alpha}^{[\ell]s} {B^{\star}}_{k,\alpha}^{[\ell]r} |jk)
 \; +\\+
 \sum_{j,k}^{D_{\ell-1}} \sum_{\alpha, \beta}^{D_\ell}
 \left( \sum_p \sum_{r,s}^{d} \langle r | {\Theta'}^{[\ell]}_p | s \rangle \right)
 B_{j,\alpha}^{[\ell]s} {B^{\star}}_{k,\beta}^{[\ell]r} |jk) \,(\alpha \beta| \chi^{[\ell]}_p)
 \; +\\+
 \sum_{j,k}^{D_{\ell-1}} \sum_{\alpha, \beta}^{D_\ell}
 \sum_s^d B_{j,\alpha}^{[\ell]s} {B^{\star}}_{k,\beta}^{[\ell]s} |jk) \,(\alpha \beta| \xi^{[\ell]}_p).
\end{multline}
By means of the transfer matrices formalism, we can rewrite the two previous equations in a more compact, and clearer, form
\begin{equation} \label{eq:xi_shorter}
 \begin{aligned}
 |\xi^{[\ell-1]}) &= \sum_q g^{[\ell]}_q \Etra_{\Theta_q}^{[\ell]} |\Phi^{+}) +
 \sum_p \Etra_{\Theta'_p}^{[\ell]} | \chi^{[\ell]}_p) +
 \Etra_{\Id}^{[\ell]} |\xi^{[\ell]}) \\
 | \chi^{[\ell-1]}_p) &= h^{[\ell]}_p \Etra_{\Theta''_p}^{[\ell]} |\Phi^{+}).
 \end{aligned}
\end{equation}
Acquiring all these data, for every $\ell$ requires an overall computational cost (apart subleading trends) of
\begin{equation}
 \# \mbox{cost} \sim L \left(\#q + 2\#p + 1 \right) \left(2 d D^3 + d^2 D^2 \right),
\end{equation}
where $\#q$ and $\#p$ are respectively the number of one body and two body terms in the Hamiltonian expression.
As you see, even in this complex scenario, the scaling behavior with $D$ and $L$ of the computational cost
remains roughly the same. Now we can conclude that
\begin{equation}
 \langle \Psi | \tilde{H}^{\rightarrow}_{\ell} | \Psi \rangle = 
 \Etra_{\Id}^{[1]} \cdot
  \Etra_{\Id}^{[2]} \cdot \ldots \cdot \Etra_{\Id}^{[\ell]} |\xi^{[\ell]}),
\end{equation}
and in particular $\langle \Psi | H | \Psi \rangle = |\xi^{[0]})$ which is a scalar number, and it is exactly
the energy of the state if $H$ is the actual Hamiltonian of the system.

The great improvement in computing expectation values we encountered for separable observable is recovered
in case of (short-range) Hamiltonian operators: the full computational cost to acquire the energy
(which will be later adopted as variational functional) scales only linearly with the system size.
If one thinks that (full-search) exact methods typically bear an exponential cost in $L$, it is easy to
understand why DMRG/MPS architectures are regarded with great interest.

\section[Pictorial representation of MPS]{Pictorial representation of Matrix\\Product States Tensor Network}

Through the present chapter, we learned how to deal with MPS: their mathematical properties that allow
algebraic manipulation (gauge group), and their physical properties that allow to control quantum entanglement.
On the other hand,
the equations we encounter start to look cumbersome and confusing, like eq.~\eqref{eq:xi_cumbersome}.
To work around this issue, we are now going to provide an alternative way to express MPS representation
that is based on diagrams and graph theory rather than standard analytical expressions. This will prove
a faster and clearer fashion to represent states, observations, and matrix multiplication; and will become
instrumental in later chapters.

Let us start back from our definition of MPS. If our system is a 1D-OBC lattice with $L$ sites, and
$\{ |s\rangle \}_s$ is the local canonical basis, then a generic state of the system is written as
$|\Psi\rangle = \sum_{s_1 \ldots s_L} \mathcal{T}_{s_1 \ldots s_L} |s_1 \ldots s_L\rangle$;
the complex tensor (with $L$ indices) $\mathcal{T}_{s_1 \ldots s_L}$ uniquely defines $|\Psi\rangle$.
Now, stating that $|\Psi\rangle$ is an MPS (with fixed bondlink $D$),
is equivalent to say that $\mathcal{T}_{s_1 \ldots s_L}$ allows the following decomposition
\begin{equation} \label{eq:MPSTN}
 \mathcal{T}_{s_1 \ldots s_L} = \sum_{ \{ j_\ell = 1 \} }^{D_{\ell} \leq D}
 A_{j_1}^{[1] s_1} \; A_{j_1, j_2}^{[2] s_2} \; A_{j_2, j_3}^{[3] s_3} \ldots
 A_{j_{L-2}, j_{L-1}}^{[L-1] s_{L-1}} \; A_{j_{L-1}}^{[L] s_{L}}.
\end{equation}
As mentioned before, all the $A^{[\ell]}$ elements are three-indices tensors (apart the first
$A^{[1]}$ and last $A^{[L]}$ MPS blocks, which have only two indices due to open boundaries),
$s_{\ell}$ being the physical index, i.e. related to the local canonical state $|s\rangle_{\ell}$,
while the two $j_{\ell}$ being the correlation space indices linking to the two neighboring MPS
blocks, namely $A^{[\ell-1]}$ and $A^{[\ell+1]}$.
Equation \eqref{eq:MPSTN} tells us that a MPS is the result of a multiple contraction of (possibly variational) tensors,
or more simply a \emph{Tensor Network}.

Let us draw eq.~\eqref{eq:MPSTN} in the following pictorial graph
\begin{center}
\begin{overpic}[width = 350pt, unit=1pt]{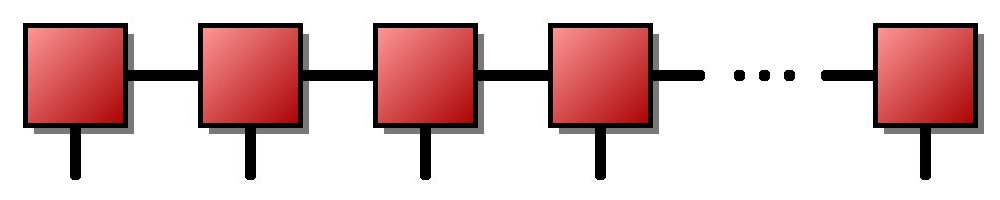}
 \put(20, 40){$A^{[1]}$}
 \put(81, 40){$A^{[2]}$}
 \put(142, 40){$A^{[3]}$}
 \put(203, 40){$A^{[4]}$}
 \put(316, 40){$A^{[L]}$}
\end{overpic}
\end{center}
where every block (graph vertex) represent a single tensor $A^{[\ell]}$, the legs/links attached to it being the indices
$s$ (vertical one), $j_{\ell-1}$ and $j_{\ell-1}$ (horizontal ones).
Connecting two tensors though a given link means
contracting the product of the two over that index. By these very simple rules, one sees that \eqref{eq:MPSTN}
is recovered, but there is no need anymore to write down either the sums or every single index $s$ or $j$ by its name.
Everything is implicit in the pictogram.

As we discussed in section \ref{sec:MPSGauge}, on each of the contracted index/connected links we can insert an
isomorphism $X_{\ell}$ together with its inverse $X^{-1}_{\ell}$. Since their global action cancels out during link contraction,
they have absolutely no effect on the global tensor $\mathcal{T}$. This is exactly the gauge of Matrix Product States,
which we represent like this
\begin{center}
\begin{overpic}[width = 320pt, unit=1pt]{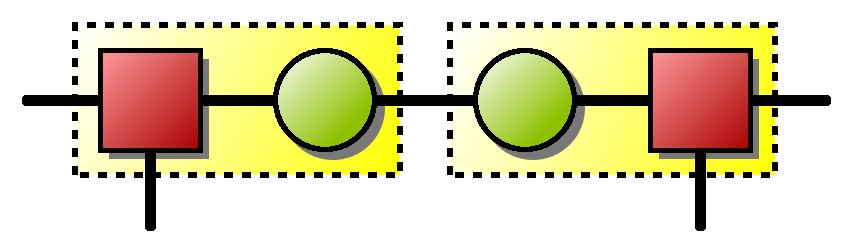}
 \put(49, 53){$A^{[\ell]}$}
 \put(116, 53){$X_{\ell}$}
 \put(188, 53){$X^{-1}_{\ell}$}
 \put(251, 53){$A^{[\ell+1]}$}
 \put(84, 69){$B^{[\ell]}$}
 \put(215, 69){$B^{[\ell+1]}$}
\end{overpic}
\end{center}
meaning that $B_{j,k}^{[\ell]s} = \sum_{\alpha}^{D_{\ell}} A_{j,\alpha}^{[\ell]s} X_{\alpha,k}$ and
$B_{j,k}^{[\ell+1]s} = \sum_{\alpha}^{D_{\ell}} X^{-1}_{j,\alpha} A_{\alpha,k}^{[\ell]s}$.
This operation is definitely equivalent to \eqref{eq:Gaugetransform},
once we have defined an isomorphism $X_{\ell}$ for every link $\ell$, $1 \leq \ell < L$.

To break by hand the freedom granted the gauge group, we defined two particular gauge choices,
namely the left and right gauges. We said that $A^{[\ell]}$ is in the left gauge if
$\sum_s {A^{\dagger}}_s^{[\ell]} A_s^{[\ell]} = \Id$, i.e.
$\sum_{s, \alpha} {A^{\star}}_{j, \alpha}^{[\ell]s} A_{\alpha, k}^{[\ell]s} = \delta_{j,k}$,
or equivalently, read in terms of transfer matrices, $(\Phi^{+}| \Etra^{[\ell]}_{\Id} = (\Phi^{+}|$; and
in pictorial representation, it becomes an equation between graphs:
\begin{equation} \label{eq:phoneLGra}
\begin{overpic}[width = 240pt, unit=1pt]{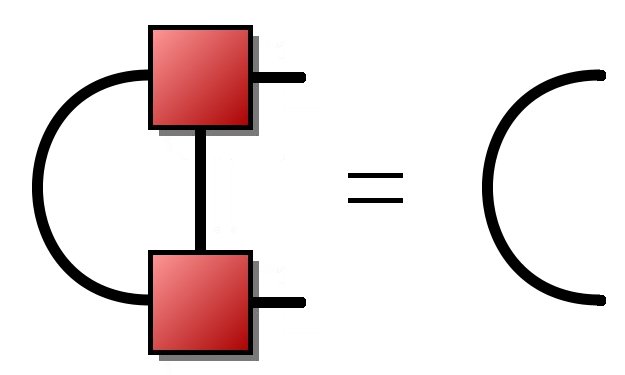}
 \put(68, 23){${A^{\star}}^{[\ell]}$}
 \put(70, 110){$A^{[\ell]}$}
\end{overpic}
\end{equation}
Similarly, the right gauge condition is the left-right specular of this graph equation.
Having such equation in graph form lets us to see immediately how to exploit the gauge by substituting
pieces of the Tensor Network and thus eliminating tensors.
In particular, assume we are to calculate the state square norm
$\langle \Psi | \Psi \rangle = \sum_{\{s_{\ell}\}} \mathcal{T}_{s_1 \ldots s_L} \mathcal{T}^{\star}_{s_1 \ldots s_L} =$
\begin{center}
\begin{overpic}[width = 260pt, unit=1pt]{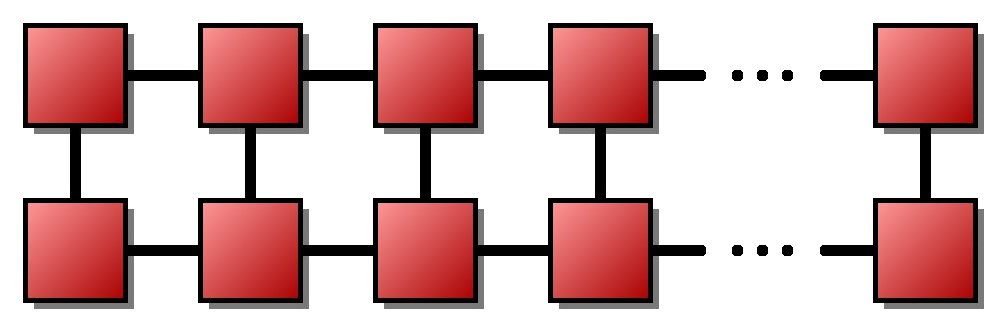}
% \put(68, 23){${A^{\star}}^{[\ell]}$}
% \put(70, 110){$A^{[\ell]}$}
\end{overpic}
\end{center}
If all the MPS tensors are in the left gauge, we can substitute \eqref{eq:phoneLGra} into the diagram,
and tensors start to literally cancel out, from left to right, until only $1 = \langle \Psi | \Psi \rangle $ remains.
Moreover, the left gauge condition ensures \eqref{eq:mappuzL}
(reduced density matrix propagation via CPT map), which corresponds to
\begin{equation} \label{eq:mappuzLGra}
\begin{overpic}[width = 240pt, unit=1pt]{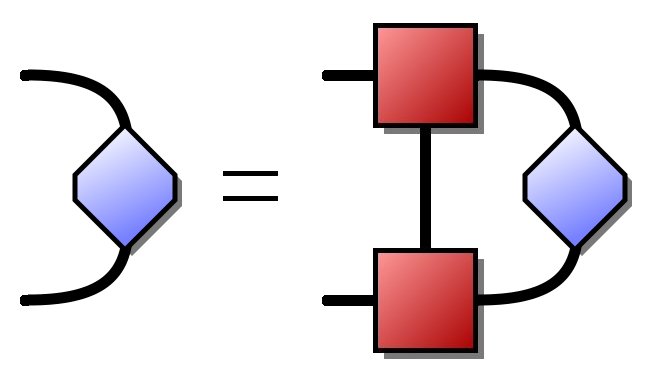}
 \put(147, 23){${A^{\star}}^{[\ell]}$}
 \put(149, 107){$A^{[\ell]}$}
 \put(207, 67){$\tilde{\rho}_{\ell}^{L}$}
 \put(37, 67){$\tilde{\rho}_{\ell-1}^{L}$}
\end{overpic}
\end{equation}

In section \ref{sec:Localsupport} we saw that having the leftmost MPS tensors in the left gauge and the
rightmost in the right gauge provides a great advantage when computing observables having local support.
Precisely, let $O = \bigotimes_{\ell = \ell_1}^{\ell_2} \Theta^{[\ell]}$, and we require that
$\sum_s {A^{\dagger}}_s^{[\ell]} A_s^{[\ell]} = \Id$ for $\ell < \ell_1$ and
$\sum_s A_s^{[\ell]} {A^{\dagger}}_s^{[\ell]} = \Id$ for $\ell > \ell_2$. Then, the part of the graph
outside $\{\ell_1..\ell_2\}$ cancels out thanks to \eqref{eq:phoneLGra} and we are left with
\begin{center}
\begin{overpic}[width = \textwidth, unit=1pt]{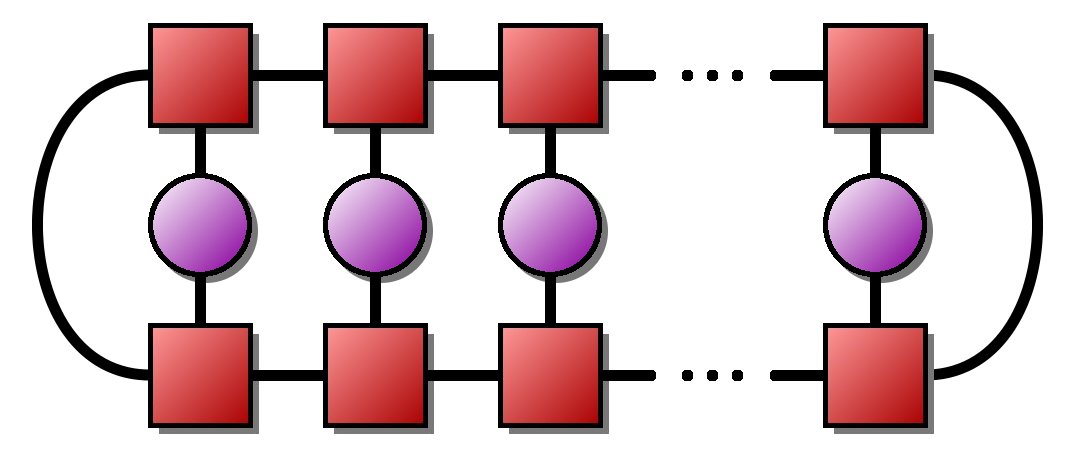}
 \put(61, 23){${A^{\star}}^{[\ell_1]}$}
 \put(63, 77){$\Theta^{[\ell_1]}$}
 \put(63, 132){$A^{[\ell_1]}$}
 \put(306, 23){${A^{\star}}^{[\ell_2]}$}
 \put(308, 77){$\Theta^{[\ell_2]}$}
 \put(308, 132){$A^{[\ell_2]}$}
\end{overpic}
\end{center}
$= (\Phi^{+}| \Etra_{\Theta_{\ell_1}}^{[\ell_1]} \cdot \ldots \cdot \Etra_{\Theta_{\ell_1}}^{[\ell_1]} | \Phi^{+} )$,
as we saw in the previous section.

To explain the algorithm we use to calculate the expectation value of a nearest-neighbor interacting Hamiltonian $H$,
we introduced the relations \eqref{eq:xi_cumbersome} and \eqref{eq:xi_shorter}, where
the transfer vector $|\xi^{[\ell]})$ was defined via a recursive scheme.
Despite the unclear look of those equations, it is possible to resume them both in a simple and intuitive graphical
equation:
\begin{equation} \label{eq:xi_fig}
\begin{overpic}[width = \textwidth, unit=1pt]{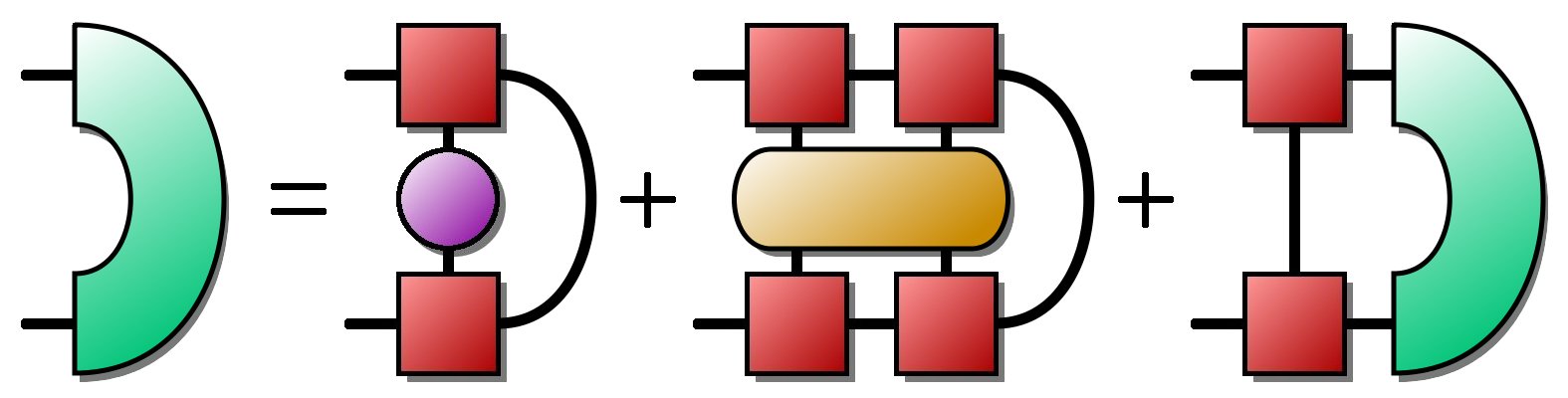}
 \put(34, 47){$\xi_{\ell-1}$}
 \put(367	, 47){$\xi_{\ell}$}
 \put(103, 77){$A^{[\ell]}$}
 \put(190, 77){$A^{[\ell]}$}
 \put(314, 77){$A^{[\ell]}$}
 \put(103, 45){$R^{[\ell]}_1$}
 \put(207, 45){$R^{[\ell+1]}_2$}
\end{overpic}
\end{equation}
where we just regrouped together the one-site operators as
$R^{[\ell]}_{1} = \sum_q g^{[\ell]}_{q} \;\Theta^{[\ell]}_{q}$, and the two-site ones:
$R^{[\ell]}_{2} = \sum_p h^{[\ell]}_{p} \;{\Theta'}^{[\ell-1]}_{p} \otimes {\Theta''}^{[\ell]}_{p}$.

\vspace{.5em}
Several equations that we will encounter in this thesis involve Tensor Network contraction
or decompositions, in most cases they allow a diagrammatic version, granting immediateness,
and clarity of understanding. So, where appropriate, we shall provide it for completeness and comfort for the reader.

\section{Minimization algorithms} \label{sec:minimizOBC}

We mentioned MPS being powerful variational tools for simulating ground state of 1D many-body systems,
and we also managed to give a prescription for evaluating the energy of the matrix product state.
It is finally time we adopt such energy $\langle \Psi | H | \Psi \rangle$ as a functional for variational simulation,
and describe an algorithm that drives our trial Matrix Product State toward the absolute minimum of this functional.
Despite the huge reduction in variational parameters we are left thanks to the MPS representation
($\sim LdD^2$ rather than $\sim d^{L}$), performing full search of the minimum within the whole parameter
space at once is still too expensive for practical purposes. Instead, we will follow a scheme similar
to original DMRG: the idea is to perform only local or quasi-local variations of the whole state
representation, namely variating a limited number of connected MPS blocks while keeping the other fixed.
Then we repeat, while choosing each time a different compact subset of MPS blocks to variate, until
convergence is eventually reached. Like traditional DMRG algorithms, usually one or two adjacent block
are variated at a time, and refrain by sweeping towards the left or the right (bouncing off the boundaries of the system,
if our problem is OBC).

\subsection{Single variational site}

In this framework, at every minimization step only one block of the MPS is being treated as variational,
say the one related to site $\ell$;
the other ones are fixed, and in practice the energy functional itself will depend on them.
We also assume that all MPS blocks to the left of $\ell$ are in the left gauge, and those on the right
are in the right gauge, so that
\begin{equation} \label{eq:284}
 |\Psi\rangle = \sum_{s_1 \ldots s_L = 1}^d 
 \left( A_{s_{1}}^{[1]} \cdot \ldots \cdot A_{s_{\ell-1}}^{[\ell-1]} \cdot C_{s_{\ell}}^{[\ell]}
 \cdot B_{s_{\ell+1}}^{[\ell+1]} \cdot \ldots \cdot B_{s_{L}}^{[L]} \right)
 | s_1 \ldots s_L \rangle,
\end{equation}
with $A^{[\ell']}$ ($\sum_s {A^{\dagger}}_{s}^{[\ell']}A_{s}^{[\ell']} = \Id$) and
$B^{[\ell']}$ ($\sum_s B_{s}^{[\ell']} {B^{\dagger}}_{s}^{[\ell']} = \Id$) fixed,
and we are searching the $C^{[\ell]}$ which minimizes $\langle \Psi | H |\Psi\rangle$.
It is always possible to gauge transform an MPS to achieve form \eqref{eq:284} by means of
repeated SVD as we saw previously, the whole singular part of the decompositions has been embedded inside $C^{[\ell]}$.

It is easy to calculate the explicit dependence on $C^{[\ell]}$ of the energy:
by means of transfer matrices formalism one can write
\begin{multline} \label{eq:cinqueleoni}
 \langle \Psi | H |\Psi\rangle = ( \Phi^{+} | \Etra^{[\ell]}_{\Id} |\xi^{\ell} ) +
 ( \xi^{\ell-1} | \Etra^{[\ell]}_{\Id} |\Phi^{+}  )
 + \sum_q g^{[\ell]}_q ( \Phi^{+} | \Etra^{[\ell]}_{\Theta} | \Phi^{+} )
  + \\ + \sum_p ( \Phi^{+} | \Etra^{[\ell]}_{\Theta'} | \chi^{[\ell]}_p ) + 
 \sum_p h^{[\ell]}_p ( \chi^{[\ell-1]}_p | \Etra^{[\ell]}_{\Theta''} | \Phi^{+} ).
\end{multline}
Let us interpret it: the first and second term embed respectively the terms of the Hamiltonian
having support to the right and left of $\ell$; the third term are those acting on $\ell$ only,
and the two last terms couple $\ell$ with its neighbors.
As all the terms contain just one $\Etra^{[\ell]}$, the functional is quadratic in the tensor $C^{[\ell]}$:
$\langle \Psi | H |\Psi\rangle = \sum_{\bullet} {C^{\star}}_{m,n}^{r} \mathcal{H}^{jks}_{mnr} {C^{\star}}_{j,k}^{s}$,
where the \emph{effective Hamiltonian} $\mathcal{H}$ is hermitian.
Also, we should take into account state normalization
$\langle \Psi | \Psi\rangle = \sum_{\bullet} {C^{\star}}_{j,k}^{s} C_{j,k}^{s}$, which is a constraint of the
problem, and therefore must be inserted in the functional with its appropriate Lagrange multiplier $\varepsilon$.

In conclusion, the resulting Lagrangian reads
$\mathcal{L}(C, C^{\star}) = \langle \Psi | H |\Psi\rangle - \varepsilon \langle \Psi | \Psi\rangle
= \langle \!\langle C | \mathcal{H} - \varepsilon \mathcal{N}| C \rangle \! \rangle = $ 
\begin{equation} \label{eq:lagrange_fig}
\begin{overpic}[width = 300pt, unit=1pt]{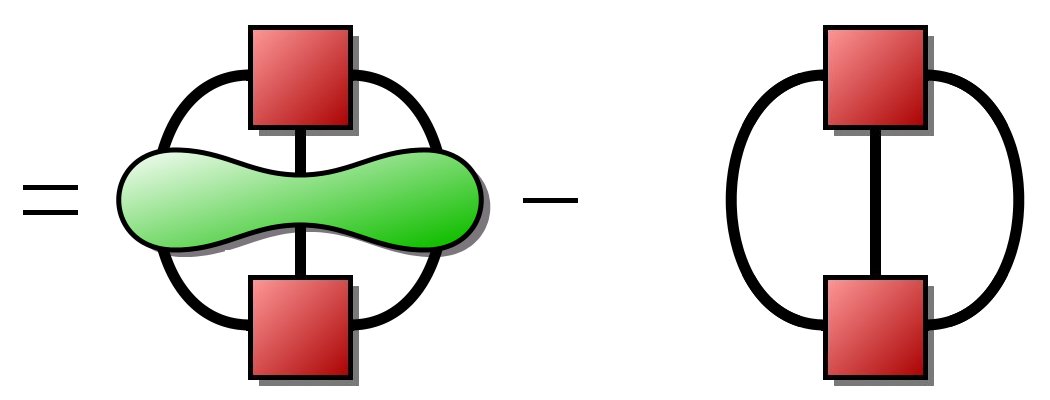}
 \put(-20, 51){\Huge $\mathcal{L}$}
 \put(181, 51){\Huge $\varepsilon$}
 \put(80, 53){$\mathcal{H}$}
 \put(78, 89){$C^{[\ell]}$}
 \put(76, 17){${C^{\star}}^{[\ell]}$}
 \put(242, 89){$C^{[\ell]}$}
 \put(240, 17){${C^{\star}}^{[\ell]}$}
\end{overpic}
\end{equation}
where $| C \rangle \! \rangle$ is intended as a $d D_{\ell} D_{\ell-1} \sim d D^2$ dimensional vector,
$\mathcal{H}$ and $\mathcal{N}$ as observables on such space, representing respectively
the effective Hamiltonian and effective square norm.
Here $\mathcal{H}$ is given by the same five terms of \eqref{eq:cinqueleoni};
in the same order of appearance they read
\begin{equation}
\begin{overpic}[width = \textwidth, unit=1pt]{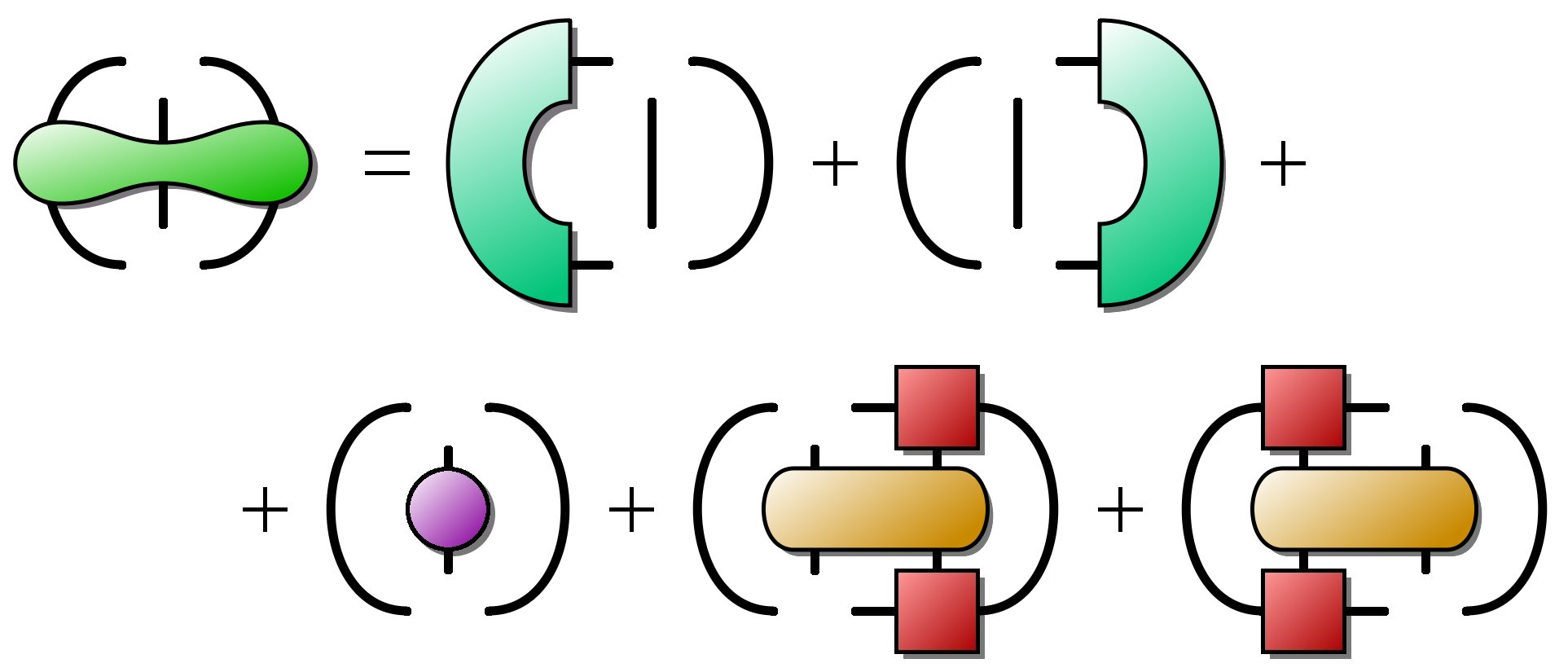}
 \put(13, 124){\scriptsize $\mathcal{H}$}
 \put(113, 124){\scriptsize $\xi_{\ell-1}$}
 \put(291, 124){\scriptsize $\xi_{\ell}$}
 \put(105, 37){\scriptsize $R_1^{[\ell]}$}
 \put(208, 37){\scriptsize $R_2^{[\ell+1]}$}
 \put(333, 37){\scriptsize $R_2^{[\ell]}$}
\end{overpic}
\end{equation}
with $|\xi^{\ell})$, as before, obtained recursively via \eqref{eq:xi_fig}, and similarly
$(\xi^{\ell-1}|$ arises from the left-right specular equation.

You see that, thanks to the gauge condition chosen for the $A^{[\ell']}$ and $B^{[\ell']}$, the
effective square norm $\mathcal{N}$ coincides with the identity operator on the $dD^2$-dimensioned effective space,
so that $\mathcal{L}(C, C^{\star}) = \langle \!\langle C | \mathcal{H} - \varepsilon \Id| C \rangle \! \rangle$
as prescribed by \eqref{eq:lagrange_fig}. Finding the minimum of a quadratic Lagrangian is now straightforward since,
by exploiting the fact that $C$ and $C^{\star}$ are independent for complex differential calculus, one has
\begin{equation} \label{eq:std_eigenvalue}
 \frac{\partial \mathcal{L}(C,C^{\star})}{\partial \langle \!\langle C |} = | 0 \rangle \! \rangle
 \quad \longrightarrow \quad
 \mathcal{H} | C \rangle \! \rangle = \varepsilon | C \rangle \! \rangle.
\end{equation}
Therefore we have to deal with a standard eigenvalue problem for $\mathcal{H}$, and among solutions
$\mathcal{H} | C \rangle \! \rangle = \varepsilon | C \rangle \! \rangle$ we have to consider the one giving
minimal value of $\langle \!\langle C | \mathcal{H} | C \rangle \! \rangle =
\varepsilon \langle \!\langle C | C \rangle \! \rangle = \varepsilon$; in other words, we have to find the
\emph{lowest eigenvalue} solution of the problem \eqref{eq:std_eigenvalue}, and we are done.
In conclusion, we mapped an eigenproblem for the whole $d^{L}$ dimensioned global system into a
$dD^2$ eigenproblem for the local tensor $C^{[\ell]}$. When $D$ is a parameter chosen by the user
and not dependent on $L$, solving the local problem \eqref{eq:std_eigenvalue} requires the same effort for every system size.
Also, the great advantage of working in the proper gauge framework is clear now, since if $\mathcal{N}$ were not to coincide
with $\Id$, we would have to deal with a generalized eigenproblem
($\mathcal{H} | C \rangle \! \rangle = \varepsilon \mathcal{N} | C \rangle \! \rangle$): more expensive and less stable.

It is clear that after one has found the minimal
$| C_{\text{min}} \rangle \! \rangle$ for \eqref{eq:std_eigenvalue} the energy
of the resulting $|\Psi_2\rangle$ is necessary decreased (or equal) from the initial guess $|\Psi_1 \rangle$
\begin{equation}
 \langle \Psi_{2} | H |\Psi_{2}\rangle = \langle \!\langle C_{\text{min}} | \mathcal{H} | C_{\text{min}} \rangle \! \rangle
 \leq \langle \!\langle C_{\text{guess}} | \mathcal{H} | C_{\text{guess}} \rangle \! \rangle = 
 \langle \Psi_{1} | H |\Psi_{1}\rangle.
\end{equation}
This is how the algorithm proceeds towards energy minimization, after this step we move to the MPS block to the immediate
right $\ell+1$ (or the immediate left $\ell-1$) and repeat. Of course, to complete the iteration step one has to
perform the proper gauge transformation so that, after finding $C_{\text{min}}^{[\ell]}$, turns it into left gauge
$\rightarrow A^{[\ell]}$ (or right gauge if we are sweeping left) so that \eqref{eq:284} immediately holds for site $\ell+1$.
But this is easy, just perform a singular value decomposition of $C^{[\ell]s}_{j,k}$
\begin{equation}
 C^{[\ell]}_{\alpha,k} = \sum_{\beta} A^{[\ell]}_{\alpha,\beta} \;\lambda_{\beta}\; U_{\beta,k}
\end{equation}
with $\alpha$ being the composite index $\{j,s\}$, and $A^{\dagger} A = U U^{\dagger} = \Id$. The gauge transformation
is then
\begin{equation}
 \begin{aligned}
 C^{[\ell]s}_{j,k} \quad &\longrightarrow \quad &
 C^{[\ell]s}_{j,\beta} \;U^{\dagger}_{\beta,k} \;\lambda^{-1}_{k} \quad &= A^{[\ell]s}_{j,k}\\
 A^{[\ell+1]s}_{j,k} \quad & \longrightarrow \quad&
 \lambda_{j} \;U_{j,\beta} \;A^{[\ell+1]s}_{\beta,k} \quad & = C^{[\ell+1]s}_{j,k},
 \end{aligned}
\end{equation}
which concludes the iteration step.

The algorithm is usually carried on until some convergence threshold in the energy $\varepsilon$ has been achieved.
In most cases
this simulation procedure converges surprisingly fast, as very few sweeps are necessary to reach a stable minimum,
even if we were to start from a completely random variational MPS.
In several simulations where the single-site framework was adopted,
computational results are in good agreement with theory and/or experiment,
yet this protocol presents some difficulties.
The monotonicity of the energy functional at every iteration step, even if it allows fast convergence,
hides the possibility of getting stuck in local minima of the variational parameters landscape:
in order to work around this issue one has to insert manually artificial fluctuations, as proposed
by S. White in DMRG context \cite{WhiteOneSite}. Similarly, the algorithm encounters trouble when
dealing with symmetries (see appendix \ref{app:symchap}),
where the user is forced to insert symmetry-breaking fluctuations by hand.

Despite on how we can solve, with big or small success, these issues in the single-block framework,
a very common way to work around them is to recover the original idea that gave birth to DMRG, i.e. dealing
with a two-site block minimization at once.

\subsection{Double variational site}

This time we want to variate two adjacent blocks at the same time, say $C^{[\ell]}$ and $C^{[\ell+1]}$,
while keeping fixed the other ones. The most clever way to do this is forgetting that $C^{[\ell]}$ and $C^{[\ell+1]}$
are two distinct MPS blocks: we consider them as a single overall tensor $M^{s_{\ell}, s_{\ell+1}}_{j,k}
= C^{[\ell]s_{\ell}}_{j,\alpha} \; C^{[\ell]s_{\ell+1}}_{\alpha,k} $ on
which the Lagrangian functional is quadratic, and adopt $M$ as our only variational element.
Of course, this allows us to momentarily describe \emph{more} entanglement
across the bond $\{\ell,\ell+1\}$ than what would be normally allowed by a $D$-bondlink MPS. Therefore,
to provide an iterative scheme, we will embed in the algorithm a method for entanglement truncation,
so that in the end the original MPS representation is recovered.

Then, let us start from our initial guess for the iteration step
\begin{equation} \label{eq:Dublock}
 |\Psi\rangle = \sum_{s_1 \ldots s_L = 1}^d 
 \left( A_{s_{1}}^{[1]} \cdot \ldots \cdot A_{s_{\ell-1}}^{[\ell-1]} \cdot
 M_{s_{\ell}, s_{\ell+1}}
 \cdot B_{s_{\ell+2}}^{[\ell+2]} \cdot \ldots \cdot B_{s_{L}}^{[L]} \right)
 | s_1 \ldots s_L \rangle,
\end{equation}
where every $M_{s_{\ell}, s_{\ell+1}}$ is given by the matrix product
$C_{s_{\ell}}^{[\ell]} \cdot C_{s_{\ell+1}}^{[\ell+1]}$
and MPS blocks $A_{s_{\ell'}}^{[\ell']}$ (resp. $B_{s_{\ell'}}^{[\ell']}$) are in the left (right) gauge.
As a whole, $M$ is a tensor with four indices, and notice that alongside the sudden increase
of allowed entanglement (from $\log D$ to $\log dD$), an increase of variational parameters,
with respect to the standard MPS case, comes out: from $2dD^2$ to $d^2 D^2$.

As we mentioned the Lagrangian is quadratic in $M$, and thanks to gauge relations for $A$ and $B$
matrices, the effective normalization $\mathcal{N}$ is again the identity operator,
since $\langle \Psi | \Psi \rangle = \sum_{j k s r} |M^{s,r}_{j,k}|^2$. Then
\begin{equation}
 \mathcal{L}(M,M^{\star}) = 
 \langle \!\langle M | \mathcal{H} | M \rangle \! \rangle -
 \varepsilon \langle \!\langle M |  M \rangle \! \rangle,
\end{equation}
where the effective Hamiltonian $\mathcal{H}$ is given by
\begin{equation}
\begin{overpic}[width = \textwidth, unit=1pt]{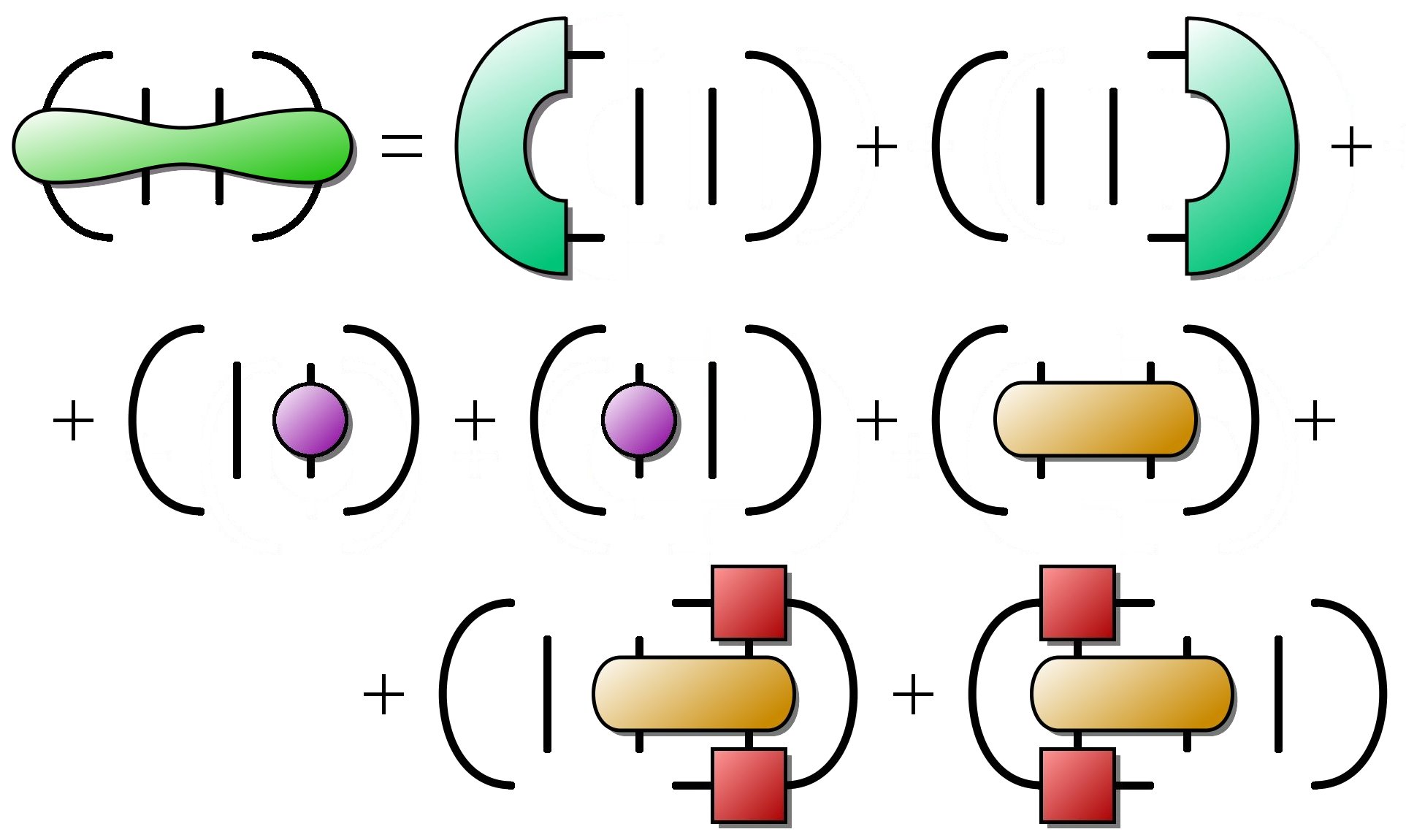}
 \put(15, 190){\scriptsize $\mathcal{H}$}
 \put(128, 190){\scriptsize $\xi_{\ell-1}$}
 \put(343, 190){\scriptsize $\xi_{\ell+1}$}
 \put(171, 113){\scriptsize $R_1^{[\ell]}$}
 \put(296, 113){\scriptsize $R_2^{[\ell+1]}$}
 \put(184, 37){\scriptsize $R_2^{[\ell+2]}$}
 \put(310, 37){\scriptsize $R_2^{[\ell]}$}
\end{overpic}
\end{equation}
Graphs 1 and 2 contain the terms of the Hamiltonian having support outside $\{\ell, \ell+1\}$; graphs 3,4 and 5 those
of support inside $\{\ell, \ell+1\}$, and the last two represent interaction of the inner sites with the environment.

Like for the single site case, the optimal $M$ is found via lowest eigenvalue problem, i.e. smallest $\varepsilon$ allowing
\begin{equation}
 \mathcal{H} | M \rangle \! \rangle = \varepsilon | M \rangle \! \rangle.
\end{equation}
Now we have to manipulate the newly found $M$ in order to recover the standard MPS form.

To do this we proceed again via singular value decomposition.
First we write the tensor $M$ as a matrix $M_{\alpha \beta}$
where the composite index $\alpha \sim \{j,s_{\ell}\}$ refer to the left bondlink index $j$ and the
left physical index $s_{\ell}$, while $\beta \sim \{k,s_{\ell+1}\}$ to the right bondlink $k$ and site $s_{\ell+1}$ indices.
Now we calculate the SVD as
\begin{equation}
 M_{\alpha \beta} = \sum_{\gamma}^{dD} A^{[\ell]}_{\alpha \gamma} \; \lambda_{\gamma}\; B^{[\ell+1]}_{\gamma \beta}
 = \sum_{\gamma}^{dD} A^{[\ell]s_{\ell}}_{j \gamma} \; \lambda_{\gamma}\; B^{[\ell+1] s_{\ell+1}}_{\gamma k}
\end{equation}
so we can stick it into \eqref{eq:Dublock} and obtain again the MPS representation. Still, the bondlink
$\ell$ has increased dimension $D_{\ell}^{\text{new}} = \min\{d D_{\ell-1}, d D_{\ell + 1}\} \sim dD$.
But now the positive values $\lambda_{\gamma}$ are the Schmidt coefficients of the partition at bond $\ell$
of the state $|\Psi\rangle$, and $\langle \Psi | \Psi \rangle = \sum_{\gamma}^{dD} \lambda_{\gamma}^2$.
Therefore, in order to recover the best approximation of this state
allowing $\log D$ entanglement, we must truncate the \emph{smallest} Schmidt coefficients
$\lambda_{\gamma}$, until only $D$ of them remain. Namely, if $\lambda_{\gamma}$ were sorted in decreasing
order ($\lambda_{\gamma} \geq \lambda_{\gamma+1}$),
we keep only the first $D$ of them, and renormalize to preserve state norm
\begin{equation} \label{eq:tronky} 
 {\tilde{\lambda}}_{\gamma} = \frac{\lambda_{\gamma}}{\sqrt{\sum_{\eta}^{D} \lambda^2_{\eta}}}
 \quad \gamma \in \{1..D \}
 \qquad \longrightarrow \qquad \sum_{\gamma}^{D} {\tilde{\lambda}}_{\gamma}^2 = 1;
\end{equation}
moreover, $A^{[\ell]}$ and $B^{[\ell]}$ will satisfy respectively the left and right gauge condition even after the truncation.

We can now write, assuming we are sweeping towards right,
$M_{s_{\ell}, s_{\ell+1}} = A^{[\ell]}_{s_\ell} \cdot C'^{[\ell+1]}_{s_\ell+1}$, where
$C'^{[\ell+1] s_\ell+1}_{\gamma, k} = {\tilde{\lambda}}_{\gamma} B^{[\ell+1] s_\ell+1}_{\gamma, k}$ so that
\begin{equation}
 |\Psi\rangle = \sum_{s_1 \ldots s_L = 1}^d 
 \left( A_{s_{1}}^{[1]} \cdot \ldots \cdot A_{s_{\ell-1}}^{[\ell-1]} \cdot
 A_{s_{\ell}}^{[\ell]} \cdot \overbrace{C_{s_{\ell+1}}^{[\ell+1]} \cdot B_{s_{\ell+2}}^{[\ell+2]}}
 \cdot \ldots \cdot B_{s_{L}}^{[L]} \right)
 | s_1 \ldots s_L \rangle.
\end{equation}
The MPS representation is now ready to perform the next algorithm iteration, just identify the new
two-site block $M_{s_{\ell+1},s_{\ell+2}} = C_{s_{\ell+1}}^{[\ell+1]} \cdot B_{s_{\ell+2}}^{[\ell+2]}$.

The double-site based algorithm we just described presents two important improvements
with respect to the single-site one. First, as two adjacent blocks are being modified at the same time,
reconstructing the correct short-range physics runs much faster, and since the Hamiltonian is made
of nearest neighboring terms the energy is extremely sensitive to the n-n physics (especially for non-critical
systems) thus leading to a faster minimization convergence.
Secondly, at the time we perform the truncation \eqref{eq:tronky}, we allow for errors in our description,
as we force the state to carry no more entanglement than the MPS representation allows.
Therefore, slight fluctuations appear, identified by eventual small increases in energy.
This is actually an advantage of the protocol, as fluctuations are
a natural way to discourage the algorithm from getting stuck in local minima of the energy landscape.

\section{Matrix Product Operators} \label{sec:MPO}

So far, we applied the Matrix Product formalism in order to build multi-indexed tensors $\mathcal{T}_{s_1 \ldots s_L}$,
which were components of a target state $|\Psi\rangle$ over a separable (canonical) vector basis
$|s_1 \ldots s_L\rangle$; but it is clear that its capabilities extend to \emph{every} algebraic construct
which can be expanded in a basis of local elements, regardless their nature.
No doubt, applying the Matrix Product concept to describe nonlocal operators seems the most natural
goal to pursue.

Matrix Product Operators (MPO) were firstly introduced in \cite{MPDO} and used, for instance,
to describe either thermal mixed states, time evolution paradigms for MPS \cite{MPOMurg} \cite{Maricarmen} \cite{Sebadrag},
or long range interaction Hamiltonians \cite{MPOFrowis}.
Their open boundary formulation is definitely similar to that of MPS:
\begin{equation} \label{eq:MPO}
 \hat{\Theta} = \sum_{s_1 \ldots s_L}^d \sum_{r_1 \ldots r_L}^d
 \left( O_{s_{1}}^{[1]r_{1}} \cdot \ldots \cdot O_{s_{L}}^{[L]r_{L}} \right)
 | s_1 \ldots s_L \rangle \langle r_1 \ldots r_L |,
\end{equation}
where, if $|s_{\ell}\rangle$ is the canonical vector basis on site $\ell$, then
$|s_{\ell}\rangle \langle r_{\ell}|$ is the canonical local operator basis.
The local-basis expansion is then performed over these canonical elements,
while adopting Matrix Product-based coefficients.
To every site $\ell$, incoming physical index $r_{\ell}$, and outcoming physical index $s_{\ell}$ we associated
a $D_{\ell-1} \times D_{\ell}$ matrix $O_{s_{\ell}}^{[\ell]r_{\ell}}$, which sums altogether to a four-indices tensor
on every site:
\begin{equation}
\begin{overpic}[width = 350pt, unit=1pt]{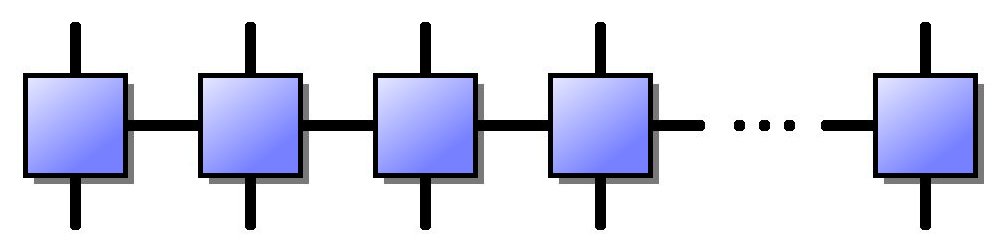}
 \put(20, 40){$O^{[1]}$}
 \put(81, 40){$O^{[2]}$}
 \put(142, 40){$O^{[3]}$}
 \put(203, 40){$O^{[4]}$}
 \put(316, 40){$O^{[L]}$}
\end{overpic}
\end{equation}
As you can see, blocks $1$ and $L$ of the MPO have only one correlation space
link index to be consistent with the OBC setting.
As for MPS, Matrix Product Operators are typically prescribed according to a maximal bondlink dimension $D$
($D_\ell < D$ $\forall \ell$, with $D$ non-scaling with $L$)
which makes the expression \eqref{eq:MPO} manageable for practical purposes
even for large system sizes $L$. Such $D$ also poses a limit in the entangling capabilities of $\hat{\Theta}$,
actually binding the amount of long-range correlation the operator can create.

MPOs are outstanding tools when the goal is to apply a transformation $\hat{\Theta}$ to a
state $|\Psi\rangle$ whose MPS representation is available. In fact, the resulting
state $\hat{\Theta} |\Psi\rangle$ is automatically expressed in Matrix Product form:
\begin{multline} \label{eq:centouno}
 \hat{\Theta} |\Psi\rangle = \sum_{\{s_{\ell}\} \{r_{\ell}\} \{t_{\ell}\}}
 \left( \vphantom{A_{t_{1}}^{[1]}} O_{s_{1}}^{[1]r_{1}} \cdot \ldots \cdot O_{s_{L}}^{[L]r_{L}} \right)
 \left( A_{t_{1}}^{[1]} \cdot \ldots \cdot A_{t_{L}}^{[L]} \right)
 \times \\ \times
 | s_1 \ldots s_L \rangle \langle r_1 \ldots r_L | t_1 \ldots t_\ell \rangle =
 \sum_{s_1 \ldots s_L} \left( B_{S_{1}}^{[1]} \cdot \ldots \cdot B_{t_{L}}^{[L]} \right) | s_1 \ldots s_L \rangle
\end{multline}
where
\begin{equation} \label{eq:centodue}
 B_{s}^{[\ell]} = \sum_{r} (A_{s}^{[\ell]} \otimes O_{s}^{[\ell]r}).
\end{equation}
\begin{center}
 \begin{overpic}[width = 350pt, unit=1pt]{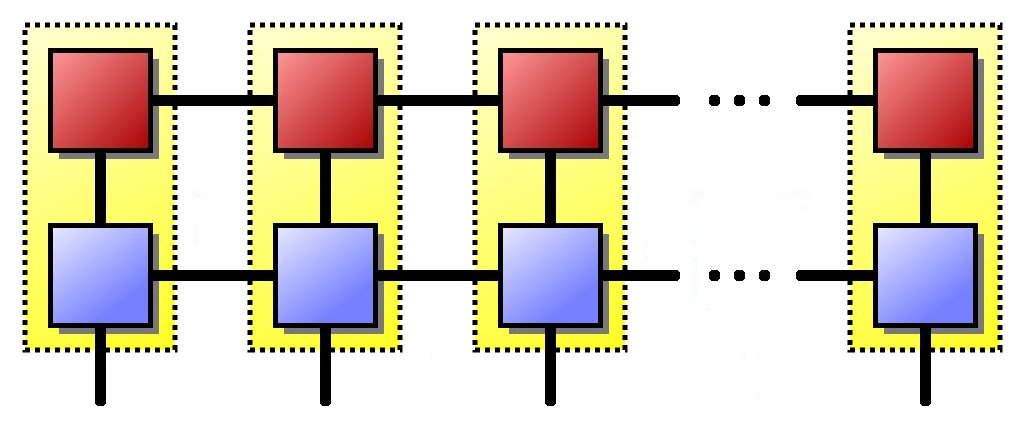}
 \put(27, 46){$O^{[1]}$}
 \put(27, 107){$A^{[1]}$}
 \put(39, 75){$B^{[1]}$}
 \put(104, 107){$A^{[2]}$}
 \put(104, 46){$\ldots$}
\end{overpic}
\end{center}
Truly, the bondlink dimension of the target MPS $\hat{\Theta} |\Psi\rangle$ is increased to $D' = D_A \cdot D_O$, the product
of the original MPS bond $D_{A}$ and that of the MPO $D_{O}$. So, it looks that application of MPO to MPS is definitely
expensive in terms of the bondlink.
This is true, and nevertheless easy to work around: it is sufficient to reduce the target MPS to the desired $D''$
properly. This is quickly done by following the usual steps:
\begin{itemize}
 \item Choose a bond, say $\{\ell,\ell+1\}$.
 \item Gauge transform the MPS so that blocks to the left (resp. right) of the chosen bond are in left (right) gauge,
 so that Schmidt coefficients of the partition emerge explicitly
 \item Truncate the smallest Schmidt coefficients and renormalize to one the remaining ones (squared), as \eqref{eq:tronky}.
 \item choose another bond and repeat, until every bondlink has been renormalized to $D''$ or less.
\end{itemize}
The error we intake when renormalizing the state is compatible with the amount of entanglement we are discarding
(which is explicitly known by comparing Von-Neumann entropies before and after truncation).
Being able to transform MPS into MPS becomes fundamental, for instance, if we want to describe a time-evolution
of a system whose starting point is a finitely correlated state: within this paradigm it is very useful to understand
how to write an MPO representation of a given Hamiltonian, and how to exponentiate it efficiently.
This is a major point of interest of ref.~\cite{MPOMurg}.

\subsection{Matrix Product Density Operators}

A relevant class of operators we are typically interested in,
is the family of density matrices, i.e. positive, unity trace, operators.
Although it is instructive and useful decomposing such operators into MPO form, it is even more interesting
to exploit their positivity (as well as positivity of any partial trace, i.e. degree of freedom reduction)
to further decompose their Matrix Product structure.
Indeed, if we consider that any $\rho \geq 0$ can be written as $\rho = X X^{\dagger}$
(and, conversely, the whole space of matrices $X$ generate the class of positive operators via $X X^{\dagger}$)
one is encouraged to build the MPO decomposition of $X$ rather than  $\rho$ itself.
Doing so not only eliminates the positivity restraint on the resulting MPO, but also gives us an edge for
dealing with state transformations, as the application $T \rho T^{\dagger}$ simplifies into $TX$,
which could be a nontrivial numerical improvement.

Precisely, in ref.~\cite{MPDO} the Matrix Product Density Operators (MPDO) are properly defined. They are those MPO,
according to \eqref{eq:MPO}, whose blocks $O^{[\ell]}$ are given by
\begin{equation} \label{eq:MPDO}
 O^{[\ell]r}_{s} = \sum_{\tau = 1}^{\tilde{d}_{\ell}} {A^{\star}}^{[\ell]r}_{\tau} \otimes {A}^{[\ell]\tau}_{s},
\end{equation}
where $\tilde{d}_{\ell}$ is at most $d D_{\ell-1} D_{\ell}$.
Decomposition \eqref{eq:MPDO} is actually splitting
the matrix product layer into two sub-layers stacked together, as
\begin{equation} \label{eq:MPDOfig}
\begin{overpic}[width = 350pt, unit=1pt]{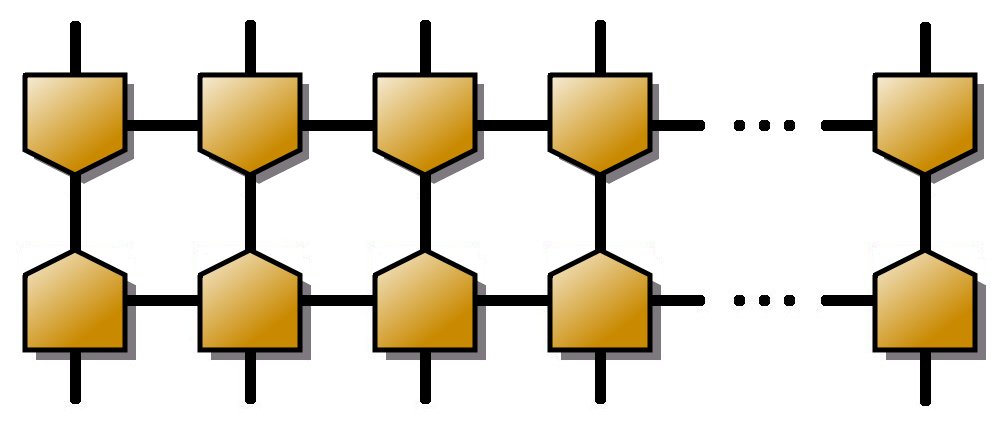}
 \put(19, 37){$A^{[1]}$}
  \put(16, 104){${A^{\star}}^{[1]}$}
 \put(81, 37){$A^{[2]}$}
  \put(78, 104){${A^{\star}}^{[2]}$}
 \put(142, 37){$A^{[3]}$}
  \put(139, 104){${A^{\star}}^{[3]}$}
 \put(203, 37){$A^{[4]}$}
  \put(200, 104){${A^{\star}}^{[4]}$}
 \put(316, 37){$A^{[L]}$}
  \put(313, 104){${A^{\star}}^{[L]}$}
\end{overpic}
\end{equation}
where the pentagonal shape of the tensors in the diagram specifies that tensors in the upper layer are
up-down specular to those in the lower layer (plus complex conjugation).

If we were to give an interpretation to the tensorial index dimension $\tilde{d}_{\ell}$ we could invoke again the valence
bond picture: indeed $\log \tilde{d}_{\ell}$ is the maximal allowed entanglement that the system can share with
an external degree of freedom coupling expressly to site $\ell$, e.g. a local thermal bath.

MPDO are useful tools for addressing one-dimensional open systems, especially where the mixing with external media
acts on the bulk itself. It is even possible to formulate master equation problems with matrix product formalism.

\section[Slater determinant MPS]{Example: exact Matrix Product State representation for Slater Determinants} \label{sec:Slater}

We would like to conclude this chapter with a simple, yet practical example of the Matrix Product
formalism applied analytically to a specific class of many body states: Slater Determinants.
In fermionic problems, Slater Determinants are the starting point of most many-body calculations
(like Hartree-Fock); they are states where $N$ fermions share no quantum correlations,
each one of them filling an orbital which is typically solution of the mean-field Hamiltonian.
Nevertheless, since such orbitals are non-necessarily localized in a chosen configuration
space, they can still manifest self-correlation entanglement in the separable basis. So it is probably
the simplest among non-trivial matrix product decomposition problems.
The following construction is somehow related to ref.~\cite{MPOFrowis}, but I developed it
during my Philosophiae Doctorateship as an independent project, supported by G.~Santoro and V.~Giovannetti.

Let us deal with spinless fermions, for simplicity: the first step we have to perform is
to match the physics of this context with the algebraic formulation adopted so far in this chapter.
To do this, we can completely forget about physical
dimensionality of the problem and boundary conditions: the only initial structure we need is
a complete set of $L$ one-body wavefunctions. We also choose a complete ordering for these.
They will represent the sites in our 1D OBC (spin) system, placed according to the chosen ordering;
the canonical local basis corresponding to $|0\rangle =$ empty level and $|1\rangle =$ filled level.
The mapping of a fermionic system into a spin system is made via standard Wigner transformation
\begin{equation} \label{eq:Wigner1}
 \begin{aligned}
 | \Omega \rangle \quad &\longrightarrow \quad | 00 \ldots 0 \rangle \\
 c_{\ell} \quad &\longrightarrow \quad \sigma^{z}_{1} \otimes \ldots \otimes \sigma^{z}_{\ell-1} \otimes \sigma^{-}_{\ell}
 \otimes \Id \otimes \ldots \otimes \Id,
 \end{aligned}
\end{equation}
where $| \Omega \rangle$ is the vacuum state, $c_{\ell}$ the destruction operator on level $\ell$,
and $\sigma$ being Pauli matrices. Now any state $|\Psi\rangle$ can be expanded in such product basis
$|\Psi\rangle = \sum_{s_1 \ldots s_L} \mathcal{T}_{s_1 \ldots s_L} |s_1 \ldots s_L\rangle$, and
we are going to apply the matrix product formalism to the components tensor $\mathcal{T}_{s_1 \ldots s_L}$,
local dimension $d = 2$.
For sake of completeness, let us even write the explicit MPS expansion in the second quantization formalism,
\begin{equation} \label{eq:2quantMPS}
 |\Psi\rangle_{\text{Fermi}} = \sum_{s_1 \ldots s_L = 1}^2
 \left( \vphantom{\sum} A_{s_{1}}^{[1]} \cdot \ldots \cdot A_{s_{L}}^{[L]} \right)
 (c_{1}^{\dagger})^{s_1} \ldots (c_{L}^{\dagger})^{s_L} |\Omega\rangle.
\end{equation}
where the construction operators $c_{\ell}^{\dagger}$
are placed in the correct order, and obviously $(c_{\ell}^{\dagger})^{0} = \Id$.
Simple as that.

A Slater determinant state $|\Sigma\rangle$ is defined as follows
\begin{equation} \label{eq:Slaz}
 |\Sigma\rangle = \tilde{c}_1^{\dagger}\,\tilde{c}_2^{\dagger}\,\ldots \tilde{c}_{N}^{\dagger}\,| \Omega \rangle
\end{equation}
where $\tilde{c}_{\alpha}^{\dagger}$ fills a one-body orbital which may have a nontrivial expansion
over the original one-body levels we chose as basis. Precisely the transformation is given by
\begin{equation}
 \tilde{c}_{\alpha} = \sum_{\ell = 1}^{L} \phi^{\star}_{\alpha}(\ell)\; c_{\ell},
\end{equation}
$\phi^{\star}_{\alpha}(\ell)$ being the first-quantization
decomposition of the Slater orbital $\alpha$ onto the original wavefunction basis $\ell$.
Orthogonality is required among $\alpha$ orbitals, i.e.
$\sum_{\ell} \phi_{\alpha}(\ell) \phi^{\star}_{\beta}(\ell) = \delta_{\alpha, \beta}$.
As we know from literature, such considerations lead us to write $|\Sigma\rangle$ in its explicit determinant form
\begin{equation} \label{eq:slasla}
 |\Sigma\rangle = \sum_{0 \leq \ell_1 < \ldots < \ell_{N} \leq L}
 \left| \begin{array}{cccc}
    \phi_{1}(\ell_1) & \phi_{1}(\ell_2) & \cdots & \phi_{1}(\ell_N) \\
    \phi_{2}(\ell_1) & \phi_{2}(\ell_2) & \cdots & \phi_{2}(\ell_N) \\
    \vdots & \vdots & \ddots & \vdots \\
    \phi_{N}(\ell_1) & \phi_{N}(\ell_2) & \cdots & \phi_{N}(\ell_N) \\
 \end{array} \right|
 c_{\ell_{1}}^{\dagger} \ldots c_{\ell_{N}}^{\dagger} |\Omega\rangle.
\end{equation}

To find a MPS representation for $|\Sigma\rangle$, will be instrumental to give a Matrix Product
Operator description for Fermi operators $\tilde{c}_{\alpha}$ over delocalized orbitals.
Since the vacuum is already in (trivial) MPS form $|0\ldots 0\rangle$ we will then find the
MPS structure of $|\Sigma\rangle$ by stacking together MPOs of $\tilde{c}_{\alpha}^{\dagger}$
according to \eqref{eq:Slaz}, as we did in \eqref{eq:centouno} and \eqref{eq:centodue}

\subsection{MPO for delocalized Fermi operators}

We are now going to provide a MPO representation of
$\tilde{c}_{\alpha}^{\dagger}$ which is compact, elegant, and very general. The only ingredient we need is knowing
the expansion of the orbital $\alpha$ in the original one-body wavefunctions basis $\phi_{\alpha}(\ell)$.
Then the goal is finding the $B^{[\ell]}$ satisfying
\begin{equation} \label{eq:CreatMPO}
 \begin{aligned}
 \tilde{c}_{\alpha}^{\dagger} &\longrightarrow \sum_{s_1 \ldots s_L = 0}^1 \sum_{r_1 \ldots r_L = 0}^1
 ( b_0 | \vphantom{\sum} B_{s_{1}}^{[1]r_1} \cdot \ldots \cdot B_{s_{L}}^{[L]r_L} | b_L)\;
 |s_1 \ldots s_L\rangle \langle r_1 \ldots r_L |\\
 &= \sum_{\{s\},\{r\} = 0}^{1}
 ( b_0 | \vphantom{\sum} B_{s_{1}}^{[1]r_1} \cdot \ldots \cdot B_{s_{L}}^{[L]r_L} | b_L)\;
 (c_{1}^{\dagger})^{s_1} \ldots (c_{L}^{\dagger})^{s_L} |\Omega\rangle \langle \Omega |
 \,c_{L}^{r_L} \ldots c_{1}^{r_1},
 \end{aligned}
\end{equation}
where we explicitly set vector boundaries to the matrix product expression
(actually $| b_L)$ is a vector and $(b_0|$ a functional) so that we will able to define
every $B_{s}^{[\ell]r}$ homogeneously, even those at the furthest sites. In particular we need $D = 2$,
and the solution we found is given by
\begin{equation} \label{eq:FermiMPOrecipe}
 \begin{array}{ccc}
  {B^{[\ell]}}_{0}^{0} = \left( \begin{array}{cc}
 1 & 0 \\ 0 & 1
 \end{array} \right) = \Id
 & \qquad
 & {B^{[\ell]}}_{0}^{1} = \left( \begin{array}{cc}
 0 & 0 \\ 0 & 0
 \end{array} \right) = 0
 \\ \\
  {B^{[\ell]}}_{1}^{0} = \left( \begin{array}{cc}
 0 & 0 \\ \phi_{\alpha}(\ell) & 0
 \end{array} \right) = \phi_{\alpha}(\ell) \, \sigma^{-}
 &&   {B^{[\ell]}}_{1}^{1} = \left( \begin{array}{cc}
 1 & 0 \\ 0 & -1
 \end{array} \right) = \sigma^{z}
 \\ \\
 | b_L ) = \left( \begin{array}{cc}
 1 \\ 0
 \end{array} \right) = |0)
 && ( b_0 | = \left( \vphantom{{B^{[\ell]}}_{1}} 0\;\;\;1 \right) = (1|,
 \end{array}
\end{equation}
where the information on $\phi_{\alpha}(\ell)$ is used on only
one of the 16 elements of the four-indices tensor $B^{[\ell]}$. Apart from that
the expression \eqref{eq:FermiMPOrecipe} is formally homogeneous in $\ell$, as we wanted.\

To show that \eqref{eq:FermiMPOrecipe} reproduces the correct action of $\tilde{c}_{\alpha}^{\dagger}$
we first notice that when performing the matrix product contraction, the terms which contain one and only
one $\sigma^{-}$ are the ones that survive: indeed $(1|0) = (1|\sigma^{z}|0) = 0$, while
$(1|\sigma^{-}|0) = 1$, but on the other hand $(\sigma^{-})^2 = 0$. So we can reduce the expression
\eqref{eq:CreatMPO} in a simple sum over the site $\ell$ upon which the $\sigma^{-}$ is activated, becoming
\begin{equation}
 \sum_{\ell} \phi_{\alpha}(\ell) \left[ \sigma_{1}^{z} \otimes \ldots \otimes \sigma_{\ell-1}^{z}
 \otimes \sigma_{\ell}^{-} \otimes \Id_{\ell+1} \otimes \Id_{L} \right] \longrightarrow
 \sum_{\ell} \phi_{\alpha}(\ell)\, c_{\ell}^{\dagger} = \,\tilde{c}_{\alpha}^{\dagger}
\end{equation}
which proves the equivalence. As an additional remark, it is easy to see that it is possible to
\emph{deactivate} the global action of such MPO by just changing a correlation boundary state,
say $(b_0|$. Precisely, if we were to set $(b_0| = (0|$ instead of $(1|$, the MPO expression \eqref{eq:CreatMPO}
would coincide with the identity $\Id$ instead of $\tilde{c}_{\alpha}^{\dagger}$; one can then regard
the correlation space boundaries as local switches that control the whole matrix product behavior,
even if it is not localized.

\subsection{MPO stack to MPS}

We can now go back to the Slater determinant state
$|\Sigma\rangle = \tilde{c}_1^{\dagger}\,\tilde{c}_2^{\dagger}\,\ldots \tilde{c}_{N}^{\dagger}\,| \Omega \rangle$
and adopt the engineering we learned to define its whole MPS exact representation \eqref{eq:2quantMPS}.
In particular we start from the vacuum state $|00\ldots 0\rangle$, which is trivially an
MPS with $D=0$, and we apply $\tilde{c}_{N}^{\dagger}$ to obtain again an MPS.
Then we refrain, by applying in order $\tilde{c}_{N-1}^{\dagger}$,  $\tilde{c}_{N-2}^{\dagger} \ldots$ and so forth,
up to $\tilde{c}_{1}^{\dagger}$: every step is performed
following the prescription of eq.~\eqref{eq:centodue}
(although we should never renormalize if we want our description to be exact).
In conclusion we have
\begin{equation} \label{eq:Slamprecipe}
 A^{[\ell]} = \sum_{q_1 \ldots q_N = 0}^{1}
 \left( {B^{[\ell,N]}}_{q_N}^{0} \otimes \ldots \otimes
 {B^{[\ell,2]}}_{q_2}^{q_3} \otimes {B^{[\ell,1]}}_{q_1}^{q_2} \right).
\end{equation}
As you see, our description uses as a whole a total bondlink dimension of $D = 2^{N}$, regardless from $L$.
Actually, since every $B^{1}_{0}$ is the null operator, we can also restrict the previous sum to
$q_1 \geq q_2 \geq \ldots \geq q_N$, since every term for which any $q_k < q_{k+t}$, with $t>0$, would give zero contribution.
In the end it is a sum of merely $N$ terms.
Similarly, we define the correlation boundary vectors:
\begin{equation}
 |b_{L}) = |0)^{\otimes N} =
 \left( \begin{array}{c} 1 \\ 0 \\ \vdots \\ 0 \end{array} \right),
 \quad \text{and} \quad
 (b_0| = (1|^{\otimes N} = \left( 0 \; \cdots\; 0 \; 1 \right).
\end{equation}
Putting these ingredients together leads us to the decomposition of our Slater Determinant $\Sigma$ in the MPS
representation, where explicit boundaries of the matrix product expression are present
\begin{equation} \label{eq:Slamp}
 |\Sigma\rangle = \sum_{s_1 \ldots s_L = 1}^2
 (b_0| A_{s_{1}}^{[1]} \cdot \ldots \cdot A_{s_{L}}^{[L]} |b_L)\;
 (c_{1}^{\dagger})^{s_1} \ldots (c_{L}^{\dagger})^{s_L} |\Omega\rangle.
\end{equation}
Let us briefly analyze the matrices we built via \eqref{eq:Slamprecipe}. It is easy to see that
$A^{[\ell]}_{0}$  it is always the identity $A^{[\ell]}_{0} = \Id_{D \times D}$, while $A^{[\ell]}_{1}$
contains the information upon orbitals, expanded in the original wavefunctions.
To make this clear, we show as an example the cases $N = 2$, for which it holds
$A^{[\ell]}_{1} = \phi_1(\ell) [\Id \otimes \sigma^{-}] + \phi_2(\ell) [\sigma^- \otimes \sigma^z ]$, i.e.
\begin{equation}
 N = 2 \quad \longrightarrow \quad
 A^{[\ell]}_{1} = \left(
 \begin{array}{cccc}
  0 & 0 & 0 & 0 \\
  \phi_1(\ell) & 0 & 0 & 0 \\
  \phi_2(\ell) & 0 & 0 & 0 \\
  0 & -\phi_2(\ell) & \phi_1(\ell) & 0 \\
 \end{array} \right).
\end{equation}
It is clear that, the only matrix products that lead to nonzero amplitude are those where two
excitations $|1\rangle$ are present, Thus the sum \eqref{eq:Slamp} reduces to
\begin{equation}
 |\Sigma_2\rangle = \sum_{\ell_1 < \ell_2} \left( \phi_1(\ell_1)\, \phi_2(\ell_2) - \phi_1(\ell_2)\, \phi_2(\ell_1) \vphantom{\sum} \right)
 (c_{1}^{\dagger})^{\ell_1} (c_{1}^{\dagger})^{\ell_2} |\Omega\rangle,
\end{equation}
where we have recovered explicitly the determinant expression.
Also, let us write down the case with three orbitals to be filled $N = 3$, in this scenario we and up with

\begin{equation} 
 A_1^{[r]} = \mbox{\begin{footnotesize} $
 \left( \begin{array}{cccccccc}
 0 & 0 & 0 & 0 & 0 & 0 & 0 & 0\\
 \phi_1(r) & 0 & 0 & 0 & 0 & 0 & 0 & 0\\
 \phi_2(r) & 0 & 0 & 0 & 0 & 0 & 0 & 0\\
 0 & -\phi_2(r) & \phi_1(r) & 0 & 0 & 0 & 0 & 0\\
 \phi_3(r) & 0 & 0 & 0 & 0 & 0 & 0 & 0\\
 0 & -\phi_3(r) & 0 & 0 & \phi_1(r) & 0 & 0 & 0\\
 0 & 0 & -\phi_3(r) & 0 & \phi_2(r) & 0 & 0 & 0\\
 0 & 0 & 0 & \phi_3(r) & 0 & -\phi_2(r) & \phi_1(r) & 0
 \end{array} \right),
 $ \end{footnotesize}}
\end{equation}
the reader is invited to check that the resulting amplitudes are correct.

\subsection{Efficiency of the description}

We want now argument that, if no further information upon the orbitals $\phi_{\alpha}(\ell)$
being filled is exploited, the exact representation we just gave is the most efficient in terms of MPS.
By this statement we mean that we are spending the smallest bondlink dimension needed
to faithfully reproduce the correct amount of correlation the state can manifest.
From section \ref{sec:valencebond}, we know that a $D$-dimensioned bondlink MPS allows up to
$\mathcal{S} \leq \log_2 D$ entanglement, i.e. Von Neumann entropy of a partition (the logarithm base of 2 is chosen
as common ground in quantum information theory), thus the optimal is $D = 2^{\mathcal{S}}$.
Now, inequality $\mathcal{S} \leq N$ is guaranteed by the existence of 
an exact MPS representation \eqref{eq:Slamp}. But if equality $\mathcal{S} = N$ can be achieved for
some choice of $\phi_{\alpha}(\ell)$, we also proved representation optimality.

To obtain it, we just adopt a special set of (doubly-periodic) plane-waves
$\phi_{\alpha}(\ell) = \exp(4\pi i \alpha \ell / L)$. For simplicity let us perform a half-system partition, and
define a new double set of orbitals $\{\phi^{[L]}_{\alpha}(\ell), \phi^{[R]}_{\alpha}(\ell)\}_{\alpha}$
from the previous ones as
\begin{equation} \label{eq:planesplit}
 \begin{aligned}
 \phi^{[L]}_{\alpha}(\ell) &= \sqrt{2}\; \Theta(L/2 - \ell) \phi_{\alpha}(\ell) \quad \mbox{and}\\
 \phi^{[R]}_{\alpha}(\ell) &= \sqrt{2}\; \Theta(\ell - L/2) \phi_{\alpha}(\ell),
 \end{aligned}
\end{equation}
with $\Theta$ being the Heaviside step function.
Even though in a general case a new set of wavefunctions generated via \eqref{eq:planesplit}
would no longer be orthonormal, it is clear that with the specific choice of
$L/2$ periodic plane waves, orthonormality is preserved:
$\sum_{\ell} \phi^{[L]}_{\alpha}(\ell) {\phi^{\star}}^{[R]}_{\beta}(\ell) = 0$,
as the supports are disjoint, and
$\sum_{\ell} \phi^{[L]}_{\alpha}(\ell) {\phi^{\star}}^{[L]}_{\beta}(\ell) =
\sum_{\ell} \phi^{[R]}_{\alpha}(\ell) {\phi^{\star}}^{[R]}_{\beta}(\ell) = \delta_{\alpha, \beta}$.
So we can define Fermi operators corresponding to this new set, satisfying the anticommutation rules
$ \{ \tilde{c}_{\alpha, \bullet}, \tilde{c}_{\beta, \bullet} \} =
\{ \tilde{c}_{\alpha, L}, \tilde{c}_{\beta, R}^{\dagger} \} = 0$ and
$ \{ \tilde{c}_{\alpha, L}, \tilde{c}_{\beta, L}^{\dagger} \} = 
\{ \tilde{c}_{\alpha, R}, \tilde{c}_{\beta, R}^{\dagger} \} = \delta_{\alpha, \beta}$.
It is clear that the original $\tilde{c}$ decompose in the new ones as
$\tilde{c}_{\alpha} = 2^{-1/2} (\tilde{c}_{\alpha,L} +  \tilde{c}_{\alpha,R})$,
thus letting us write the whole Slater determinant state as:
\begin{equation}
 |\Sigma\rangle = \frac{1}{2^{N/2}} \left( \tilde{c}^{\dagger}_{1,L} +  \tilde{c}^{\dagger}_{1,R} \right)
  \ldots \left( \tilde{c}^{\dagger}_{N,L} +  \tilde{c}^{\dagger}_{N,R} \right) |\Omega\rangle,
\end{equation}
Of this state, we want to calculate the density matrix reduced to half the system, say the left one,
so we trace over the right-half degrees of freedom
$\rho_{L}^{\Sigma} = \trace_{R}\left[ |\Sigma\rangle\langle\Sigma|\right]$.
With this goal, we set $|\Sigma'\rangle = \tilde{c}_{1} |\Sigma\rangle$ and consider:
\begin{equation}
 \begin{aligned}
 \rho^{\Sigma}_L &= \frac{1}{2}\; \trace_{R} \left[ ( \tilde{c}^{\dagger}_{1,L} +  \tilde{c}^{\dagger}_{1,R} )
 \;|\Sigma'\rangle \langle \Sigma'|\; ( \tilde{c}_{1,L} +  \tilde{c}_{1,R} ) \right] \\
 &= \frac{1}{2} \left( \tilde{c}^{\dagger}_{1,L} \trace_{R} \left[ |\Sigma'\rangle \langle \Sigma'|\right] \tilde{c}_{1,L}
 + \trace_{R} \left[ |\Sigma'\rangle \langle \Sigma'| \right] \right) \\
 &= \frac{1}{2} \;\Id_{2 \times 2} \otimes \trace_{R} \left[ |\Sigma'\rangle \langle \Sigma'| \right]
 = \frac{\Id_{2 \times 2}}{2} \otimes \rho^{\Sigma'}_L,
\end{aligned}
\end{equation}
where we used the cyclicity of the trace over right support operators $\tilde{c}_{\alpha,R}$, and clearly
$\tilde{c}_{1,R} |\Sigma'\rangle = 0$; then we noticed that $\rho^{\Sigma'}_L$ and
$(\tilde{c}^{\dagger}_{1,L} \,\rho^{\Sigma'}_L\, \tilde{c}_{1,L})$ have orthogonal supports.
Now we repeat the same argument on $|\Sigma'\rangle$, and proceed by induction. In conclusion,
we can claim that $\rho^{\Sigma}_L$ is (isometrically equivalent to) $2^{-N} \Id_{2^{N} \times 2^{N}}$,
the maximally mixed state on a $2^{N}$ dimensioned space, whose Von Neumann entropy is just $N$.
This concludes the proof.

An intuitive, but not naive, interpretation of such result can be given in the following terms.
As fermions occupying the various orbitals must be mutually uncorrelated due
to the Slater determinant state nature, the only possible entanglement
the system can manifest under a real-space partition is given by the self-correlation of every orbital, separately accounted.
In fact, in the studied case, we presented $N$ uncorrelated completely delocalized orbitals, each one carrying
the entanglement of a unit (i.e. the amount of entanglement shared by a spin singlet), so $N$ is naturally
the total amount.

\subsubsection*{OVERALL REMARKS}
\begin{itemize}
 \item The $\sigma^{z}$ matrix in equation \eqref{eq:FermiMPOrecipe} is the one and only responsible for
   establishing the correct anticommutation relations of Fermi statistics. That said,
   it is straightforward to modify \eqref{eq:FermiMPOrecipe} so that the corresponding MPO
   is describing a \emph{Bose operator} instead:
   you just need to replace ${B^{[\ell]}}_{1}^{1} = \Id$ and leave the rest unchanged
   (also extensible to abelian anyons by using phase gates ${B^{[\ell]}}_{1}^{1} = e^{i \varphi \sigma^{z}}$).
 \item We mentioned that the present design is modeled on spinless fermions, but actually is naturally extensible
   to fermion with spins. The only difference is that at the very beginning, when we are selecting a complete
   basis of orbitals, we need to specify a complete basis of spin-orbitals instead, and then choose a complete ordering.
   Any ordering is fine and does not compromise the MPS cost in terms of $D$ as long as the particles are uncorrelated
   and fixed in number.
\end{itemize}

\subsection{Tensor grid representation of one-body\\wavefunction basis change}

Let us recall that, when we derived the MPO representation for $\tilde{c}_{\alpha}^{\dagger}$, we also mentioned
that it is possible to control its overall action by adjusting the left correlation boundary vector $(b_0|$:
namely the MPO coincides with $\tilde{c}_{\alpha}^{\dagger}$ if $(b_0|=(1|$, while it is just the identity $\Id$ for
$(b_1| = (0|$. In other words
\begin{equation}
 \sum_{\{s\},\{r\} = 0}^{1}
 ( q | \vphantom{\sum} B_{s_{1}}^{[1]r_1} \cdot \ldots \cdot B_{s_{L}}^{[L]r_L} | b_L)\;
 {c_{1}^{\dagger}}^{s_1} \ldots {c_{L}^{\dagger}}^{s_L} |\Omega\rangle \langle \Omega |
 \,c_{L}^{r_L} \ldots c_{1}^{r_1} = \tilde{c}_{\alpha}^{\dagger \,q},
\end{equation}
with $q \in \{0,1\}$.
Also recall that the fermionic orbitals $\phi_{\alpha}(\ell)$
we filled to build the Slater determinant state were an orthonormal
set: let us complete it to an orthonormal basis $\{\phi_{\alpha}(\ell)\}_{\alpha}$, with
$\alpha \in \{1..L\}$. The dimension must be $L$ by the
assumption that the original set of $L$ wavefunctions was complete.
For any of those $\phi_{\alpha}(\ell)$ the corresponding MPO is given by \eqref{eq:FermiMPOrecipe}.

Now we stack together the MPOs, like we did for the Slater state, but instead using only $N$ of them,
we stack the complete set, ordered from $\alpha = 1$ on top to $\alpha = L$ at the bottom;
moreover, instead of using the standard left correlation boundary vector $(b_0|_\alpha = (1|$
we set a generic $(b_0|_\alpha = (q_{\alpha}|$.
It is obvious that the operator arising from this construction is equivalent to
\begin{equation}
  \tilde{c}_{1}^{\dagger\, q_1} \; \tilde{c}_{2}^{\dagger\, q_2} \;
  \tilde{c}_{3}^{\dagger\, q_3} \ldots \tilde{c}_{L}^{\dagger\, q_L}.
\end{equation}
Finally, we apply such operator to the vacuum $|\Omega\rangle$.
The meaning of all this construction is that we actually defined an application
on the binary strings of $\{q_{\alpha}\}_{\alpha}$ to the real Fermi space, as
\begin{equation} \label{eq:Wigner2}
 (q_1 \ldots q_L| \longrightarrow \tilde{c}_{1}^{\dagger\, q_1} \; \tilde{c}_{2}^{\dagger\, q_2}
  \ldots \tilde{c}_{L}^{\dagger\, q_L} |\Omega\rangle.
\end{equation}
By linearity, this map extends to all the space generated by $(q_1 \ldots q_L|$, which corresponds to
the whole correlation bondlink space (as $(q_1 \ldots q_L|$ is its canonical product basis). The
map is clearly bijective and thus invertible.
But you notice that the inverse of \eqref{eq:Wigner2} is basically a Wigner transformation from the Fermi space
to its spin representation where this time the $\phi_\alpha$ have been chosen as basis of one-body wavenfunctions,
so it is formally similar to \eqref{eq:Wigner1}, but the basis has \emph{changed}
(the old one is associated to the $c$, the new one to the $\tilde{c}$).

In conclusion, we could use all this MPO stack formalism to represent a many-body state transformation $|\Psi\rangle$
corresponding to a change of the chosen basis of one-body wavefunctions. I.e. assuming that we can expand
\begin{equation}
 |\Psi\rangle = \sum_{\{s\} = 0}^{1} \mathcal{T}_{s_1 \ldots s_L}^{\text{old}}
 {c}_{1}^{\dagger\, s_1} \ldots {c}_{L}^{\dagger\, s_L} |\Omega\rangle =
 \sum_{\{q\} = 0}^{1} \mathcal{T}_{q_1 \ldots q_L}^{\text{new}}
 \tilde{c}_{1}^{\dagger\, q_1} \ldots \tilde{c}_{L}^{\dagger\, q_L} |\Omega\rangle,
\end{equation}
then the two components tensors $\mathcal{T}^{\text{old}}$ and $\mathcal{T}^{\text{new}}$ satisfy the
equation:
\begin{equation} \label{eq:Grid}
 \begin{overpic}[width = \textwidth, unit=1pt]{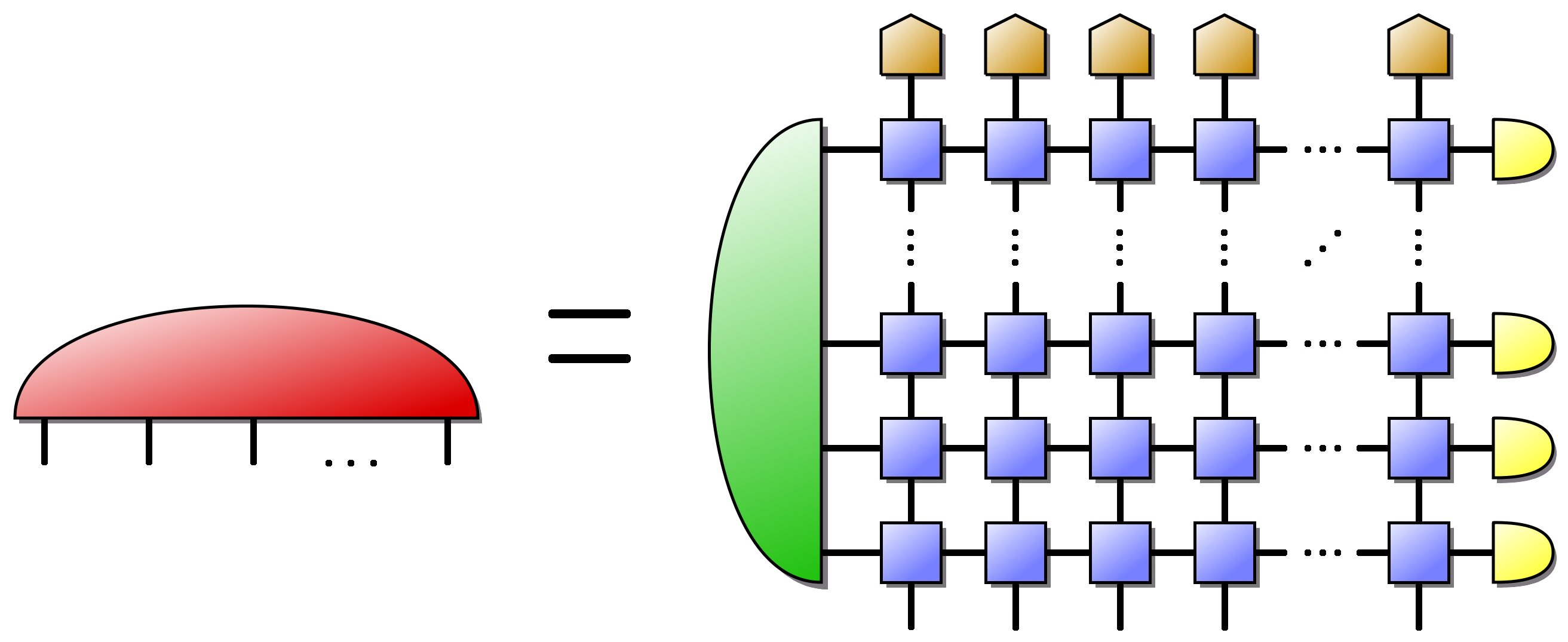}
  \put(54, 65){$\mathcal{T}^{\text{old}}$}
  \put(178, 72){$\mathcal{T}^{\text{new}}$}
  \put(225, 144.5) {\scriptsize $0$}
  \put(251, 144.5) {\scriptsize $0$}
  \put(276, 144.5) {\scriptsize $...$}
  \put(376, 119) {\scriptsize $0$}
  \put(376, 71) {\scriptsize $0$}
 \end{overpic}
\end{equation}
where the blue tensors in the grid are exactly $B^{[\ell, \alpha]}$ of \eqref{eq:FermiMPOrecipe},
with $\ell$ being the coordinate in the horizontal axis, and $\alpha$ the one in the vertical axis
(the origin is the lower-left corner). The upper and rightmost edge tensors are trivially
$|0\rangle$ and $|0)$.

An interesting remark to this result, is that the grid Tensor Network that appear in \eqref{eq:Grid} can be
efficiently contracted, despite having several closed loops in its geometry (see section \ref{sec:Closedloop}),
thanks to particle-conservation symmetry relations.

\subsection{Extensions to Configuration Interaction}

In quantum chemistry settings, the simplest path to move beyond the mere mean field paradigm
is adopting Configuration Interaction. In those descriptions, Hartree-Fock solutions are adopted as a
canonical vector basis of orbitals for further calculations.
According to such viewpoint, one is interested to express correlations by superposing
few to several Slater determinant states, which typically \emph{share} some of the HF orbitals as well as \emph{differ}
for other ones. If the energy minimization problem were to be performed over the whole space of Slater states
the result would be exact, still this would be an extremely hard problem: thus generally the amount of
orbitals for which the involved Slaters differ, is kept to a small, manageable number.

Having this scheme in mind, we would like to extend our previous Slater MPS (MPO stack) representation to embed
also Configuration Interaction states, where different orbital excitations are coherently added.
The ultimate ingredient of this perspective would be writing Matrix Product representation for every operator
generated by the Fermi ones $\tilde{c}^{\dagger}_{\alpha}$ through sums and multiplications. Of course the related
zoology is huge, so we will limit our discussion the simplest nontrivial case.

Consider for instance the expression
\begin{equation}
 \Theta_{2+2} = \alpha \;\tilde{c}^{\dagger}_1 \,\tilde{c}^{\dagger}_2 +
 \beta \;\tilde{c}^{\dagger}_3 \,\tilde{c}^{\dagger}_4.
\end{equation}
We want to describe $\Theta_{2+2}$ as a Matrix Product Operator, and as you can guess there is no unique way to perform
the extension from the normal Fermi operator case.
Depending on whether we focus on the adaptability of the description or the economy on the bondlink
dimension we end up with different proposals.

\vspace{.5em}
\emph{\textbf{Standard Guess -}} this path exploits the standard technique to sum coherently Matrix
Product objects, and is strongly based on \eqref{eq:FermiMPOrecipe}; thus is highly suitable for further generalization,
but at the cost of a sub-optimal bondlink dimension.
Let us now adopt $D = 8$ and consider
\begin{equation} \label{eq:blockdj}
 {B^{[\ell,2+2]}}_{j}^{i} = \sum_k \left( \begin{array}{c|c}
 {B^{[\ell,1]}}_{k}^{i} \otimes {B^{[\ell,2]}}_{j}^{k} & 0 \\ \hline
 0 & {B^{[\ell,3]}}_{k}^{i} \otimes {B^{[\ell,4]}}_{j}^{k}
 \end{array} \right),
\end{equation}
where the $B^{[\ell,\alpha]}$ tensors are those defined in \eqref{eq:FermiMPOrecipe} for $\tilde{c}^{\dagger}_{\alpha}$.
The basic idea behind this construction is to use a correlation space which is the
\emph{Cartesian sum} of the two original correlation spaces, and a matrix product object which is the block
diagonal composition. Similarly we define the correlation boundary vectors, which contain information on $\alpha$ and $\beta$:
\begin{equation}
 | b_L ) =
 \mbox{\scriptsize{$ \left( \begin{array}{c} \alpha \\ 0 \\ 0 \\ 0 \\ \beta \\ 0 \\ 0 \\ 0 \end{array} \right) $}}
 = \alpha
 \mbox{\footnotesize{$ \left( \begin{array}{c} 1 \\ 0 \\ 0 \\ 0 \end{array} \right) $}}
 \oplus \beta
 \mbox{\footnotesize{$ \left( \begin{array}{c} 1 \\ 0 \\ 0 \\ 0 \end{array} \right) $}} 
 = \mbox{\footnotesize{$ \left( \begin{array}{c} \alpha \\ \beta \end{array} \right) $}}\otimes
 \mbox{\footnotesize{$ \left( \begin{array}{c} 1 \\ 0 \\ 0 \\ 0 \end{array} \right) $}},
\end{equation}
where we used distributivity of the tensor product $\otimes$ with respect to the Cartesian sum $\oplus$.
Similarly, $(b_0| = ($\mbox{\tiny{1 1}}$) \otimes ($\mbox{\tiny{0 0 0 1}}).

\vspace{.5em}
\emph{\textbf{Cheap Guess -}} this path focuses on keeping the lowest correlation bondlink dimension possible, and
actually requires $D = 6$.
\[
 {B^{[\ell,2+2]}}_{0}^{0} = \Id_{6 \times 6} \qquad \quad {B^{[\ell,2+2]}}_{0}^{1} = 0
\]
\[
 {B^{[\ell,2+2]}}_{1}^{0} = \left( \begin{array}{cccccc}
 0 & 0 & 0 & 0 & 0 & 0 \\
 \sqrt{\alpha} \phi_1(\ell) & 0 & 0 & 0 & 0 & 0 \\
 \sqrt{\beta} \phi_3(\ell) & 0 & 0 & 0 & 0 & 0 \\
 \sqrt{\beta}\phi_4(\ell) & 0 & 0 & 0 & 0 & 0 \\
 \sqrt{\alpha} \phi_2(\ell) & 0 & 0 & 0 & 0 & 0 \\
 0 & - \sqrt{\alpha} \phi_2(\ell) & -\sqrt{\beta} \phi_4(\ell) &
 \sqrt{\beta} \phi_3(\ell) & \sqrt{\alpha} \phi_1(\ell) & 0
 \end{array} \right)
\]
\begin{equation} \label{eq:cheapyMPO6}
 \mbox{and}\qquad
 {B^{[\ell,2+2]}}_{1}^{1} = \left( \begin{array}{cccccc}
 1 & 0 & 0 & 0 & 0 & 0 \\
 0 & -1 & 0 & 0 & 0 & 0 \\
 0 & 0 & -1 & 0 & 0 & 0 \\
 0 & 0 & 0 & -1 & 0 & 0 \\
 0 & 0 & 0 & 0 & -1 & 0 \\
 0 & 0 & 0 & 0 & 0 & 1
 \end{array} \right),
\end{equation}
while boundaries are as before $|b_L) = |0)$ and $(b_0| = \left( \mbox{\tiny{0 \ldots 0 1}} \right) = (5|$.
By multiplying the $B^{[\ell,2+2]}$ matrices it is easy to see that we are reproducing the correct action of
the operator, i.e.
\begin{multline} \label{eq:sommy1}
 \sum_{\ell_1 < \ell_2} \left\{ \alpha \left( \vphantom{\sum} \phi_1(\ell_1) \phi_2(\ell_2) 
 - \phi_2(\ell_2) \phi_1(\ell_1) \right) \right. + \\ +
 \left. \beta \left(  \vphantom{\sum} \phi_3(\ell_1) \phi_4(\ell_2) - \phi_4(\ell_2) \phi_3(\ell_1) \right) \right\}
 \;c^{\dagger}_{\ell_1} \;c^{\dagger}_{\ell_2}.
\end{multline}
Like previously, we argued if this Matrix Product representation is optimal
in terms of correlation bondlink dimension: we found that a state of the form $\Theta_{2+2}|\Omega\rangle$ has a real-space
partition entropy of entanglement \emph{at most} equal to $5/2$. This implies that a faithful MPS description would
require a $D \geq \sqrt{32}$, so that $D = 6$ is the smallest allowed integer, and thus optimal.

The present proposal presents various options for generalization, although finding the analytical MPO expression
for a generic operator which is cheapest in terms of $D$ is definitely a hard task.
With this last speculation we conclude this analytical example of Matrix Product formalism for
interesting states in condensed matter physics and quantum chemistry.

\newpage

Throughout this chapter we dealt uniquely with open boundary condition problems, and developed
a formalism of Matrix Product States based on the OBC framework. Of course, such a description
can be adjusted to fit naturally periodic boundary conditions as well,
taking care of the correct amount of entanglement.
In the next chapter we will introduce a periodic description for finitely-correlated states
and thus Matrix Product States, with its proper formulation and tricks of the trade;
this will be instrumental in the proper definition of a thermodynamical limit.

\chapter[Periodic and infinite MPS]{Periodic and infinite Matrix Product States} \label{chap:PBCMPS}

One of the major issues for standard DMRG architectures in 1D problems is dealing with Periodic Boundary Conditions (PBC).
It was soon clear that traditional DMRG ideas could not be applied to PBC with the same success and simulation
precision, but it was only with the advent of MPS representations that this trouble become clear and argumented.
Indeed, while in OBC the DMRG procedures describes all and only the finitely correlated states, i.e. those states whose
entanglement is bounded by a finite value (which typically does not scale with system size $L$) in PBC the correspondence
is not exact any longer. Nevertheless, finitely correlated states play again a very important role in describing
ground states of short-range interacting models, as they manifest the correct entanglement area-law.
Indeed, even in PBC finitely correlated states naturally lead to a matrix product representation,
but the formulation \cite{Rossinisupersolid, Rossinistiff} is slightly different from their OBC counterpart.

\section[Valence bond picture for PBC-MPS]{Valence bond picture for Periodic MPS}

In section \ref{sec:valencebond} we introduced the valence bond picture to argument and contextualize
MPS with open boundaries; its is straightforward to extend such description to a periodic system.
To every site we associate a pair of spins, each one $D$ dimensioned ($D$ chosen by the user, often
sensibly larger than the local degree of freedom dimension $d$). We prepare this virtual state so that
every pair of neighboring sites share a maximally entangled state through the $D$-dimensioned spins
$|\Phi^{+}\rangle = D^{-\frac{1}{2}}\sum_{\alpha}^{D} | \alpha \alpha \rangle$  (entangled bond).
Notice the difference with the OBC case, where we had $L$ sites and thus $L-1$ physical bonds:
in PBC every site has two neighbors (there is neither first nor last site, or, if you prefer, sites $1$ and $L$
are neighbors), so the amount bonds is $L$.
The virtual-to-physical mapping is defined identically to the OBC case:
\begin{equation}
 \mathcal{A}^{[\ell]} = \sum_{s = 1}^{d} \sum_{j,k = 1}^{D} A^{[\ell]s}_{j,k} |s\rangle_{\ell}
 \,\langle j,k |^{\text{aux}}_{\ell}.
\end{equation}
As before, which we are going to apply it to the composite entangled bond state
$\bigotimes_{\ell} \mathcal{A}^{[\ell]} (\bigotimes_{\ell'} |\Phi^{+}\rangle^{\text{aux}}_{\ell', \ell'+1})$.
Immediately, one can see that the resulting state can be expressed as
\begin{equation} \label{eq:PBCMPS}
 |\Psi\rangle = \sum_{s_1 \ldots s_L = 1}^d 
 \trace \left[ A_{s_{1}}^{[1]} \cdot A_{s_{2}}^{[2]} \cdot A_{s_{3}}^{[3]}
 \cdot \ldots \cdot A_{s_{L}}^{[L]} \right]
 | s_1 \ldots s_L \rangle,
\end{equation}
where the $A_{s_{L}}^{[L]}$ (resp $A_{s_{1}}^{[1]}$) are no longer vectors (dual vectors) in the correlation space,
but matrices, $0 \neq D_{L \equiv 0} \leq D$, like for every other site $\ell$.
The trace operator in \eqref{eq:PBCMPS} makes the inner matrix product
cyclic, so there is no starting nor ending point of the 1D ring. Also let us represent $| \Psi \rangle =$
\begin{equation}
\begin{overpic}[width = \textwidth, unit=1pt]{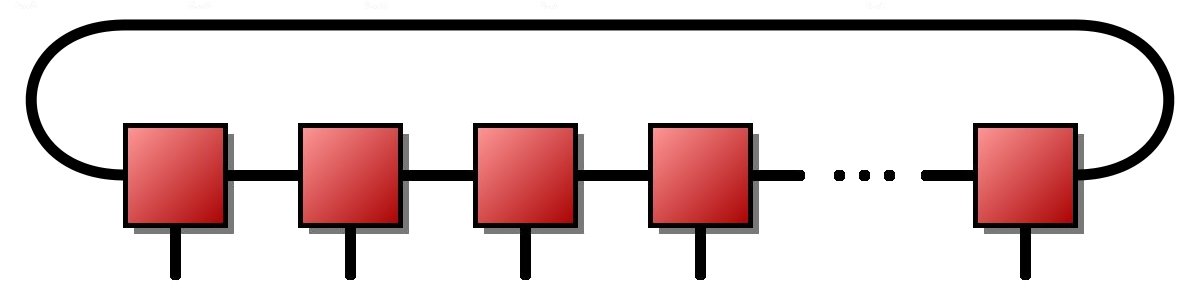}
 \put(49, 36){$A^{[1]}$}
 \put(106, 36){$A^{[2]}$}
 \put(163, 36){$A^{[3]}$}
 \put(220, 36){$A^{[4]}$}
 \put(318, 36){$A^{[L \equiv 0]}$}
\end{overpic}
\end{equation}
diagrammatic version of \eqref{eq:PBCMPS}. If we now are interested in estimating the entanglement of a connected subset
of sites, we can use the same argument for OBC and get a similar conclusion.
In fact, if we want to part the system in an given interval of sites and its complementary,
we need to break two entangled
bonds of the virtual state $\bigotimes_{\ell'} |\Phi^{+}\rangle^{\text{aux}}_{\ell', \ell'+1}$.
And since the resulting state $|\Psi\rangle$ has entanglement bounded bounded by the first one, we have
\begin{equation}
 \mathcal{S}_{\text{VN}}\left( {\rho}_{\ell_1 \ldots \ell_2} \right) \equiv
 -\trace \left[ {\rho}_{\ell_1 \ldots \ell_2} \log {\rho}_{\ell_1 \ldots \ell_2} \right] \leq 
 \log D_{\ell_1} + \log D_{\ell_2} \sim 2 \log D,
\end{equation}
which is twice as in the OBC case, where we could split the system while breaking just one entangled bond.

An interesting point concerning periodic systems is dealing with translational invariance symmetry. As most models
have translationally invariant (TI) Hamiltonians $H$, exploiting the expected TI of the ground state
becomes fundamental for every simulation method.
This is mostly true for PBC, where TI is meaningful and
spontaneously broken only in exceptional cases (when ground space degeneracies arise),
whereas in OBC the breaking is naturally induced by the presence of boundaries.

It is immediate to see that if the tensors in the MPS representation do not depend on the site, i.e.
$A^{[\ell]} \longrightarrow A$ regardless from $\ell$, then the state $|\Psi\rangle$ is translationally invariant:
\begin{multline}
 \mathbb{T} |\Psi\rangle = \sum_{s_1 \ldots s_L = 1}^d 
 \trace \left[ A_{s_{1}} \cdot A_{s_{2}} \cdot A_{s_{3}}
 \cdot \ldots \cdot A_{s_{L}} \right]\,
 \mathbb{T} | s_1 \ldots s_L \rangle
 = \\ =
 \sum_{s_1 \ldots s_L = 1}^d 
 \trace \left[ A_{s_{2}} \cdot A_{s_{3}}
 \cdot \ldots \cdot A_{s_{L}} \cdot A_{s_{1}} \right]
 | s_2 \ldots s_L s_1 \rangle = |\Psi\rangle,
\end{multline}
where $\mathbb{T}$ is the elementary translation operator.
The original state is obtained again by using trace cyclicity and a relabeling of the indices $s$. A more pressing
problem is the inverse: given a translational state $\mathbb{T} |\Psi\rangle = |\Psi\rangle$ allowing a periodic MPS
representation, does it have also a \emph{homogeneous} representation, i.e. where matrices are not site dependent?
We will constructively, and positively, answer such question right away.

\section[TI-MPS admit a homogeneous description]{Translational MPS admit a\\homogeneous description} \label{sec:MPShomo}

Assume we are starting from a site-dependent MPS representation $A^{[\ell]}$ of a state $|\Psi\rangle$
as in \eqref{eq:PBCMPS},
we will build another MPS rep. $B$ where matrices do not depend on the site any longer.
Let us write
\begin{equation} \label{eq:howtohomo}
 B_{s} = L^{- \frac{1}{L}} \left(
 \begin{array}{ccccc}
  0 & A^{[1]}_s &&&\\
  & 0 & A^{[2]}_s && \\
  && \ldots && \\
  &&& 0 & A^{[L-1]} \\
  A^{[L]} &&&& 0 \\
 \end{array}
 \right),
\end{equation}
we will now show that the MPS built with these matrices is equivalent to the original one. In fact
\begin{multline*}
 \sum_{s_1 \ldots s_L = 1}^d 
 \trace \left[ B_{s_{1}} \cdot B_{s_{2}} \cdot \ldots \cdot B_{s_{L}} \right]
 | s_1 \ldots s_L \rangle
 = \\ =
 \frac{1}{L} \sum_{q = 0}^{L-1} \sum_{s_1 \ldots s_L = 1}^d 
 \trace \left[ A^{[1+q]}_{s_1} \cdot \ldots \cdot A^{[L+q]}_{s_L} \right]
 | s_1 \ldots s_L \rangle =
\end{multline*}
\begin{equation}
 = \frac{1}{L} \sum_{q = 0}^{L-1} \sum_{s_1 \ldots s_L = 1}^d 
 \trace \left[ A^{[1]}_{s_{1-q}} \cdot \ldots \cdot A^{[L]}_{s_{L-q}} \right]
 | s_1 \ldots s_L \rangle =
 \frac{1}{L} \sum_{q = 0}^{L-1} \mathbb{T}^{q} |\Psi\rangle = |\Psi\rangle,
\end{equation}
because $\mathbb{T} |\Psi\rangle = |\Psi\rangle$ by hypothesis.
In conclusion, we succeeded in building a homogeneous representation for a generic finitely correlated state
on a 1D PBC ring, but not without expenses. Notice, indeed, that the bondlink dimension we end up with
is $D' = \sum_{\ell} D_{\ell} \sim LD$, with $D_{\ell}$ being the original bondlink dimensions of $A^{[\ell]}$;
a linear scaling law with the system size $L$ arises. At the same time, we are describing the same
amount of entanglement as before, so the $B$ representation is definitely sub-optimal.

Unfortunately, this is common ground when dealing with PBC Matrix Product descriptions
(for both states and operators). Regarding this issue, ref.~\cite{MPSreview} proposes a trivial example involving
the W-state $\sum_q \mathbb{T}^{q} |0\ldots01\rangle$, which has minimal MPS bond dimension of 2, that
necessarily increases to $L$ if we want to give a homogeneous MPS description.

\begin{figure}
 \begin{center}
 \begin{overpic}[width = 300pt, unit=1pt]{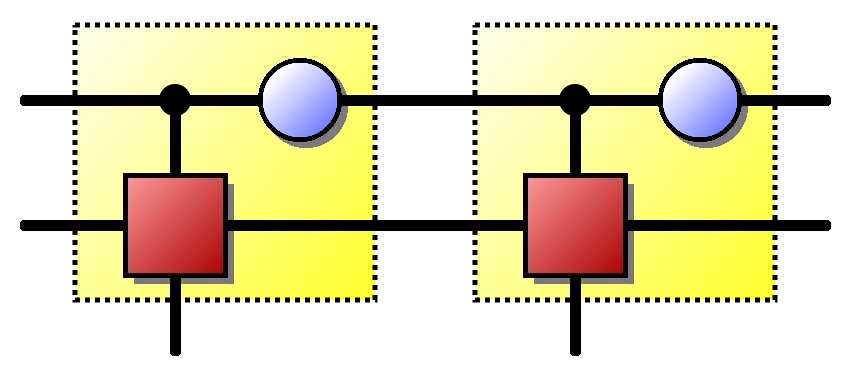}
  \put(51, 48){$A^{[q]s}_{j,k}$}
  \put(99, 94){$\Delta^{+}$}
  \put(-3, 48){$j$}
  \put(146, 104){$q'$}
  \put(146, 58){$k$}
  \put(-3, 95){$q$}
  \put(68, 7){$s$}
  \put(36, 108){$B$}
 \end{overpic}
 \end{center}
\caption{Construction of the homogeneous MPS $B$ \eqref{eq:howtohomo}. 
the resulting bondlink can be seen as composed by the channel carrying $j \in \{1..D\}$ and that carrying
$q \in \{1..L\}$ so its total dimension is $LD$. the starting MPS $A^{[\ell]s}_{jk}$ is now seen as a four-legs
tensor, as the site label $\ell$ becomes a tensor index. The black dot is the triple Kronecker delta,
while $\Delta^{+}_{q,q'} = \delta_{q',q+1 \,\text{mod}\, L}$.}
\end{figure}
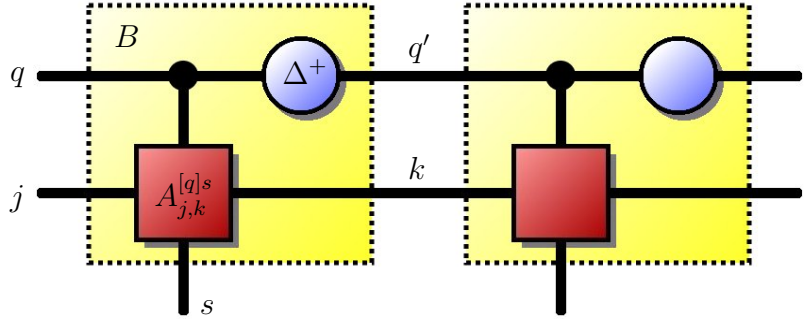

\section{Expectation values for periodic MPS} \label{sec:PBCObs}

Often in numerical settings, dealing with periodicity is more tricky and expensive than corresponding OBC versions
of the same problems: for MPS this peculiarity manifests immediately in an increase of computational costs
needed to acquire expectation values of observables.

In section \ref{sec:OBCMpsObs} we defined a procedure to evaluate physical quantities of a MPS in OBC
with a number of operations that scales nicely, as \eqref{eq:MPScostcompact}, with main simulation parameters:
system size $L$ and correlation bondlink $D$. In the best case scenario, where the separable
observable $O$ acted on an interval of $\ell$ sites, we estimated the cost to scale like $\sim \ell D^3 d$.

Scaling laws are not so nice for periodic MPS, and the compactness of operator supports does not help,
due to the presence of a global closed loop in the graph (see also section \ref{sec:Closedloop}).
Precisely, assume that $O = \bigotimes_{\ell = \ell_1}^{\ell_2} \Theta^{[\ell]}$,
and we are looking for the expectation value
\begin{multline}
 \langle \Psi | O | \Psi \rangle = \sum_{s_1 \ldots s_n} \sum_{r_1 \ldots r_n}
 \trace\left[ A_{s_{1}}^{[1]} \cdot \ldots \cdot A_{s_{L}}^{[L]} \right]
 \times \\ \times
 \trace\left[ {A^{\star}}_{r_{1}}^{[1]} \cdot \ldots \cdot {A^{\star}}_{r_{L}}^{[L]} \right]
 \langle r_1 \ldots r_L | \bigotimes_{\ell = \ell_1}^{\ell_2} \Theta^{[\ell]} | s_1 \ldots s_L \rangle.
\end{multline}
We can still rewrite this equation in a simpler Matrix Product form thanks to the formalism of transfer matrices,
defined identically as before
$ \Etra_{X}^{[\ell]} \equiv \sum_{s,r}^d \langle r | X | s \rangle
( A_{s}^{[\ell]} \otimes {A^{\star}}_{r}^{[\ell]} )$.
The difference is that this Matrix Product is also cyclic, i.e.
\begin{equation} \label{eq:TranstringPBC}
 \langle \Psi | O | \Psi \rangle =
 \trace \left[ \Etra_{\Theta}^{[\ell_1]} \cdot \ldots \cdot 
 \Etra_{\Theta}^{[\ell_2]} \cdot \Etra_{\Id}^{[\ell_2+1]} \cdot \ldots \cdot \Etra_{\Id}^{[\ell_1-1]} \right].
\end{equation}
Multiplying two transfer matrices costs $\sim D^6$, an expense that can be reduced to $2dD^5 + d^2D^4$
by calculating in the order $\mathbb{M}_s = (A_s \otimes \Id) \mathbb{E}$, then
$\mathbb{Q}_r = \sum_{s,r} \langle r | X | s \rangle \mathbb{M}_s$, and finally
$\mathbb{E}' = \sum_r (\Id \otimes A^{\star}_r) \mathbb{Q}_r$.
Unfortunately, this is the only improvement that can be made in general.

Equation \eqref{eq:TranstringPBC} has no right and left boundary vectors, which were instrumental to remove
a $D^2$ scaling power out of the cost. Moreover, the gauge group can be no longer exploited to eliminate
terms from the product of matrices; this can be argumented as follows.
We would like, for instance,
to transform the $\mathbb{E}^{[\ell]}_{\Id}$ into the identity so that it disappears from
\eqref{eq:TranstringPBC}.
However, the MPS gauge group transforms the transfer matrix $\mathbb{E}^{[\ell]}_{X}$ according to
\begin{equation}
 \mathbb{E}^{[\ell]}_{X} \longrightarrow \left( Y_{\ell-1}^{-1} \otimes Y_{\ell-1}^{\star\,-1} \right) \cdot
 \mathbb{E}^{[\ell]}_{X} \cdot \left( Y_{\ell} \otimes Y_{\ell}^{\star} \right).
\end{equation}
But the input matrix $\mathbb{E}^{[\ell]}_{X}$ could be entangling, and the transformation is local and invertible,
so there is no chance that a generic $\mathbb{E}^{[\ell]}_{X}$
can be mapped into a non-entangling operator (like $\Id$) this way.

In conclusion, if we want to acquire the \emph{exact} expectation value of a product observable on a
PBC MPS, the computational cost is
\begin{equation} \label{eq:PBCMPScost}
 \# \mbox{cost} \sim L \left(2 d D^5 + d^2 D^4 \right).
\end{equation}
Honestly, it is absolutely convincing that for large $L$ the system will be less sensitive
to finite size effects, thus manifesting an emergent physics which is very similar to OBC physics in the bulk.
We could exploit somehow this limit to reduce costs while acquiring controlled errors;
nevertheless, it is useful to understand how to work with MPS in the thermodynamical limit before
elaborating this idea.

\section{Thermodynamical limit MPS} \label{sec:MPSTD}

The chance of extending a Matrix Product State description so that it is actually representing an infinite system
$L \to \infty$, follows directly from the notion of MPS homogeneity we discussed in section \ref{sec:MPShomo}.

Let $A^{s}_{j,k}$ be the elementary tensor block of a PBC homogeneous Matrix Product State. For every system size
$L$, this $A$ defines a unique state in the $2^L$ dimensioned Hilbert space, thus forming a sequence of states
$|\Psi_L\rangle$.
The thermodynamical state is defined through physically relevant quantities, namely
expectation values $\langle O \rangle_{\infty}$ of compact support observables
which should coincide with the limit of $\langle O \rangle_{L}$ for $L \to \infty$.
\begin{equation}
  \begin{overpic}[width = 340pt, unit=1pt]{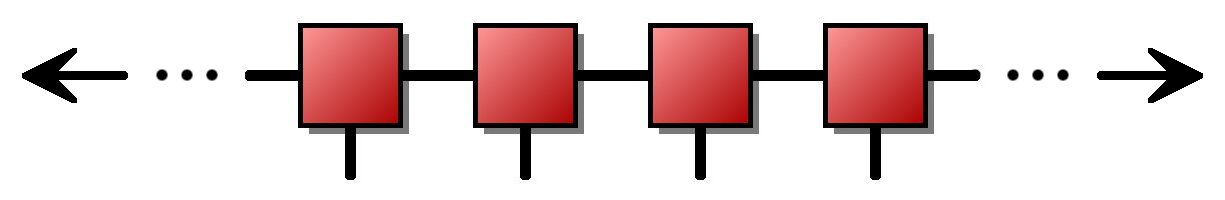}
  \put(13, 16){$-\infty$}
  \put(92, 31){$A$}
  \put(141, 31){$A$}
  \put(190, 31){$A$}
  \put(238, 31){$A$}
  \put(309, 16){$+\infty$}
 \end{overpic}
\end{equation}
Of course such expectation value limit must be well-defined in order for the thermodynamical state to be consistent;
thus we need to understand under which conditions upon $A$ this uniqueness is achieved.

Therefore, consider the expectation value $\Psi ( O )$
of a compact support observable $O$, which acts on $\ell$ adjacent sites
\begin{multline} \label{eq:TDlimlim}
 \Psi_{\infty} ( O ) = \lim_{L \to \infty} \Psi_{L} ( O )
 \equiv \lim_{L \to \infty} \frac{\langle \Psi_L | O | \Psi_L \rangle}{\langle \Psi_L | \Psi_L \rangle} = \\
 = \lim_{L \to \infty}
 \frac{\trace \left[ \Etra_{\Id} \ldots \Etra_{\Id} \,\widetilde{\Etra}_{O}\, \Etra_{\Id}
 \ldots \Etra_{\Id} \right]}{ \trace \left[ \Etra_{\Id} \ldots \Etra_{\Id}
 \vphantom{\widetilde{\Etra}_{O}} \right] } =
 \lim_{L \to \infty} \frac{\trace \left[ \Etra_{\Id}^{L-\ell} \;\widetilde{\Etra}_{O}\right]}{
 \trace \left[ \vphantom{\widetilde{\Etra}_{O}} \Etra_{\Id}^{L} \right]}
\end{multline}
where $\widetilde{\Etra}_{O}$ is the $\ell$-sites composite transfer matrix of the operator $O$:
\[
 \widetilde{\Etra}_{O} =
 \sum_{\mbox{{\scriptsize $\begin{array}{c}
        s_1 \ldots s_{\ell} = 1 \\
        r_1 \ldots r_{\ell} = 1
       \end{array} $}}}^{d}
 \!\!\!\!\!\! \langle r_1 \ldots r_{\ell} | O | s_1 \ldots s_{\ell} \rangle\;\,
 \trace\left[ \left( A_{s_1} \otimes A^{\star}_{r_1} \right) \cdot \ldots \cdot
 \left( A_{s_{\ell}} \otimes A^{\star}_{r_{\ell}} \right) \right].
\]
In eq.~\eqref{eq:TDlimlim} you see that,
since we are keeping $A$ fixed, for a generic $L$ the MPS state $| \Psi_L \rangle$ will not be normalized,
so we have to introduce manually the square norm in the expression
$\langle \Psi_L | \Psi_L \rangle = \trace [ \Etra_{\Id}^{L} ]$.

It is clear that the Thermodynamical limit state $\Psi_{\infty} (\cdot)$ must not depend on how we perform the
limit itself, nor which boundary conditions we used at finite sizes. So we must obtain the same result even starting
from an open boundary setting, as long as $A$ still describes the bulk, and distance between the support of $O$
and boundaries diverges. So we will introduce arbitrary correlation-space boundaries
$(b_{\text{left}} |$ and $|b_{\text{right}})$ by hand, and write
\begin{equation} \label{eq:TDlim}
 \Psi_{\infty} ( O ) = \lim_{L \to \infty}
 \frac{ (b_{\text{left}} | \;\Etra_{\Id}^{nL} \;\widetilde{\Etra}_{O}\; \Etra_{\Id}^{mL}\; |b_{\text{right}})}{
 (b_{\text{left}} | \;\Etra_{\Id}^{(n+m)L+\ell}\; |b_{\text{right}}) },
\end{equation}
we will require that this limit does not depend on $(b_{\text{left}} |$, $|b_{\text{right}})$, $n$ or $m$
($n$ and $m$ being any two positive integers); it also must coincide with the limit in equation \eqref{eq:TDlimlim}.

\vspace{.5em}
\emph{\textbf{Lemma -}} uniqueness of limit \eqref{eq:TDlimlim}, \eqref{eq:TDlim}
holds \emph{iff} some spectral requirements
upon the transfer matrix of the identity operator $\Etra_{\Id}$ are satisfied:
\begin{itemize}
 \item among eigenvalues $\lambda_{\alpha}$ of $\Etra_{\Id}$, there is one $\lambda_0$ strictly bigger than all others
 in modulus, i.e. $|\lambda_{\alpha \neq 0}| < |\lambda_{0}|$,
 \item $\lambda_0$ eigenvalue is simple, meaning that only one related eigenvector $|e_0^{\rightarrow})$ exists
 i.e. the eigenspace of $\lambda_0$ has dimension 1.
\end{itemize}
Let us prove this statement.
The transfer matrix $\Etra_{\Id}$ is not necessarily diagonalizable, but as it is on complex field,
we can expand it in its generalized eigenvector basis (which is not orthogonal in general),
so it appears in the Jordan block form
\begin{equation} \label{eq:etrajordan}
 \Etra_{\Id} = \left( \begin{array}{ccccc|cc}
  \lambda_0 & 1 &&&&\\
  & \lambda_0 & 1 &&&\\
  && \cdots & \cdots &&\\
  &&& \lambda_0 & 1 &\\
  &&&&\lambda_0\\ \hline
  &&&&& \lambda_1 & \cdots\\
  &&&&&& \cdots
 \end{array} \right)
\end{equation}
where we highlighted the generalized eigenspace of $\lambda_0$. The conditions we required upon $\Etra_{\Id}$ tells
us that $|\lambda_{\alpha \neq 0}| < |\lambda_0|$,  and that there is only one Jordan block
corresponding to $\lambda_0$, as it appears in \eqref{eq:etrajordan}.
Then, given a random vector $|v)$, it is possible to demonstrate that
\begin{equation} \label{eq:totheigen}
 \lim_{q \to \infty} \frac{\Etra^{q}_{\Id} |v)}{
 \sqrt{(v| {\Etra^{\dagger}_{\Id}}^q \Etra^{q}_{\Id} |v) }} = |e^{\rightarrow}_0),
\end{equation}
with $|e^{\rightarrow}_0)$ normalized $(e^{\rightarrow}_0|e^{\rightarrow}_0) = 1$,
provided that $|v)$ has no null component over the generalized eigenspace of $\lambda_0$.
To show \eqref{eq:totheigen} first expand $|v)$ in the generalized eigenbasis $|e_{\alpha, \partial}^{w})$,
where $\alpha$ is the eigenvalue index, $\partial$ refers to the Jordan block where the basis element belongs,
and $w$ its position within the block. Then the composite application of several $\Etra^{\dagger}_{\Id}$
gives
\begin{equation} \label{eq:Polyexpand}
 \Etra^{q}_{\Id} |v) = \sum_{\alpha} \lambda_{\alpha}^{q}\;
 \sum_{\partial, w} \mathcal{P}^{[\Delta_{\partial}-w]}_{\lambda_{\alpha}}(q)
 \; |e_{\alpha, \partial}^{w}),
\end{equation}
where $\mathcal{P}^{[x]}_{\lambda}(q)$ is a polynomial function of $q$, of degree $x$,
and whose coefficients depending on $\lambda$; $\Delta_{\partial}$ is the size of Jordan block $\partial$.
When we take the limit \eqref{eq:totheigen} of the latter expression, all the components belonging
to gen. eigenspaces different from the $\lambda_0$ one vanish, as $(\lambda_{\alpha}/|\lambda_0|)^q \to 0$ for
any $\alpha \neq 0$.
Then, also components over generalized eigenvectors of $\lambda_0$ which are not the true unique eigenvector $|e^{\rightarrow}_0)$
disappear, since their polynomial multiplier is of lower degree, i.e.
$\mathcal{P}^{[\Delta_{0}-w]}_{\lambda_{0}}(q) / |\mathcal{P}^{[\Delta_{0}]}_{\lambda_{0}}(q) |\to 0$
for any $w \neq 0$.

Therefore the composite action of $\Etra^{q}_{\Id}$ on a generic vector $|v)$ maps it
(after normalization) to $|e_0^{\rightarrow})$. With a similar argument, it can be shown
that $(v| \Etra^{q}_{\Id} \to (e_0^{\leftarrow}|$, where $(e_0^{\leftarrow}|$ is the only 'left-eigenvector'
(eigenfunctional) of $\Etra^{q}_{\Id}$. Notice that by construction $(e_0^{\leftarrow}|$ is not necessarily the
dual of $|e_0^{\rightarrow})$ via Riesz representation theorem, but the two vectors are not orthogonal either,
so that $(e_0^{\leftarrow}|e_0^{\rightarrow}) \neq 0$. After all these considerations, we can
extract the desired conclusion, i.e.
\begin{equation} \label{eq:TDexpect}
 \begin{aligned}
 \Psi_{\infty} ( O ) &= \lim_{L \to \infty} \frac{\trace \left[ \Etra_{\Id}^{L-\ell} \;\widetilde{\Etra}_{O}\right]}{
 \trace \left[ \vphantom{\widetilde{\Etra}_{O}} \Etra_{\Id}^{L} \right]}
 = \lim_{L \to \infty} \frac{\sum_v \trace \left[ |v)(v| \Etra_{\Id}^{L} \;\widetilde{\Etra}_{O} \Etra_{\Id}^{L} \right]}{
 \sum_u \trace \left[ |u)(u| \vphantom{\widetilde{\Etra}_{O}} \Etra_{\Id}^{2L+\ell} \right]} \\
 &= \lim_{L \to \infty}
 \frac{ (b_{\text{left}} | \;\Etra_{\Id}^{nL} \;\widetilde{\Etra}_{O}\; \Etra_{\Id}^{mL}\; |b_{\text{right}})}{
 (b_{\text{left}} | \;\Etra_{\Id}^{(n+m)L+\ell}\; |b_{\text{right}}) }\\
 &= 
 \frac{(e_0^{\leftarrow}| \widetilde{\Etra}_{O} |e_0^{\rightarrow})}{(e_0^{\leftarrow}|\Etra^{\ell}_{\Id}|e_0^{\rightarrow})}
 = \lambda_{0}^{-\ell}\;\frac{(e_0^{\leftarrow}| \widetilde{\Etra}_{O} |e_0^{\rightarrow})}{(e_0^{\leftarrow}|e_0^{\rightarrow})},
 \end{aligned}
\end{equation}
where the limit correlation boundaries are defined by
$\lambda_0 |e_0^{\rightarrow}) = \Etra_{\Id} |e_0^{\rightarrow})$, and
$\lambda_0 (e_0^{\leftarrow}| = (e_0^{\leftarrow}| \Etra_{\Id}$. As you see, the result is consistent
regardless if we are following an OBC or PBC scheme.
The reverse implication in the Lemma is trivial by counterexample.
%Conversely, if more than one eigenvector of
%$\Etra_{\Id}$ sharing the same $|\lambda|$ existed, eq.~\eqref{eq:Polyexpand} could give any superposition
%of such vectors (depending on the starting vector) for $q \to \infty$, and the limit would then be ill-defined, thus
%the requirements in the Lemma are also a necessary condition.

The capability of expressing the expectation value for every observable as in \eqref{eq:TDexpect},
can as well be formulated in terms of density matrices $\rho_{\ell}$.
That is, a thermodynamical limit
quantum state can be properly defined by the sequence of reduced density matrices $\{\rho_{\ell}\}_{\ell}$ for any
finite size $\ell$,
having the property that when tracing partially larger-sized ones, we recover smaller-sized ones:
\begin{equation} \label{eq:partrace}
 \{\rho_{\ell}\}_{\ell} : \longrightarrow \quad
 \trace_{1..\ell_0}\left[ \vphantom{X^1} \rho_{\ell} \right] =
 \trace_{\ell-\ell_0 .. \ell} \left[ \vphantom{X^1} \rho_{\ell} \right] = \rho_{\ell - \ell_0}
 \quad \forall\, \ell_0 < \ell.
\end{equation}
Here we are implicitly considering translational invariance as well, which is automatically
granted by homogeneity of the $A$ in our case.
Then, following the prescriptions of the lemma and what we learned from \eqref{eq:TDexpect}, we can
write down the expression for the reduced density matrix $\rho_{\ell}$ of an arbitrary number of sites as
\begin{equation} \label{eq:TDrho}
 \rho_{\ell} = 
 \!\!\!\!\! \sum_{\mbox{{\scriptsize $\begin{array}{c}
        s_1 \ldots s_{\ell} = 1 \\
        r_1 \ldots r_{\ell} = 1
       \end{array} $}}}^{d}
 \!\!\!\!\!\! \frac{ (e_0^{\leftarrow}| \left( A_{s_1} \otimes A^{\star}_{r_1} \right)
 \ldots \left( A_{s_{\ell}} \otimes A^{\star}_{r_{\ell}} \right)
 | e_0^{\rightarrow}) }{\lambda_0^{\ell} (e_0^{\leftarrow}|e_0^{\rightarrow})}
 \;|s_1 \ldots s_{\ell}\rangle \langle r_1 \ldots r_{\ell}|,
\end{equation}
where the partial trace property \eqref{eq:partrace} is an automatic consequence of the fact that
$\sum_{r,s} \delta_{r,s} (A_{s} \otimes A^{\star}_{r}) |e_0^{\rightarrow}) = \lambda_0 |e_0^{\rightarrow})$.

An interesting and physically relevant
comparison with Matrix Product Density Operators, we introduced in section \ref{sec:MPO}, can be made
once we gave the pictorial representation of \eqref{eq:TDrho}:
\begin{equation} \label{eq:TDrhofig}
  \begin{overpic}[width = \textwidth, unit=1pt]{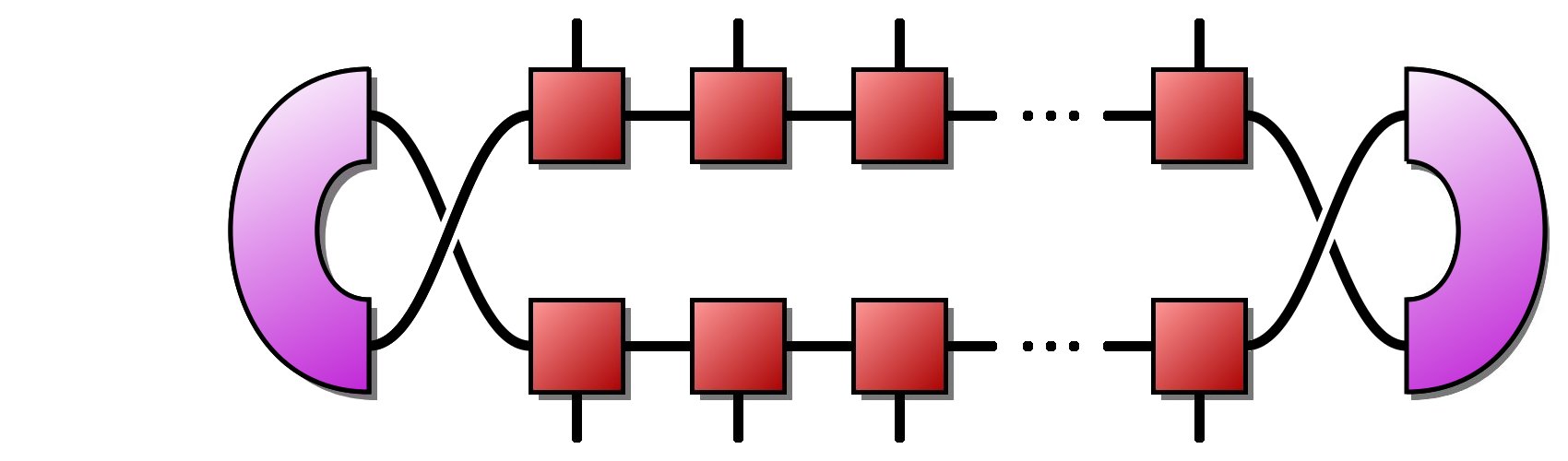}
  \put(10, 54){$\rho_\ell = \xi \times$}
  \put(62, 54){$e_0^{\leftarrow}$}
  \put(366, 54){$e_0^{\rightarrow}$}
  \put(139, 24){$A$}
  \put(138, 83){$A^{\star}$}
  \put(179, 24){$A$}
  \put(178, 83){$A^{\star}$}
  \put(219, 24){$A$}
  \put(218, 83){$A^{\star}$}
  \put(294, 24){$A$}
  \put(293, 83){$A^{\star}$}
 \end{overpic}
\end{equation}
where $\xi = [\lambda_0^{\ell} (e_0^{\leftarrow}|e_0^{\rightarrow}) ]^{-1}$ is put for correct state normalization,
and correlation boundary vectors are the solutions of the eigenproblem
\begin{equation}
  \begin{overpic}[width = \textwidth, unit=1pt]{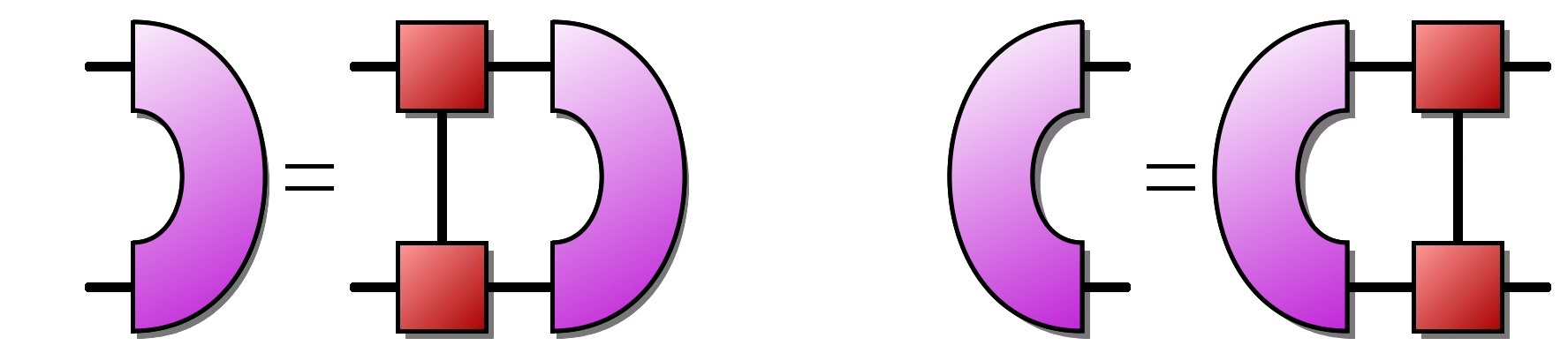}
  \put(10, 41){$\lambda_0 \times$}
  \put(48, 41){$e_0^{\rightarrow}$}
  \put(153, 41){$e_0^{\rightarrow}$}
  \put(105, 68){$A$}
  \put(104, 12){$A^{\star}$}
  \put(211, 41){$\lambda_0 \times$}
  \put(241, 41){$e_0^{\leftarrow}$}
  \put(307, 41){$e_0^{\leftarrow}$}
  \put(357, 12){$A^{\star}$}
  \put(358, 68){$A$}
 \end{overpic}
\end{equation}
for maximal modulus eigenvalue $|\lambda_0|$. The resemblance of \eqref{eq:TDrhofig} with \eqref{eq:MPDOfig} is
evident, in particular our MP-thermodynamical state is written in a peculiar MPDO form where
the mixing dimension $\tilde{d}_{\ell} = 1$ (meaning no external degrees of freedom that couple locally)
except for the boundaries where $\tilde{d}_{\ell} = D$.
Moreover, the two designs match exactly \emph{iff} $e_0^{\rightarrow}$, read as a matrix
(being a two-indices tensor, consider one of the two as incoming matrix index, and the other one outcoming),
is \emph{positive}. In this case we can write $e_0^{\rightarrow} = X X^{\dagger}$, and embed the $X$ within
the MP-block $A_{s}^{[\ell]} \to B_{s}^{[\ell]} = A_{s} X$ as it was a gauge transformation, and
we find precisely the formalism \eqref{eq:MPDOfig}.

Truly, we know a peculiar case when both matrices $e_0^{\rightarrow}$ and $e_0^{\leftarrow}$ are necessarily
positive: that is when $A$ is either in the left or the right gauge. For instance, let us assume that $A$ is in the
left gauge, then the uniqueness condition translates into the requirement that the CPT map \eqref{eq:mappuzL}
is mixing (see appendix \ref{app:mixin}).
If this requirement is satisfied, then automatically: $\lambda_0 = 1$, $e_0^{\leftarrow} = \Id_{D \times D}$
and $e_0^{\rightarrow} = \Lambda$, but this implies that $(e_0^{\leftarrow}|e_0^{\rightarrow}) =
(\Phi^{+}|\Lambda) = \trace[\Lambda] = 1$, telling us that $\xi = 1$ $\forall \ell$.
So the denominator in equation \eqref{eq:TDrho} vanishes, simplifying to
\begin{equation} \label{eq:TDrhogau}
 \rho_{\ell} = 
 \!\!\!\!\! \sum_{\mbox{{\scriptsize $\begin{array}{c}
        s_1 \ldots s_{\ell} = 1 \\
        r_1 \ldots r_{\ell} = 1
       \end{array} $}}}^{d}
 \!\!\!\!\!\! ( \Phi^{+}| \left( A_{s_1} \otimes A^{\star}_{r_1} \right)
 \ldots \left( A_{s_{\ell}} \otimes A^{\star}_{r_{\ell}} \right)
 | \Lambda)
 \;|s_1 \ldots s_{\ell}\rangle \langle r_1 \ldots r_{\ell}|.
\end{equation}
Similarly, the argument can be applied to a right-gauged $A$, leading to
a version of \eqref{eq:TDrhogau} where correlation boundaries are exchanged.

\subsection{Size matters?}
Being able to address infinite states in a formally exact fashion,
while employing a finite amount of resources, sounds really nice and useful for numerical issues.
Honestly, in many simulation algorithms and architectures, the computational times often scale non-trivially
with system size $L$, which makes approaches to the thermodynamical limit clumsy trials:
expensive and imprecise. With Matrix Product States (as well as with other classes of self-similar tensor
network structures, like CPS, TTN and MERA, we are going to describe in the next chapters)
the infinite problem is perfectly addressed with finite-effort numerics: the amount of calculus
scales solely with $D$, the correlation bondlink dimension.

A good question, now, is whether a Matrix Product Simulation is actually capable of describing,
with a manageable $D$, physical states with a good precision, even at the thermodynamical limit.
We understood that a $D$-bondlink MPS can represent
exactly the whole class of finitely correlated states, with entanglement bound by $\log D$;
this extends to thermodynamical limit as well.
Now consider a ground state of a 1D non-critical system: its partition entropy ought
to satisfy the area-law of entanglement, stating that as $L$ grows, $\mathcal{S}_{\text{VN}}$ scales
like $L^{0}$, going towards a finite value $\varepsilon$ in the TD-limit. Therefore, even
the appropriate bondlink dimension $D = 2^{\varepsilon}$ stays
finite, meaning that a thermodynamical MPS description can be definitely made.
More precisely, in literature there are several classes of quantum states which are physically relevant,
and an exact (typically optimal) MPS representation has been found (see e.g. the review \cite{MPSreview}, where
MPS for AKLT, Majumdar-Gosh, GHZ, W, and Cluster states are presented).

On the contrary, this argument suggests us to think that representing critical ground states,
whose $\mathcal{S}_{\text{VN}} \sim (c \log L) / 6$ require infinite bondlink $D$ for a successful
description of the thermodynamical limit. This looks to be true, and despite this would be
troublesome for numerical implementation, the issue of dealing with infinite-bondlink Matrix
Product States has been studied from analytical perspectives, like in refs.~\cite{InfyMPS1, InfyMPS2}
where an equivalence with conformal field theory (CFT) has been established.

To provide more clearance on the relationship between Matrix Product representations
and criticality in 1D, we are going to discuss about correlations in MPS.

\section{Matrix Product States and correlations} \label{sec:MPScorr}

One dimensional quantum systems are quite peculiar: they can manifest no quantum phase transition at finite temperature,
nor they can exhibit long-range order parameters. Yet, the relationship between criticality and non-criticality
of a system is a matter of utmost importance. As quantum entanglement is not a physical observable,
the basic way to recognize and identify the presence of a quantum phase
transition is through scaling laws for two-point correlations.

Let us go back to the thermodynamical limit (TD) MPS, defined homogeneously by the matrices $A_s$,
and define the correlation function
of two (separate) local observables $\Theta$ and $\Theta'$, acting at arbitrary distance $\ell+1$:
\begin{equation}
 \mathfrak{C}_{\ell+1}(\Theta, \Theta') \equiv \langle \Theta^{[\ell_0]} \otimes \Theta'^{[\ell_0 + \ell + 1]} \rangle
 - \langle \Theta^{[\ell_0]} \rangle \langle \Theta'^{[\ell_0 + \ell + 1]} \rangle,
\end{equation}
since the system is translationally invariant, $\ell_0$ is irrelevant.
To workaround normalization issues
we will just adopt an $A$ in the left gauge, as we did for \eqref{eq:TDrhogau}, then we end up with
\begin{equation} \label{eq:325}
 \begin{aligned}
  \mathfrak{C}_{\ell+1}(\Theta, \Theta') &= (\Phi^{+}| \Etra_{\Theta} \;\Etra_{\Id}^{\ell}\; \Etra_{\Theta'} |\Lambda)
  - (\Phi^{+}| \Etra_{\Theta} |\Lambda)\; (\Phi^{+}| \Etra_{\Theta'} |\Lambda)\\
  &= (\Phi^{+}| \Etra_{\Theta} \cdot \left[ \Etra_{\Id}^{\ell} - |\Lambda) (\Phi^{+}| \right] \cdot \Etra_{\Theta'} |\Lambda)\\
  &= (v^{\text{left}}_{\Theta} | \left[ \Etra_{\Id}^{\ell} - |\Lambda) (\Phi^{+}| \right] | v^{\text{right}}_{\Theta'} ).
 \end{aligned}
\end{equation}
Let us now study the expression in \eqref{eq:325} from an algebraic viewpoint.
As we previously stated, both left and right (double) correlation boundary vectors,
$|\bullet)$ and $(\bullet|$, can be seen as matrices, whose vector version is the so-called
Liouville representation:
\begin{equation}
 O = \sum_{i,j} \alpha_{i,j} |i\rangle \langle j| \quad \longrightarrow \quad |O) = \sum_{i,j} \alpha_{i,j} |ij),
\end{equation}
where, in particular $\Id \to |\Phi^{+})$. Then $\Etra_{\Id}$ is a Complete Positivity, Trace preserving map
when applies to the matricial element on its right, as we know from \eqref{eq:mappuzL}; similarly, it is
a Complete Positive and Unital (i.e. it maps the identity into itself) map when applying to the left.

We begin requiring that the CPT map has \emph{mixing} property,
meaning that has a single attraction point, that would be $\Lambda$.
CPT maps are always contractive, as proven in appendix \ref{app:CPTspec}, but the
relaxation requirement tells us also that $\Etra^{\infty} = |\Lambda)(\Phi^{+}|$ so that we can write
\begin{equation} \label{eq:linfiz}
  \mathfrak{C}_{\ell+1}(\Theta, \Theta')
  = (v^{\text{left}}_{\Theta} | \left[ \Etra_{\Id}^{\ell} - \Etra_{\Id}^{\infty} \right] | v^{\text{right}}_{\Theta'} ),
\end{equation}
which has a clear physical meaning: the uncorrelated product of expectation values is, as should be, equivalent
to the operator product at infinite distance.
We can now exploit the expansion \eqref{eq:APolyexpand} for multiple application of a CPT mixing map
and write the correlator as:
\begin{equation} \label{eq:linfiz2}
   \mathfrak{C}_{\ell+1}(\Theta, \Theta') =
   \sum_{\alpha = 2} \lambda_{\alpha}^{\ell} \;\mathcal{P}^{[\Delta_{\alpha}]}_{\lambda_{\alpha}}(\ell)
\end{equation}
where $\lambda_{\alpha \geq 2}$ are the eigenvalues of $\Etra_{\Id}$ other than 1, and for
which it holds $|\lambda_{\alpha \geq 2}| < 1$, while
$\mathcal{P}^{[x]}_{\xi}(\ell)$ are polynomial functions of degree $x$ with coefficients
depending on $\xi$, and $\Delta_{\alpha}$ is the size of the largest Jordan block belonging to
the generalized eigenspace of $\alpha$. Two features of \eqref{eq:linfiz2} are worthy of remark:
\begin{itemize}
 \item Since every $\lambda_{\alpha}$ has modulus strictly smaller than 1, $\mathfrak{C}_{\ell}$ goes
 necessarily to $0$ at $\ell \to \infty$, regardless of $\Theta$ and $\Theta'$, which is telling
 us that the \emph{state manifests no long range order parameter}. Notice that this property depends
 strictly on the mixing condition employed for $\Etra_{\Id}$.

 \item \emph{Correlations decay exponentially}. In particular it is possible to dominate $\mathfrak{C}_{\ell}$
 with decreasing exponentials:
 precisely, let us order the eigenvalues $\{\lambda_{\alpha}\}_{\alpha}$ so that they decrease in
 modulus. Then
 \begin{equation} \label{eq:bibol}
  \lim_{\ell \to \infty} \frac{|\mathfrak{C}_{\ell+1}(\Theta, \Theta')|}{(|\lambda_2| + \varepsilon)^{\ell}} = 0,
 \end{equation}
 for any $\varepsilon \geq 0$; but if $\varepsilon$ is chosen small enough (i.e. $\varepsilon < 1 - |\lambda_2|$)
 the denominator in \eqref{eq:bibol} decays exponentially, and thus the numerator decays faster.
\end{itemize}

It is still possible for two points correlators of MPS, as \eqref{eq:linfiz2}, to \emph{resemble}
power-law decay rates, by playing with several exponentials with $|\lambda_{\alpha}|$ very close to 1,
but only for short ranges, and many eigenvalues are required so $D$ must be chosen appropriately.
When the TD-limit MPS state is investigated at long ranges, its ultimate non-critical nature becomes
clear, and dominant: ruled by the eigenvalues $\lambda_{\alpha}$ of the Identity transfer matrix $\Etra_{\Id}$.

\section[Faster expectation values]{Faster expectation values for PBC-MPS at large sizes} \label{sec:PBCimprove}

When using Matrix Product States as a variational tailored wavefunction ansatz for classical simulations
of quantum systems, is fundamental that we make economy on every computational step of the algorithm.
Calculating expectation values of observables, and in particular Hamiltonians, is one of the
numerical ingredients which require most computational effort so it is important to optimize its scheme beforehand.

In section \eqref{sec:PBCObs} we acknowledged that acquiring expectation values in periodic Matrix
Product States is quite more expensive than in the open boundary case, carrying an overall $D^2$ multiplier
to the cost (from $\propto dLD^3$ to $\propto dLD^5$). In a paradoxical way, as we approach the
thermodynamical limit $L \to \infty$ the cost drops again to $\propto dLD^3$
(plus solving a fixed point equation, usually subleading), since boundaries
of the TD state are not correlated through external channels.
Therefore it is natural to think that if the size $L$ of the periodic system we are considering
is sensibly large, the outcoming state shall be close enough to the TD-limit, and thus
the system will feel little of the periodicity, identified by small amplitudes for finite-size effects.
A way to exploit this fact to improve the evaluation algorithm was proposed in ref.~\cite{Pippan},
we now sketch the same idea with a slightly different formulation.

Assume we want to calculate the composite Transfer Matrix
\begin{equation}
 \widetilde{\Etra}^{\ell} = \Etra_{\Theta_1}^{[1]} \cdot \Etra_{\Theta_2}^{[2]}
 \cdot \ldots \cdot \Etra_{\Theta_\ell}^{[\ell]},
\end{equation}
where we chose a tensor product observable $\bigotimes_{\ell'}^{\ell} \Theta^{[\ell']}_{\ell'}$ for simplicity,
but the following arguments apply to an entangling operator as well.
We will also state that the whole MPS segment $1 \ldots \ell$, which $\widetilde{\Etra}^{\ell}$ is calculated through,
is in the left (or right) gauge, but not necessarily homogeneous. In particular if sites $1 \ldots \ell$ are
not the whole system, $\ell < L$,
it is always possible to take the singular part to its complementary, and satisfy the gauge condition.
We want to achieve $\widetilde{\Etra}^{\ell}$ but without performing singularly any product
$\Etra \cdot \Etra$ which costs $\sim D^5$. To this purpose, let us consider the singular value decomposition of
$\widetilde{\Etra}^{\ell} = $, i.e.
\begin{equation}
 \widetilde{\Etra}^{\ell}_{\alpha \beta} = \sum_{\gamma} U_{\alpha \gamma}\,
 \sigma_{\gamma}\, V_{\beta \gamma} = U \cdot \Delta^{[\sigma]} \cdot V^{\dagger},
\end{equation}
in a formal sense, where $U^{\dagger} U = V^{\dagger} V = \Id$, and
$\Delta^{[\sigma]}_{\alpha \beta} = \delta_{\alpha \beta}\, \sigma_{\alpha}$
is diagonal and positive ($\sigma_{\alpha} \geq 0$). Now, \emph{if} the $\Etra_{\Theta_{\ell'}}^{[\ell']}$ were
all positive and homogeneous, the singular values $\sigma$ would coincide with the eigenvalues and $U = V$,
which leads to $\sigma_{\alpha} = \lambda_{\alpha}^{\ell}$, where
$\lambda_{\alpha}^{\ell}$ are the eigenvalues of $\Etra$.
But this tells us that the ratio between two singular values
$\sigma_{\alpha} / \sigma_{\beta} = (\lambda_{\alpha} / \lambda_{\beta})^{\ell}$ decays as much fast
as the segment $\ell$ is long. Telling us that for $\ell$ long enough very few singular values $\sigma$
are relevant before reaching the numerical precision of the calculator.
This argument holds for positive $\Etra$, but extends naturally to hermitian matrices,
and by linearity and continuity it reasonably works for every matrix, for some $\ell$ large enough,
and holds even for dishomogeneous matrix products (proven numerically in \cite{Pippan}).

We could equivalently state that the \emph{range} of $\widetilde{\Etra}^{\ell}$ has actual dimension $p$ smaller
than $D^2$, where $p$ is the number of singular values $\sigma$ that are being kept, while the other are discarded as they
are of equal or smaller order of magnitude than computational precision.
During simulations, $p$ becomes a parameter, and can be kept smaller and smaller as the size $\ell \sim L$ increases.
Then the stochastic procedure for calculating $\widetilde{\Etra}^{\ell}$ goes like this:
\begin{enumerate}
 \item Choose a random matrix $X$, of dimension $D^2 \times p$. If the random number generator is satisfactory,
 there will be zero probability that one column vector will be linearly dependent on the other $p-1$ ones,
 due to the fact that those $p-1$ may generate a set of zero probability measure (absolutely continuous with Lebesgue measure).
 \item Apply and calculate $\Etra_{\Theta_1}^{[1]} \cdot \ldots \Etra_{\Theta_\ell}^{[\ell]} \cdot X
 = \widetilde{\Etra}^{\ell} \cdot X = Y $, obviously starting from the right. Every step costs $2dD^3p$ operations.
 $Y$ is again a $D^2 \times p$ matrix, but the vector columns will span only the range of $\widetilde{\Etra}^{\ell}$
 which we will now assume is exactly $p$-dimensional. As before, chances are that all column vectors of $Y$ will
 be linearly independent, so they will span the whole range $\widetilde{\Etra}^{\ell}$, because dimensions match.
 \item Orthonormalize the columns of $Y$, either via a Graham-Schmidt or a QR decomposition.
 We obtain $Z = Y \cdot T$, with $T$ typically triangular, containing the whole singular part of $Y$.
 The matrix $Z$ is still $D^2 \times p$ dimensioned, and isometric: $Z^{\dagger} Z = \Id$.
 Moreover, since the columns of $Z$ span $\text{Rng}(\widetilde{\Etra}^{\ell})$, we have
 $Z \, Z^{\dagger} = P_{\text{Rng}(\widetilde{\Etra}^{\ell})}$ the projector over the range. But then
 \begin{equation}
  \widetilde{\Etra}^{\ell} = P_{\text{Rng}(\widetilde{\Etra}^{\ell})} \cdot \widetilde{\Etra}^{\ell}
  = Z \cdot Z^{\dagger} \cdot \widetilde{\Etra}^{\ell}
 \end{equation}
 \item Apply and calculate $Z^{\dagger} \cdot \Etra_{\Theta_1}^{[1]} \cdot \ldots \Etra_{\Theta_\ell}^{[\ell]}
 = Z^{\dagger} \cdot \widetilde{\Etra}^{\ell} = W$, from left to right; the cost is $\sim 2dD^3p$ per step.
 We are done now, since
 \begin{equation}
  \widetilde{\Etra}^{\ell} = Z \cdot W
 \end{equation}
 and we have both $Z$ and $W$ matrices. We do not even have to multiply them together, and instead keep them separated:
 whenever we will have to use $\widetilde{\Etra}^{\ell}$ as a part of a whole MPS-network contraction,
 contracting over the $p$-dimensioned index space in the middle will be the last operation to be performed.
\end{enumerate}
In conclusion, we can resume these simple steps as follows:
\begin{equation} \label{eq:Pippantech}
  \begin{overpic}[width = \textwidth, unit=1pt]{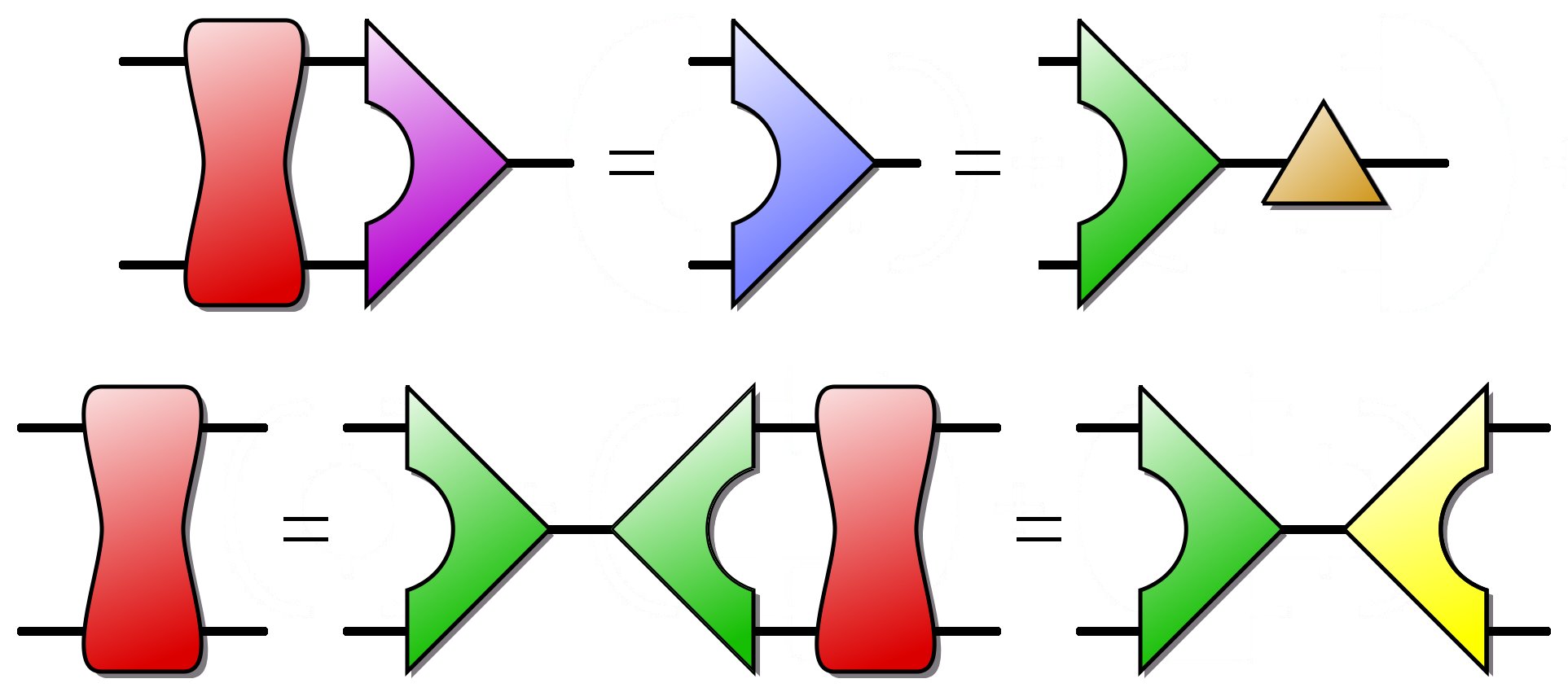}
  \put(56, 129){$\widetilde{\Etra}^{\ell}$}
  \put(107, 129){$X$}
  \put(199, 129){$Y$}
  \put(284, 129){$Z$}
  \put(326, 127){$T$}
  \put(31, 36){$\widetilde{\Etra}^{\ell}$}
  \put(213, 36){$\widetilde{\Etra}^{\ell}$}
  \put(161, 36){$Z^{\star}$}
  \put(117, 36){$Z$}
  \put(300, 36){$Z$}
  \put(345, 36){$W$}
 \end{overpic}
\end{equation}
\[
  \mbox{where $Z$ is isometric, i.e.} \qquad
  \begin{overpic}[width = 180pt, unit=1pt]{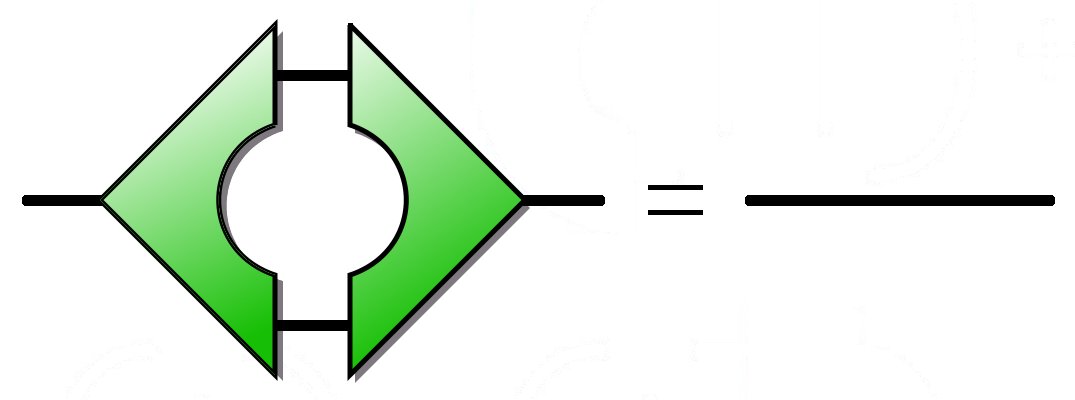}
  \put(24, 31){\footnotesize $Z^{\star}$}
  \put(71, 31){\footnotesize $Z$}
 \end{overpic}
\]
If we wish to check explicitly the behavior of singular values of $\widetilde{\Etra}^{\ell}$ we just have
to perform a SVD on $W$, because
\begin{equation} \label{eq:trentini}
 \widetilde{\Etra}^{\ell} = Z \cdot W = Z \cdot (\tilde{U} \cdot \Delta^{[\sigma]}_{W} \cdot V^{\dagger} ) =
 (Z \cdot \tilde{U}) \cdot \Delta^{[\sigma]}_{W} \cdot V^{\dagger};
\end{equation}
but $(Z \tilde{U}) (Z \tilde{U})^{\dagger} = \Id$, meaning that \eqref{eq:trentini} is a singular value decomposition
for $\widetilde{\Etra}^{\ell}$ and singular values of a matrix are uniquely defined; so 
$\Delta^{[\sigma]}_{W} = \Delta^{[\sigma]}$ and we can see how fast they decay for large $\ell$.
By adopting this process, we spent a total amount of elementary computational operations equal to
\begin{equation} \label{eq:Pipancost}
 \# \mbox{cost} \sim 2Lp \left(2 d D^3 + d^2 D^2 \right),
\end{equation}
a nice improvement, even because at reasonable lengths ($L \sim 50$), $p$ can be usually chosen
orders of magnitude smaller than $D^2$ with practically no loss in simulation precision
(see figure \ref{fig:Rossin02}).

\section[PBC-MPS Minimization]{Minimization algorithms\\with periodic MPS}

For open boundary conditions MPS we presented a fast-converging and numerically manageable algorithm to
find the ground state of a generic (short-range interacting) Hamiltonian. In section \ref{sec:minimizOBC}
we discussed that the essence of such algorithm is minimizing one MPS block at a time, keeping the
other fixed; the basic step is composed by a partial contraction of the MPS
network with the Hamiltonian operator (effective Hamiltonian), followed by a $dD^2$-dimensioned eigenvalue problem
($d^2D^2$ in the two-blocks simultaneous minimization case).

In the periodic boundary case we have more than one naturally available path. If the Hamiltonian
is translationally invariant, then a good guess would be using the set of homogeneous MPS as
variational wavefunctions, because a translational ground state must exist. This idea would lead to
a all-at-once minimization of the MPS state, but unfortunately the Lagrangian would not be quadratic
in the MPS homogeneous block and the problem to solve would be way harder than an eigenvalue problem.
Moreover, forcing the variational state to be translational would let us not identify easily
Hamiltonians bearing a spontaneous translational symmetry breaking.

For these reasons, in this section we will instead describe an algorithm that not assumes translationality
in the variational state (and thus using dishomogeneous MPS) and minimizes blocks one at a time to
preserve a quadratic structure for the Lagrangian \cite{Rossinisupersolid, Rossinistiff}.
Then let us start again from a nearest neighbor Hamiltonian
$H = \sum_{\ell = 1}^{L} R_2^{[\ell]}$ where for comfort we regrouped in $R_2^{[\ell]}$ both one-body and two-body terms, i.e.
\begin{equation} \label{eq:quacchio}
 R_2^{[\ell]}  = \sum_{q} g^{\ell}_{q} \;\Id^{[\ell-1]}\ \otimes \Theta^{[\ell]}_{q} +
 \sum_{r} h^{\ell}_{r} \;{\Theta'}^{[\ell-1]}_{r} \otimes {\Theta''}^{[\ell]}_{r}.
\end{equation}
Let us assume that we are going to minimize the MPS tensor block $A^{[\ell]}$ associated to site $\ell$.
First, we split the Hamiltonian as $H = R_2^{[\ell]} + R_2^{[\ell+1]} + \bar{R}_2$, where $\bar{R}_2$ contains
all the terms of \eqref{eq:quacchio} that have support in the complementary of site $\ell$.
Then we calculate two composite of transfer matrices, which shall be the ingredients of our Lagrangian functional, namely
\begin{itemize}
 \item The transfer matrix $\widetilde{\Etra}_1$ from site $\ell+2$ to $\ell-2$ of the Identity operator, i.e.
 $\widetilde{\Etra}_1 = \Etra_{\Id}^{[\ell+2]} \cdot \ldots \cdot \Etra_{\Id}^{[\ell-2]}$
 \item The transfer matrix $\widetilde{\Etra}_2$ of $\bar{R}_2$ from site $\ell+1$ to $\ell-1$.
 Even though $\bar{R}_2$ is not a separable operator it is possible, with some engineering,
 to calculate $\widetilde{\Etra}_2$ spending quite the same computational cost \eqref{eq:Pipancost},
 apart a non-scaling prefactor.
\end{itemize}
Thanks to the technique \eqref{eq:Pippantech}, acquiring these transfer matrices is efficient; actually we
prefer to store in memory $Z_1$, $W_1$ and $Z_2$, $W_2$ (where $\widetilde{\Etra}_{\alpha} = Z_{\alpha} W_{\alpha}$),
since $p \ll D^2$ so keeping $4\,(p \times D^2)$ matrix elements is less expensive than $2\,(D^2 \times D^2)$.

Then the Lagrangian for $A^{[\ell]}$ reads:
$\mathcal{L}(A^{[\ell]}, {A^{\star}}^{[\ell]}) = \langle \Psi | H |\Psi\rangle - \varepsilon \langle \Psi | \Psi\rangle
= \langle \!\langle A^{[\ell]} | \mathcal{H} - \varepsilon\,
 \mathcal{N} | A^{[\ell]} \rangle \! \rangle = $ 
\begin{equation}
\begin{overpic}[width = 308pt, unit=1pt]{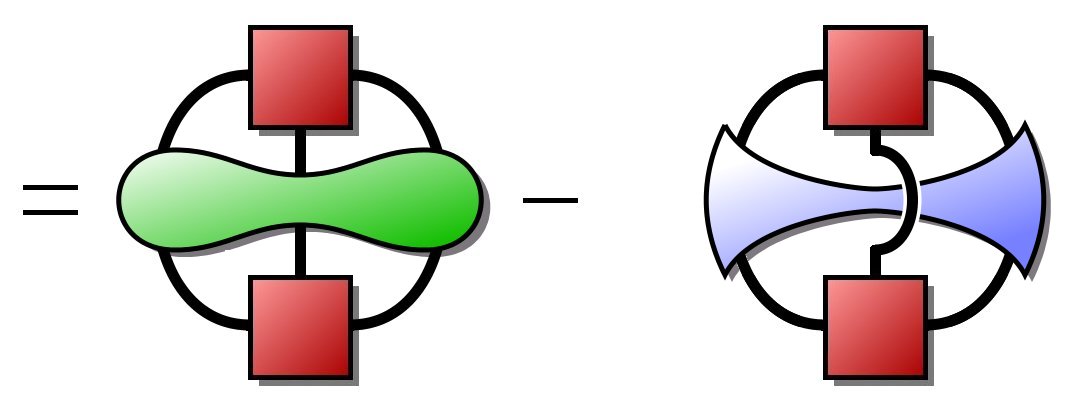}
 \put(-20, 51){\Huge $\mathcal{L}$}
 \put(181, 51){\Huge $\varepsilon$}
 \put(80, 53){$\mathcal{H}$}
 \put(215, 53){$\mathcal{N}_0$}
 \put(78, 89){$A^{[\ell]}$}
 \put(76, 17){${A^{\star}}^{[\ell]}$}
 \put(242, 89){$A^{[\ell]}$}
 \put(241, 17){${A^{\star}}^{[\ell]}$}
\end{overpic}
\end{equation}
$\mathcal{N} = \mathcal{N}_0 \otimes \Id_{d \times d}$ is the effective square-norm operator, where $\mathcal{N}_0$ is
given by
\begin{equation}
\begin{overpic}[width = 340pt, unit=1pt]{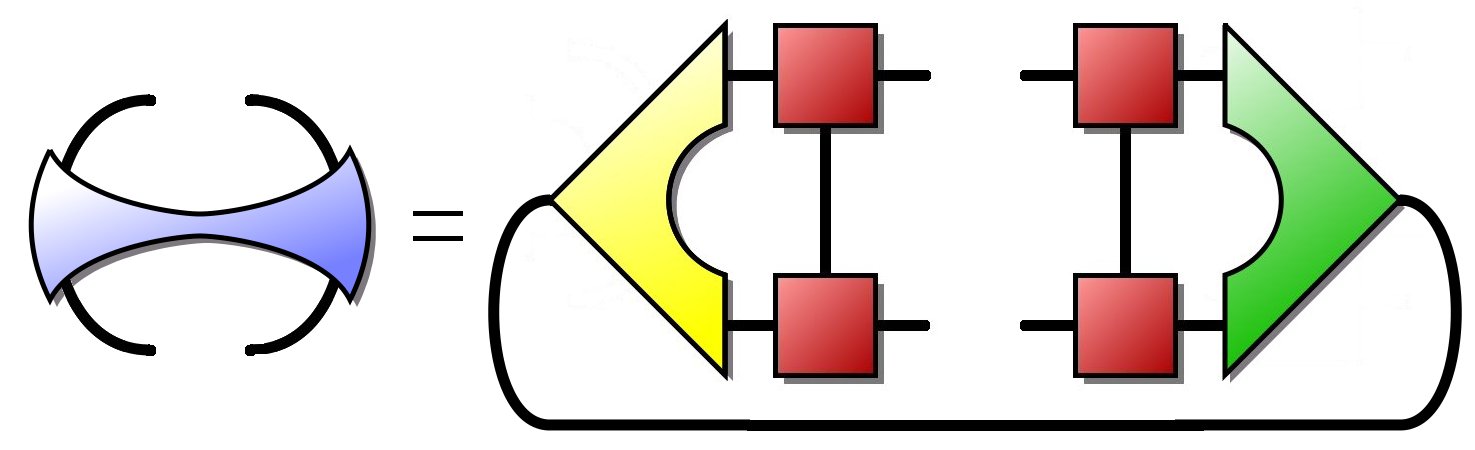}
 \put(14, 49){\footnotesize $\mathcal{N}_0$}
 \put(137, 55){\footnotesize $W_1$}
 \put(300, 55){\footnotesize $Z_1$}
\end{overpic}
\end{equation}
while the effective Hamiltonian $\mathcal{H}$ is obtained as follows
\begin{equation}
\begin{overpic}[width = \textwidth, unit=1pt]{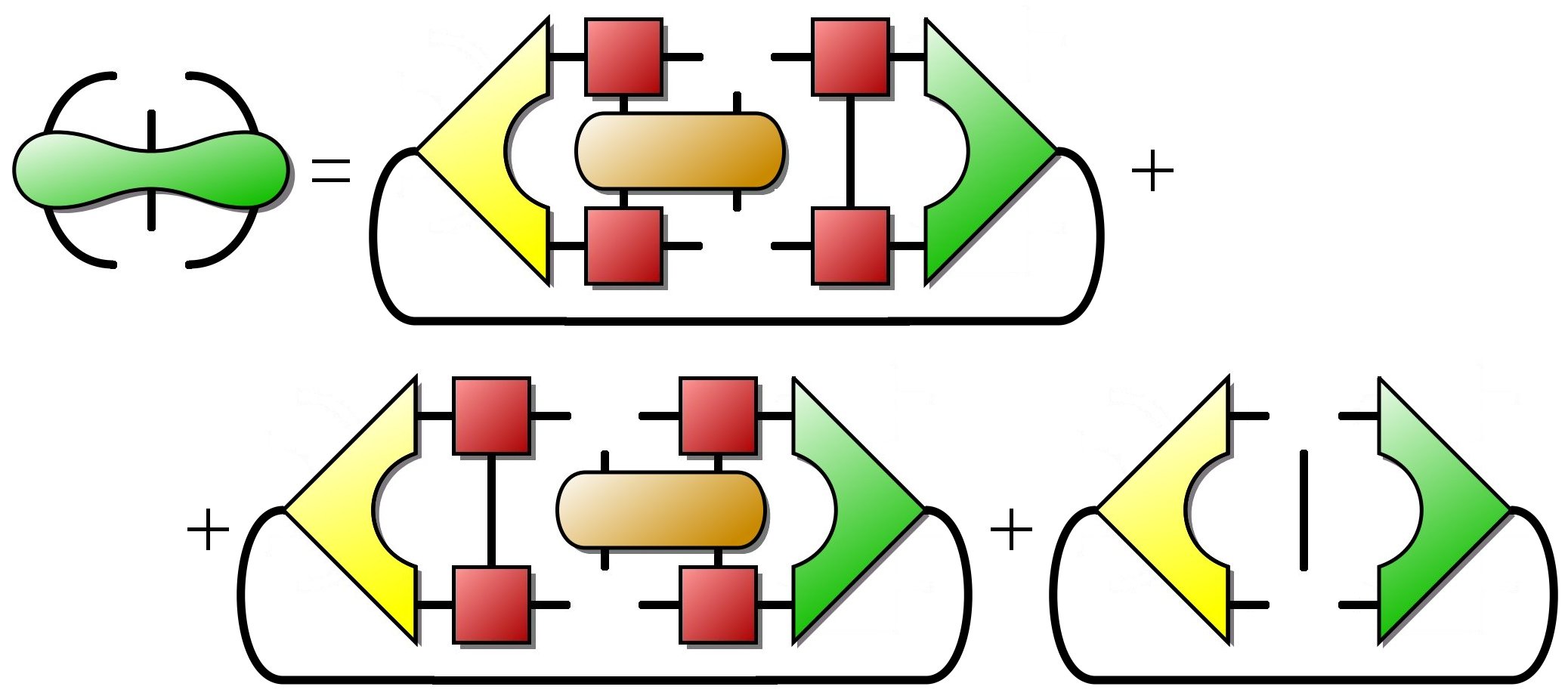}
 \put(14, 128){\footnotesize $\mathcal{H}$}
 \put(111, 133){\footnotesize $W_1$}
 \put(163, 132){\footnotesize $R_2^{[\ell]}$}
 \put(244, 133){\footnotesize $Z_1$}
 \put(78, 43){\footnotesize $W_1$}
 \put(154, 42){\footnotesize $R_2^{[\ell+1]}$}
 \put(211, 43){\footnotesize $Z_1$}
 \put(280, 43){\footnotesize $W_2$}
 \put(357, 43){\footnotesize $Z_2$}
\end{overpic}
\end{equation}
We immediately see a difference from the corresponding Lagrangian in the OBC case \eqref{eq:lagrange_fig}:
the effective (square) norm operator is no longer the identity $\mathcal{N} \neq \Id$.
If we recall correctly in the OBC case it was a property strictly depending on the choice of a gauge
condition for the other MPS blocks. But when the MPS design is Periodic,
in general there is no trick with gauge transformations in order to map $\mathcal{N}$ into $\Id$.
This also means that to find the optimal $A^{[\ell]}$ one has to solve a
\emph{generalized eigenvalue problem}, instead of a simple one as in \eqref{eq:std_eigenvalue}.
Precisely, the Euler-Lagrange equation of our problem is:
\begin{equation} \label{eq:gen_eigenvalue}
 \frac{\partial \mathcal{L}(A^{[\ell]}, {A^{\star}}^{[\ell]}) }{\partial \langle \!\langle
 A^{[\ell]} |} = | 0 \rangle \! \rangle
 \quad \longrightarrow \quad
 \mathcal{H} | A^{[\ell]} \rangle \! \rangle = \varepsilon \,\mathcal{N} | A^{[\ell]} \rangle \! \rangle,
\end{equation}
whose solution with minimal $\varepsilon$ is the optimal one, since
$\varepsilon$ is actually the energy:
$\varepsilon = \langle \!\langle A^{[\ell]} |\mathcal{H} | A^{[\ell]} \rangle \! \rangle
 / \langle \!\langle A^{[\ell]} |\mathcal{N} | A^{[\ell]} \rangle \! \rangle
= \langle \Psi | H |\Psi\rangle / \langle \Psi | \Psi\rangle$. After we found the solution, we
put the optimized $A^{[\ell]}$ into the MPS and choose another site for the variational paradigm.
Then repeat and sweep until convergence is achieved.

\begin{figure} 
 \begin{center}
  \includegraphics[width=260pt]{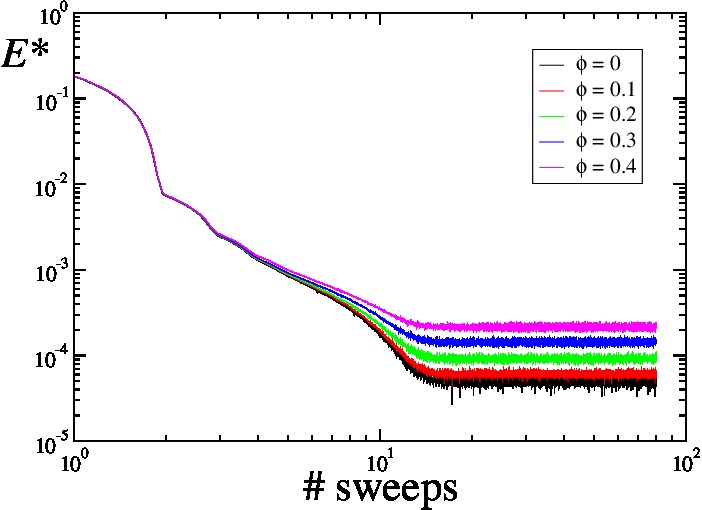}
 \end{center}
 \caption{ \label{fig:Rossin01}
Offset $E^{*}$ from the ground state energy of a variational periodic MPS a function
of the number of algorithm sweeps. The model considered is a 1D spin-$\frac{1}{2}$ XXY ring, with anisotropy
$\Delta = 0.5$ and twisted periodic boundary conditions ($\phi$ is the twisting phase). Here a bondlink $D = 18$ was used.
The graphic was kindly contributed by D.~Rossini \cite{Rossinistiff}.
 }
\end{figure}

\subsection{Stabilizing the generalized eigenproblem}

Dealing with the generalized eigenproblem \eqref{eq:gen_eigenvalue}
is no small trouble. Even with the most advanced linear
algebra techniques, numerical costs are much greater than those required for addressing
standard eigenproblems (for compatible dimensions). For the latter, eigenvector related to minima
or maxima of the spectrum are often found quickly thanks power method-inspired procedures.
Several algorithms for the generalized eigenproblem, like the Jacobi-Davidson, are also based on
power-method principles, but they should be wielded with care.
These protocols are very efficient when the inverse matrix of $\mathcal{N}$
is available; and, when it is not, converge faster the more $\mathcal{N}$ is easy to invert.
Obviously, an operator $\mathcal{N}$ is not suitable for inversion when there are eigenvalues
which are much smaller (closer to zero) than other ones, because numerical methods perceive the relative eigenspaces
as if they were a kernel. In this sense, we can relate the 'fast-invertibility' of a (positive) matrix
with the requirement  that the relative spreading of its eigenvalues is small, i.e.
$(\lambda_{\text{max}} - \lambda_{\text{min}} )/ \lambda_{\text{min}} \ll 1$,
and also means that $\mathcal{N}$ is
somehow 'close' to the identity, as the multiples of $\Id$ are the
only positive operators having relative spread zero.

Now, can we perform some gauge transformation that takes $\mathcal{N}$ as close as possible to the identity?
We stated that there exist no general solution to this question, although, for large system sizes
$\mathcal{L}$ we can argue that the system shall manifest small finite-size effects, and an affinity
with the OBC version should be met. In section \ref{sec:PBCimprove} we saw that the largest eigenvalues of
a composite transfer matrix decay more fast the larger is $L$. In particular, if every $\Etra_{\Id}$ was
mixing-CPT, we could write
\begin{equation} \label{eq:casino}
 \widetilde{\Etra}_{\Id}^{\ell} \sim |\Lambda_L) (\Lambda_R| + \varepsilon^{\ell} \,O_{\ell}
\end{equation}
where $\| O_{\ell} \|$ is bounded regardless from $\ell$,
and $\varepsilon$ is somehow related to the largest (in modulus) eigenvalue $\lambda_{\alpha}$ of $\Etra$ smaller
than 1: $\varepsilon \sim |\lambda_{\alpha}| < 1$. It is clear that for $\ell \to \infty$ the second term in
\eqref{eq:casino} vanishes, but also for $\ell$ finite but large $\varepsilon^{\ell} \,O_{\ell}$
is just a small perturbation to the first, leading, term.
$\Lambda_L$ and $\Lambda_R$, read as matrices, are positive thanks to CPT condition.

Relying on this concept we perform the following operations. consider
$\widetilde{\Etra}_3 = \Etra_{\Id}^{[\ell+1]} \cdot \ldots \cdot \Etra_{\Id}^{[\ell-1]} = 
\Etra_{\Id}^{[\ell+1]} \cdot \widetilde{\Etra}_1 \cdot \Etra_{\Id}^{[\ell-1]}$; if $L$ is large
and all the fixed MPS-blocks are in the left gauge, then
$\widetilde{\Etra}_3$ is written as \eqref{eq:casino} and in particular $(\Lambda_R| = (\Phi^{+}|$,
Then, since $\Lambda_L$ is positive, we can write
(via SVD, for example) $\Lambda_L = X X^{\dagger}$, or equivalently
$|\Lambda_L) = (X \otimes X^{\star})|\Phi^{+})$. Now we can perform a gauge transformation upon
$A^{[\ell-1]}$ so that it adsorbs the $X$ operator:
$A_s^{[\ell-1]} \to A_s^{[\ell-1]} \cdot X$. After the transformation, we are left with
\begin{equation}
 \widetilde{\Etra}_3 \sim |\Phi^{+})(\Phi^{+}| + \varepsilon^{\ell} \,O'_{\ell},
\end{equation}
which, in turn, makes the effective square-norm operator to read as follows:
\begin{equation}
 \mathcal{N} \sim \Id + \varepsilon^L \; \widetilde{\mathcal{N}}_{0}^{[L]},
\end{equation}
whose relative spread of the eigenvalues scales with $\varepsilon^{L}$, where $\varepsilon < 1$.
We ended up with an effective normalization operator which is actually the identity apart
a small perturbation, which decays exponentially with the length $L$. When we apply the
Jacobi-Davidson method in this framework, we find the generalized eigenproblem solution
much faster than if we do it naively, as proven in numerical simulations
\cite{Rossinisupersolid, Rossinistiff}.

\section{MPS and Tensor Networks}
With this last consideration, we conclude our discussion on how Matrix Product States
(either in their open or periodic boundary contexts) relate to simulation paradigms as variational
tailored wavefunctions, with surprising efficiency even at high precision calculus,
and wide manipulation features that let them \emph{overpower} the old-fashioned DMRG design.

As you could imagine, though, their application does not limit to numerical settings. MPS
are kept in great regard even for analytical calculation purposes. The possibility of building
interesting parent Hamiltonians for \emph{any} Matrix Product State \cite{MPSreview}, and
their continuous-space enhancement to describe a finitely-correlated
quantum field theory \cite{MPScontinuous, MPSHolographic}, are just
two of the many developments achieved in the last ten years.

More than anything, MPS have been the main inspiration that led physicists to investigate
thoroughly in the simulation capabilities of Tensor Networks in general. PEPS, TTN, MERA,
and other well-known Tensor Networks designs, actually gathered interest only after
the MPS representation of DMRG states was fully understood. In the following chapters we will
try to understand how such Tensor Network-based structures work; how the ideas behind MPS technique
can be redesigned in order, for instance, to extend the success of this method
even in settings where standard MPS/DMRG fails, like higher-dimensionality systems.

\begin{figure} 
 \begin{center}
  \includegraphics[width=\textwidth]{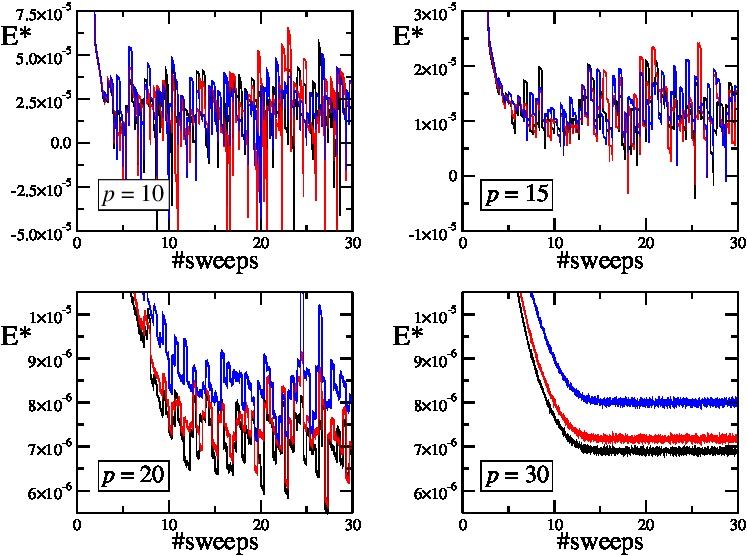}
 \end{center}
 \caption{ \label{fig:Rossin02}
Convergence rates of the energy offsets with sweeps, using the same 1D periodic XXY model of figure \ref{fig:Rossin01}, 
for various transfer matrices truncation parameter $p$, as explained in section \ref{sec:PBCimprove}
and equation \eqref{eq:Pippantech}. Notice that $p = 30$ achieves an excellent precision, even though
the original double-bondlink had dimension $D^2 = 324$, thus a factor 10 in algorithm speed.
The graphic was kindly contributed by D.~Rossini \cite{Rossinistiff}.
 }
\end{figure}

\chapter{General features of Tensor Networks} \label{chap:TN}

Every many-body/multipartite quantum state, when described as amplitude
components over a separable product basis, is uniquely represented by a complex Tensor.
However, the number of identifiers, of descriptors, one has to express to
locate specifically that state, within the whole manifold of system states,
is huge: in principle it scales exponentially with the number of elementary constituents
of the system, regardless of their nature.
Tensor networks (TN) are the trial to express the same state with a number of numerical descriptors
(be they variational or parametric) which is small, meaning that they must scale nicely with
the system size $L$, and that lead to the original wide amount of descriptors by means of
simple \emph{linear} algebraic operations. It is really not fundamental whether the desired analytical state
is reproduced exactly as much is important to recover the real-physics features the state should exhibit.

It is really impressive to acknowledge how much interesting physics can be generated for an approach
that sounds so naive, even in contexts where other analytical or numerical methods are totally clueless.
In this chapter, inspired by what we learned about MPS (the father-archetype of TN) we will
try to understand what common properties and features the Tensor Network architectures share, and
also present some useful examples and comparisons.

\section{Definition of Tensor Network state}

Let us start from a generic multipartite system $\mathcal{H}^{\otimes L} = \bigotimes_{\ell = 1}^{L} \mathcal{H}$,
where $\mathcal{H}$ is the Hilbert representation of the elementary degree of freedom; here we are taking
the constituents to bear a homogeneous representation, as it is the usual physical setting,
but this is not really a requirement of
the TN description. An orthonormal basis for the single degree of freedom must be chosen as canonical one,
which we shall represent as $|s\rangle$; then it is standard procedure to expand any given state $|\Psi\rangle$ of
$\mathcal{H}$ over the product canonical basis:
\begin{equation} \label{eq:tenspan}
 |\Psi\rangle = \sum_{s_1 \ldots s_L = 1}^{d} \mathcal{T}_{s_1 \ldots s_L} \;|s_1\,s_2 \ldots s_L\rangle
\end{equation}
where $d$ is the dimension of the elementary $\mathcal{H}$, $L$ the total number of constituents,
and $\mathcal{T}_{s_1 \ldots s_L}$ is a complex tensor (i.e. a multi-indexed collection of complex numbers),
with $L$ indices, each one allowing $d$ different values.
The normalization condition reads $\sum_{\{s_\alpha\}_\alpha}\mathcal{T}_{s_1 \ldots s_L}
\mathcal{T}_{s_1 \ldots s_L}^{\star} = 1$. As long as we ordered completely the $L$ degrees of freedom,
\eqref{eq:tenspan} is meaningful for every dimensionality of the physical system, and for every nature
(spin / bosonic / fermionic) of the constituents as well. In fact, the complete ordering allows us to write
always a second-quantization version of \eqref{eq:tenspan}, as follows
\begin{equation} \label{eq:tenspanII}
 |\Psi\rangle = \sum_{s_1 \ldots s_L = 1}^{d} \mathcal{T}_{s_1 \ldots s_L} \;
 (c_{1}^{\dagger})^{s_1} (c_{2}^{\dagger})^{s_2} \ldots (c_{L}^{\dagger})^{s_L} |\Omega\rangle.
\end{equation}
where $c_{\alpha}^{\dagger}$ are either Bose or Fermi operators, and in each case they satisfy the proper
commutation or anticommutation relations. Spin-orbitals $\alpha$ are now the elementary components of the system,
their on-site filling being the local canonical basis.
Expression \eqref{eq:tenspanII} represents the more general many-body
state, and $\mathcal{T}_{s_1 \ldots s_L}$ contains all its physics and information.

At the same time $\mathcal{T}$ is a huge array, with $d^L$ elements, hard to manipulate in every sense.
But assume that the state $|\Psi\rangle$ is such that $\mathcal{T}$ can be obtained, via
contracting over an additional index $q$, from a pair of tensors $\mathcal{T}'$ and $\mathcal{T}''$, like
\begin{equation}
 \mathcal{T} = \sum_q \mathcal{T}'_{s_1 \ldots s_{\ell},q} \;\mathcal{T}''_{s_{\ell+1} \ldots s_L,q}. 
\end{equation}
Then we would need a number of descriptors equal to $D (d^{\ell} + d^{L-\ell}) \sim D d^{L/2}$,
with a lot less information needed when the number $D$ of allowed values for index $q$
(or index $q$ dimension) is smaller than $d^{L/2}$.
The same argument can be applied again and refrained for $\mathcal{T}'$ or $\mathcal{T}''$;
every time we split a tensor into smaller tensors partially contracted together.
Every time, there is a chance (and typically happens) that we lose description capacity,
meaning that the resulting set of states allowing the new decomposition
is often than before. But this is not an issue as long as the states we are cutting out
of our ansatz are those which are not physically relevant, and we keep those that contain the true physics
of the problem we want to study.

This is the central point of the Tensor Network paradigm, inspired to MPS.
We write $\mathcal{T}$ as product of multiple tensorial objects, where $L$ indices are left open,
they are the $d$-dimensioned physical indices, while an arbitrary number of
'fictitious-space' indices $\{q\}$ (of arbitrary dimension $D_q$) are contracted.
Like for MPS, for a given scheme of contraction, which from now on we will call \emph{Network},
there is \emph{always} a choice of virtual links $D_q$ large enough so the whole $\mathcal{H}^{\otimes L}$
is described. But that would be a waste of effort, since it is very unlikely that a physical state
would require that very amount of information shared among tensors in the network.
Which yields a definition of TN-state that takes into account parametric bounds to
our description in terms of $(i)$ the number of tensors, $(ii)$ the number of indices per tensor,
and $(iii)$ allowed values per non-physical index:

\vspace{.5em}
\emph{\textbf{Definition of Tensor Network}} - A multipartite state $|\Psi\rangle$ is a Tensor Network state
$\{V,k,D\}$,
with maximal tensor number $V$, correlation link-number $k$, and link dimension $D$ if
\begin{itemize}
 \item It exists a decomposition of $\mathcal{T}_{s_1 \ldots s_L}$ as a contracted product of tensors:
  \begin{equation} \label{eq:quaqua}
   \mathcal{T}_{s_1 \ldots s_L} = \sum_{q_1 = 1}^{D_1} \ldots \sum_{q_B = 1}^{D_B}
   \left( \prod_{n = 1}^{V'} T^{[v]\,q_{w(1,n)} \ldots q_{w(b_n,n)}}_{s_{v(1,n)} \ldots s_{v(a_n,n)}} \right),
  \end{equation}
  where $V'$ is the total number of elementary tensors $=$ nodes in the network,
  $B$ is the total number of contracted indices $=$ connected links in the network,
  Every $s_\alpha$ appears once as tensor index in the expanded expression
  \eqref{eq:quaqua}, and every $q_\alpha$ appears twice
 \item The number of tensors in the decomposition is bounded by $V' \leq V$, 
 the bondlink dimensions are bounded by $D_\alpha \leq D$ for every $\alpha \in \{1..B\}$, and
 the total amount of indices of a tensor is bound by $a_n + b_n \leq k$ for every $n \in \{1..V \}$.
\end{itemize}

As an immediate consequence, the overall number of complex value descriptors
(i.e. variational parameters) required for such representation is
\begin{equation}
 \# \mbox{elements} \leq \,V \, \left( \vphantom{A^{[1]}} \sup\{ D, d \} \right)^{k} .
\end{equation}
Although this is a well-formulated definition, there is no doubt that the formalism of
\eqref{eq:quaqua} is cumbersome and confusing. For most purposes involving Tensor Network
states is actually preferable to involve a diagrammatic representation, much similar to the one
we adopted for Matrix Product States. In these diagrams, Tensor Network states are represented
as graphs: Tensors being vertices, indices being links, either left open if they are physical
indices, or connected if they are fictitious. Let us consider the following example:
\begin{equation} \label{eq:Netpic}
\begin{overpic}[width = \textwidth, unit=1pt]{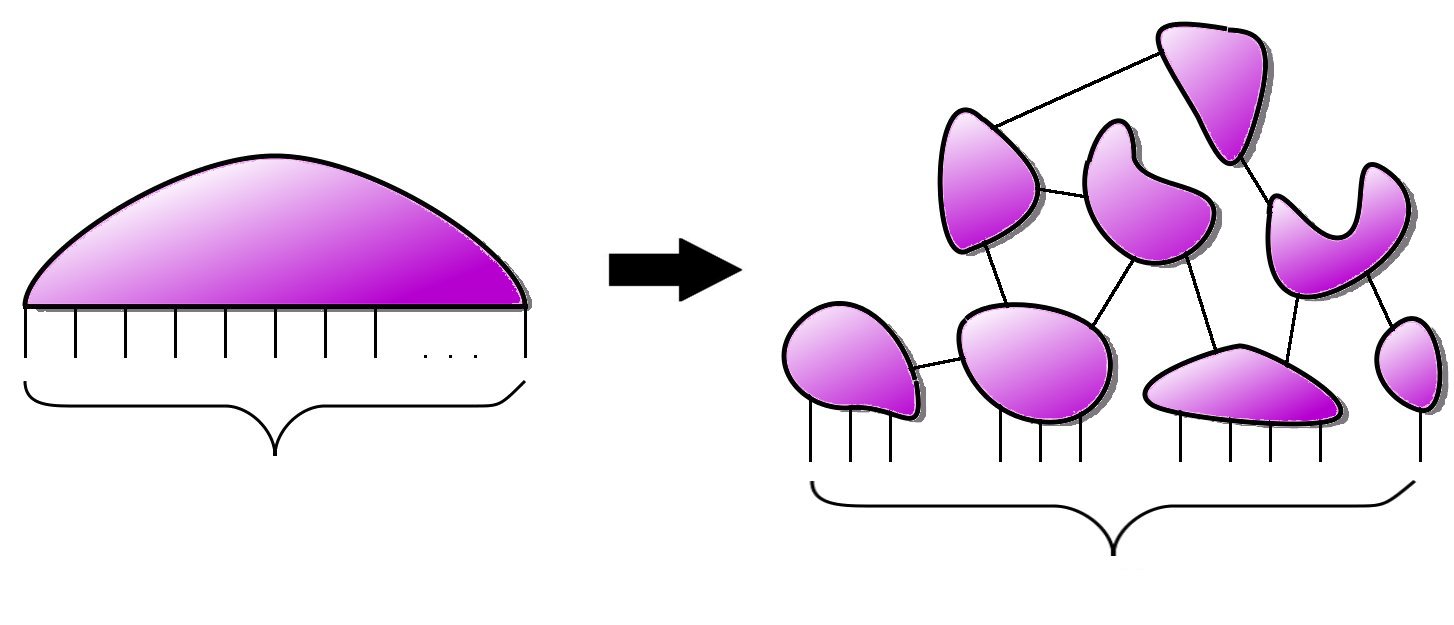}
 \put(47, 98){$\mathcal{T} \longleftrightarrow |\Psi\rangle$}
 \put(70, 30){$L$}
 \put(294, 3){$L$}
 \put(221, 67){\footnotesize $T^{[1]}$}
 \put(272, 67){\footnotesize $T^{[2]}$}
 \put(258, 113){\footnotesize $T^{[3]}$}
 \put(303, 106){\footnotesize $\ldots$}
\end{overpic}
\end{equation}
This network is made out of $V = 8$ tensors, with maximal correlation number $k = 6$; the
analytical tailored expression of the corresponding (spin) state is given by
\begin{multline} \label{eq:Netugly}
 |\Psi\rangle = \sum_{\{s_\alpha\} = 1}^{d} \sum_{\{q_\beta \} = 1}^{D_{\beta} \leq D}
 \left( T^{[1]q_1}_{s_1 s_2 s_3} \; T^{[2]q_1 q_2 q_3}_{s_4 s_5 s_6} \; T^{[3]q_2 q_4 q_5}
 \; T^{[4]q_3 q_4 q_6}
 \right. \\ \left.
 \; T^{[5]q_5 q_7} \; T^{[6]q_6 q_8}_{s_7 s_8 s_9 s_{10}}
 \; T^{[7]q_7 q_8 q_9} \; T^{[8]q_9}_{s_{11}}
 \right) |s_1 \ldots s_L\rangle
\end{multline}
which is messy, and not immediate as \eqref{eq:Netpic} although they represent the same
parametric set of states.

Most classes of Tensor Network commonly considered in literature,
are \emph{scalable} with system size. We intend that the network is
obtained by repeating some fixed local pattern of vertices contraction to build a
self-similar structure, so it can be adjusted to fit any $L$ bu just adding new
tensors according to the same pattern as before. When doing so, it is important that
$k$ and $D$ can be kept fixed, and the number $V$ of tensors (recall that $V$ is proportional to the number
of variational parameters) scales nicely, e.g. linearly, with $L$. This was the
case of Matrix Product States, where precisely $V = L$.
$D$ is also occasionally referred to as \emph{refinement parameter} \cite{EisMERAPEPS},
as its value directly influences the capability of the TN ansatz.

\section{Entanglement of Tensor Network states} \label{sec:entanet}

Quantum entanglement is the primary responsible for argumenting that Tensor Network are actually a
good technique to describe physical multipartite states. Ground states are characterized by small correlations,
compared to a generic random state in the Hilbert. So probing them with trial wavefunctions
that admit a simple description, and yet
capable to reproduce just the needed amount of entanglement seems a suitable choice.
Tensor Network states have the outstanding feature that their entanglement is perfectly controlled
by the network topology itself, as we will show in this section.

\vspace{.5em}
\emph{\textbf{Entanglement bounds of a Tensor Network state}} - Assume $|\Psi\rangle$ is a state
which allows a Tensor Network representation, as in \eqref{eq:quaqua}. Let us choose
any partition of the physical sites $s_j$ into two disjoint subsets $s^{[A]}_j$ and $s^{[B]}_j$.
Then the Von Neumann entanglement entropy associated to this partition satisfies the following inequality:
\begin{equation} \label{eq:TNEbound}
 \mathcal{S}_{\text{VN}} \left( {\rho}^{[A]}  \right) \leq
 \min_{ \text{part.} \to \{ D_\alpha \}} \left( \sum_{\alpha} \log D_\alpha \right)
\end{equation}
where the minimum is taken over all the partitions of the network graph into two subgraphs,
with the condition that the first subgraph embeds all the $A$ sites and the other one the $B$ sites.
The $D_\alpha$ are the bondlink dimensions of the links
we need to break in order to disconnect the two subgraphs.

To make an example, let us consider again a Tensor Network state like \eqref{eq:Netpic} \eqref{eq:Netugly},
and assume that we are to estimate the entanglement shared between the six leftmost sites 
$A=\{1..6\}$ and the five rightmost sites $B=\{7..11\}$. Then identify all the possible ways to
part the network into two subnetworks, respectively containing $A$ and $B$. Three smart choices are given by:
\begin{equation} \label{eq:Netsplit}
\begin{overpic}[width = \textwidth, unit=1pt]{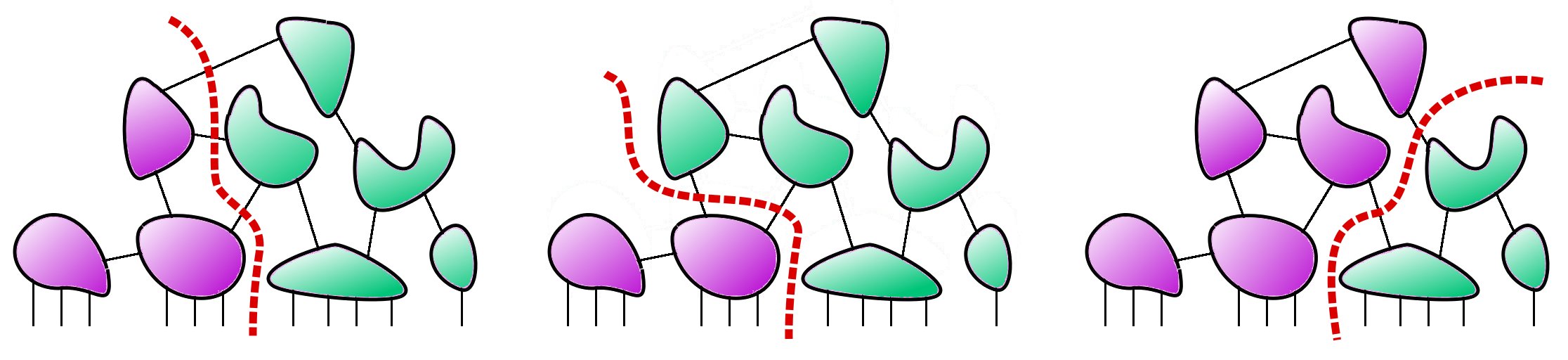}
\end{overpic}
\end{equation}
Finally detect all the network bondlinks $\alpha$ we should break to separate the two subgraphs
(violet and green), and
sum their $\log D_{\alpha}$ to obtain a bound on the entanglement.
In particular, the three graph partitions picted in \eqref{eq:Netsplit} tell us that
\begin{equation}
 \mathcal{S}_{\text{VN}}^{[A \leftrightarrow B]} \leq
 \min\left\{ \log(D_3 D_4 D_5), \log( D_2 D_3), \log( D_7 D_8) \right\}
\end{equation}
and, since any $D_\alpha \leq D$, we conclude that $\mathcal{S}_{\text{VN}} < D^2$. The reader can easily check
that no other graph partition would lead to a tighter bound.

Let us prove the statement \eqref{eq:TNEbound}, by adopting an argument very similar to the
valence bond picture for MPS. Consider any network graph partition into two subgraph, in accordance
with the lattice sites partition $A \leftrightarrow B$, as above.
Then, let us write a starting virtual state
\begin{equation}
 |\Phi^{+}_{\otimes}\rangle = \bigotimes_{\alpha \in \text{cut}} |\Phi^{+}_{D_\alpha}\rangle
 \quad \mbox{where} \quad |\Phi^{+}_{p}\rangle = \frac{1}{\sqrt{p}} \sum_{q = 1}^{p} |qq\rangle.
\end{equation}
Each maximally state $|\Phi^{+}_{D_\alpha}\rangle$ contributes to the entanglement of
$|\Phi^{+}_{\otimes}\rangle$ separately, since they lie in different degrees of freedom,
and each one contributing with $\log D_{\alpha}$.
But now we can find a linear mapping taking the state $|\Phi^{+}_{\otimes}\rangle$ into the original $|\Psi\rangle$
\begin{equation} \label{eq:locazzin}
 |\Psi\rangle = \left( M_1^{[A]} \otimes M_2^{[B]} \right)|\Phi^{+}_{\otimes}\rangle
\end{equation}
the mapping $M_1^{[A]}$ is given by the contraction of the subgraph on $A$, and therefore is linear,
and so is $M_1^{[B]}$, telling us that \eqref{eq:locazzin} can be viewed as a quantum transformation,
not necessarily invertible. But as it is a tensor product, it is also local, and thus can only degrade
entanglement, not enhance it. Therefore, the entanglement of $|\Psi\rangle$ must be less than that
of $|\Phi^{+}_{\otimes}\rangle$ which is exactly $\sum_{\alpha} \log D_{\alpha}$. The same
argument can be repeated for any subgraph partition, thus concluding the proof.

\section{Operators and link exchange-statistics} \label{sec:linkexchange}

Since we now have a diagrammatic representation for linearly-connected tailored variational wavefunctions,
we want now to exploit this idea also to include the action of operators. Especially operators that
act locally or at short-ranges appear as new tensorial pieces to add to the network structure,
connecting to those physical links that were left open in the TN-state design: for instance,
a three-site operator (acting on sites number 2,3 and 4) applied to \eqref{eq:Netsplit} reads
\begin{equation} \label{eq:Netops}
\begin{overpic}[width = 165pt, unit=1pt]{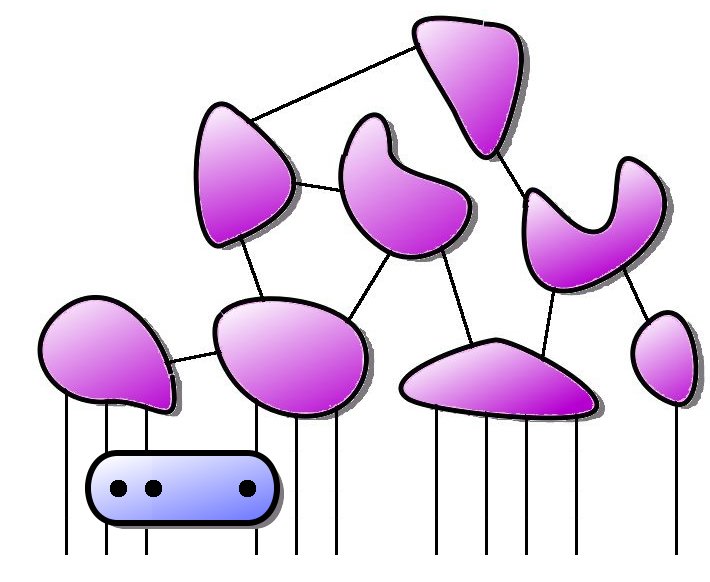}
  \put(-70, 60){\large $\Theta_{2,3,4} \;|\Psi\rangle = $}
\end{overpic}
\end{equation}
As was pointed out in refs.~\cite{Corboz, FOC1, FOC2}, attaining this representation is undoubtedly trivial for
a spin or boson system, but we should handle the issue with care when our Tensor Network state
is describing a system of fermions. Indeed, when we introduced the second quantization version \eqref{eq:tenspanII}
of Tensor decomposition, ordering the sites was a crucial point.
It is clear that changing the ordering of those sites would not only
rearrange the components of $\mathcal{T}$ but also change some signs appropriately due to Fermi statistic.
We are about to show that this feature can be embedded as an inherent property of the network links,
that manifests under crossing (exchanging) of the links themselves.

For instance, we will start from a two-site operator $\Theta_{1,2}$; we will also assume for
simplicity that this operator preserves \emph{parity}, i.e.
\begin{multline}
 \Theta_{1,2} = \alpha_0 \,\Id + \alpha_1 \,c_1 c_2 + \alpha_2 \,c_1^{\dagger} c_2
 + \alpha_3 \,c_1 c_2^{\dagger} +  \\ +
 \alpha_4 \,c_1^{\dagger} c_2^{\dagger} +
 \alpha_5 \,c_1^{\dagger} c_1 + \alpha_6 \,c_2^{\dagger} c_2 +
 \alpha_7 \,c_1^{\dagger} c_1 c_2^{\dagger} c_2
\end{multline}
requiring parity preservation has the advantage that (quasi-) local operators preserve
support locality when representing fermions as spins.
%\begin{equation} \label{eq:parypres}
% \Theta_{1,2}
% \left( \begin{array}{cccc}
% \alpha_0 & 0 & 0 & \alpha_1 \\
% 0 & \alpha_0 + \alpha_6 & \alpha_3 & 0 \\
% 0 & \alpha_2 & \alpha_0 + \alpha_5 & 0 \\
% \alpha_4 & 0 & 0 &\alpha_0 + \alpha_5 +  \alpha_6 + \alpha_7
% \end{array} \right)
%\end{equation}
Now, assume the tensorial representation of $\Theta_{1,2}$ is known and available, encoded through
descriptors $\{ \alpha_0 \ldots \alpha_4\}$.
We wonder: how do we express the action of $\Theta_{1,3}$ instead? The standard procedure
is swapping the second and third (or equivalently, first and second) sites
both before and after performing $\Theta_{1,2}$, as
\begin{equation} \label{eq:CrossX}
\begin{overpic}[width = \textwidth, unit=1pt]{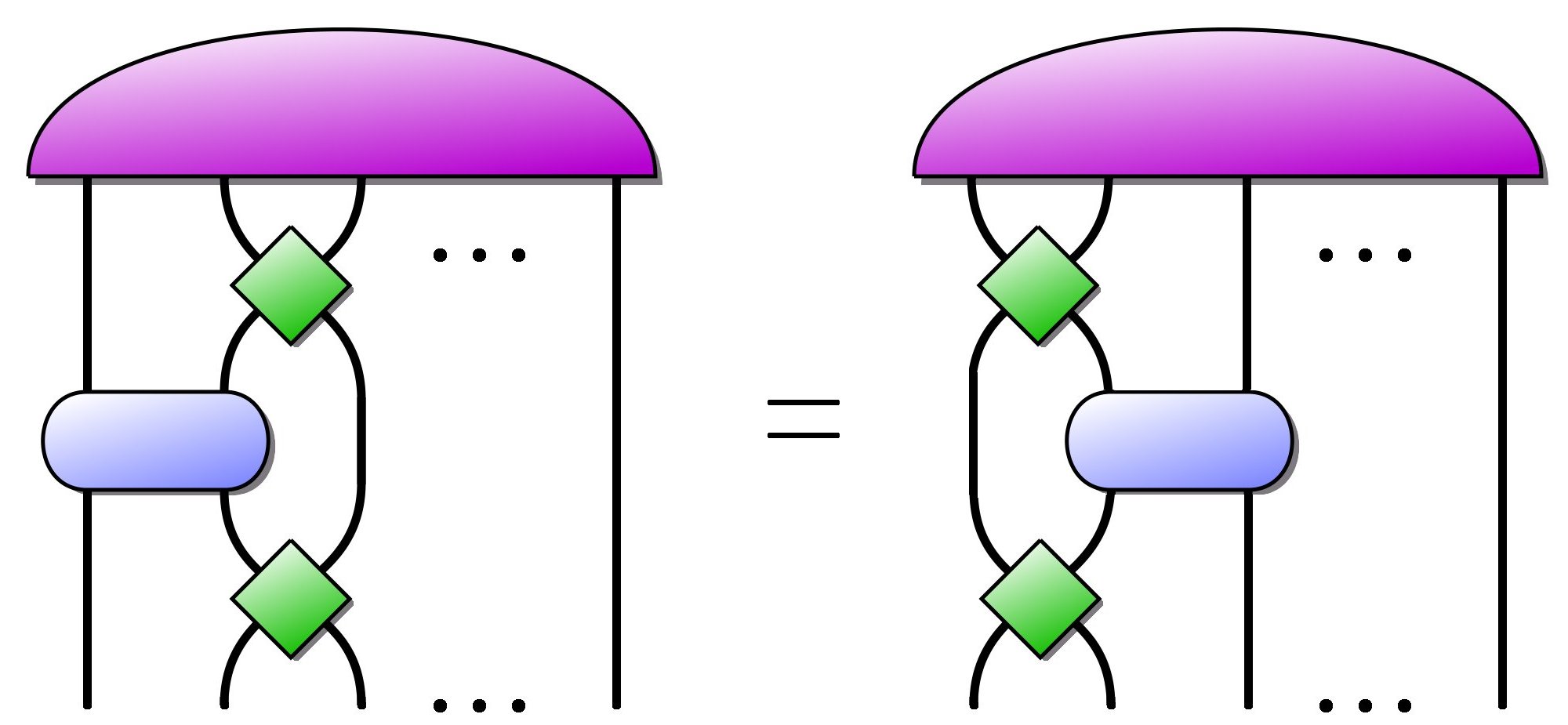}
  \put(35, 66){$\Theta$}
  \put(68.5, 27){$S$}
  \put(255, 27){$S$}
  \put(68.5, 105){$S$}
  \put(255, 105){$S$}
  \put(290, 66){$\Theta$}
\end{overpic}
\end{equation}
According to anticommutation rules, it is during the exchanging process that the fermionic statistic
should emerge. Indeed the operator $S$ in equation \eqref{eq:CrossX} is the standard
quantum information Swap gate only for spin and bosonic Tensor Network. For fermionic Tensor Networks
it reads instead:
\begin{equation} \label{eq:Swapfermi}
\begin{overpic}[width = 140pt, unit=1pt]{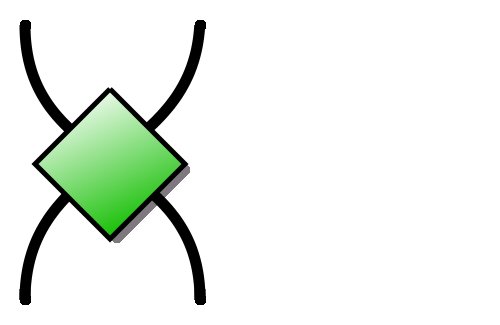}
  \put(27, 41){$S$}
  \put(70, 41){$= \quad
  \left( \begin{array}{cccc}
   1 &&&\\
   & 0 & 1&\\
   & 1 & 0&\\
   & & & -1 \\
  \end{array} \right)$}
\end{overpic}
\end{equation}
where the antisymmetrization sign $-1$ appears in correspondence to 
$c^{\dagger}_1 c^{\dagger}_2 |\Omega\rangle = - c^{\dagger}_2 c^{\dagger}_1 |\Omega\rangle$,
the doubly occupied canonical state.
all the other 2-sites canonical vectors, namely $|\Omega\rangle$, $c^{\dagger}_1 |\Omega\rangle$ and
$c^{\dagger}_2 |\Omega\rangle$, being either zero or one particle states,
feel no difference from the spin/Bose setting,

In conclusion we acknowledged that Tensor Networks ansatze can be successfully applied to fermionic
systems with no basic difference in efficiency or computational costs. The only issue we need to
take care of, is the exchange of network links within diagrams, that can be rearranged from the original
ordering by means of \eqref{eq:Swapfermi}. We could state that the network links satisfy
an exchange statistics themselves, that identifies the nature of particles the TN-picture is representing.

As an additional remark, let us point out that it is straightforward to generalize all these arguments
to include \emph{abelian anyon} (like those of fractional quantum Hall effect) algebras as well.
Indeed, let us assume that we are representing, with our trial Tensor Network state, a 2D system,
which is where anyons arise. Then if anyonic operators undergo the  exchange rule
$c^{\dagger}_1 c^{\dagger}_2 = e^{i \phi_0} c^{\dagger}_2 c^{\dagger}_1$, then the
the bondlink exchange statistics of the TN representation \eqref{eq:Swapfermi} is replaced by
\begin{equation}
 S_{\text{anyonic}} = \left(
 \begin{array}{cccc}
   1 &&&\\
   & 0 & 1&\\
   & 1 & 0&\\
   & & & e^{i\phi_0} \\
  \end{array} \right).
\end{equation}
The problem of extending such formulation to include also non-abelian anyons was discussed in ref.~\cite{Corbozanyon}.
A similar scheme to deal with fermionic statistics is by using directed-link
Tensor Network designs introduced recently, called \emph{Fermi operator circuits} (FOC) \cite{FOC1, FOC2}.

\section{Gauge group of Tensor Network states} \label{sec:TNgauge}

Inspired by our discussion of section \ref{sec:MPSGauge} where we defined a group of MPS transformations
under which the physical state is invariant, we want to extend this paradigm to any given
Tensor Network geometry. As for Matrix Product States, a gauge transformations group provide a
clever way to manipulate analytically or numerically tensors within the network structure. The purposes of
exploiting this group are many, for example, to perform faster contractions.

The generalization of \eqref{eq:Gaugetransform} to an arbitrary TN-state is natural.
Consider, for instance, a shared (connected at both sides) bondlink $\nu$, corresponding to a virtual degree of
freedom, of dimension $D_{\nu}$. Tensors $T^{[\alpha]}$ and $T^{[\beta]}$ are the two nodes
in the network sharing link $\nu$. Then we can define the composite tensor emerging from the contraction of the two
\begin{equation}
 T^{[\alpha \leftrightarrow \beta] \{q_\alpha, q_\beta \}}_{\{s_\alpha, s_\beta\}}
 = \sum_{q_{\nu} = 1}^{D_{\nu}}
 T^{[\alpha] \{q_\alpha\},q_\nu}_{\{s_\alpha\}} \;
 T^{[\beta] \{q_\beta \},q_\nu}_{\{s_\beta\}}
\end{equation}
We are also assuming that no homogeneity or geometric constraint is requested for the TN to hold,
so that the state $|\Psi\rangle$ actually depends on $T^{[\alpha]}$ and $T^{[\beta]}$ only through
$T^{[\alpha \leftrightarrow \beta]}$, i.e. even if each of the former two changes but the latter
is unaltered, then $|\Psi\rangle$ is also unchanged.

Now let us choose an isomorphism $X$ on $\mathbb{C}^{D_{\nu}}$, i.e.
an invertible linear application on the $D_{\nu}$-dimensional complex vector space:
$X X^{-1} = X^{-1} X = \Id_{D_{\nu} \times D_{\nu}}$. Then it is clear that
\begin{equation} \label{eq:TNgauinva}
 \begin{aligned}
 T^{[\alpha \leftrightarrow \beta] \{q_\alpha, q_\beta \}}_{\{s_\alpha, s_\beta\}}
 &= \sum_{r_1 \ldots r_3 = 1}^{D_{\nu}}
 T^{[\alpha] \{q_\alpha\},r_1}_{\{s_\alpha\}} \;
 X^{\phantom{-1}}_{r_1 r_2} X^{-1}_{r_3 r_3}
 T^{[\beta] \{q_\beta \},r_3}_{\{s_\beta\}} \\
 &= \sum_{q_{\nu} = 1}^{D_{\nu}}
 W^{[\alpha] \{q_\alpha\},q_\nu}_{\{s_\alpha\}} \;
 W^{[\beta] \{q_\beta \},q_\nu}_{\{s_\beta\}},
 \end{aligned}
\end{equation}
where $W^{[\alpha]}$ is the contraction of $T^{[\alpha]}$ with $X$, while
$W^{[\beta]}$ is obtained by linking  $X^{-1}$ to $T^{[\beta]}$, as in the following diagram
\begin{equation} \label{eq:TNgaupic}
\begin{overpic}[width = 230pt, unit=1pt]{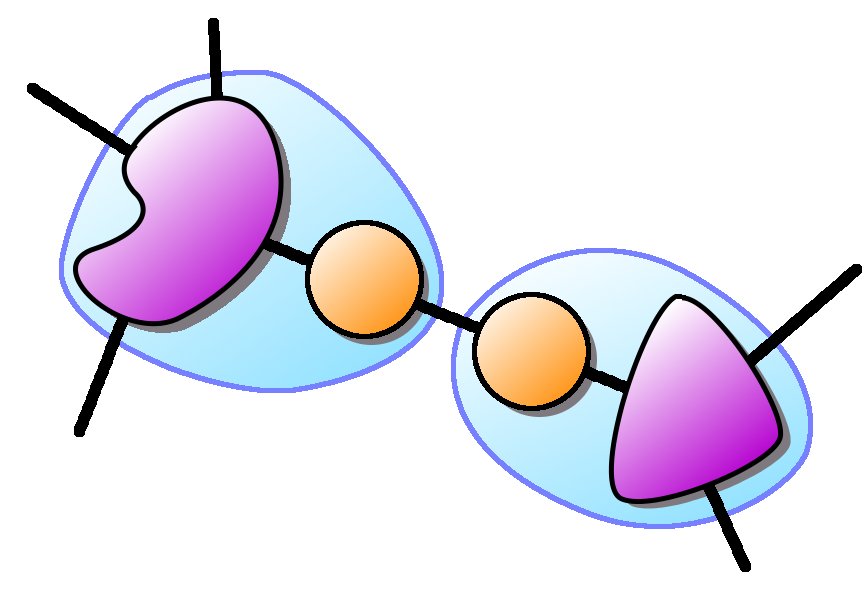}
  \put(47, 100){$T^{[\alpha]}$}
  \put(91, 80){$X$}
  \put(100, 123){$W^{[\alpha]}$}
  \put(116, 18){$W^{[\beta]}$}
  \put(131, 61){$X^{-1}$}
  \put(176, 46){$T^{[\beta]}$}
\end{overpic}
\end{equation}
Since $T^{[\alpha \leftrightarrow \beta]}$ is invariant under this transformation, $|\Psi\rangle$
is invariant as well. Rephrasing \eqref{eq:TNgauinva} in other words, we could say that, as
$\nu$ is a fictitious, external degree of freedom over which we are performing a (partial) trace,
the physical system (set of real degrees of freedom) is insensitive to any local invertible transformation $X$
acting upon $\nu$, which is what \eqref{eq:TNgauinva} does. There is no need to say that
transformation \eqref{eq:TNgaupic} has a natural group structure, arising from the fact that
the set of Isomorphisms $\{X\}$ is closed under composition.

This very argument can be applied and repeated for every closed (doubly connected, i.e. non-physical) link in
the network structure. Indeed, we associate an invertible matrix $X_{\nu}$ and a \emph{direction}
to every virtual TN-bondlink and transform tensors 
$T^{[\alpha]} \rightarrow W^{[\alpha]}$ according to
\begin{equation} \label{eq:TNgaugetransform}
 W^{[\alpha] \{q\},\{p\}}_{\{s\}} = 
 \sum_{\{v\},\{w\}}
 \left( \prod_j^{k_{\alpha}^{\text{in}}} X^{-1}_{q_j, v_j} \right)
 \left( \prod_j^{k_{\alpha}^{\text{out}}} X_{p_j, w_j} \right)
 T^{[\alpha] \{v\},\{w\}}_{\{s\}}
\end{equation}
where $k_{\alpha}^{\text{in}}$ (resp. $k_{\alpha}^{\text{out}}$) is the number of links whose chosen
direction is incoming to (outcoming from) node $\alpha$.
Precisely, we are contracting $T^{[\alpha]}$ to $X_{\nu}$ at link $\nu$, if the direction given to $\nu$ is
outcoming from node $\alpha$, and to $X^{-1}_{\nu}$ otherwise. As stated, no transformation is
allowed on the physical links, i.e.:
\begin{equation} \label{eq:TNgaupic2}
\begin{overpic}[width = 260pt, unit=1pt]{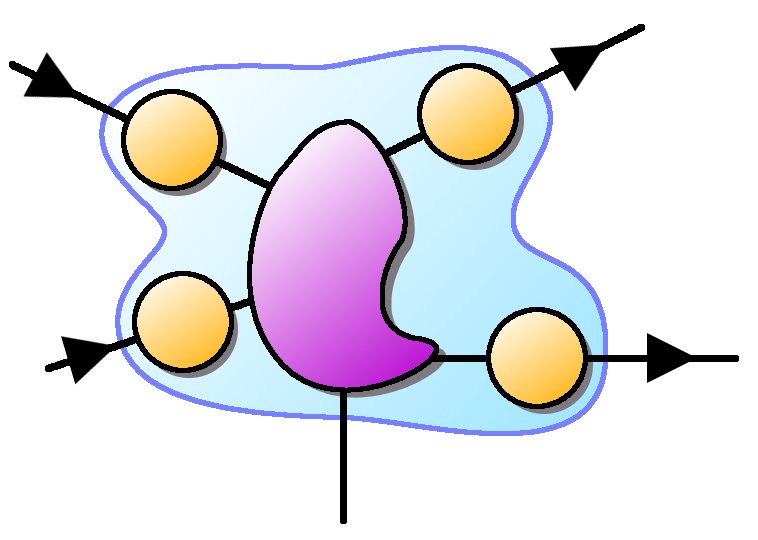}
  \put(53, 70){$X^{-1}_1$}
  \put(49, 133){$X_2^{-1}$}
  \put(104, 92){$T^{[\alpha]}$}
  \put(184, 105){$W^{[\alpha]}$}
  \put(126, 10){$\longleftarrow$ physical index}
  \put(155, 141){$X_3$}
  \put(178, 57){$X_{4}$}
\end{overpic}
\end{equation}
It is worth mentioning that the group we found, which coincides with the Cartesian sum
$\mathcal{G} \equiv \bigoplus_{\nu}^{n} \mbox{Iso}(\mathbb{C}^{D_{\nu}})$
(with $n$ the total number of closed links), is
the \emph{most general} state-invariant transformation, provided that the TN-geometry and the bondlink
dimensions $D_{\nu}$ are fixed.

\section{No closed loop $\Rightarrow$ efficient contraction} \label{sec:Closedloop}

As we learned how to match a quantum operator with the Tensor Network representation of a state,
we need to work out how to achieve expectation values, to get physical information on the state itself.
For MPS we argumented the importance of a contraction algorithm which is numerically efficient:
as we typically adopt energy as a benchmark for simulation convergence towards the ground state,
it is likely that we will need to evaluate $\langle \Psi | H | \Psi \rangle$ many times.

An optimal contraction scheme must be modeled on the topology of Tensor Network we are considering.
unfortunately, even when such scheme is optimized, its efficiency could still be unsatisfactory for practical purposes,
and that obviously depends on the network design.
Precisely, even if we have a $L$-scalable Tensor Network structure $\{V(L), k, D \}$ 
with a number of variational parameters $V$ proportional to $L$, the optimal
contraction cost could still scale exponentially with $L$. This is, in particular, the unfortunate case
of Product Entangled Pair states, which we will introduce in section \ref{sec:PEPS}.

However, we can identify a sub-class of Tensor Networks which is still quite general
and whose contraction efficiency is guaranteed. We are talking about Tensor Networks that
\emph{lack closed loops} in the graph structure. Acquiring an expectation value out of
a Tensor Network state $\{V, k, D\}$ without close loops has a computational cost bound by
\begin{equation} \label{eq:Noloopcost}
 \# \mbox{cost} \sim 2\:V \, D^{2k},
\end{equation}
which scales linearly with size if $V \propto L$. Here we present an example of the
comparison between TN-structures without or with closed loops:
\begin{equation} \label{eq:loop}
\begin{overpic}[width = \textwidth, unit=1pt]{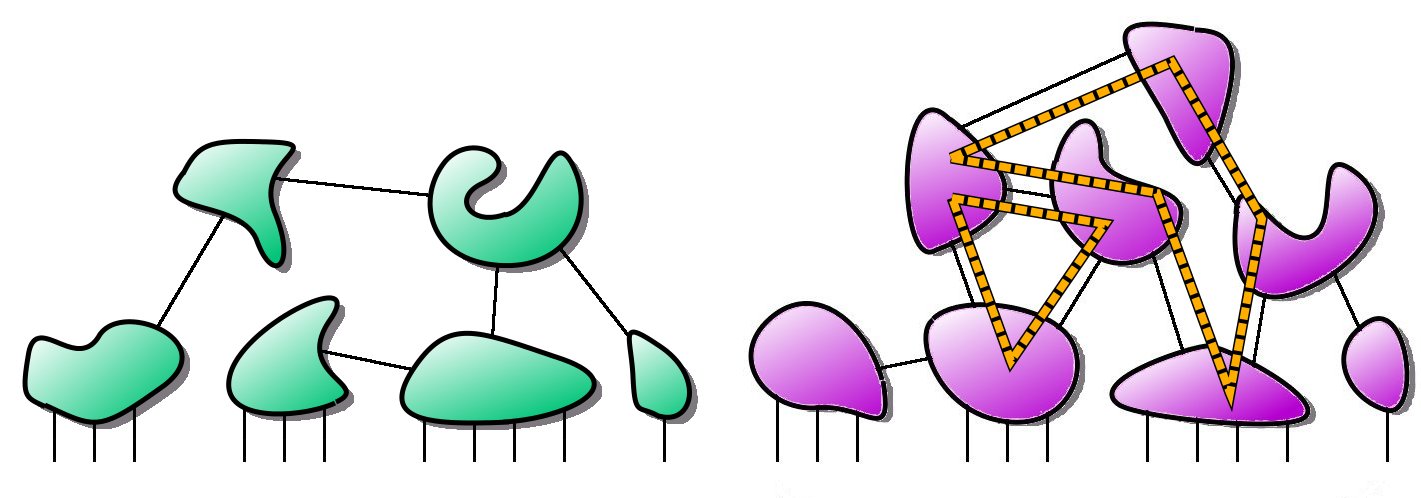}
  %\put(35, 66){$\Theta$}
\end{overpic}
\end{equation}
where orange dashed lines highlight the minimal loops.
The contraction scheme that leads to \eqref{eq:Noloopcost} is definitely intuitive
and often close to be the optimal one for a given network. Let us sketch it briefly.

Let $\Theta_{\otimes} = \bigotimes \Theta^{[\ell]}_{\ell}$ be the separable product operator
whose expectation values $\langle \Psi | \Theta_{\otimes} | \Psi \rangle$ we are interested in.
Then, for every node $\mu$ in the network, we define a new 'double'-tensor
\begin{equation}
 \mathcal{W}^{[\mu] q_1 \ldots q_m}_{p_1 \ldots p_m} = 
% \sum_{s_{v(1)} \ldots s_{v(\mu)} = 1}^{d} \sum{r_{v(1)} \ldots s_{r(n)} = 1}^{d}
 \!\!\!\!\! \sum_{\mbox{{\scriptsize $\begin{array}{c}
        s_{v(1)} \ldots s_{v(n)}\\
        r_{v(1)} \ldots r_{v(n)}
       \end{array} \!\!\! = 1$}}}^{d}
 \!\!\!\!\!\!
 T^{[\mu]q_1 \ldots q_r}_{ s_{v(1)} \ldots s_{v(n)} }
 T^{\star\:[\mu]p_1 \ldots p_r}_{ r_{v(1)} \ldots r_{v(n)} }
 \;\prod_{j = 1}^{n} \langle r_{v(j)} | \Theta_{v(j)} | s_{v(j)} \rangle.
\end{equation}
where $n$ is the number of physical links connected to node $\mu$, which correspond to sites
$v(1) \ldots v(n)$, while $m$ is the number of virtual links; of course $m+n \leq k$.
The tensor $\mathcal{W}^{[\mu]}$ is obtained by pairing together tensors $T^{[\mu]}$ and
its complex conjugate $T^{\star\:[\mu]}$, while contracting every physical link though the action of
$\Theta_{\otimes}$; notice that $\mathcal{W}^{[\mu]}$ has correlation number $2m \leq 2k$.
Obtaining every $\mathcal{W}^{[\mu]}$ has a computational cost of at most $D^{2k}$.

Now, it is easy to see that the $\mathcal{W}^{[\mu]}$ form again a Tensor Network, which is
identical to the original one except for having no open (physical) links,
which vanished. 
So it contracts completely into a real number, equal to
$\langle \Psi | \Theta_{\otimes} | \Psi \rangle$.
The resulting bondlink dimension is the original one squared: $D' = D^2$.
We can now exploit that this new network
has no closed loop, then we start contracting from \emph{terminal nodes}, i.e. nodes that have correlation
number 1 (a finite graph without close loops has always two terminal nodes at least).
Contracting a terminal node has a cost in elementary operations that scales like $D^{2k}$,
and i have to repeat this operation $\sim V$ times. Thus the overall cost of the
full contraction is $\sim V D^{2k}$, and by adding the expense to obtain the double-tensors
$\mathcal{W}^{[\mu]}$ we recover \eqref{eq:Noloopcost}.

It is important to notice that such procedure was explained disregarding the ordering
of sites, so it works for Tensor Network states of spins and bosons, but it is not so
trivial for fermions, since link-exchange operators are not separable operators. Nevertheless,
it was pointed out in \cite{VerstraTree} that this very scheme can be generalized to
the Fermi case, with equally-scaling efficiency rates, by adopting $\mathbb{Z}_2$-symmetric
tensors in the network, with practically no loss of generality (see appendix \ref{app:symchap}).

\subsection{Peripheral gauge} \label{sec:peripheral}

We gain a remarkable computational speed-up by manipulating with gauges
a Tensor Network state without closed loops in its structure. Assume, for instance,
that we want to acquire the expectation value of an operator $\Theta$ having support
on a restricted set of sites $\{\alpha_{t}\}$. Now, we define the nucleus $\Pi$ as the smallest connected subgraph of the
network containing all the external links $\{\alpha_{t}\}$. Then it always exist
a gauge for which the number of operations to obtain $\langle \Theta \rangle$ is
\begin{equation} \label{eq:Periphcost}
 \# \mbox{cost} \sim 2\:\#(\Pi) \, D^{2k},
\end{equation}
instead of $2V D^{2k}$, where $\#(\Pi)$ is the number of tensors in subgraph $\Pi$. This is definitely
an interesting improvement, especially when the support of $\Theta$ is very local with respect
to the network geometry, i.e. when $\#(\Pi) \ll V$.

This is the same principle that, in OBC-MPS context, lead us to \eqref{eq:MPScostcompact}; and the definition
is totally similar. Let us briefly sketch the idea on how to convert the tensors \emph{not} belonging
to $\Pi$ into the \emph{peripheral} gauge, that eliminates them automatically (with no need of numerical calculus)
when acquiring $\langle \Theta \rangle$. First, let us associate to every tensor in $\Pi^{\text{c}}$, the complementary
subgraph of $\Pi$, a distance: the graph distance to the nucleus $\Pi$. Then, starting to the tensors having highest
distance, we perform Singular Value Decompositions
\begin{equation}
 T^{[\alpha]}_{\{q\},\{p,s\}} = \sum_{\gamma} U^{[\alpha]}_{\{q\},\gamma}
 \;\lambda_{\gamma}\; V^{[\alpha]}_{\gamma,\{p,s\}},
\end{equation}
where $q$ (resp. $p$) indices are related to links that decrease (increase) distance, 
i.e. that lead towards (far from) the nucleus $\Pi$, and $s$ are physical link indices.
Then we damp all the singular part $ U\cdot\lambda$ of every tensor in
$\Pi^{\text{c}}$ into the only linked tensor with shorter distance
(this is a gauge transformation mapping $T \to V$). This is done recursively, until
all the singular part has been embedded into $\Pi$. And since all the periphery $\Pi^{\text{c}}$ is isometric,
it cancels out analytically when contractions with $\langle \Psi |$ are made.

It is easy to see that only for non-closed loop Tensor Network a peripheral gauge can be
defined regardless of the support of $\Theta$. if $\Pi^{\text{c}}$ contained a loop then
there is no gauge that contracts automatically the branch. This is what happened exactly
for periodic MPS, whose single loop is sufficient to break down the definition of a peripheral gauge.

\section{Energy minimization techniques}

Performing the variational step towards the ground state is often the bottleneck
of any simulation algorithm. In Tensor Network ansatze, a good procedure of
minimization makes the practical difference between better and worse network designs.
However, there are some common ideas involving variational algorithms around TN:
first, the globular structure of a state allows us to treat various tensors in
the network as independent variables, and therefore we are highly encouraged
to minimize one (or few) of them at a time, while keeping the other ones fixed.
This, of course, is possible as long as we do not require some homogeneity constraint
among tensors: we might, for instance, wish to fix some tensors to be identical,
e.g. due to the presence of a global symmetry (like homogeneous periodic
MPS under Translational invariance); in this case the minimization of various network nodes
would be simultaneous. Then we should treat the two cases separately.

\vspace{.5em}
\emph{\textbf{No homogeneity constraint -}} if we are free to adjust every
tensor independently of the others, then addressing the problem is simple because the
Lagrangian functional is always quadratic on the single tensor we choose to variate
$\mathcal{L} = \langle \Psi | H |\Psi\rangle - \varepsilon \langle \Psi | \Psi\rangle$.
In practice, we could contract the whole network except for the tensor $T$ of interest,
which appears twice in the expression, once as $T$ and once as $T^{\star}$,
similarly to the PBC-MPS case. Then
$\mathcal{L}(T, T^{\star}) = \langle \!\langle T | \mathcal{H} - \varepsilon \mathcal{N}| T \rangle \! \rangle = $
\begin{equation} \label{eq:qulan}
\begin{overpic}[width = 320pt, unit=1pt]{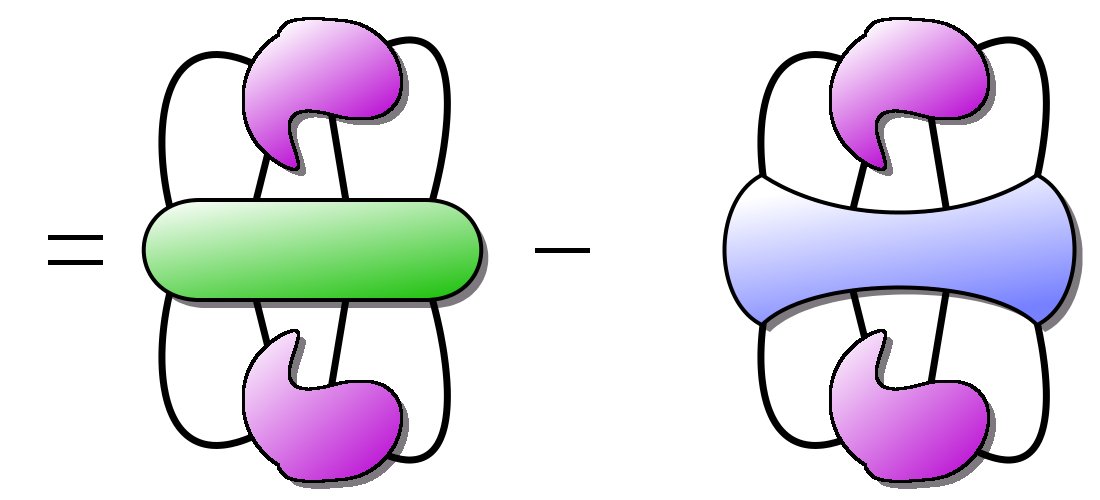}
 \put(-15, 64){\Huge $\mathcal{L}$}
 \put(184, 64){\Huge $\varepsilon$}
 \put(85, 68){$\mathcal{H}$}
 \put(90, 121){$T$}
 \put(88, 16){${T^{\star}}$}
 \put(262, 121){$T$}
 \put(260, 16){${T^{\star}}$}
 \put(257, 68){$\mathcal{N}$}
\end{overpic}
\end{equation}
where the effective Hamiltonian $\mathcal{H}$ is hermitian, and the effective square norm is
a positive operator $\mathcal{N}$. This is due to the fact that partially contracting
the Tensor Network around an operator $O$ is formally equivalent to
\begin{equation}
 \mathcal{M}(O) = \sum_{\{s\}} \left(
 \sum_{\{q\}} \prod M^{\{q\}}_{\{s\}} \right) \cdot O \cdot
 \left( \sum_{\{p\}} \prod M^{\dagger\, \{p\}}_{\{s\}} \right)
\end{equation}
but this is a completely positive map (see appendix \ref{app:cptchap}), which
preserves operator positivity and hermiticity.
Then we are allowed to address the Lagrangian problem \eqref{eq:qulan} as a generalized
eigenvalue problem $\mathcal{H} |T\rangle = \varepsilon \mathcal{N} |T\rangle$, whose
minimal $\varepsilon$ solution defines the optimal $T$.

\vspace{.5em}
\emph{\textbf{With homogeneity constraint -}} when two or more tensors in the network
are chosen a priori to be identical, but not fixed, then the problem becomes more complicated
as the Lagrangian reads
\begin{equation} \label{eq:nonqulan}
 \mathcal{L}(T, T^{\star}) =
 \left( \bigotimes_{j}^{w} \;\langle \!\langle T |_j  \right) 
 [ \mathcal{H} - \varepsilon \mathcal{N} ]
 \left( \bigotimes_{j}^{w} | T \rangle \! \rangle_j \right).
\end{equation}
where $w$ is the number of times the same variational tensor is repeated in the network structure.
The problem \eqref{eq:nonqulan} can be approached with various numerical methods \cite{MatteoMera};
those that gathered most interest in literature are the following
\begin{itemize}
 \item \emph{Gradient methods} - this idea involves guessing random
  variations to the original tensor by using the Lagrangian gradient
  $\partial \mathcal{L}(T,T^{\star}) / \partial T$ (or the conjugate gradient, which
  converges usually faster) as favored direction of random walk.
 \item \emph{Linearized problem} - here we try to treat $\mathcal{L}$ as a linear functional
 $\mathcal{L} = \langle \!\langle Q_{(T,T^{\star})} | T \rangle\!\rangle$ whose solution is
 found immediately by polar decomposition. Doing so is very cheap and fast-paced
 but convergence is not guaranteed in general.
 \item \emph{Imaginary time evolution} - Another way of proceeding is starting from a random
 state and cooling down the system by applying $e^{-\beta H}$ to it, for $\beta \to \infty$ the
 state thermalizes at zero temperature, i.e. is the ground state. But writing an unperturbative
 exponential of the Hamiltonian operator $H$ is usually hard task, and not always suitable
 to match the Tensor Network geometry.
\end{itemize}

This mostly concludes the general features of Tensor Network states that we wanted to discuss
in this chapter. Before in-depth investigating the peculiar properties of the famous class
of Hierarchical Tensor Networks (Tree networks, Multiscale entanglement renormalization ansatz)
we focus on some other archetypal examples that are worth mentioning.

\section[PEPS]{Example I:\\Product Entangled Pair States (PEPS)} \label{sec:PEPS}

Product Entangled Pair states are the natural generalization at dimensionality $> 1$
of Matrix Product States. The idea behind their formulation is to adapt the valence bond picture of MPS
to 2D or 3D systems. We discussed in section \ref{sec:valencebond} that an MPS can be seen as an on-site
transformation applied to a starting, virtual state made by several maximally entangled pairs $|\Phi^{+}\rangle =
\sum_{\alpha}^D |\alpha \alpha \rangle$, (with $D$ is arbitrary): every pair belonging
to a physical bond in the lattice.

It is clear that this very construction is meaningful for any dimensionality, and for every lattice
we can think of as long as nearest neighboring sites are well-defined. For instance, when we apply this
framework on a square lattice we obtain:
\begin{equation} \label{eq:PepsVB}
\begin{overpic}[width = 300pt, unit=1pt]{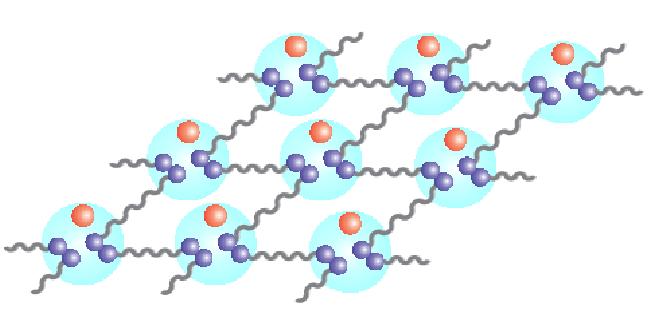}
 %\put(-15, 64){$\mathcal{L}$}
\end{overpic}
\end{equation}
(picture taken from MPQ-Garching)
where red dots are physical sites. Every site has four bonds, and every bond carries an entangled pair.
Then a linear transformation maps the $D^4$-dimensioned
fictitious degree of freedom into the $d$-dimensioned physical one, that defines uniquely the PEPS.

Since PEPS are fully parametrized by such local transformation, they are actually a Tensor Network
$\{L,k,D\}$, with a number of tensors equal to the total number of sites $L$, correlation number
equal to the number of neighbors per site plus 1 for the physical link
(e.g. $k = 5$ for a square lattice), and arbitrary bondlink dimension $D$.
Their TN representation reads
\begin{equation} \label{eq:PepsTN}
\begin{overpic}[width = 270pt, unit=1pt]{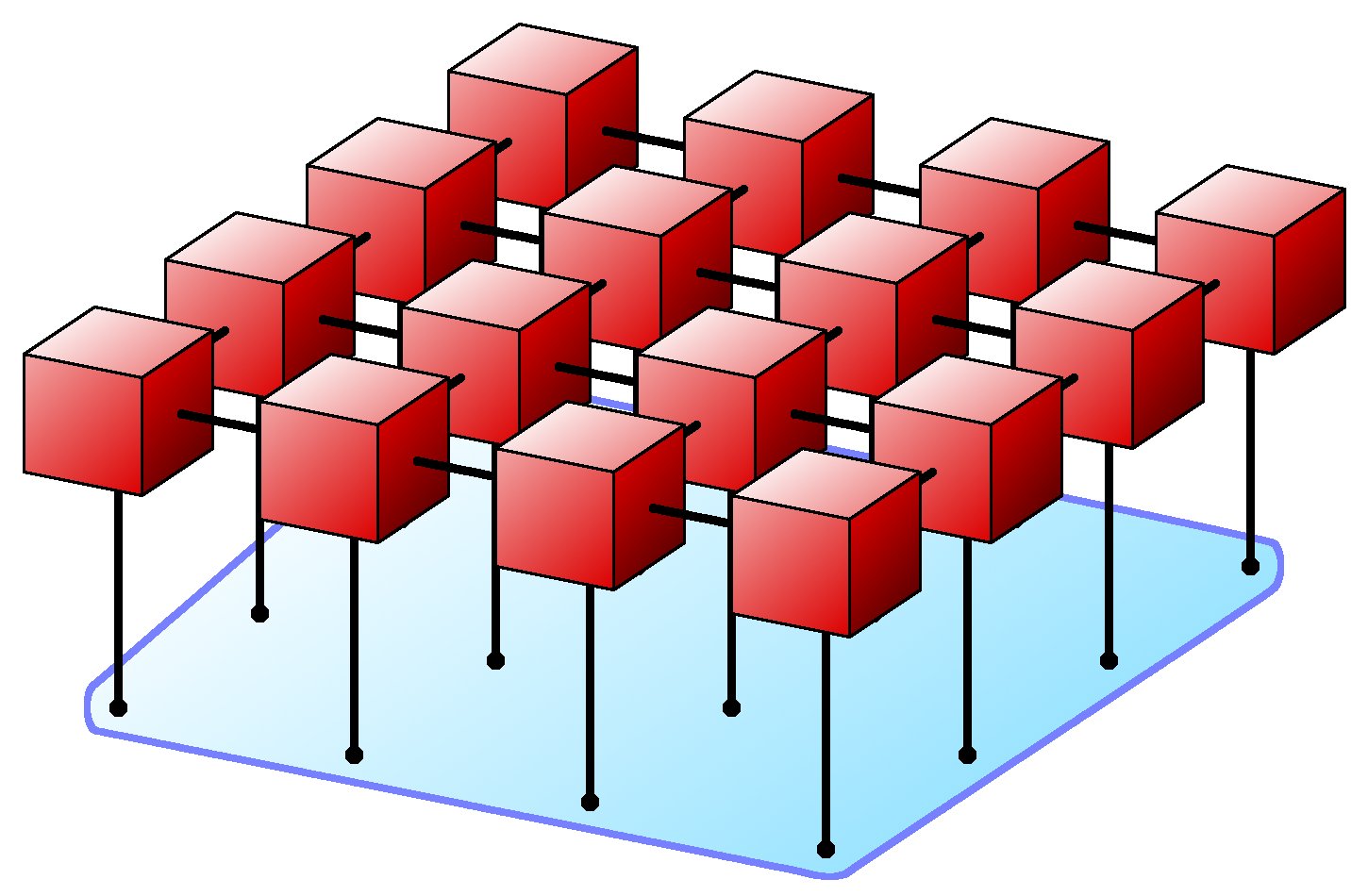}
 %\put(-15, 64){$\mathcal{L}$}
\end{overpic}
\end{equation}
for a square lattice, where vertical links are physical indices. Red boxes are tensors
$A^{[\mu]s}_{q_1,q_2,q_3,q_4}$, where $\mu$ is the site, $s$ the local canonical level index,
and $q_1 \ldots q_4$ the four correlation indices.

The reason why PEPS gathered so much interest in the latest years \cite{PowaPEPS, SchuchPEPS},
is that they obey the exact entanglement area law for their geometric dimensionality;
e.g. a 2D PEPS satisfy a 2D area law of entanglement.
It is quite easy to check it, by using the TN-entanglement bound rules we derived in section
\ref{sec:entanet}. Consider, for instance, a connected subset of sites in the square lattice,
and the entanglement entropy related to the reduced density matrix: we mentioned that this
entanglement is bound by the minimal number of links one has to cut to separate the graph
into two subgraphs (times $\log D$). Then one sees immediately that this minimal number
coincides with the \emph{perimeter} of the subset of sites. But that is
the 2D area law: for a region scaling like a surface $\sim \ell^2$, the entanglement,
\begin{equation}
 \mathcal{S}_{\text{VN}} \sim \ell \log D,
\end{equation}
scales with the characteristic lenghtscale $\ell$ of the region. This suggests
that PEPS are optimal tools to approximate ground states of short-ranged Hamiltonians,
even because, similarly to their one-dimensional cousins MPS, they also yield a completeness theorem.

PEPS have also a huge drawback when compared to MPS, which is their computational complexity \cite{HardPEPS}.
Namely, if we want to contract \emph{exactly} a PEPS, e.g. for acquiring an expectation value,
the number of elementary operators needed scales exponentially with the system size, no matter the
contraction scheme. This is due to the presence of an extremely high amount of 
loops in the network, not an easy trouble to work around.
Of course one can deal with the problem approximately, by, for example, renormalizing the width
of the resulting composite bondlink during contraction; still, we must be careful that the
approximation we are adopting should not break down the area law principle. To this
goal, several proposals have been introduced in literature \cite{PEPSgo, Maricarmen}.

\section[Correlator Product States]{Example II:\\Correlator Product States (CPS)}

Correlator product states \cite{CPStrong, CPSplaq, CPScompletegraph,CPS2D} are a simple, but clever, way
to adjust correlations in many body states, by means of a whole abelian algebra made of ranged
canonical weight-factor operators. They have been acknowledged and used long since;
in some sense, they can be regarded as the completely variational generalization of Jastrow factors.

They are called 'states', but it would be more suitable to regard correlator product elements
as operators. Indeed they hardly stand alone as a mere Tensor Network structure, because as
they must be finite-support and commuting, they are forbidden to establish global symmetries,
even the simplest ones like particle number conservation.
On the positive side, they excel
in preserving symmetries already present in the state prior to their application as operators,
so they are often used in conjunction to a starting ansatz:
\begin{equation} \label{eq:CPS}
 |\Psi_{\text{CPS}}\rangle = \sum_{s_1 \ldots s_L = 1}^{d}
 \left( \prod_{\alpha} C^{[\alpha]}_{ s_{v(\alpha,1)} \ldots s_{v(\alpha,k)} } \right)
 \Phi^{0}_{s_1 \ldots s_L} |s_1 \ldots s_L\rangle.
\end{equation}
In this representation, $|\Phi^{0}\rangle = \sum \Phi^{0}_{s_1 \ldots s_L} |s_1 \ldots s_L\rangle$ is
exactly the starting ansatz state, capable of controlling desired symmetries; e.g.
nonzero $\Phi^{0}_{s_1 \ldots s_L}$ components should be only those with the right particle number.
It is clear that correlator product factors $C^{[\alpha]}$ can not break such symmetry,
since sectors with zero amplitude will stay zero amplitude.
Nevertheless, $C^{[\alpha]}$ can build correlations, which is especially useful when
$|\Phi^{0}\rangle$ is a somehow uncorrelated trial state, like a Slater determinant
(in particular, see the MPS representation for Slater determinants in section \ref{sec:Slater}).

Notice two facts involving \eqref{eq:CPS}: first, the action of correlator product factors $C^{[\alpha]}$
can be applied in any ordering, because they are all diagonal in the canonical basis,
so they must commute as they share a basis of eigenvectors altogether.
Secondly, \eqref{eq:CPS} is clearly a Tensor Network, since all the relations between the various
$C^{[\alpha]}$ and $\Phi^{0}$ are linear; so we can use arguments treated in this chapter to study CPS
(e.g. to guess entanglement bounds). You could argue that such picture does not match exactly the definition
of TN we gave in \eqref{eq:quaqua}, since the same index is shared by more tensors than merely
two, but this is easily worked around by adding triple Kronecker delta nodes in the network
$\delta^{[3]}_{i,j,k} = \delta_{i,j} \delta_{j,k}$.

Precisely, consider a two-site diagonal operator $\Theta$, defined as
\begin{equation}
  \Theta = \sum_{s_1, s_2 = 1}^{d} C^{[1\leftrightarrow 2]}_{s_1 s_2} \; |s_1 s_2\rangle \langle s_1 s_2|
  = \!\!\!\sum_{s_1, s_2, r_1, r_2, k_1 k_2}^{d}\!\!\!
  \delta^{[3]}_{s_1,r_1,k_1} \delta^{[3]}_{s_2,r_2,k_2} C^{[1\leftrightarrow 2]}_{k_1 k_2}
  |s_1 s_2\rangle \langle r_1 r_2|
\end{equation}
then we can exploit its diagonal nature to decompose its operator representation even
in the diagrammatic scheme
\begin{equation} \label{eq:DiagOp}
\begin{overpic}[width = 280pt, unit=1pt]{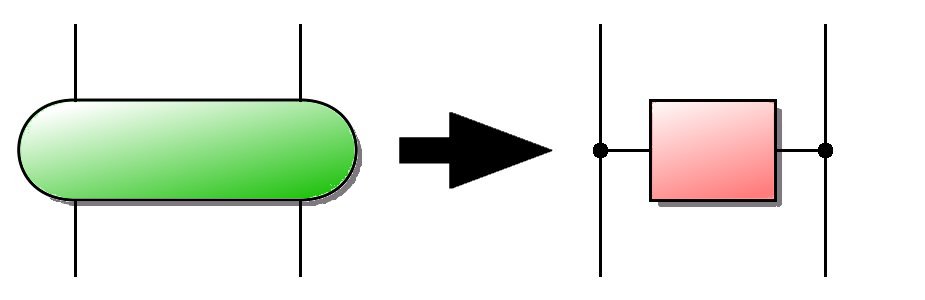}
 \put(52, 40){$\Theta$}
 \put(201, 40){$C^{[1\leftrightarrow 2]}$}
 \put(26, 6){$s_1$}
 \put(26, 81){$r_1$}
 \put(94, 6){$s_2$}
 \put(94, 81){$r_2$}
\end{overpic}
\end{equation}
where the black nodes are $\delta^{[3]}$ tensors; so the network structure of \eqref{eq:CPS} is well-defined.
Of course, correlation capabilities of CPS depend strongly on the
specifics of chosen $C^{[\alpha]}$ factors:
\begin{itemize}
 \item \textbf{Involved sites} $(k)$ - any number of sites can be chosen to appear as indices in a
 single $C^{[\alpha]}$ factor.
 The amount of variational descriptors and thus the speed of simulation algorithm
 is extremely sensitive to this parameter (usually scaling exponentially with $k$).
 \item \textbf{Range} $(\ell)$ - fixed the number of involved sites per correlator $C^{[\alpha]}$ we can still
 choose the maximal distance among such sites for which we require the corresponding $C^{[\alpha]}$ to be present
 in the product.
 \item \textbf{Shape} - if we are dimension higher than 1, not only the number of involved sites per
 factor, but also their shape in the lattice is relevant. For example, a common fashion for square lattice is work with
 plaquette correlator product \cite{CPSplaq}, where every $C^{[\alpha]}$ involve the $k=4$ vertices of a lattice square
 (or a rescaled lattice square).
\end{itemize}
At any rate, CPS are are considered successful variational tools, for their ridiculously small number
of variational parameters, high manipulability, and the capability of adjusting any single
correlation by adding just one suitable factor to the product.
Choosing a suitable starting state $|\Phi^{0}\rangle$ is also a delicate issue for the CPS ansatz.
For instance, a MPS (with modest bondlink $D$) could be a suitable choice;
this lead to the construction of hybrid MPS$\leftrightarrow$CPS Tensor Network designs,
also known in literature as \emph{Renormalization Algorithm with Graph Enhancement} (RAGE) \cite{rage1, rage2}.

\subsection{CPS $\Rightarrow$ MPS/PEPS} \label{sec:MPJastrow}

We would like to point out an interesting fact about CPS. a Correlator Product having a finite
range $\ell$ of factors which does not scale with system size $L$ clearly describes a finitely-correlated
state: this tells us that finite-range CPS can be put in relation with MPS and PEPS, or,
more precisely, with Matrix Product Operators and Product Entangled Pair Operators
(PEPO - operatorial version of PEPS). A result of my personal thesis work, is that when $\ell$ is finite
is always possible to represent a CPS as an analytical subclass of MPS/PEPS with finite correlation
bondlink $D$ being a function of $\ell$. Such construction obviously depends on the involved
sites per factor and their shape, so we are here going to present the mapping
CPS $\to$ MPS in the simplest setting: 1D, and binary factors $k = 2$, kept at all ranges $\ell'$
from neighbors up to an arbitrary finite bound $2 \leq \ell' \leq \ell$.

Let us give this prescription starting from $\ell = 1$ and then increasing  $\ell$:
\begin{equation} \label{eq:Jas1}
 \sum_{s_1 \ldots s_L} \left( \prod_{j}^{\tilde{L}} C^{[1,j]}_{s_j s_{j+1}} \right)
 |s_1 \ldots s_L \rangle \langle s_1 \ldots s_L |,
\end{equation}
where we are assuming that $C^{[1,j]}$ may depend on the pair of sites $\{j, j+1\}$ on which it is acting,
to keep major variational freedom. Also, boundary conditions are small issue: they only set
the upper bound $\tilde{L}$ for the $j$ over which the product is taken (to $L$ for PBC, $L-\ell'$ for OBC).
Let us draw the diagrammatic representation of \eqref{eq:Jas1}, given by
\begin{equation} \label{eq:Jas1pic}
\begin{overpic}[width = \textwidth, unit=1pt]{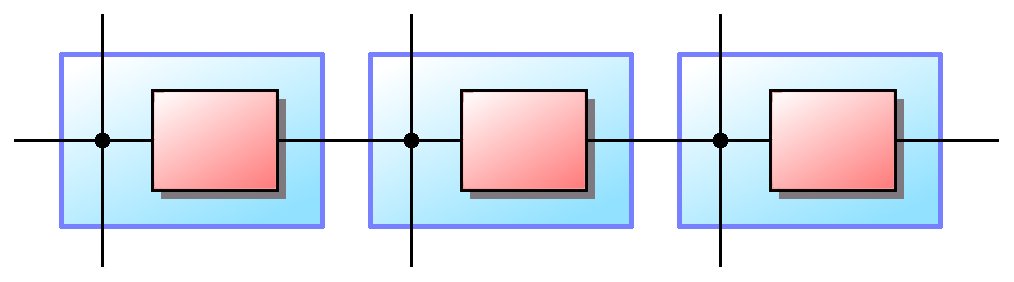}
 \put(72, 50){$C^{[1,j]}$}
 \put(82, 92){$O^{[j]}$}
 \put(185, 50){$C^{[1,j+1]}$}
 \put(304, 50){$C^{[1,j+2]}$}
\end{overpic}
\end{equation}
As you see, it is possible to embed the correlator within an MPO structure $O^{[j]}$ where
the product matrices correspond to the blue boxes in \eqref{eq:Jas1pic}, or equivalently
\begin{equation}
 O^{[j] r}_{s} = \sum_{s_{j+1}} \delta_{r,s}\; C^{[1,j]}_{s, s_{j+1}} \,|s)(s_{j+1}|
 = \sum_{\alpha, \beta} \; \delta^{[3]}_{\alpha,r,s}\; C^{[1,j]}_{\alpha, \beta} \;|\alpha)(\beta|.
\end{equation}
Now let us move to $\ell = 2$, where next-nearest neighboring correlator factors
$C^{[2,j]}$are introduced. Then the overall correlator product operator is expressed by:
\begin{equation} \label{eq:Jas2pic}
\begin{overpic}[width = \textwidth, unit=1pt]{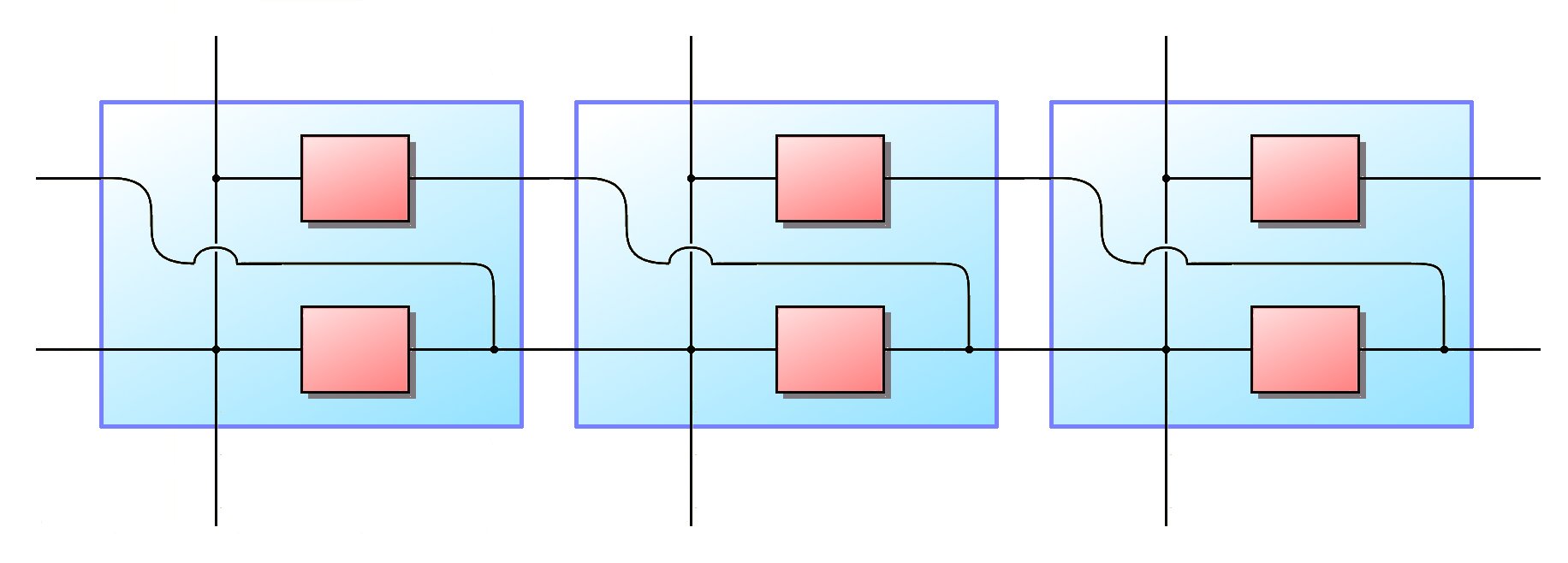}
 \put(78, 49){\footnotesize$C^{[1,j]}$}
 \put(78, 92){\footnotesize$C^{[2,j]}$}
 \put(103, 119){\footnotesize$O^{[j]}$}
\end{overpic}
\end{equation}
It is easy to check that the circuitry of links and delta nodes performs the right
index plugging at the correlator factors. Indeed, focus on $C^{[2,j]}$ in the picture,
while it is obvious that its left link corresponds to $s_j$, you can similarly
follow the right link and verify that it is exactly $s_{j+2}$.
So the Matrix Product Operator block $O^{[j]}$ is also correct, as it contains
all the inner structure (deltas) and variational information (correlators) needed.

Equation \eqref{eq:Jas2pic} has a vertical pattern that is suitable to be copied and repeated,
every time we add a layer, it is equivalent to add a new set of product correlators,
effectively enhancing the maximal range $\ell$ by one. So if our purpose is to describe
the action of
\begin{equation} \label{eq:Jasel}
 \sum_{s_1 \ldots s_L}
 \left( \prod_{\ell' = 1}^{\ell} \; \prod_{j}^{\tilde{L}} C^{[\ell',j]}_{s_j s_{j+\ell'}} \right)
 |s_1 \ldots s_L \rangle \langle s_1 \ldots s_L |,
\end{equation}
then the corresponding Matrix Product Operator representation would read
\begin{equation}
 O^{[j] r}_{s} = \sum_{\{\alpha\}, \{\beta\}}
 \left( \prod_{\ell' = 1}^{\ell} C^{[\ell',j]}_{s,\beta_{\ell'}} \right)
 \delta^{[3]}_{\alpha_1,s,r} \left( \prod_{\ell' = 1}^{\ell-1} \delta_{\alpha_{\ell'+1}, \beta_{\ell'}} \right)
 |\alpha_1 \ldots \alpha_{\ell})(\beta_1 \ldots \beta_{\ell}|
\end{equation}
For better comprehension, let us sketch it for $\ell = 4$:
\begin{equation} \label{eq:Jastrowpic}
\begin{overpic}[width = 200pt, unit=1pt]{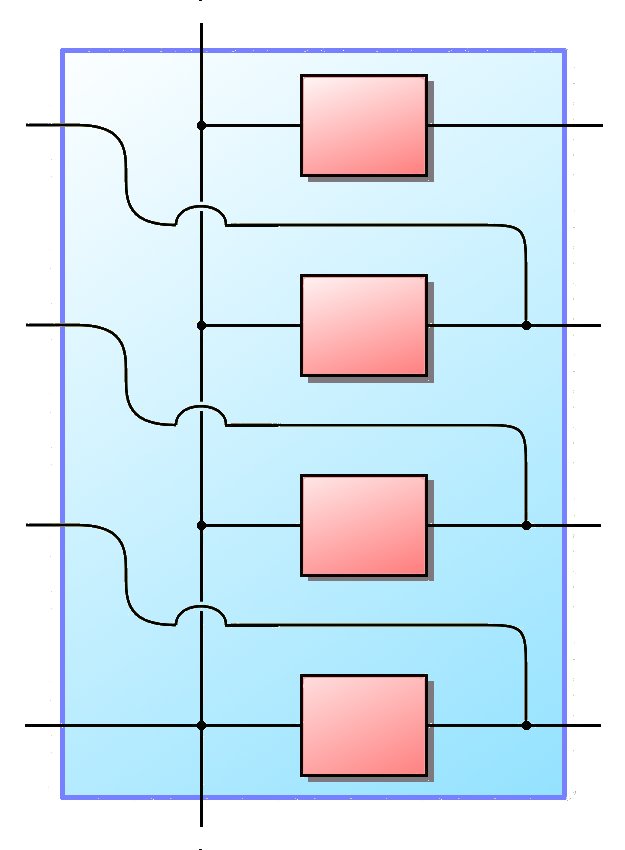}
 \put(106, 36){$C^{[1,j]}$}
 \put(106, 100){$C^{[2,j]}$}
 \put(106, 164){$C^{[3,j]}$}
 \put(106, 228){$C^{[4,j]}$}
 \put(146, 260){$O^{[j]}$}
\end{overpic}
\end{equation}
The total correlation bondlink dimension used for this representation is $d^{\ell}$,
it should be convincing that this representation is also optimal (if no further
restraints on the $C$ factors are requested), since we are transferring through the
MPO bondlink the minimal information to reproduce the factors exactly.

In conclusion, we can establish a general entanglement bound on 1D binary Correlator
Product States, which is
\begin{equation}
 \mathcal{S}_{\text{VN}} \leq \ell \log d,
\end{equation}
characterizing the correct 1D area law of entanglement if the maximal range $\ell$ does not scale with
the total length $L$.

The scheme we just presented has been originally conceived by me during my doctorateship work, and supported
by R.~Fazio and F.~Becca.

\section{Towards hierarchical Tensor Networks}

Tensor Network architectures can be put in tight relation with numerical renormalization groups.
In fact, consider a tensor $T^{[\cdot]}$ attached to some of the physical indices; it can be as well
interpreted as the action of a linear transformation acting on the local density matrix,
mapping it into the virtual links space. As the overall effective dimension is typically reduced, a numerical renormalization
is taking place. Tensor Networks entanglement bounds guarantee that the amount of correlation in
the TN tailored variational state matches the entanglement that can be built via renormalization process.

We want now to analyze and describe detailed properties of another renown class of Tensor Networks,
corresponding to the original Wilson's real-space numerical RG. We are talking about
Tree networks, and of course also their recent generalization: Multiscale Entanglement Renormalization Ansatz (MERA).
These two network geometries share the intriguing property of embedding a scale-invariance in
their pattern, so they are ideal candidates for dealing with critical models.
We will classify them together as \emph{hierarchical} Tensor Network states because network
nodes are linked according to hierarchical relations.

The interest revolving around these methods, and the following in-depth analytical study,
are such that it is appropriate to devote an entire chapter to describe their features and peculiarities.

\chapter{Trees and MERA} \label{chap:TTNMERA}

When the Matrix Product State representation as tailored wavefunction paradigm version
of the DMRG was realized and understood, it was an important breakthrough.
But it was with the advent of Tree Tensor Networks (TTN)
\cite{VidalTree1, IoTree}
and Multiscale Entanglement Renormalization Ansatz (MERA)
\cite{MERAzero, MERAalgo, MERAevo, MERAgauge, MERAtopo, QuMERAchan, BERA}
that the computational physicists' community started
talking about Tensor Networks in general. TTN and MERA, similarly to MPS, have a
network pattern which is not only very simple and highly adaptive but also self-similar.
Nevertheless, the relevant difference between TTN/MERA and MPS is that, while
self-similarity in MPS arises every time a new site is added, for (binary) TTN/MERA
it happens every time the number of sites is \emph{doubled}
(or tripled for ternary TTN/MERA, and so forth).
This suggests us that, as we will see, TTN/MERA bear somehow a resemblance to real-space renormalization
processes, and also that they should be especially suitable for describing systems in which
a scale invariance is emergent.
Most of the results provided in this chapter are original work developed by me and was supported
by V.~Giovannetti, M.~Rizzi, S.~Montangero and R.~Fazio.

\section{Real-Space numerical renormalization}

In section \ref{sec:dmrgtomps} we argumented thoroughly how a Matrix Product State can be explicitly
built from the data of the recursive (standard) density matrix renormalization transformations. Let us
follow a similar path now, but with a different starting algorithm in mind. For consistence we will
work in 1D.

We are again assuming that numerical renormalization process is being performed,
but not according to the traditional DMRG framework, where at every step the degree of freedom of
a single site is added to the picture, and the joint density matrix
is then renormalized, i.e. old block $\cup$ added site $\to$ new block
($D \otimes d \to D$).
Instead, here we are assuming that the renormalized degree of freedom of the old block is
doubled, or equivalently, coupled with a copy of itself. Now we select
a new state within this space (according to whatever benchmark we prefer),
describe its density matrix, and then renormalize,
i.e. old block $\cup$ another old block $\to$ new block
($D \otimes D \to D$).
What inspires this proposal is that during the same time we perform a density matrix
renormalization step, 1D lattice sites in the real-space are also renormalized, in a coarse-graining fashion.

Let us now recall equation \eqref{eq:wata}, used in a standard-DMRG, while we are right-propagating
the scheme, and keeping $D \times D$ renormalized density matrices
$\tilde{\rho}_{\ell} = \sum_{j=1}^{D} p_j \;|L_j\rangle \langle L_j|$.
The basic step of the algorithm could be expressed through $A^{[\ell]}$ as
\begin{equation}
 |L_j\rangle_{\ell}^{L} = \sum_{k = 1}^D \sum_{s = 1}^d 
 A_{k,j}^{[\ell] s} \; |L_k\rangle_{\ell-1}^{L} \otimes |s\rangle_{\ell}.
\end{equation}
and $A^{[\ell]}$ became the elementary tensor of MPS description.
Let us write the corresponding recursive expansion for real-space numerical RG:
if no translational invariance assumption is made (for instance, we could be in an OBC setting)
we must keep track of every $\tilde{\rho}^{[h]}_{\ell}$, where $h$ tells us
how many times the renormalization has been performed already, and $\ell$ carries the
information on which original sites the density matrix is actually describing.
With these considerations, and defining
$\tilde{\rho}^{[h]}_{\ell} = \sum_{j=1}^{D} p_j \;|M_j\rangle^{[h]}_{\ell} \langle M_j|$, we get
\begin{equation}
 |M_j \rangle_{\ell}^{[h]} = \sum_{k_1, k_2 = 1}^{D_{h-1}}
 \Lambda_{k_1, k_2}^{[h,\ell]\, j} \;\, |M_{k_1} \rangle_{2\ell-1}^{[h-1]} \otimes
 |M_{k_1} \rangle_{2\ell-1}^{[h-1]}.
\end{equation}
where $h$ goes from $0$ to $\bar{h} = \log L$, with $L$ the system size, while
$\ell$ ranges from $1$ to $2^{-h}L$ (notice that the number of allowed $\ell$ values halves at every
layer $h$ increase, which is exactly the coarse graining).
Finally, $|M_j \rangle_{\ell}^{[h]}$ goes from $1$ to the renormalization dimension $D_{h}$,
eventually depending on $h$.

You should convince yourself that with this procedure, the state is completely and
uniquely characterized by the set of $\Lambda^{[h,\ell]}$, which is a three indices complex tensor.
Since the global state $|\Psi\rangle$, when written in the canonical separable basis, has components
determined by linear contraction relations between the $\Lambda^{[h,\ell]}$, it is a Tensor Network state.
Precisely, it reads
\begin{equation}
 |\Psi\rangle =
 \left( \bigotimes_{\ell = 1}^{2^{\bar{h} -1}} \Lambda^{[1,\ell]} \right) \cdot
 \left( \bigotimes_{\ell = 1}^{2^{\bar{h} -2}} \Lambda^{[2,\ell]} \right)
 \ldots
 \left( \bigotimes_{\ell = 1}^{2} \Lambda^{[\bar{h}-1,\ell]} \right)
 |\mathcal{C}\rangle,
\end{equation}
where we wrote the $\Lambda$ as $D_h^{2} \times D_{h+1}$ matrices. Equivalently,
in the diagrammatic Tensor Network representation it appears like this
\begin{equation} \label{eq:TN2pic}
\begin{overpic}[width = \textwidth, unit=1pt]{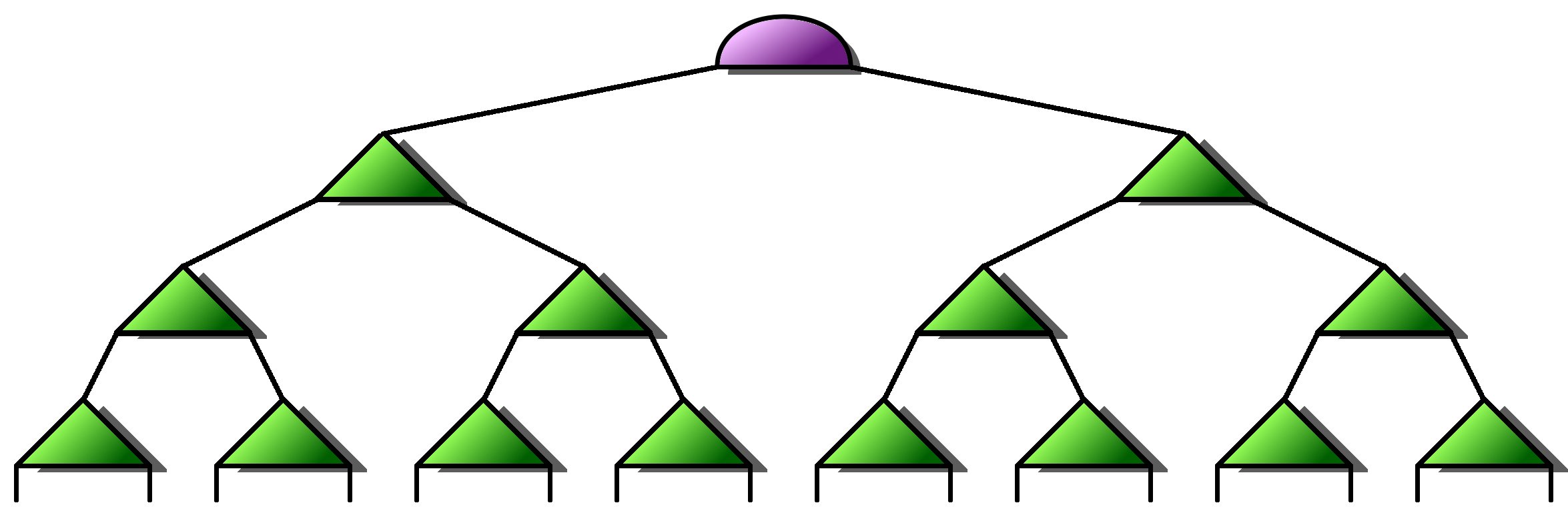}
 \put(-2, 27){\footnotesize $\Lambda^{[1,1]}$}
 \put(16, 55){\footnotesize $\Lambda^{[2,1]}$}
 \put(115, 54){\footnotesize $\Lambda^{[2,2]}$}
 \put(63, 85){\footnotesize $\Lambda^{[3,1]}$}
 \put(170, 116){\footnotesize $\mathcal{C}$}
\end{overpic}
\end{equation}
as you see it is a tree graph, with branching number $b = 2$. The tensor $\mathcal{C}$ standing at
the top of the structure is the only one which is topologically different, as it has
only two links, and due to its placement is often referred to as \emph{hat} (or \emph{root}) tensor.

From equation \eqref{eq:TN2pic} the coarse graining action performed by renormalizers
$\Lambda$ becomes immediate and clear: every \emph{layer} of $\Lambda$ tensors maps an
adjacent pair of (eventually already renormalized) sites into a single renormalized site,
thus effectively halving the overall size of the system. Once $L$ has been reduced to $2$
we describe it as a simple binary state $|\mathcal{C}\rangle$.

Due to hermiticity of every $\tilde{\rho}^{[h]}_{\ell}$ we would like for their
respective eigenvectors $|M_j \rangle_{\ell}^{[h]}$ to be an orthonormal set, at every $h$ and $\ell$.
As for MPS, where a similar requirement lead to a gauge symmetry breaking, this
restraint translates into a condition that every $\Lambda$ must satisfy, namely
\begin{equation} \label{eq:TTNisometr}
 \delta_{j_1, j_2} = \sum_{k_1, k_2}^{D_{h-1}} \Lambda^{\star\,[h,\ell]\,j_1}_ {k_1, k_2}
 \Lambda^{[h,\ell]\,j_2}_ {k_1, k_2}, \qquad \forall\;\{h,\ell\},
\end{equation}
where $\vphantom{A}^{\star}$ stands for complex conjugation.
In other words, every $\Lambda$, read as a $D^2_{h-1} \times D_{h}$ matrix, must be (left-)
isometric, i.e. $\Lambda^{\dagger} \Lambda = \Id$. This is indeed a gauge symmetry breaking.
Truly, this is exactly the peripheral gauge we defined in section \ref{sec:peripheral},
when the nucleus $\Pi$ corresponds to the hat tensor $\mathcal{C}$.
We remarked that, for a given Tensor Network with no closed loops (as a tree graph is), it is always possible
to find the gauge transformation that maps it into the peripheral gauge, no matter
the starting state: this tells us that the isometricity condition \eqref{eq:TTNisometr} for $\Lambda$
carries no loss of generality at all.

You can guess that the binary character $b = 2$ of the tree network in \eqref{eq:TN2pic}
is due to the fact that we renormalized just two copies of the old block
density matrices into a new one. Of course, mapping
an arbitrary number $b$ of copies of the old block into a single one leads
to tree Tensor Networks with the corresponding branching number $b$.
For example, we could have a ternary tree Tensor Network when $b = 3$:
\begin{equation} \label{eq:TN3pic}
\begin{overpic}[width = \textwidth, unit=1pt]{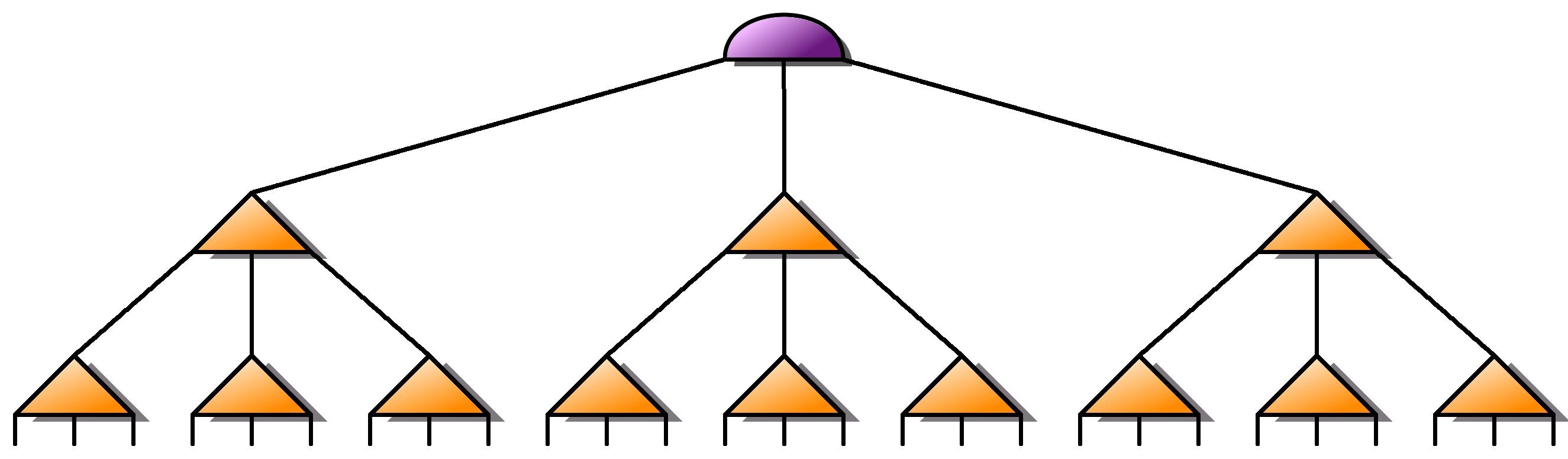}
 \put(0, 27){\footnotesize $\Lambda^{[1,1]}$}
 \put(33, 57){\footnotesize $\Lambda^{[2,1]}$}
 \put(166, 58){\footnotesize $\Lambda^{[2,2]}$}
 \put(173, 102){\footnotesize $\mathcal{C}$}
\end{overpic}
\end{equation}
where the total number of layers $\bar{h}$ would be $\bar{h} = \log_3 L$.
Throughout this chapter we will present various results, deriving our calculations in accordance to
the binary tree Tensor Network (2TTN) design. We want to remark, however, that most of the
claims and statements extend naturally to TTN with higher branching number: typically
one has just to substitute $2$ with $b$ where appropriate 
(usually as a logarithm basis) to get the right result.

Moreover, notice that we disregarded any assumption on boundary conditions so far.
Indeed, unlikely from Matrix Product States, where different boundary conditions
lead to substantially different network topologies (no closed loop in OBC, but with closed loop in PBC),
in TTN representations of a state, like \eqref{eq:TN2pic} and \eqref{eq:TN3pic}, the
network topology is insensitive to the presence of a boundary. This, as we will see,
will naturally bring a unique and well-built definition of the thermodynamical limit.

\section[Tree entanglement and MERA]{Tree network entanglement instability and the introduction of MERA}
\label{sec:MERAintro}
 
If we wish for our real-space self-similar Tree network to be actually capable of describing
1D critical states, we must fist check that the amount of entanglement it can hold
satisfies the typical entanglement area-law violation:
\begin{equation}
 \mathcal{S}_{\text{VN}} (\rho_{\ell}) = \frac{c}{3} \log \ell + \mbox{const.}
\end{equation}
where $\rho_{\ell}$ is the density matrix of $\ell$ adjacent sites, and in PBC conditions.
Also we are assuming that we are approaching the thermodynamical limit, or at least that the region
$\ell$ is too small to capture finite size effects: $\ell \ll L$.

In order to evaluate upper bounds to the entanglement of a TTN, we are going to employ
the Tensor Network entanglement arguments we provided in section \ref{sec:entanet}. There
we found that the entanglement of a partition $\Gamma_{\ell}$ is bounded by $\sim D^{\chi}$ where
$\chi$ is the minimal number of broken links needed to separate the network graph into
two subgraphs, respectively containing $\Gamma_{\ell}$ and its complementary $\Gamma_{\ell}^{\text{c}}$.
Let us perform the count for binary TTN. First, notice that, since the Tree network design is not
translationally-invariant defined, we will expect that $\mathcal{S}_{\text{VN}} (\rho_{\{\ell_1,\ell_1+\ell\}})$
will not depend only on the number of sites $\ell$ in the interval, but also on its placement $\ell_1$.

Now we proceed recursively with layers. If $\ell_1$ is odd, then I have to cut no link in the lowest
layer, if it is even, i cut one link; either way, i move up one layer and site $\ell_1$ is mapped
into renormalized site $\lceil \ell_1 / 2 \rceil$ (where $\lceil x \rceil$ is the smallest integer
which is larger than $x$), and we can repeat the procedure. The same argument holds for the other
region boundary $\ell_2 = \ell_1 + \ell$, even though the link breaking is needed when $\ell_2$ is odd
and not when it is even.
In conclusion we have that $\mathcal{S}_{\text{VN}} (\rho_{\{\ell_1,\ell_1+\ell\}}) \leq \# \mbox{cuts} \cdot \log D$, where
\begin{equation} \label{eq:fluctucut}
 1 \leq \# \mbox{cuts} \leq 2 \bar{h} = 2 \log_b \ell,
\end{equation}
depending on $\ell_1$.
For clarity, let us show an example of the described procedure when $L \geq 16$ ($\bar{h} \geq 4$),
are we are considering $\rho_{3,11}$:
\begin{equation} \label{eq:TN2cut}
\begin{overpic}[width = \textwidth, unit=1pt]{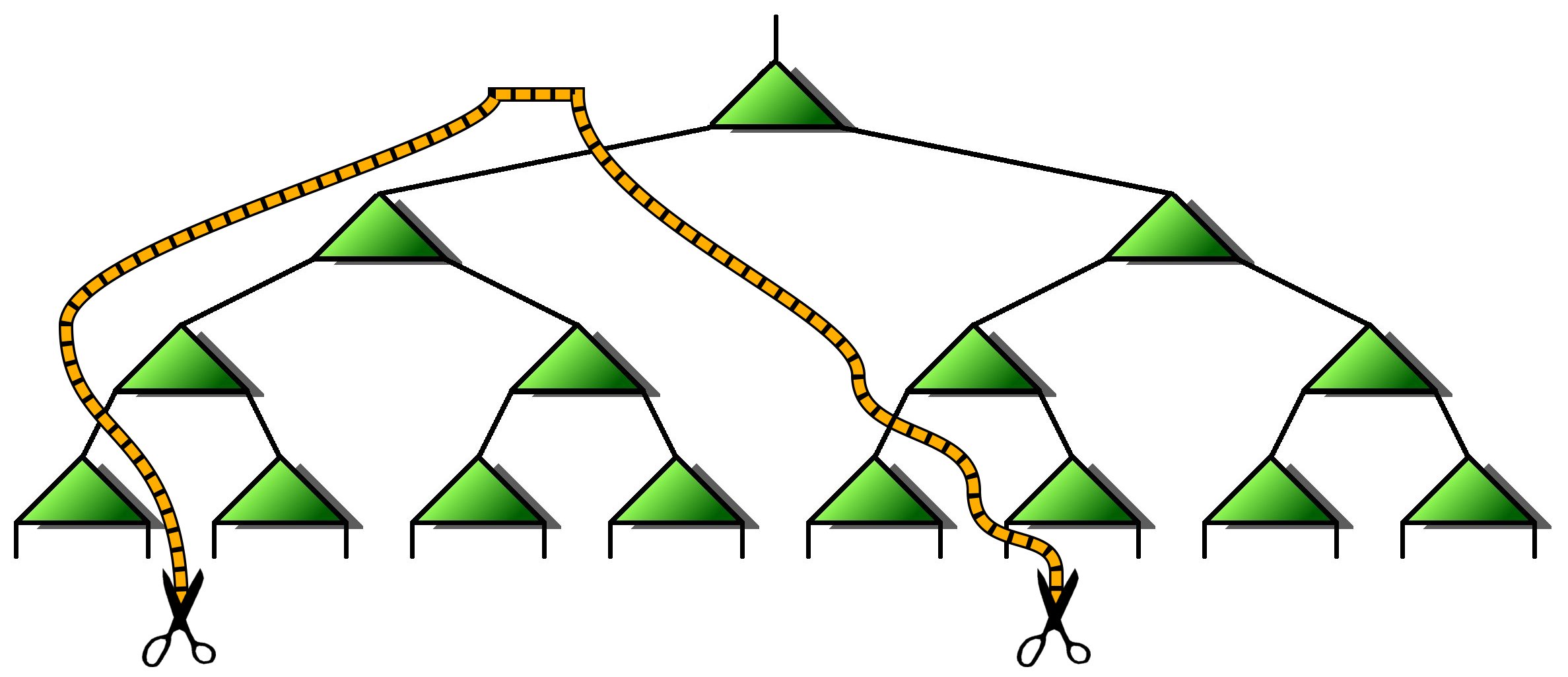}
% \put(0, 27){\footnotesize $\Lambda^{[1,1]}$}
\end{overpic}
\end{equation}
where we have drawn only a branch of the whole Tree network.
In this example the entanglement bound is given by $\sim 4 \log D$
(actually by $\sim [3 \log D + \log d]$ since one of the links to be cut is a physical one).

Result \eqref{eq:fluctucut} is quite relevant: it tells us that it is true that in a TTN state
one can choose a sequence of subsystems of growing size $\ell$ exhibiting the correct area-law violation scaling,
but it is also true that it exists another sequence which is nearly separable.
In other words, we can state that the entangling capacity of a TTN \emph{fluctuates} widely with translations,
ranging between almost no entanglement and critical entanglement. Such \emph{translational instability}
is embedded within the nature of Tree networks themselves, and must be handled with care.
In particular, in order to work around this entanglement instability,
and representing a variational quantum state showing a more smooth area law violation,
two methods are adopted:

\vspace{.5em}
\emph{\textbf{Incoherent translational mixture -}} This path focuses on calculating the
translational average every time we wish to acquire an expectation value on the TTN state.
In practice, let $\Theta_{\{\ell_\alpha\}_\alpha}$ be an observable having support on
sites $\{\ell_\alpha\}_\alpha$, then what we are actually interested in is
\begin{equation}
 \langle \bar{\Theta} \rangle = \frac{1}{L} \sum_{\ell = 1}^{L} \langle \Theta_{\{\ell + \ell_\alpha\}_\alpha} \rangle.
\end{equation}
By doing so we always integrate out translational fluctuations, as we are considering solely the zero
Fourier mode.
This operatorial average procedure is theoretically equivalent to
consider a state which is the incoherent mixture of all possible translations of the original TTN state.
Namely we define
\begin{equation} \label{eq:incoherentrasl}
 \bar{\rho}_{\{\ell_\alpha\}_\alpha} = \frac{1}{L} \sum_{\ell = 1}^{L} \rho_{\{\ell+\ell_\alpha\}_\alpha}.
\end{equation}
for any choice of the support $\{\ell_\alpha\}_\alpha$ and we have represented a state, which is translational by construction:
$\bar{\rho}_{\{\ell_\alpha\}_\alpha} = \bar{\rho}_{\{\ell'+\ell_\alpha\}_\alpha}$.
Notice that we are \emph{not} considering the coherent superposition of the TTN state translations,
because the interference graphs can not be contracted efficiently due to the presence of several closed loops.
We can now combine the entanglement argument of Tensor Networks with the concavity property
of Von Neumann entropy, and obtain an overall bound on the entanglement for state \eqref{eq:incoherentrasl}:
\begin{equation} \label{eq:traslmix}
 \mathcal{S}_{\text{VN}}\left( \bar{\rho}_{\ell} \right) \leq \log_2 \ell \cdot \log D
\end{equation}
which satisfies the right logarithmic behavior, and is perfectly smooth under translations.

\vspace{.5em}
\emph{\textbf{Adding new tensor elements: MERA -}} The idea of MERA arose from the request
of saving some information on correlations shared by neighboring sites scheduled to
be renormalized into different, separate blocks. For instance, in \eqref{eq:TN2pic} site
7 and site 8, despite being neighbors, are not going to be renormalized together until
the hat is reached: this means that most of the entanglement they share is likely to be lost
in the renormalization process, and poorly described by the corresponding Tensor Network design.
Therefore, the scheme proposed by G. Vidal \cite{MERAzero} consists into performing a
(unitary) operation $X$ coupling these sites, whose purpose is to \emph{disentangle} the
two as much as possible before renormalizing them separately. This way, the information
concerning the original entanglement of the pair is stored within this operation $X$,
and renormalization needs not to concern about it.
Disentangling operations and real-space renormalizators are then applied alternately.

It is clear that, since unitary disentanglers are linear operations, quantum states
achievable by this process are again Tensor Network states. They are known
as Multiscale Entanglement Renormalization Ansatz states, and show a hierarchical and real-space self-similar
pattern just like TTN states.
Here we present a partial frame of a binary MERA Tensor Network
\begin{equation} \label{eq:MERA2pic}
\begin{overpic}[width = \textwidth, unit=1pt]{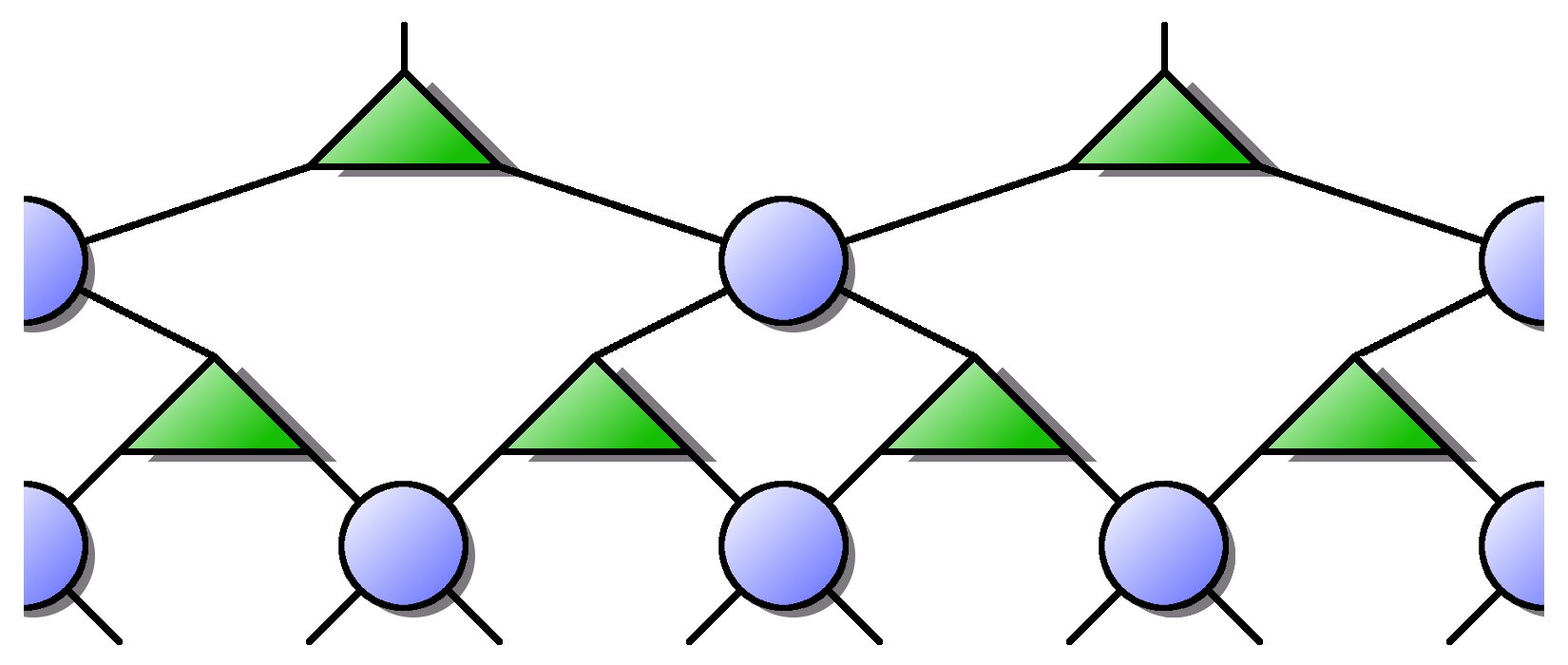}
 \put(43, 56){\footnotesize $\Lambda^{[h,\ell]}$}
 \put(90, 25){\footnotesize $X^{[h,\ell]}$}
\end{overpic}
\end{equation}
where for every layer $h$ and horizontal position $\ell$, tensors are chosen to be
unitary/isometric, i.e. $X^{\dagger} X = \Id_{D^2 \times D^2}$ and 
$\Lambda^{\dagger} \Lambda = \Id_{D \times D}$. Of course, the corresponding
MERA version can be modeled on a Tree Network of arbitrary branching number;
in literature, both binary and ternary MERA have been considerably used for simulation purposes.
For completeness, let us draw the diagrammatic pattern of a ternary MERA:
\begin{equation} \label{eq:MERA3pic}
\begin{overpic}[width = \textwidth, unit=1pt]{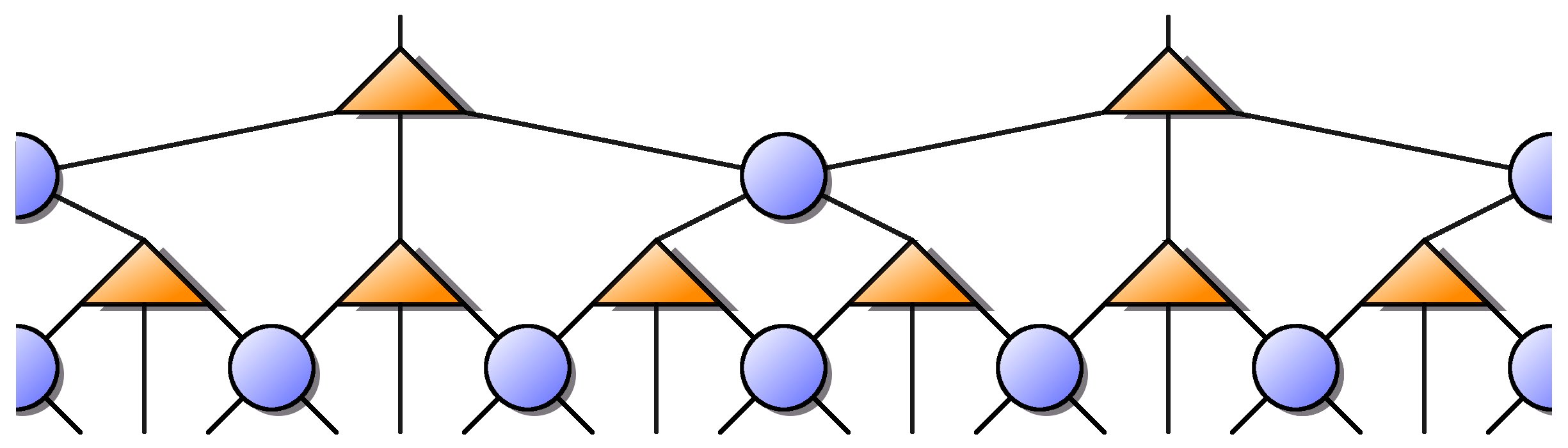}
 \put(43, 50){\footnotesize $\Lambda^{[h,\ell]}$}
 \put(58, 32){\footnotesize $X^{[h,\ell]}$}
\end{overpic}
\end{equation}
We will show now that introducing the new Tensor elements $X$ plays a fundamental
role in regularizing the entanglement scaling law under translation, even though
MERA are again, like TTN, a non-translational network design.

Let us assume, as before, that we are to evaluate the entanglement entropy of $\rho_{\{\ell_1, \ell_1+\ell\}}$,
estimating an upper bound via the Tensor Network link-cut argument, and proceed recursively layer-by-layer
(a MERA full layer is composed by stacking together disentanglers $X^{[h,\bullet]}$
and coarse-grainers $\Lambda^{[h,\bullet]}$).
For every layer, and each one of the two region boundaries ($\ell_1$ and $\ell+\ell_1$), I have
to cut either one or two links, depending on the region location with respect to
the MERA network geometry, as you see from this example
\begin{equation} \label{eq:MERAcut}
\begin{overpic}[width = 280pt, unit=1pt]{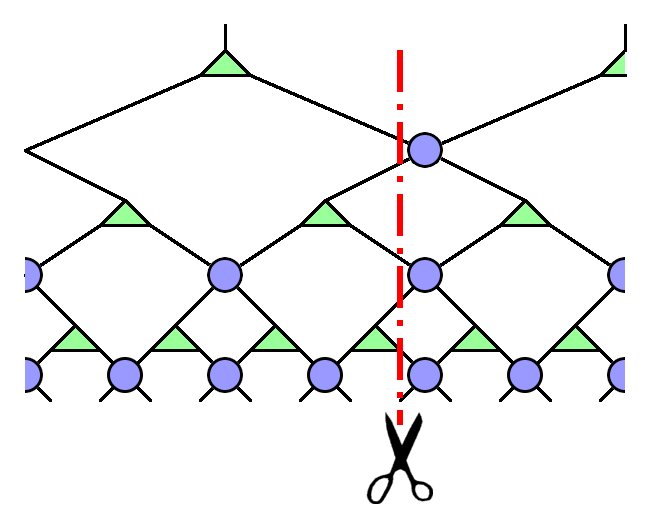}
 %\put(43, 50){\footnotesize $\Lambda^{[h,\ell]}$}
\end{overpic}
\end{equation}
This leads to a bounding function for the entanglement 
$\mathcal{S}_{\text{VN}} (\rho_{\ell_1,\ell_1+\ell}) \leq \#\mbox{cuts} \cdot \log D$, where
\begin{equation} \label{eq:MERAfluctucut}
 2 \log_b \ell \leq \# \mbox{cuts} \leq  4 \log_b \ell,
\end{equation}
depending on $\ell_1$. It is clear from \eqref{eq:MERAfluctucut} that entanglement-bound fluctuations are present
even in the MERA case. However, differently from \eqref{eq:fluctucut} they are hardly a problem: indeed even in
the worst case scenario $\mathcal{S}_{\text{VN}} (\rho_{\ell_1,\ell_1+\ell}) \leq 2 \log_b \ell \cdot \log D$: the
entanglement is ruled by a logarithmic violation of the 1D area-law, as we wished. This is the main reason
why MERA are kept in high regard in 1D critical systems simulations.

\vspace{.5em}
Despite the enhanced accuracy given by disentangling elements of a MERA, we would like to show,
throughout this chapter, that MERA and TTN representations manifest really a common behavior,
as they can both easily capture critical properties of strongly-correlated states, e.g. in terms of critical exponents,
primary fields, and so forth. Acknowledging that, intuition suggests that renormalizer operations $\Lambda$ are those
responsible for keeping track of long-range properties of the TTN/MERA state, while disentanglers $X$
are mostly used to adjust the local variational structure.
To avoid translational invariance-breaking issues, we will make large use of translational averages of
operators and correlators in the following sections, thus substantially adopting the 
incoherent mixture paradigm of \eqref{eq:traslmix} whenever possible, even for MERA.

Similarly, notice that TTN can be definitely seen as a subclass of MERA. They actually correspond
to MERA networks where all the disentangling tensors are set equal to the identity, i.e.
$X^{[h,\ell]} = \Id_{D^2 \times D^2}$ $\forall h,\ell$.

One additional remark: we stated that Tree networks are suitable for any boundary condition we might inquire,
this is due to the presence of dense graph frontiers in their geometry. On the other hand
the MERA Tensor Network has a natural attitude for preferring Periodic boundary conditions. Indeed,
as you see from \eqref{eq:MERA2pic} and \eqref{eq:MERA3pic}, each site has a network path through
the lowest layer linking it to both its neighbors, recalling somehow the valence bond picture idea
for Periodic MPS. Precisely, one can represent the whole 1D binary MERA state as follows:
\begin{equation} \label{eq:MERAPBC}
\begin{overpic}[width = 250pt, unit=1pt]{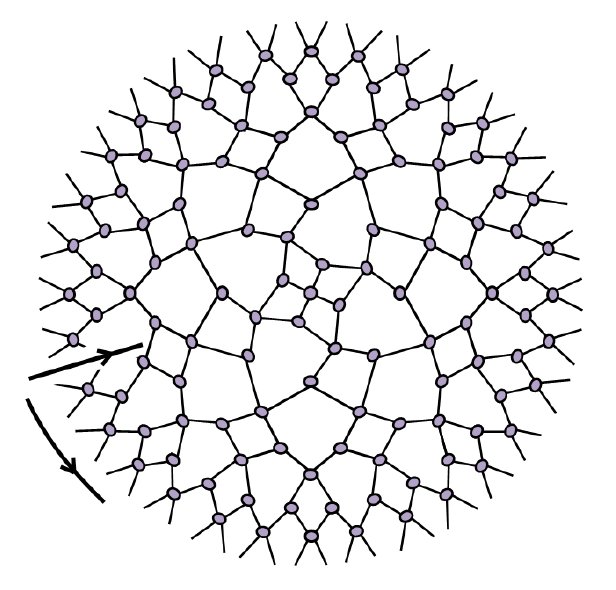}
 \put(14, 46){$\ell$}
 \put(46, 102){$h$}
\end{overpic}
\end{equation}
(figure taken from ref.~\cite{{TNevenbly}})
where the ring of physical sites stands at the edge of the circle.
3-link nodes are $\Lambda$ tensors, 4-link ones are $X$ tensors, and the tensor in the
very center of the diagram is a 4-legs hat tensor $|\mathcal{C}_4\rangle$.
You can easily check that \eqref{eq:MERA2pic} is exactly the local pattern of \eqref{eq:MERAPBC}.
Of course, a generalization of MERA for OBC can clearly be designed by adjusting its geometry,
we will treat this point in section \ref{sec:BERA}.

\section{Causal cone property} \label{sec:causalcone}

A remarkable feature, important both from analytical and computational point of view, shared
by TTN and MERA alike is the so-called causal cone property. It is a generalization of the peripheral
gauge paradigm, allowing us to contract at no expense whole branches of the Tensor Network, that
extends also to MERA geometry, despite having closed loops in their graph,
thanks to the isometricity requirements for $X$ and $\Lambda$ tensors.

Assume we want to achieve the expectation value over a 1D TTN/MERA state for an
observable $\Theta_{\{\ell_1 .. \ell_2 \}}$ whose support is an interval of sites, say $\{\ell_1 .. \ell_2 \}$.
Then, by adopting the standard network contraction method, one performs
$\langle \Psi | \Theta | \Psi \rangle =
\sum_{\{r\},\{s\}} \mathcal{T}^{\star}_{\{r\}} \langle r_{\ell_1} \ldots r_{\ell_2}|\Theta|
s_{\ell_1} \ldots s_{\ell_2} \rangle \mathcal{T}_{\{s\}}$. But now,
consider the lower layer of disentangler $X^{[1,\ell]}$ (for a MERA), which are unitary by requirement $X^{\dagger} X = \Id$.
Those $X$ whose (both) physical indices do not belong to the support interval $\{\ell_1 .. \ell_2 \}$
are contracted with their adjoint and automatically vanish. Precisely, we got rid of a whole layer of disentanglers
except for those directly connected to the support. The same argument holds for renormalizers $\Lambda$
belonging to the lower layer, which disappear alike, as sketched in this diagram:
\begin{equation} \label{eq:ccone1}
\begin{overpic}[width = \textwidth, unit=1pt]{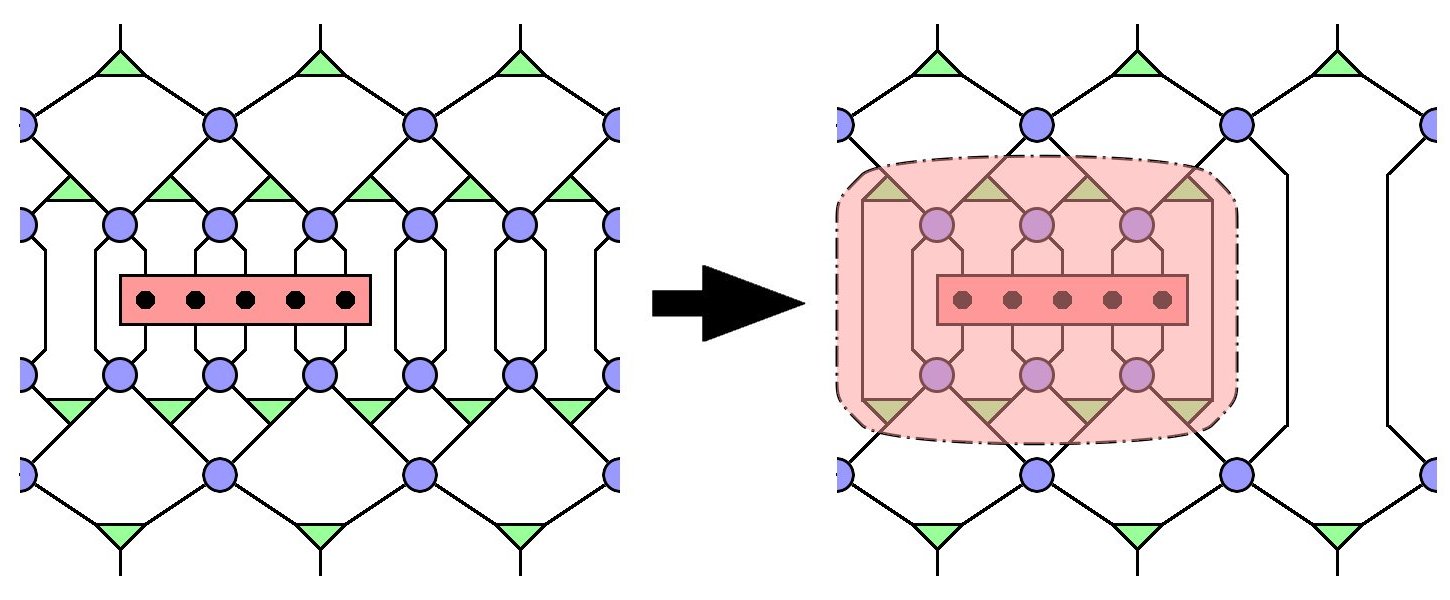}
 \put(0, 145){$a)$}
 \put(220, 145){$b)$}
\end{overpic}
\end{equation}
As we eliminated an entire (double) layer from the Tensor Network state and its conjugate,
we could say that we mapped the original expectation value problem into a new one:
\begin{equation}
 \langle \Psi | \Theta | \Psi \rangle = \langle \Psi' | \mathcal{A}_{\{\ell_1 .. \ell_2\}}\left(\Theta\right) | \Psi' \rangle
\end{equation}
where the quantum state $|\Psi'\rangle$
is the original TTN/MERA state without the bottom layer, and the new observable
$\mathcal{A}_{\{\ell_1 .. \ell_2\}}\left(\Theta\right)$ is the big composite pink tensor
of equation (\ref{eq:ccone1}.b). Acquiring $\mathcal{A}_{\{\ell_1 .. \ell_2\}}\left(\Theta\right)$
starting from $\Theta$, $X$ and $\Lambda$ tensors is a finite operation, whose complexity
scales with the size of the support $\ell_2 - \ell_1$, \emph{not} with the size of the system $L$.
Moreover, when mapping $\Theta \to \mathcal{A}_{\{\ell_1 .. \ell_2\}}\left(\Theta\right)$ the
actual support typically shrinks: you see how in \eqref{eq:ccone1} we map a 5-site
observable into a 4-site one.
We can obviously refrain this argument, every time mapping up
the effective observable in a recursive fashion
\begin{equation} \label{eq:mapup2}
 \Theta^{[h]} \to \Theta^{[h+1]} = \mathcal{A}^{[h]}_{\{\ell_1^{h} .. \ell_2^{h}\}}(\Theta^{[h]})
\end{equation}
until we reach the hat tensor $|\mathcal{C}\rangle$, for which it holds
$\langle \Psi | \Theta | \Psi \rangle =  \langle \mathcal{C} | \Theta^{[\bar{h}-1]} | \mathcal{C}\rangle$.
Each mapping contraction \eqref{eq:mapup2} has a limited computational cost, and we have
to perform it a number of times equal to the total number of layers $\bar{h}$, but we know that
$2^{\bar{h}} = L$, the total system size. In conclusion, for binary TTN/MERA the computational cost
for evaluating compact-support observables scales as
\begin{equation}
 \# \mbox{cost} \propto \log_2 L.
\end{equation}

Obviously, tensors canceled out by isometricity relations can not influence in any way the result
of the expectation value, so all the physical properties of sites in $\{\ell_1 \ldots \ell_2 \}$,
and so their reduced density matrix, can be determined by a small subset of tensors in the network.
Namely, graph nodes we can reach, starting from physical links $\{\ell_1 \ldots \ell_2 \}$,
by means of sole vertical network propagation (only moves that increase the layer index are accepted).
If we want to use a language formalism borrowed from relativity, those 'influent'
tensors form the \emph{causal cone} of sites $\{\ell_1 \ldots \ell_2 \}$.
Here we show you an example for a binary MERA with $L = 16$ sites:
\begin{equation} \label{eq:MERAcc}
\begin{overpic}[width = 340pt, unit=1pt]{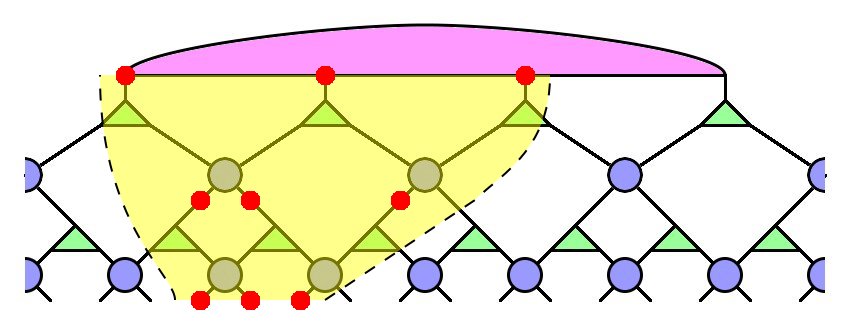}
 \put(167, 106){$\mathcal{C}$}
\end{overpic}
\end{equation}
the yellow region is the causal cone of sites $\{4,5,6\}$, it contains only
4 disentanglers $X$, and 5 renormalizators $\Lambda$, out of 25 tensors in the whole network.

Causal cone 'horizontal' sizes (width) are dynamic while moving across the layers, as already pointed out.
In practice, given support size and placement of the pre-mapping observable, by simple diagram
contraction rules one can determine the size of the mapped observable, which depend also
on the chosen network geometry and branching number $b$. Precisely for $b = 2$:
\[
 \mbox{binary TTN} \left[
 \begin{aligned}
  2 \ell - 1 &\longrightarrow \ell \\
  2 \ell &\longrightarrow \left\{
  \begin{array}{cc}
   \ell &\mbox{if $\ell_1$ odd} \\
   \ell+1 &\mbox{if $\ell_1$ even}
  \end{array} 
  \right.
 \end{aligned}\right.
\]
\begin{equation} \label{eq:ccsizerulez}
 \mbox{binary MERA} \left[
 \begin{aligned}
  2 \ell - 1 &\longrightarrow \ell+1 \\
  2 \ell &\longrightarrow \left\{
  \begin{array}{cc}
   \ell+1 &\mbox{if $\ell_1$ odd} \\
   \ell+2 &\mbox{if $\ell_1$ even}
  \end{array} 
  \right.
 \end{aligned}\right.
\end{equation}
We immediately see that for binary TTN geometry, a cone width of 1 is \emph{stable} through the causal path.
Accordingly, a size of 3 or greater shrinks in an almost-exponential fashion with layers.
Finally, an observable supporting 2 adjacent sites may either be mapped into a one-site observable,
or a two-site one, depending on how these sites match the network structure: we will say, then,
that 2-sites is a \emph{metastable} size, as it may collapse to 1 but not necessarily, nor immediately.
This remains true for higher branching number $b$ TTN states.

For a MERA state it can even happen that we have to grow the causal cone width, e.g. if the observable
was one-site is mapped into a two-site one. In binary MERA geometry, 3 is the only stable size,
while 2 and 4 are metastable. When moving to a ternary MERA geometry (or a higher branching $b$ one),
we see that these characteristic widths are smaller: precisely size 2 is stable,
sizes 1 and 3 metastable (but 3 becomes unstable if $b > 3$).

\section{Ascending and descending maps}

In this section we will focus on those maps that propagate operators upwards trough the layers,
like the one of equation \eqref{eq:mapup2}. According to their nature, we will call them
\emph{ascending} maps, and relate them to completely positive maps described in appendix \ref{app:cptchap}.

For simplicity, let us start with a one-site operator $\Theta_{\ell}$ acting on a binary tree state.
We acknowledged that size 1 is a stable causal cone width for this TTN geometry, so
the mapped operator $\mathcal{A}_{\{\ell\}}(\Theta)$ will still be one-site. Let us express analytically
the action of this mapping, which in turn depends on the parity of $\ell$
\begin{equation} \label{eq:TNmap1}
 \mathcal{A}_{\{\ell\}}(\Theta) = \left\{
 \begin{aligned}
  \mathcal{A}_{L}(\Theta) &\equiv \Lambda^{\dagger} (\Theta \otimes \Id) \Lambda
  %= \sum \Lambda^{\star\,q}_{i,k} \Theta_{i,j} \Lambda^{r}_{j,k} |r\rangle \langle q| \\
  \qquad \mbox{if $\ell$ odd} \\
  \mathcal{A}_{R}(\Theta) &\equiv \Lambda^{\dagger} (\Id \otimes \Theta) \Lambda
  %= \sum \Lambda^{\star\,q}_{k,i} \Theta_{i,j} \Lambda^{r}_{k,j} |r\rangle \langle q| \\
  \qquad \mbox{if $\ell$ even}
 \end{aligned} \right.
\end{equation}
where the subscript $L/R$ refers to which of the two lower links of the node tensor $\Lambda$
(left or right) is touching site $\ell$. Of course, tensor $\Lambda$ also depends on the layer $h$ and horizontal position
$\ell$ that locate its placement within the network, and thus similarly will do maps $\mathcal{A}_{L}$ and
$\mathcal{A}_{R}$ (however we shall treat implicitly this dependence to avoid carrying on too many indices).

It becomes immediately clear from \eqref{eq:TNmap1} that both $\mathcal{A}_{R}$ and $\mathcal{A}_{L}$
maps are Completely Positive and Unital. This is due to the fact that we already expressed
the Kraus expansion of the mapping, indeed $\Lambda^{i}_{j,k}$ form a Kraus set over $k$ of $D \times D$ matrices
$\Lambda^{[L]}_{k} = \sum_{i,j} \Lambda^{i}_{j,k} |i\rangle \langle j|$,
and another set over $j$ of matrices
$\Lambda^{[R]}_{j} = \sum_{i,k} \Lambda^{i}_{j,k} |i\rangle \langle k|$. Therefore
\begin{equation}
 \mathcal{A}_{L}(\Theta) = \sum_{k} {\Lambda^{[L]}_k}^{\dagger} \Theta  {\Lambda^{[L]}_k}
 \quad \mbox{and} \quad
 \mathcal{A}_{R}(\Theta) = \sum_{j} {\Lambda^{[R]}_j}^{\dagger} \Theta  {\Lambda^{[R]}_j},
\end{equation}
but due to isometricity condition on renormalizators, we have
\begin{equation}
\sum_{k} {\Lambda^{[L]}_k}^{\dagger} \Lambda^{[L]}_k = 
\sum_{j} {\Lambda^{[R]}_j}^{\dagger} \Lambda^{[R]}_j  = \Lambda^{\dagger} \Lambda
= \mathcal{A}_{R}(\Id) = \mathcal{A}_{L}(\Id) = \Id,
\end{equation}
which is exactly the Kraus set requirement \eqref{eq:Kraus}.
Interestingly enough, it is relevant that this map should be completely positive;
the reason becomes clear one we introduce the formalism of \emph{descending maps}
for density matrices.
In fact, since $\Theta$ has 1-site support, the reduced density matrix of that site
is the sole responsible for determining expectation values:
\begin{equation}
 \langle \Psi | \Theta | \Psi \rangle = \trace \left[ \Theta \cdot \rho_{\ell} \right].
\end{equation}
The causal cone argument tells us that this is also equivalent to contracting
$\mathcal{A}_{\{\ell\}}(\Theta)$ with the TTN state without the bottom layer, on which
$\mathcal{A}_{\{\ell\}}(\Theta)$ acts again as a single site operator, so that
\begin{equation} \label{eq:uptodown}
 \langle \Psi' | \mathcal{A}_{\{\ell\}}(\Theta) | \Psi' \rangle = 
 \trace \left[ \mathcal{A}_{\{\ell\}}(\Theta) \cdot \rho'_{\ell'} \right] =
 \trace \left[ \Theta \cdot \mathcal{D}_{\{\ell\}}(\rho'_{\ell'}) \right],
\end{equation}
where $\rho'_{\ell'}$ is the reduced density matrix of $|\Psi'\rangle$ at site $\ell' = \lceil \ell/2 \rceil$,
and $\mathcal{D}_{\{\ell\}}$ is the map adjoint to $\mathcal{A}_{\{\ell\}}$ with respect
to the trace scalar product for matrices $(A|B) = \trace[A^{\dagger} B]$.
It is clear that since \eqref{eq:uptodown} holds for the whole algebra of 1-site
observables $\Theta$ it must necessarily be that
\begin{equation} \label{eq:meaningful}
 \rho_{\ell} = \mathcal{D}_{\{\ell\}}(\rho'_{\ell'}).
\end{equation}
Luckily $\mathcal{D}_{\{\ell\}}$, being adjoint of a CP-unital map,
is a completely-positive trace-preserving (CPT) map, which takes density matrices to density matrices
and thus \eqref{eq:meaningful} is perfectly meaningful.
As pointed out in ref.~\cite{QuMERAchan}, applying $\mathcal{D}_{\{\ell\}}$
can be actually seen as the action of a quantum channel \cite{Nielsen}.
The equivalence in \eqref{eq:uptodown} can be interpreted as some TTN version
of the Heisenberg $\leftrightarrow$ Shr\"odinger scheme duality:
we can map ascending-wise operators, or descending-wise density matrices,
the expectation value is achieved either way and the number of contractions is the same.
Similarly to $\mathcal{A}_{\{\ell\}}$, $\mathcal{D}_{\{\ell\}}$ also depends on the parity of target site $\ell$, namely
\begin{equation} \label{eq:TNmap2}
 \begin{aligned}
  \mathcal{D}_{L}(\rho') &\equiv \trace_2 \left[ \Lambda \,\rho'\, \Lambda^{\dagger} \right]
  = \sum_{k} {\Lambda^{[L]}_k} \,\rho'\,  {\Lambda^{[L]}_k}^{\dagger} \\
  \mathcal{D}_{R}(\rho') &\equiv \trace_1 \left[ \Lambda \,\rho'\, \Lambda^{\dagger} \right]
  = \sum_{j} {\Lambda^{[R]}_j} \,\rho'\, {\Lambda^{[R]}_j}^{\dagger},
 \end{aligned}
\end{equation}
both CPT, and expressed in Kraus formalism. Performing the algebraic contraction of any of the maps
defined in \eqref{eq:TNmap1} and \eqref{eq:TNmap2} requires a number of elementary operations scaling
like $\sim D^{4}$, as you can easily guess from the following diagram representation:
\begin{equation} \label{eq:AscDesc1}
\begin{overpic}[width = \textwidth, unit=1pt]{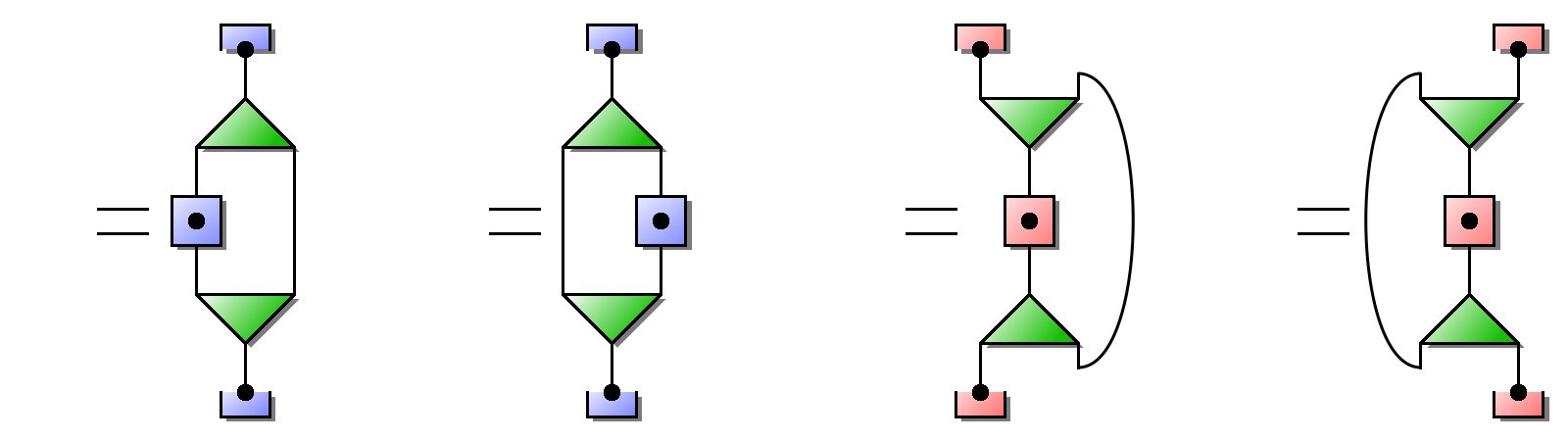}
 \put(0, 51){\large $\mathcal{A}_L$}
 \put(99, 51){\large $\mathcal{A}_R$}
 \put(203, 51){\large $\mathcal{D}_L$}
 \put(300, 51){\large $\mathcal{D}_R$}
\end{overpic}
\end{equation}

We presented the ascending/descending map formulation for one-site operator/density matrix,
but we are not surprised that it extends naturally for any size, according to the causal cone
width rules \eqref{eq:ccsizerulez}. Clearly, ascending maps will typically reduce the size
of the effective operator, and at the same time, descending maps with a given target size
will be function of density matrices with smaller size. Let us consider the case of two adjacent sites $\{\ell_1, \ell_1+1\}$,
depending whether $\ell_1$ is even or odd we have:
\begin{equation} \label{eq:TNmap3}
 \begin{aligned}
  \rho_{\{2\ell-1, 2\ell\}} &= \mathcal{S}(\rho'_{\ell}) = \Lambda \,\rho'_{\ell}\, \Lambda^{\dagger}\\
  \rho_{\{2\ell, 2\ell+1\}} &= \mathcal{D}_R \otimes \mathcal{D}_L (\rho'_{\{\ell, \ell+1\}})
 \end{aligned}
\end{equation}
where $\mathcal{S}$ is the CPT map obtained by just multiplying the density matrix by $\Lambda$ and
$\Lambda^{\dagger}$; and notice that it holds
$\mathcal{D}_L(\cdot) = \trace_R [ \mathcal{S}(\cdot)]$ and
$\mathcal{D}_R(\cdot) = \trace_L[\mathcal{S}(\cdot)]$.
The map appearing in the bottom line of \eqref{eq:TNmap3} is separable, so it can only
decrease entanglement, and likely it will since it is not an invertible transformation and
typically satisfy the mixing property (see appendix \ref{app:mixin}).
Let us sketch the diagrams for the descending operators of \eqref{eq:TNmap3} as well:
\begin{equation} \label{eq:TNmap4}
\begin{overpic}[width = 330pt, unit=1pt]{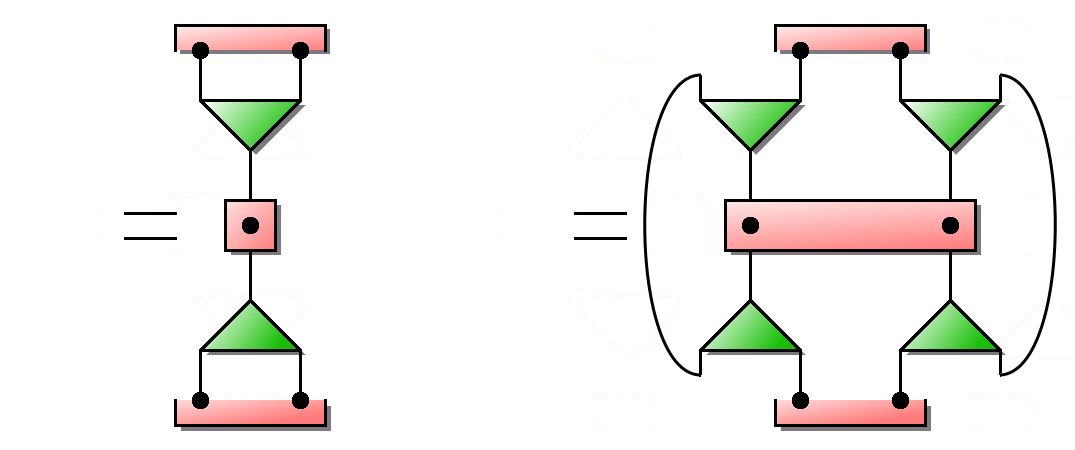}
 \put(20, 65){\large $\mathcal{S}$}
 \put(115, 65){\large $\mathcal{D}_R \otimes \mathcal{D}_L$}
\end{overpic}
\end{equation}
the ascending ones are simply defined by taking the adjoint of \eqref{eq:TNmap4}.
Due to the fact that TTN has a renormalization flow which is quasi-local at every step,
it is easy to check that maps $\mathcal{S}$, $\mathcal{D}_L$ and $\mathcal{D}_R$ are the
\emph{only} ingredients we need to build all the maps for any given size. For instance, let us write the
generic descending map for the binary TTN, in order to obtain $\rho_{\{\ell_1,\ell_1+\ell\}}$
\[
 \mbox{odd length} \left[
 \begin{aligned}
  \rho_{\{\ell_1, \ell_1 + 2\ell \}} & =
  \mathcal{S} \otimes \ldots \otimes \mathcal{S} \otimes \mathcal{D}_L (\rho'_{\{\lceil \ell_1/2 \rceil,
  \lceil \ell_1/2 \rceil + \ell \}}) \quad & \mbox{for $\ell_1$ odd}\\
  \rho_{\{\ell_1, \ell_1 + 2\ell \}} & =
  \mathcal{D}_R \otimes \mathcal{S} \otimes \ldots \otimes \mathcal{S} (\rho'_{\{\ell_1/2,
  \ell_1/2 + \ell \}}) \quad & \mbox{for $\ell_1$ even}\\
 \end{aligned} \right.
\]
\begin{equation} \label{eq:TNmap5}
 \mbox{even length} \left[
 \begin{aligned}
  \rho_{\{\ell_1, \ell_1 + 2\ell - 1\}} & =
  \mathcal{S} \otimes \ldots \otimes \mathcal{S} (\rho'_{\{\lceil \ell_1/2 \rceil,
  \lceil \ell_1/2 \rceil + \ell \}}) \quad & \mbox{for $\ell_1$ odd}\\
  \rho_{\{\ell_1, \ell_1 + 2\ell - 1\}} & =
  \mathcal{D}_R \otimes \mathcal{S} \otimes \ldots \otimes \\ & \otimes \mathcal{S} \otimes \mathcal{D}_R
  (\rho'_{\{\ell_1/2, \ell_1/2 + \ell + 1 \}}) \quad & \mbox{for $\ell_1$ even}\\
 \end{aligned} \right.
\end{equation}
We agree that working with all these even versus odd possibilities looks definitely messy and confusing.
To work around this issue (even though partially) we suggest to employ the
translational average $\leftrightarrow$ incoherent translational mixture formalism. That is what we
will discuss in the next section.

\vspace{.5em}
\emph{\textbf{Ascending/descending maps for MERA -}}
Before moving on, it should be pointed out that, as we developed an in-depth formalism
of completely positive ascending and descending maps for binary tree network, this can
be similarly done for other TTN geometries and for MERA as well. Precisely, in MERA
geometries, maps also embed the local action of disentanglers $X$ as well as renormalizers $\Lambda$.
As an example, let us write the descending maps for 3-site density matrices
(recall that 3 is the stable causal cone width) $\mathcal{D}^{\text{MERA}}_{3 \to 3, L}$
and $\mathcal{D}^{\text{MERA}}_{3 \to 3, R}$, respectively reading
\begin{equation} \label{eq:TNmap6}
\begin{overpic}[width = 330pt, unit=1pt]{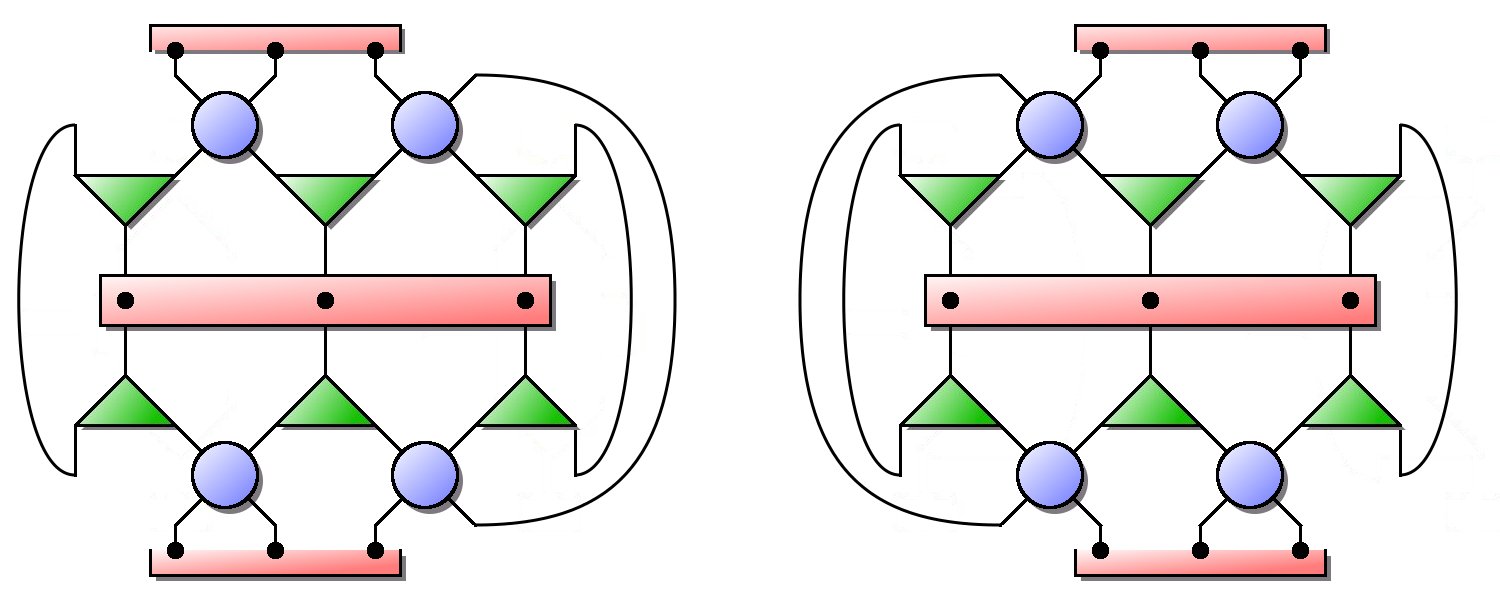}
 \put(160, 60){,} 
\end{overpic}
\end{equation}
defined in this fashion so that
\begin{equation} \label{eq:TNmap7}
\rho^{\text{MERA}}_{\{\ell_1, \ell_1 + 2\}} = \left\{
 \begin{aligned}
\mathcal{D}^{\text{MERA}}_{3 \to 3, L} (\rho^{\text{MERA}}_{\{ \ell_1/2 , \ell_1/2 + 2\}}) \quad & \mbox{for $\ell_1$ even} \\ \\
\mathcal{D}^{\text{MERA}}_{3 \to 3, R} (\rho^{\text{MERA}}_{\{ \lfloor \ell_1 \rfloor, \lfloor \ell_1 \rfloor + 2\}})
\quad & \mbox{for $\ell_1$ odd}
 \end{aligned} \right.
\end{equation}
Extending the descending map formalism to density matrices of any size is performed similarly to \eqref{eq:TNmap5}
for MERA, but with a main difference: the maps can not be any longer expressed as tensor products of local objects
due to the presence of disentanglers.

\section[translational averages]{Horizontal homogeneity and\\translationally averaged maps}

In TTN/MERA Tensor Network, interesting new properties arise when we adopt the additional
assumption that tensors belonging to the same layer $h$ of the network structure
(and of the same type $X$, $\Lambda$ for a MERA) are identical, that they are actual copies of
the same variational tensor, appearing $L/2^h$ times at different nodes in the network.
We will call this requirement \emph{horizontal} homogeneity, because we are still letting
the tensors belonging to different layers free to be variationally different.

In section \ref{sec:MPShomo} we showed how tensor homogeneity in periodic Matrix Product States is
a condition strictly related with translational invariance symmetry. Neither for Tree networks
nor for MERA, the relationship between homogeneity and translational invariance is so sharp
and well-defined, since the network geometry itself is visibly not translational. Nevertheless,
horizontal homogeneity is not only the best approximation for emulating a translational invariance,
but also, and more importantly, allows us to build a well-defined ascending/descending map formalism
for the translational averages $\leftrightarrow$ incoherent translational mixture scheme.
This will let us write simplified versions of the (say, descending) map definitions we provided
in last section, namely \eqref{eq:TNmap2} \eqref{eq:TNmap3} and \eqref{eq:TNmap5}, as they
will depend no longer from the placement index $\ell_1$, but solely on the size $\ell$.

Then let $\bar{\rho}^{[\mu]}_{\ell}$ be the one-site density matrix of the TTN, averaged over translations
\begin{equation} \label{eq:transpla}
 \bar{\rho}^{[\mu]}_{\ell} = \frac{1}{2^{\mu}} \sum_{\ell_1 = 1}^{2^{\mu}} \rho^{[\mu]}_{\{\ell_1, \ell_1+\ell-1 \}},
\end{equation}
where now the layer index $\mu$ is counted starting from the hat and moving down (i.e. $\mu = \bar{h} - h$), for comfort.
This will prove a labeling advantage when we will discuss the thermodynamical limit $\mu \to +\infty$.

As before, we start from smaller sizes and then grow with increasing $\ell$, to enhance immediateness and comprehension.
Consider then $\bar{\rho}^{[\mu]}_{1}$, by exploiting \eqref{eq:TNmap2} we can write
\begin{multline}
 \bar{\rho}^{[\mu]}_{1} =  \frac{1}{2^{\mu}} \sum_{\ell_1 = 1}^{2^{\mu}} \rho^{[\mu]}_{\{\ell_1\}}
 =  \frac{1}{2^{\mu}} \sum_{\ell_1 = 1}^{2^{\mu}/2} \rho^{[\mu]}_{\{ 2 \ell_1 - 1\}} + \rho^{[\mu]}_{\{ 2 \ell_1\}} \\
 =  \frac{1}{2^{\mu-1}} \sum_{\ell_1 = 1}^{2^{\mu-1}}
 \frac{1}{2} \left( \mathcal{D}_{L}^{[\mu]} \left( \rho^{[\mu-1]}_{\{ \ell_1 \}} \right) +
 \mathcal{D}_{R}^{[\mu]} \left(\rho^{[\mu-1]}_{\{ \ell_1 \}} \right) \right)
\end{multline}
where $\mathcal{D}_{L}^{[\mu]}$ and $\mathcal{D}_{L}^{[\mu]}$ are those descending maps defined in
\eqref{eq:AscDesc1} when tensor $\Lambda$ belonging to layer $\mu$ is being used.
Then by linearity of the descending maps we can re-sum the argument density matrices to obtain again the
appropriate translational mixture
\begin{multline} \label{eq:avemap1}
 \bar{\rho}^{[\mu]}_{1} = 
 \frac{1}{2} \,\mathcal{D}_{L}^{[\mu]}
       \left( \frac{1}{2^{\mu-1}} \sum_{\ell_1 = 1}^{2^{\mu-1}} \rho^{[\mu-1]}_{\{ \ell_1 \}} \right) +
 \frac{1}{2} \,\mathcal{D}_{L}^{[\mu]}
       \left( \frac{1}{2^{\mu-1}} \sum_{\ell_1 = 1}^{2^{\mu-1}} \rho^{[\mu-1]}_{\{ \ell_1 \}} \right) = \\
 \frac{\mathcal{D}_{L}^{[\mu]} + \mathcal{D}_{R}^{[\mu]}}{2} \left( \bar{\rho}^{[\mu-1]}_{1} \right) = 
 \mathcal{D}^{[\mu]} \left( \bar{\rho}^{[\mu-1]}_{1} \right),
\end{multline}
where we defined the average map $\mathcal{D} \equiv (\mathcal{D}_L + \mathcal{D}_R )/2$.
This new map is again CPT, as it is the equally-weighted mixture of the two original maps, its Kraus decomposition
is given by the union of the two Kraus decompositions of $\mathcal{D}_L$ and $\mathcal{D}_R$
(every matrix multiplied by a $2^{-1/2}$ factor). Eq.~\eqref{eq:avemap1}
yields to a single layer-to-layer recursive relation
$\bar{\rho}^{[\mu]}_{\ell} = \mathcal{D}^{[\mu]} ( \bar{\rho}^{[\mu-1]}_{\ell} )$ instead of the whole family
of equations \eqref{eq:meaningful} and \eqref{eq:TNmap2}, thanks to horizontal homogeneity requirement.

Let us move to two adjacent sites, a causal cone size which is metastable for TTN geometry: its stability will be
regularized when performing translational averages, as we will see.
Then
\begin{multline} \label{eq:avemap2}
 \bar{\rho}^{[\mu]}_{2} = \frac{1}{2^{\mu}} \sum_{\ell_1 = 1}^{2^{\mu-1}}
 \left( \rho^{[\mu]}_{\{2 \ell_1 - 1, 2 \ell_1\}} + \rho^{[\mu]}_{\{2 \ell_1, 2 \ell_1 + 1\}} \right) = \\
 \frac{1}{2^{\mu}} \sum_{\ell_1 = 1}^{2^{\mu-1}}
 \mathcal{S}^{[\mu]} \left( \rho^{[\mu-1]}_{\{\ell_1 \}} \right) + 
 \mathcal{D}_R^{[\mu]} \otimes \mathcal{D}_L^{[\mu]} \left( \rho^{[\mu-1]}_{\{\ell_1, \ell_1 +1 \}} \right) = \\
 \frac{1}{2}\, \mathcal{S}^{[\mu]} \left( \bar{\rho}^{[\mu-1]}_{1} \right) +
 \frac{1}{2}\, \mathcal{D}_R^{[\mu]} \otimes \mathcal{D}_L^{[\mu]} \left( \bar{\rho}^{[\mu-1]}_{2} \right)
\end{multline}
Watching this result, one can argue that in order to achieve $\bar{\rho}^{[\mu]}_{2}$ by descending recursive
relations, we should need both $\bar{\rho}^{[\mu-1]}_{1}$ and $\bar{\rho}^{[\mu-1]}_{2}$.
But indeed, $\bar{\rho}^{[\mu-1]}_{2}$ contains by itself all the information concerning $\bar{\rho}^{[\mu-1]}_{1}$:
this is obvious by definition \eqref{eq:transpla}, which tells us that
$\trace_L [\bar{\rho}^{[\mu-1]}_{2}] = \trace_R [\bar{\rho}^{[\mu-1]}_{2}] = \bar{\rho}^{[\mu-1]}_{2}$.
The present consideration allows us to write $\bar{\rho}^{[\mu]}_{2} = 
 \mathcal{D}_{2 \to 2}^{[\mu]} ( \bar{\rho}^{[\mu-1]}_{2} )$ where
\begin{equation} \label{eq:avemap2bis}
 \mathcal{D}_{2 \to 2}^{[\mu]}(\cdot) = \frac{1}{2}
 \mathcal{D}_R^{[\mu]} \otimes \mathcal{D}_L^{[\mu]} (\cdot) +
 \frac{\cos^2 \theta}{2} \mathcal{S}^{[\mu]} \left(\trace_L[\cdot]\right) +
 \frac{\sin^2 \theta}{2} \mathcal{S}^{[\mu]} \left(\trace_R[\cdot]\right).
\end{equation}
$\cos^2 \theta$ and $\sin^2 \theta$ are just two free real parameters which must be both positive,
to guarantee complete positivity of $\mathcal{D}_{2 \to 2}^{[\mu]}$, and whose sum must be 1 to
ensure trace preserving property. Then for any real angle $\theta$, eq.~\eqref{eq:avemap2bis} performs the right mapping.

By natural extension, we can generalize descending mappings for any density matrix size,
which can be always written in terms of $\mathcal{S}$, $\mathcal{D}_L$ and $\mathcal{D}_R$.
We report here the result for completeness:
\begin{equation} \label{eq:avemap3}
 \begin{aligned}
 \mathcal{D}_{\ell \to 2\ell-1} &= \frac{1}{2}
 \left( \mathcal{D}_R \otimes \mathcal{S} \otimes \ldots \otimes \mathcal{S} +
 \mathcal{S} \otimes \ldots \otimes \mathcal{S} \otimes \mathcal{D}_L \right)\\
 \mathcal{D}_{\ell \to 2\ell-2} &= \frac{1}{2}
 \left( \mathcal{D}_R \otimes \mathcal{S} \otimes \ldots \otimes \mathcal{S} \otimes \mathcal{D}_L +
 \left[\mathcal{S} \otimes \ldots \otimes \mathcal{S}\right] \circ \trace_{\theta} \right),
 \end{aligned}
\end{equation}
where $\trace_{\theta}$ means tracing away the rightmost site times $\cos^2 \theta$ plus
tracing out the leftmost site times $\sin^2 \theta$ (angle $\theta$ arbitrary), as in \eqref{eq:avemap2bis}.
All these maps are convex combinations of the maps previously defined, so they are indeed CPT.

Ascending maps can be naturally defined in a translational invariant fashion as well; it is easy to see
that the composition relations \eqref{eq:avemap3} hold for the adjoint maps 
$\mathcal{A}_{2\ell-2 \to \ell}$ and $\mathcal{A}_{2\ell-2 \to \ell}$ identically.

\section[Thermodynamical limit TTN]{Complete homogeneity and\\definition of thermodynamical limit}

We mentioned that the main purpose of introducing Tree and MERA Tensor Networks is to simulate
strongly-correlated systems. It is clear, though, that this ansatz
would be more suitable the more we are capable of reproducing a conformal symmetry, where
translational invariance, and more importantly \emph{scale} invariance hold.

In previous section we dealt with translational invariance by merging together the horizontal
homogeneity requirement and the incoherent translational mixture framework. Now we are going to
force scale invariance into the system by assuming complete homogeneity: i.e. requesting that
\emph{every} renormalization tensor $\Lambda$ in the network is identical to the others,
even throughout the layers. We will see that this assumption will naturally lead to the definition
of a thermodynamical limit, in which every physical quantity is perfectly controlled
thanks to isometricity relations, and where the scale invariance becomes manifest.

This can be explained heuristically, by considering the family of tree graphs.
Every tree graph has a self-similar pattern, but only for an infinite tree the self-similarity
becomes exact: every branch is topologically identical to each of its sub-branches because
they are all infinitely-long. Such symmetry, though, would be broken if one were
able to distinguish somehow the graph nodes, thus the assumption of complete tensor homogeneity.

Let us then consider the sequence of completely homogeneous TTN states $|\Psi^{[\mu]}(\Lambda)\rangle$, defined solely
by a single tensor $\Lambda$, repeated at every network node, and indexed by the total amount of layers $\mu$.
We are going to characterize the limit of this sequence for $\mu \to \infty$. As we discussed
in section \eqref{sec:MPSTD} the thermodynamical limit state is defined by the family of density matrices
$\bar{\rho}^{\infty}_{\ell}$ for every size $\ell$, with the requirement that elements of this set
undergo the correct partial trace relation. In our TTN setting we are first considering translational
averages, and then approaching the limit, so that
\begin{equation}
 \rho^{\infty}_{\ell} = \lim_{\mu \to \infty} \bar{\rho}^{[\mu]}_{\ell}.
\end{equation}
We will again proceed starting from size 1. Horizontal homogeneity tells us that \eqref{eq:avemap1} holds,
but at the same time, vertical homogeneity implies that $\mathcal{D}$ does not depend on the
layer we are considering. Therefore
\begin{equation}
 \bar{\rho}^{[\mu]}_{1} = \mathcal{D} \left( \bar{\rho}^{[\mu-1]}_{1} \right)
 = \mathcal{D}^2 \left( \bar{\rho}^{[\mu-2]}_{\ell} \right) = \ldots =
 \mathcal{D}^{\mu} \left( \bar{\rho}^{[\text{hat}]}_{1} \right),
\end{equation}
where $\bar{\rho}^{[\text{hat}]}_{1}$ is the translationally averaged one-site density matrix
of the hat tensor. Then we have that
\begin{equation}
 \rho^{\infty}_{1} = \lim_{\mu \to \infty} \mathcal{D}^{\mu} \left( \bar{\rho}^{[\text{hat}]}_{1} \right),
\end{equation}
i.e. one-site physical properties of the thermodynamical limit state are given by applying $\mathcal{D}$ to the
hat tensor. We will now make an additional assumption, which is indeed a constraint on $\Lambda$ but
formulated it in terms of $\mathcal{D}$: we will require that $\mathcal{D}$ is a mixing CPT map (see \ref{app:mixin}).
This condition on the eigenspace decomposition of $\mathcal{D}$ not only states that $\mathcal{D}$ has
a unique fixed point $\rho_{f}$, but also that $\rho_{f}$ is the only attraction pole,
so that every state is mapped into $\rho_{f}$ after infinite applications of $\mathcal{D}$.
Then $\rho^{\infty}_{1}$ is exactly the fixed point of $\mathcal{D}$:
\begin{equation}
 \rho^{\infty}_{1} = \mathcal{D} \left( \rho^{\infty}_{1} \right)
\end{equation}
As you guessed, normalization is kept correctly under control at every finite $\mu$, and thus in the
limit of an infinite amount of layers $\mu$. Indeed, we argumented that isometricity of $\Lambda$ is
equivalent to the peripheral gauge, for which the state normalization condition is automatically dumped onto
the hat tensor alone, $1 = \langle \Psi^{[\mu]}(\Lambda) |\Psi^{[\mu]}(\Lambda)\rangle =
\langle \mathcal{C} | \mathcal{C}\rangle$; trace preservation property of descending maps does the rest.

Let us proceed to size 2: the two-adjacent sites density matrix in the thermodynamical limit
$\rho^{\infty}_{2}$ should satisfy \eqref{eq:avemap2} for $\mu \to \infty$, which reads
\begin{equation} \label{eq:dm2td}
 \rho^{\infty}_{2} =
 \frac{1}{2}\, \mathcal{S} \left( \rho^{\infty}_{1} \right) +
 \frac{1}{2}\, \mathcal{D}_R \otimes \mathcal{D}_L \left( \rho^{\infty}_{2} \right).
\end{equation}
This is a quasi-recursive relation which has also a direct dependence upon
$\rho^{\infty}_{1}$. In particular by nesting \eqref{eq:dm2td} within itself, we can rewrite the
recursive expression into a series, namely
\begin{equation} \label{eq:dm2td2}
 \rho^{\infty}_{2} = \left[ \sum_{\tau = 0}^{\infty} \frac{1}{2^{\tau+1}}
 \left( \mathcal{D}_R \otimes \mathcal{D}_L \right)^{\tau} \right]
 \circ \mathcal{S}(\rho^{\infty}_{1}),
\end{equation}
where the expression inside the square parentheses is a CPT map, because the positive weights $1/2^{\tau+1}$ sum to 1.
At the same time, we want our $\rho^{\infty}_{2}$ to be consistent with equation \eqref{eq:avemap2bis}, which
becomes a fixed point equation in the thermodynamical limit:
\begin{multline} \label{eq:dm2td3}
  \rho^{\infty}_{2} = \mathcal{D}_{2 \to 2} \left( \rho^{\infty}_{2} \right)
 \equiv \frac{1}{2} \,\mathcal{D}_R \otimes \mathcal{D}_L \left( \rho^{\infty}_{2} \right) + \\ +
 \frac{\cos^2 \theta}{2} \,\mathcal{S} \left( \trace_1 \left[ \rho^{\infty}_{2} \right] \right) +
 \frac{\sin^2 \theta}{2} \,\mathcal{S} \left( \trace_2 \left[ \rho^{\infty}_{2} \right] \right).
\end{multline}
In order to show that \eqref{eq:dm2td} and \eqref{eq:dm2td3} are actually compatible equations (for every $\theta$),
it suffices to prove that $\trace_1 [\rho^{\infty}_{2}] = \trace_2 [\rho^{\infty}_{2}] = \rho^{\infty}_{1}$.
To show this, simply consider the partial trace, say with respect to the rightmost site,
of \eqref{eq:dm2td}. Then we get
\begin{equation}
 \trace_2 \left[ \rho^{\infty}_{2} \right] = \frac{1}{2}\,\mathcal{D}_L \left( \rho^{\infty}_1 \right)
 + \frac{1}{2}\, \mathcal{D}_R \left( \trace_2 \left[ \rho^{\infty}_{2} \right] \right);
\end{equation}
but $\mathcal{D}_L ( \rho^{\infty}_1 ) = [ 2 \mathcal{D} - \mathcal{D}_R ]( \rho^{\infty}_1 )$ and due to the
fixed point property is also equal to $ 2 \rho^{\infty}_1 - \mathcal{D}_R ( \rho^{\infty}_1 )$. Therefore we can write
\begin{equation} \label{eq:bacro}
 \mathcal{D}_R \left( \rho^{\infty}_{1} - \trace_2 \left[ \rho^{\infty}_{2} \right] \right) =
 2 \, \left( \rho^{\infty}_{1} - \trace_2 \left[ \rho^{\infty}_{2} \right] \right),
\end{equation}
which is an eigenvalue equation. But since a CPT map has spectral radius 1, it can not have 2 as eigenvalue;
thus the only solution of \eqref{eq:bacro} is $\rho^{\infty}_{1} = \trace_2 \left[ \rho^{\infty}_{2} \right]$.
Similarly, we can trace out the left site in \eqref{eq:dm2td}
and obtain the other equality: $\rho^{\infty}_{1} = \trace_1 \left[ \rho^{\infty}_{2} \right]$.
This guarantees that $\rho^{\infty}_{2}$ is unique and well-defined.

Thermodynamical limit density matrices for sizes 1 and 2 are the only ones that require dealing
with a fixed point equation to be achieved: all the other $\rho^{\infty}_{\ell \geq 3}$
can be directly calculated by applying a finite number of maps to $\rho^{\infty}_{2}$.
Precisely we are referring to size-increasing descending maps of equation \eqref{eq:avemap3},
that not only generate the whole family of $\rho^{\infty}_{\ell}$, but ensure that this family
satisfies the partial trace requirement. For example:
\begin{equation}
   \rho^{\infty}_5 = \mathcal{D}_{3 \to 5} \circ \mathcal{D}_{2 \to 3} \left( \rho^{\infty}_{2} \right),
\end{equation}
and by partially tracing this equation, say on the two rightmost sites, we obtain
\begin{multline}
 \trace_{4,5} \left[ \rho^{\infty}_5 \right] =
 \mathcal{D}_{2 \to 3} \left( \trace_3 \left[ \mathcal{D}_{2 \to 3} (\rho^{\infty}_2)  \right] \right) = \\
 \mathcal{D}_{2 \to 3} \circ \mathcal{D}_{2 \to 2} \left( \rho^{\infty}_2 \right)
 = \mathcal{D}_{2 \to 3}\left( \rho^{\infty}_2 \right) = \rho^{\infty}_3,
\end{multline}
which is the right consistency check.

In conclusion, we characterized properly and completely the thermodynamical limit TTN, whose uniqueness is
ensured by the mixing requirement of maps $\mathcal{D}_{1 \to 1} = \mathcal{D}$ and $\mathcal{D}_{2 \to 2}$.
Also notice that in this limit, any residual dependence on the hat tensor $|\mathcal{C}\rangle$ that may
linger at finite sizes, vanishes. We can interpret this consideration as a hint that $\Lambda$ is the
only responsible for capturing, and keeping track, of all the bulk properties of $|\Psi (\Lambda)\rangle$.
We will further argument this claim in the next section, where we will calculate two-point
correlators and show that the manifest critical behavior of TTN states is ruled by spectral
properties of maps, and thus by $\Lambda$.

\section{Correlations and criticality}

A critical ground state is a scale-invariant state, whose unmistakable signature is a power-law
decay rate, with the distance, of two-point correlation functions. We will now investigate such
correlations within a TTN state, to prove its criticality. The translational framework scheme will allow us define
correlations depending on the two-point distance alone and not on the location; and then we will
drive the results towards the thermodynamical limit, to better match conformal symmetry.

Then, we start defining a correlation function similarly to what we did for 
thermodynamical MPS in section \ref{sec:MPScorr}, but now we average over translations
\begin{equation} \label{eq:avecordef}
 \bar{\mathfrak{C}}_{\ell}(\Theta, \Theta')
  \equiv \frac{1}{L} \,\sum_{\ell_0 = 1}^{L} \langle \Theta_{[\ell_0]} \otimes \Theta'_{[\ell_0 + \ell]} \rangle
  - \langle \Theta_{[\ell_0]} \rangle \langle \Theta'_{[\ell_0 + \ell + 1]} \rangle,
\end{equation}
for any pair of single-site observables $\Theta$, $\Theta'$. We start by probing this correlator
on the finite but fully homogeneous TTN state $|\Psi^{[\mu]}(\Lambda)\rangle$, with $\mu$ being the total tree
graph height. Then it is clear that
\begin{equation} \label{eq:corrextend}
 \bar{\mathfrak{C}}^{[\mu]}_{\ell}(\Theta, \Theta') =
 \trace \left[ \left( \Theta \otimes \Theta' \right) \cdot \left( \bar{\sigma}^{[\mu]}_{\ell} -
 \eta^{[\mu]}_{\ell} \right) \right]
\end{equation}
which extends by linearity the definition of two-point correlator for any two-site observable $\Gamma$,
not only tensor product ones.
The pair of two-sites density matrices in expression \eqref{eq:corrextend} is defined as:
\begin{equation} \label{eq:sigmaeta}
 \begin{aligned}
  \bar{\sigma}^{[\mu]}_{\ell} &= \frac{1}{L} \sum_{\ell_0 = 1}^{L} \sigma^{[\mu]}_{\ell_0, \ell_0+\ell} = 
   \frac{1}{2^{\mu}} \sum_{\ell_0 = 1}^{2^{\mu}} \trace_{\{\ell_0+1 \ldots \ell_0 + \ell - 1\}}
   \left[ \rho^{[\mu]}_{\{\ell_0, \ell_0+\ell\}} \right] \\
  \eta^{[\mu]}_{\ell} &= \frac{1}{2^{\mu}} \sum_{\ell_0 = 1}^{2^{\mu}}
  \rho^{[\mu]}_{\{\ell_0\}} \otimes \rho^{[\mu]}_{\{\ell_0 + \ell\}}
 \end{aligned}
\end{equation}
Notice the difference between these two objects:
$\bar{\sigma}^{[\mu]}_{\ell}$ is the reduced density matrix of two sites at distance $\ell$, averaged over translations.
$\eta^{[\mu]}_{\ell}$, instead, is the translational average of the product of one-site density matrices, whose sites
stand exactly at distance $\ell$. Honestly, $\eta^{[\mu]}_{\ell}$ is a separable density matrix by definition
\eqref{eq:sigmaeta}, but not necessarily a tensor product matrix, i.e.
$\eta^{[\mu]} \neq \bar{\rho}^{[\mu]} \otimes \bar{\rho}^{[\mu]}$. One can say that
$\eta^{[\mu]}$ keeps track of \emph{classical correlations} between sites, but it is free of quantum entanglement.
At the same time we have that
$\trace_1[\bar{\sigma}^{[\mu]}_{\ell}] = \trace_2[\bar{\sigma}^{[\mu]}_{\ell}] = 
\trace_1[\eta^{[\mu]}_{\ell}] = \trace_2[\eta^{[\mu]}_{\ell}] =
\bar{\rho}^{[\mu]}_{1}$ for any $\ell$.

By exploiting the descending translational map formalism of eq.~\eqref{eq:avemap3}, we can as well develop
recursive relations for the quantities $\bar{\sigma}^{[\mu]}_{\ell}$ and $\eta^{[\mu]}_{\ell}$.
For simplicity and clarity we will write only the equations for an even distance $\ell$, as follows:
\begin{equation}
  \bar{\sigma}^{[\mu]}_{2 \ell} = \Dsla \left( \bar{\sigma}^{[\mu-1]}_{\ell} \right)
  \qquad \mbox{and} \qquad
  \eta^{[\mu]}_{2 \ell} = \Dsla \left( \eta^{[\mu-1]}_{\ell} \right),
\end{equation}
where the 2-sites descending CPT map $\Dsla$ is defined as
\begin{equation}
 \Dsla = \frac{1}{2} \left( \mathcal{D}_L \otimes \mathcal{D}_L +
 \mathcal{D}_R \otimes \mathcal{D}_R \right).
\end{equation}
Notice that the propagation map for the two density matrix elements is the same $\Dsla$,
although it applies to different arguments, so the difference is nonzero in general. In fact,
assume that we are calculating the averaged correlator \eqref{eq:avecordef} at a two-point
distance $\ell$ which is a power of 2, say $2^{q}$
\begin{equation} \label{eq:TNcorr1}
 \bar{\mathfrak{C}}^{[\mu]}_{\ell = 2^{q}}(\Gamma) =
 \trace \left[ \,\Gamma \cdot 
 \Dsla^{\,q} \left( \bar{\sigma}^{[\mu-q]}_{1} -
 \eta^{[\mu-q]}_{1} \right) \right],
\end{equation}
where, by definition \eqref{eq:sigmaeta}, we have $\bar{\sigma}^{[\nu]}_{1} = \bar{\rho}^{[\nu]}_{2}$.
On the other hand, we have that $\eta^{[\nu]}_{1}$ satisfies a recursive
equation formally similar to \eqref{eq:avemap2}, but with a different inhomogeneous term
\begin{equation} \label{eq:etavemap}
 \eta^{[\nu]}_{1} =
 \frac{1}{2}\, \mathcal{D}_R \otimes \mathcal{D}_L \left( \eta^{[\nu-1]}_{1} \right) +
 \frac{1}{2}\, \mathcal{D}_R \otimes \mathcal{D}_L \left( \eta^{[\nu-1]}_{0} \right).
\end{equation}
where for $\eta^{[\nu]}_{0}$ it holds $\eta^{[\nu]}_{0} = \Dsla (\eta^{[\nu-1]}_{0})$.

After setting up all these ingredients, we can drive equation \eqref{eq:TNcorr1} towards the thermodynamical
state easily, where every element gains a well-defined limit. Precisely:
\begin{equation} \label{eq:TNcorr2}
 \bar{\mathfrak{C}}^{[\infty]}_{2^{q}}(\Gamma) =
 \trace \left[ \,\Gamma \cdot 
 \Dsla^{\,q} \left( \Delta\sigma \right) \right],
\end{equation}
where $\Delta\sigma$ is a null trace matrix, given by the series
\begin{equation} \label{eq:eta2td2}
 \Delta\sigma = \left[ \sum_{\tau = 0}^{\infty} \frac{1}{2^{\tau+1}}
 \left( \mathcal{D}_R \otimes \mathcal{D}_L \right)^{\tau} \right]
 \circ \left( \mathcal{S}(\rho^{\infty}_{1}) -
 \mathcal{D}_L \otimes \mathcal{D}_R ( \sigsla ) \right).
\end{equation}
Here we assumed that the CPT map $\Dsla$ satisfies the mixing requirement,
with $\sigsla = \lim_{\mu \to \infty} \eta^{[\mu]}_0$ being its unique fixed point;
clearly if $\Dsla$ is mixing, then $\mathcal{D}$ is mixing as well, since
$\trace_2[\Dsla(\cdot)] = \Di(\cdot)$.

It is now worth to analyze result \eqref{eq:TNcorr2} and derive some interesting conclusions
involving this correlator. Immediately we notice that when increasing the distance to
infinity the correlation function drops to zero as it should, since infinitely distant sites
must not be entangled. This can be checked by accepting that $\Dsla$ is mixing
and using relation \eqref{eq:mixing}; then we get
\begin{equation} \label{eq:TNcorrzero}
 \lim_{q \to \infty} \bar{\mathfrak{C}}^{[\infty]}_{2^q}(\Gamma) =
 \lim_{q \to \infty} \trace \left[ \,\Gamma \cdot 
 \Dsla^{\,q} \left( \Delta\sigma \right) \right] =
 \trace \left[ \,\Gamma \cdot \sigsla \right]\, \trace[\Delta\sigma] = 0,
\end{equation}
because $\Delta\sigma$ is a traceless matrix by definition.

To investigate how the correlator scales at finite distances, we first
switch to the ascending map formalism for \eqref{eq:TNcorr2}; this is done by just taking the
adjoint map $\Asla$ of the superoperator $\Dsla$. Then
\begin{equation} \label{eq:TNcorr3}
 \bar{\mathfrak{C}}^{[\infty]}_{2^{q}}(\Gamma) =
 \trace \left[ \Asla^{\,q} \left( \Gamma \right) \cdot 
  \Delta\sigma \right],
\end{equation}
where $\Asla$ is completely positive and unital.
The linear map $\Asla$ is not necessarily diagonalizable, nevertheless it exists at least one eigenoperator
$\Gamma_{\alpha}$ for each of its eigenvalues $\lambda_{\alpha}$. Then when evaluating the correlator
on that eigenoperator we have
\begin{equation} \label{eq:TNcorr4}
 \bar{\mathfrak{C}}^{[\infty]}_{2^{q}}(\Gamma_{\alpha}) =
 \trace \left[ \Asla^{\,q} \left( \Gamma_{\alpha} \right) \cdot 
  \Delta\sigma \right] = \lambda^{q}_{\alpha} \,\trace \left[ \Gamma_{\alpha} \cdot 
  \Delta\sigma \right]
  = \lambda^{q}_{\alpha} \;\bar{\mathfrak{C}}^{[\infty]}_{1}(\Gamma_{\alpha}).
\end{equation}
Now recall that $q$ is the logarithm of the distance, which leads to
\begin{equation} \label{eq:TNcorr5}
 \bar{\mathfrak{C}}^{[\infty]}_{\ell}(\Gamma_{\alpha}) =
 \lambda^{\log_2 \ell}_{\alpha} \;\bar{\mathfrak{C}}^{[\infty]}_{1} =
 \ell^{\,\xi_{\alpha}} \;\bar{\mathfrak{C}}^{[\infty]}_{1}.
\end{equation}
We obtained an exact power-law decay for out fixed distance two-point correlator,
with the critical exponent $\xi_{\alpha} = \log_2 \lambda_{\alpha}$. This is one of
the main results I achieved during my doctorateship study \cite{IoTree}:
an infinite homogeneous tree Tensor Network defines always a critical state,
where the two-point critical exponents are the logarithms of the spectrum of $\Dsla$,
and their respective eigenoperators correspond to primary fields,
i.e. exactly-scaling operators.

Notice that since $|\lambda_{\alpha}| \leq 1$ as CPT maps are contractive, the critical
exponents have always a negative real part $\Re(\xi_{\alpha}) \leq 0$. Thus, 
although \eqref{eq:TNcorr5} might oscillate, it always decays in modulus, never explodes.

For a generic observable $\Gamma$, we can write a formal expression of its correlation
function which exploits its expansion in the generalized eigenbasis
$\{\alpha, \partial, w\}$ of $\Dsla$, like
we did for \eqref{eq:Polyexpand}. This allows us to write
\begin{equation} \label{eq:Polypoly}
 \bar{\mathfrak{C}}^{[\infty]}_{\ell}(\Gamma) =
 \sum_{\alpha} \ell^{\,\xi_{\alpha}} \cdot
 \sum_{\partial, w} \mathcal{P}^{[\Delta_{\partial}-w]}_{\lambda_{\alpha}}(\log_2 \ell)
\end{equation}
Where $\mathcal{P}^{[x]}$ are polynomials of degree $x$, which eventually
arise from the Jordan block structure of $\Dsla$.
As you see, the power-law always dominates the logarithmic part and rules
the physical critical behavior of the thermodynamical TTN state at long distances.

\begin{figure} 
 \begin{center}
  \begin{overpic}[width = 260pt, unit=1pt]{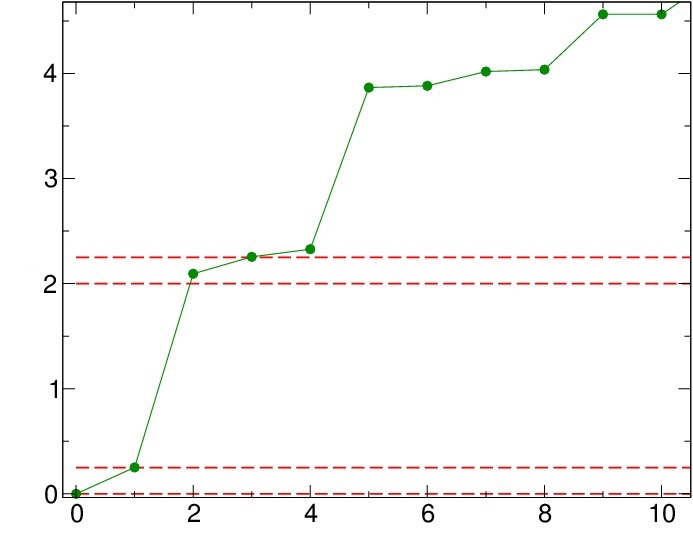}
 \put(0, 110){\large $\xi_{\alpha}$}
 \put(134, -4){\large $\alpha$}
  \end{overpic}
 \end{center}
 \caption{ \label{fig:Rizz02}
Critical exponents $\xi_{\alpha}$ (in modulus) for an infinite Ising chain, calculated with homogeneous MERA algorithm.
Here a bondlink $D = 4$ was used. The red dashed lines report the theoretical values.
The graphic was kindly contributed by M.~Rizzi and S.~Montangero \cite{MatteoMera}.
 }
\end{figure}

\vspace{.5em}
\emph{\textbf{Criticality of MERA -}}
The results and observations we just presented hold for a MERA geometry as well,
although since size 1 is not a stable causal cone width, it is necessary to involve
bigger-sized observables and density matrices. Precisely, in a binary MERA, even if we started from a
1+1 (two-point) operator, after applying few ascending maps, we increase its support size,
typically mapping it into a 3+3 operator (three adjacent sites, and other three adjacent sites, these two
parties standing at arbitrary distance), and after that size becomes stable.
According to this framework, critical exponents of 3+3 correlators
(and thus also of 1+1 correlators, which are a subclass of the 3+3 ones) are given by the
logarithms of the spectrum of
\begin{equation}
 \Dsla^{\text{MERA}} = \frac{1}{2} \left( \mathcal{D}^{\text{MERA}}_{3 \to 3, L} \otimes
 \mathcal{D}^{\text{MERA}}_{3 \to 3, L} + \mathcal{D}^{\text{MERA}}_{3 \to 3, R} \otimes
 \mathcal{D}^{\text{MERA}}_{3 \to 3, R} \right),
\end{equation}
where descending maps $\mathcal{D}^{\text{MERA}}_{3 \to 3, L}$ and $\mathcal{D}^{\text{MERA}}_{3 \to 3, R}$
are those defined in \eqref{eq:TNmap6}. This tells us that homogeneous MERA states are also critical.

It is commonly believed that, since MERA have larger stable
causal cone sizes, they are more suitable to represent a state where the physics at very short ranges
is sensibly different from the critical behavior at mid-to-long ranges: indeed homogeneous TTN states are forced
to hold \eqref{eq:TNcorr5} at every lenghtscale, even when $\ell \sim 2$, while for a MERA state
density matrices up to size 4 are loosely related related to their long-range physics.

\begin{figure} 
 \begin{center}
  \begin{overpic}[width = 260pt, unit=1pt]{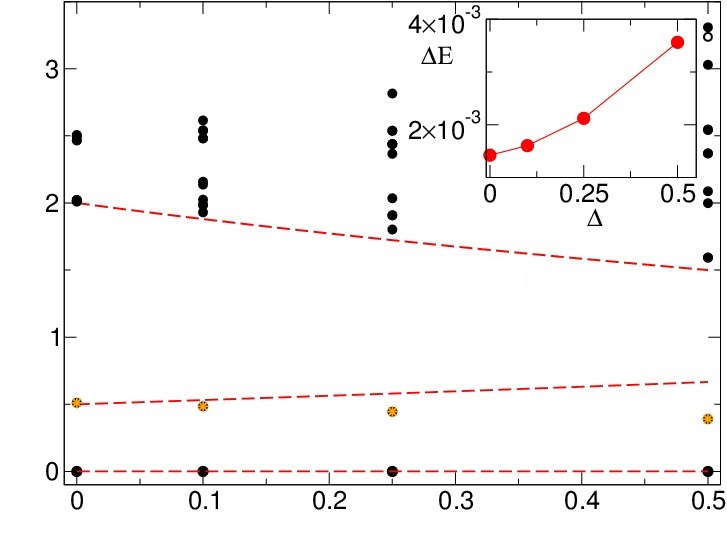}
 \put(2, 100){\large $\xi_{\alpha}$}
 \put(134, -2){\large $\Delta$}
  \end{overpic}
 \end{center}
 \caption{ \label{fig:Rizz01}
Critical exponents $\xi_{\alpha}$ for an infinite XXY chain, calculated with a homogeneous MERA algorithm,
as a function of the anisotropy $\Delta$.
Here a bondlink $D = 4$ was used. The red dashed lines report the theoretical values.
Errors on the ground state energy (per site) appear in the inset.
The graphic was kindly contributed by M.~Rizzi and S.~Montangero \cite{MatteoMera}.
 }
\end{figure}

\section[Importance of translational fluctuations]{The importance of translational\\fluctuations in TTN}

Addressing translational problems with a non-translational variational ansatz
might sound a sub-optimal choice for both analytical study and numerical simulations.
Yet, TTN and MERA, as we just showed, manifest natural scaling properties that suit so smoothly
critical systems, that are excellent variational candidates. Therefore, we are encouraged to
wonder whether the capability of hierarchical Tensor Networks to reproduce a complete conformal symmetry,
when $L \to \infty$, goes beyond the mere paradigm of incoherent translational mixture.
In other words: it is possible that in the limit of infinite layers $\mu$ (or sites), a tree Tensor Network
\emph{recovers} the same translational invariance that is forbidden to achieve at finite sizes?
We dealt with this question by considering how the translational fluctuations renormalize and
scale in tree networks, and developed a peculiar conclusion:
fluctuations are fundamental in TTN: they are necessary to describe interesting strongly-correlated physics,
especially in the thermodynamical limit.
If fluctuation vanish, the TTN state becomes trivial, and separable.
We will now sketch a derivation of this relation between fluctuations and entanglement.

In infinite systems, translational invariance is typically addressed as a hierarchy of equations,
each one of them referring to a characteristic size $\ell$, and stating that the density matrices
of that size are homogeneous in the lattice. Obviously, this implies that the same relation holds for any smaller
size $\ell' \leq \ell$, thus the hierarchical relationship.

Let us start from a one-site observable $\Theta$, and consider the
translational fluctuation upon a finite 1D lattice
\begin{equation}
 \left(\Delta \Theta^{[\mu]} \right)^2 = \frac{1}{2^{\mu}} \sum_{j = 1}^{2^{\mu}}
 \langle \Theta_j \rangle^2 - \left( \frac{1}{2^{\mu}} \sum_{j = 1}^{2^{\mu}} \langle \Theta_j \rangle \right)^2.
\end{equation}
where $L = 2^{\mu}$ is the system size. When we probe such fluctuation
upon a homogeneous binary TTN state, the expression becomes
\begin{multline}
 \left(\Delta \Theta^{[\mu]} \right)^2 =
 \trace\left[ \Theta \otimes \Theta \cdot
 \left(  \eta^{[\mu]}_0 - \bar{\rho}^{[\mu]}_1 \otimes \bar{\rho}^{[\mu]}_1 \right) \right] = \\ =
 \trace\left[ \Theta \otimes \Theta \cdot
 \left(  \Dsla^{\mu} \left(\eta^{[0]}_0 \right) -
 \Di^{\mu} \otimes \Di^{\mu} \left( \bar{\rho}^{[0]}_1 \otimes \bar{\rho}^{[0]}_1 \right) \right) \right],
\end{multline}
where we used the fact that $\trace[X]^2 = \trace[X \otimes X]$, while
$\eta^{[\mu]}_0$ is the same defined in \eqref{eq:sigmaeta}.
Moving to the thermodynamical limit is trivial now, and it reads
\begin{equation} \label{eq:zeropai}
 {\Delta \Theta}^2_{\infty}
 = \lim_{\mu \to \infty} \left({\Delta \Theta^{[\mu]}} \right)^2 =
 \trace\left[ \Theta \otimes \Theta \cdot \left(
 \sigsla - \rho^{\infty}_{1} \otimes \rho^{\infty}_{1}
 \right) \right].
\end{equation}
with $\sigsla$ and $\rho^{\infty}_{1}$ being respectively the fixed point of $\Dsla$ and $\Di$,
which are uniquely defined once we assume that both maps $\Dsla$ and $\Di$ are mixing.
We can now state that the thermodynamical TTN state is 'size-1 translational' if the quantity we just calculated vanishes;
and it can be shown that this happens only if the density matrices $\sigsla$
and $\rho^{\infty}_{1} \otimes \rho^{\infty}_{1}$ coincide.

Indeed, let $A$ and $B$ be any two one-site observables, then
\begin{multline} \label{eq:zeroby}
 \trace \left[ \left( A \otimes B + B \otimes A \right)\cdot \left(
 \sigsla - \rho^{\infty}_{1} \otimes \rho^{\infty}_{1}
 \right) \right] = \\ =
 {\Delta (A+B)}^2_{\infty} - {\Delta A}^2_{\infty} - {\Delta B}^2_{\infty} = 0,
\end{multline}
which must be zero as as we are requiring that every translational fluctuation is negligible.
Now let $F$ be the swap operator $F |\alpha\rangle \otimes |\beta\rangle = |\beta\rangle \otimes |\alpha\rangle$,
 $F = F^{-1} = F^{\dagger}$.
Clearly $F(A \otimes B)F = B \otimes A$ and
$F(\rho^{\infty}_{1} \otimes \rho^{\infty}_{1} )F = \rho^{\infty}_{1} \otimes \rho^{\infty}_{1}$.
Moreover $\Dsla_{\nu}$ is left invariant under the action of the swap gate, i.e.
$F \Dsla_{\nu}( F A F) F = \Dsla_{\nu}(A)$. This implies that $F \sigsla_{\nu}^{f} F$ is a fixed
point of $\Dsla_{\nu}$, but since it is a mixing map the fixed point must be unique, which
leads to $\sigsla_{\nu}^{f} = F \sigsla_{\nu}^{f} F$.
The previous manipulations with the swap operator allow us to write the following equivalence
\begin{equation}
 \trace \left[  A \otimes B \cdot \left(
 \sigsla - \rho^{\infty}_{1} \otimes \rho^{\infty}_{1}
 \right) \right] = 
 \trace \left[  B \otimes A \cdot \left(
 \sigsla - \rho^{\infty}_{1} \otimes \rho^{\infty}_{1}
 \right) \right].
\end{equation}
and since the sum of these terms is zero by \eqref{eq:zeroby}, they must be both zero separately, for
any operator $A$ and $B$. But since tensor product operators generate the whole algebra of 2-site operators
we must conclude that
\begin{equation}
 \sigsla = \rho^{\infty}_{1} \otimes \rho^{\infty}_{1},
\end{equation}
so that $\Dsla_{\nu}$ and $\Di_{\nu} \otimes \Di_{\nu}$ must have the same fixed point.

When this condition is verified then $\rho^{\infty}_{1}$ is automatically the fixed point of
of $\Di_{L}$ and $\Di_{R}$ as well, because
\begin{multline} \label{eq:ciseci}
 0 = \left(\Dsla - \Di \otimes \Di \right)(\rho^{\infty}_{1} \otimes \rho^{\infty}_{1})
 = \\ =
 \frac{1}{4} (\Di_{L} - \Di_{R}) (\rho^{\infty}_{1}) \otimes(\Di_{L} - \Di_{R}) (\rho^{\infty}_{1}).
\end{multline}
Therefore, a sufficient and necessary condition for translational invariance
to hold at size 1, is that $\Di_{L}(\rho^{\infty}_{1}) = \Di_{R}(\rho^{\infty}_{1})
= \Di(\rho^{\infty}_{1}) = \rho^{\infty}_{1}$.

So far, so good. Let us proceed further and require translational invariance at size 2.
The same derivation can be applied to two-adjacent sites observables $\Theta_{j,j+1}$,
its result is that a common fixed point must be shared by maps $\Di_{2,R}$ and $\Di_{2,L}$, where
\begin{equation}
 \begin{aligned}
  \Di_{2,R} &\equiv \Di_{R} \otimes \Di_{L}\\
  \Di_{2,L} &\equiv \eS \circ \left( \cos^2\!\theta\;\, \trace_R + \sin^2\! \theta\;\, \trace_L \right).
 \end{aligned}
\end{equation}
It is clear that such fixed point must coincide with $\rho_2^{\infty}$, since by \eqref{eq:dm2td3} we know that
$\Di_{2 \to 2} = \frac{1}{2} (\Di_{2,R} + \Di_{2,L})$, and if we require that $\Di_{2 \to 2}$ is mixing
its unique fixed point $\rho_2^{\infty}$ must be the shared fixed point of $\Di_{2,R}$ and $\Di_{2,L}$.
At the same time, due to \eqref{eq:ciseci}, the fixed point of $\Di_{2,R}$ is
$\rho^{\infty}_{1} \otimes \rho^{\infty}_{1}$, and by the common fixed point property we must have that
\begin{equation}
 \eS (\rho^{\infty}_{1} ) = \rho^{\infty}_{1} \otimes \rho^{\infty}_{1}
 = \Di_{2 \to 2} (\rho^{\infty}_{1} \otimes \rho^{\infty}_{1}) = \rho^{\infty}_{2},
\end{equation}
meaning that the two-sites reduced density matrix is a separable state, actually a tensor product state.
By extension, every reduced density matrix is separable, and the system can manifest no quantum correlations
at all, since $\Delta\sigma$ defined in equation \eqref{eq:eta2td2} would be the null operator.

Resuming this whole discussion, we proved that by just requiring that the thermodynamical
TTN-state manifests translational invariance property at size 2, we automatically end with a completely
factorized state, all entanglement is broken down. In this framework, we could state that translational
fluctuations are needed in a tree network if we want to describe strongly correlated physics;
they are unavoidable, even in the thermodynamical limit.

It is curious to realize that the previous demonstration does \emph{not} hold for a MERA geometry:
this could actually be one of the first remarkable arguments for preferring MERA to Trees.
Indeed for a MERA topology, according to this theoretical picture, we could hope to reproduce accurately
conformal symmetry without implications of triviality for thermodynamical limit entanglement.

\section{Parent Hamiltonians of TTN states}

Finding ground states of physically-meaningful Hamiltonians has been for decades a maximal interest topic,
it is also the very purpose of this whole Tensor Network variational ansatz itself. But even
dealing with the reverse problem can be challenging and fruitful: given a quantum (many-body) state
$|\Psi\rangle$, can we identify, characterize of even build a non-trivial Hamiltonian for which $|\Psi\rangle$ is
the ground state? Of course, the research of such a \emph{parent} Hamiltonian $H$, must be addressed
in accordance to some physically sensible constraint: we might for instance require for $H$ to
be short-ranged, to be translational, or maybe capable of coupling only a limited number of particles
per single interaction term. The more are the requirements, the harder is the problem.
In my research work, I focused on Tree Tensor Network as 1D many body states,
and analyzed how to explicitly build a \emph{non-trivial, short-ranged
and translational} Hamiltonian which is parent for the TTN state, in PBC. In this section we will
sketch the construction.

Let us start from the formal definition of the Hamiltonian $\mathcal{H}$ we want to achieve,
whose elementary terms $H$ have limited size support $\nu \ll L$:
\begin{equation} \label{eq:hamexpand}
 \mathcal{H} = \sum_{\ell_0 = 1}^{L} H_{\ell_0 + 1, \ldots, \ell_0 + \nu}.
\end{equation}
In order for this Hamiltonian to be parent for our homogeneous-TTN state
$| \Psi^{[\mu]}(\Lambda) \rangle$, we must ensure that its expectation values coincides with
the minimum of the spectrum of $\mathcal{H}$, which is also the variational minimum
of the expectation values $\langle \mathcal{H} \rangle$ on the whole Hilbert space of states:
\begin{equation}
 \langle \Psi^{[\mu]}(\Lambda) | \mathcal{H} | \Psi^{[\mu]}(\Lambda) \rangle = 
 \min_{|\Phi\rangle} \left\{ \frac{\langle \Phi |\mathcal{H}| \Phi \rangle}{\langle \Phi | \Phi \rangle} \right\}.
\end{equation}
At the same time, we could read $\mathcal{H}$ as an unnormalized translational average.
This allows us to summon again the incoherent translational mixture formalism for TTN density matrices,
which reads
\begin{multline}
 \langle \Psi^{[\mu]}(\Lambda) | \mathcal{H} | \Psi^{[\mu]}(\Lambda) \rangle =
 \sum_{\ell_0 = 1}^{L} \langle \Psi^{[\mu]}(\Lambda) | H_{\ell_0 + 1, \ldots, \ell_0 + \nu} | \Psi^{[\mu]}(\Lambda) \rangle
 = \\ =
 \sum_{\ell_0 = 1}^{L} \trace\left[ H \cdot \rho_{\{\ell_0 + 1, \ldots, \ell_0 + \nu\}}^{[\mu]} \right] =
 2^{\mu}\; \trace\left[ H \cdot \bar{\rho}_{\nu}^{[\mu]} \right].
\end{multline}
Now, suppose that exists a finite small (= non-scaling) $\nu$ for which $\bar{\rho}_{\nu}^{[\mu]}$
has non-maximal rank. If that is true, then $\bar{\rho}_{\nu}^{[\mu]}$ has some nontrivial kernel, with
strictly positive dimension, completely generated by an orthogonal set of vectors $|\kappa_{w}\rangle$.
Then we say that the elementary Hamiltonian term $H$ is built as follows
\begin{equation} \label{eq:Parentbuild}
 H = \sum_{w} \omega_w \,| \kappa_w \rangle \langle \kappa_w |,
\end{equation}
with arbitrary positive weights $\omega_w \geq 0$. With this prescription, we obtain a Hamiltonian $\mathcal{H}$
which is positive, as it is the sum of positive terms, nontrivial, if at least one $\omega_w$ is strictly
greater than zero, and for which it holds
\begin{equation}
 \langle \Psi^{[\mu]}(\Lambda) | \mathcal{H} | \Psi^{[\mu]}(\Lambda) \rangle =
 2^{\mu} \sum_{w} \omega_w \, \langle \kappa_w | \bar{\rho}_{\nu}^{[\mu]} | \kappa_w \rangle = 0,
\end{equation}
since every $|\kappa_w\rangle$ is in the kernel of $\bar{\rho}_{\nu}^{[\mu]}$. But since
$0$ is necessarily the minimum of the spectrum of $\mathcal{H}$, as it must be a positive operator,
$| \Psi^{[\mu]}(\Lambda) \rangle$ is clearly a ground state for $\mathcal{H}$.
This is the idea of our construction.

The central point of our proof, therefore, now becomes to demonstrate that for the (homogeneous) TTN state
there is always some finite non-scaling size $\nu$ for which the averaged $\nu$-sites density matrix
$\bar{\rho}_{\nu}^{[\mu]}$ has non-full rank. To show this we will exploit the fact that descending maps
can grow the size of density matrices, while the corresponding increase in entanglement
is well-kept under control by isometricity condition on $\Lambda$ (and $X$ as well, in MERA).
We will now discuss dimensionality relations leading to characterization of the smallest
size $\nu$ for which $\bar{\rho}_{\nu}^{[\mu]}$ is necessarily non-full rank. This result will obviously
depend on the Tree or MERA geometry we are employing. In the following instances,
we will consider Trees and MERA having a renormalization dimension $D$ which is equal to the physical
local dimension $d$ so that TN-homogeneity is meaningful up to the physical lattice.

\vspace{.5em}
\emph{\textbf{Binary Tree -}}
in tree geometries it is not possible to establish any bound upon the entanglement of
$\bar{\rho}_{2}^{[\mu]}$, for 2 is a (meta-) stable causal cone width.
Let us move to size three: by adopting the formalism \eqref{eq:avemap3} we know that
\begin{equation} \label{eq:TNrank1}
 \begin{aligned}
 \bar{\rho}_{3}^{[\mu]} &= \Di_{2 \to 3} \left( \bar{\rho}_{2}^{[\mu-1]} \right) =
 \frac{1}{2}\left[ \Di_R \otimes \eS + \eS \otimes \Di_L \right]
 \left( \bar{\rho}_{2}^{[\mu-1]} \right)\\
 &= \mathcal{I} \otimes \eS \left[ \frac{1}{2}\, \Di_R \otimes \mathcal{I} \left( \bar{\rho}_{2}^{[\mu-1]} \right) \right] +
 \eS \otimes \mathcal{I} \left[ \frac{1}{2}\, \mathcal{I} \otimes \Di_L \left( \bar{\rho}_{2}^{[\mu-1]} \right) \right]
 \end{aligned}
\end{equation}
where $\mathcal{I}$ is the identical map, taking every operator into itself. Notice that even though the mapping
$\eS$ increases size, as it is the application of an isometry, it is left-invertible, and thus preserves the rank;
by natural extension $\eS \otimes \mathcal{I}$ preserves the rank just as well. Now, the expression within the first
square parentheses of \eqref{eq:TNrank1} is a $d^2 \times d^2$ matrix, whose rank is obviously bound by
the amount of rows or columns (whichever the smallest), i.e. $d^2$. Finally, it is clear that the maximal rank of
a sum of two matrices $A+B$ cannot overcome the sum of ranks of $A$ and $B$ separately.
These considerations tell us that
\begin{equation} \label{eq:TNrank2}
 \text{Rnk}\left[ \bar{\rho}_{3}^{[\mu]} \right] \leq 2 \,d^2
\end{equation}
where $\bar{\rho}_{3}^{[\mu]}$ is a $d^3 \times d^3$ matrix, and thus its rank is non-maximal whenever
$d^3 > 2\,d^2$, which happens for a local physical dimension of $d \geq 3$.

When we are considering a ring of 2-level systems, e.g. a spin-{$\frac{1}{2}$} chain, size
three is not enough for ensuring non-maximality of the density matrix rank. Let us move to
size four then:
\begin{equation} \label{eq:TNrank3}
 \begin{aligned}
  \bar{\rho}_{4}^{[\mu]} &= \frac{1}{2} \, \Di_R \otimes \eS \otimes \Di_L \left(\bar{\rho}_{3}^{[\mu-1]} \right)
  + \frac{1}{2} \, \eS \otimes \eS \left(\bar{\rho}_{2}^{[\mu-1]} \right)\\
  &= \mathcal{I} \otimes \eS \otimes \mathcal{I}
  \left[ \frac{1}{2}\, \Di_R \otimes \mathcal{I} \otimes \Di_L \left(\bar{\rho}_{3}^{[\mu-1]} \right)\right]
  + \eS \otimes \eS \left[ \frac{1}{2} \,\bar{\rho}_{2}^{[\mu-1]} \right].
 \end{aligned}
\end{equation}
As previously motivated, both $\mathcal{I} \otimes \eS \otimes \mathcal{I}$ and
$\eS \otimes \eS$ maps preserve the rank; and of course the ranks of their respective arguments
is bound by their row/column dimension. Which leads to
\begin{equation} \label{eq:TNrank4}
 \text{Rnk}\left[ \bar{\rho}_{4}^{[\mu]} \right] \leq d^3 + d^2
\end{equation}
which is always strictly smaller that $d^4$ for any nontrivial dimension $d \geq 2$.
In conclusion, for binary trees, it is always possible to write a nontrivial, translational,
short-range Hamiltonian, according to the prescription \eqref{eq:Parentbuild} where every
term $H$ involves \emph{at most} four (adjacent) sites, which become three for a local dimension
$d$ greater than 2.

\vspace{.5em}
\emph{\textbf{Higher branching TTN -}}
when the Tree geometry has a branching number $b$ higher than 2, the minimal size $\nu$ of
$\bar{\rho}_{\nu}^{[\mu]}$, for which it is possible to ensure non-maximality of the rank, grows.
As an example, we can mention that for
a ternary tree this size $\nu$ is 5. But the growth rate, as a function of $b$, is somehow irregular, erratic;
sometime the local dimension $d$ is influent, sometime it is not.
Nevertheless, for each geometry, a finite $\nu$ definitely exists, and it is always equal or less than $2b$.
In fact for a $b$-branching tree network it holds
\begin{equation} \label{eq:TNrank5}
 \text{Rnk}\left[ \bar{\rho}_{2b}^{[\mu]} \right] \leq b \;d^{b+1}
\end{equation}
which is always less than its row/column dimension $d^{2b}$. This interaction-range bound is not optimal,
but it still does not scale with system size.

\vspace{.5em}
\emph{\textbf{MERA -}} The presence of the disentanglers also increases the minimal size
of non-full rank $\bar{\rho}_{\nu}^{[\mu]}$. Precisely, for a binary MERA we get
\begin{equation} \label{eq:TNrank6}
 \text{Rnk}\left[ \bar{\rho}_{5}^{[\mu]} \right] \leq 2\;d^4
 \qquad \mbox{and} \qquad
 \text{Rnk}\left[ \bar{\rho}_{6}^{[\mu]} \right] \leq d^4 + d^5;
\end{equation}
which tells us that we have to accept an interaction range $\nu = 5$ for a local dimension
$d \geq 3$, and move to $\nu = 6$ otherwise.
If our MERA geometry is ternary, we have to push to seven sites, since
\begin{equation} \label{eq:TNrank7}
 \text{Rnk}\left[ \bar{\rho}_{7}^{[\mu]} \right] \leq 3\,d^5.
\end{equation}
At any rate, the effective range of our nontrivial translational parent Hamiltonian
might not be so short, but in the end the construction \eqref{eq:Parentbuild} is always possible
in practice.

\subsection{Unfrustration and degeneracy}

It is meaningful to point out some general properties of the parent Hamiltonians generated with
the protocol \eqref{eq:Parentbuild} just described. First, we would like to highlight that
these Hamiltonians are necessarily frustration-free, meaning that the TTN is ground state of
every single interaction term.

Indeed, let us consider the expansion \eqref{eq:hamexpand}. We built every single $H$ term
to be a positive operator so that the full Hamiltonian $\mathcal{H}$ would be positive as well.
But since the TTN state has zero expectation value of $\mathcal{H}$, it must be
\begin{equation}
 0 = \langle \Psi^{[\mu]}(\Lambda) | \mathcal{H} | \Psi^{[\mu]}(\Lambda) \rangle =
 \sum_{\ell_0 = 1}^{L} \langle \Psi^{[\mu]}(\Lambda) | H_{\ell_0 + 1, \ldots, \ell_0 + \nu}
 | \Psi^{[\mu]}(\Lambda) \rangle.
\end{equation}
Now, the only way for a sum of positive terms to be zero is that every term is separately zero:
\begin{equation}
\langle \Psi^{[\mu]}(\Lambda) | H_{\ell_0 + 1, \ldots, \ell_0 + \nu}
 | \Psi^{[\mu]}(\Lambda) \rangle = 0 \qquad \forall \,\ell_0,
\end{equation}
i.e. $\mathcal{H}$ is unfrustrated. This one consideration is curiously related to
refs.~\cite{FrustrEisert1, FrustrEisert2}, where it was shown that for structured frustration-free
Hamiltonians, it is possible to build analytically a ground state via Tensor Network designs.
In some sense, our result is the other face of the same coin.

\begin{figure} 
 \begin{center}
\begin{overpic}[width = 320pt, unit=1pt]{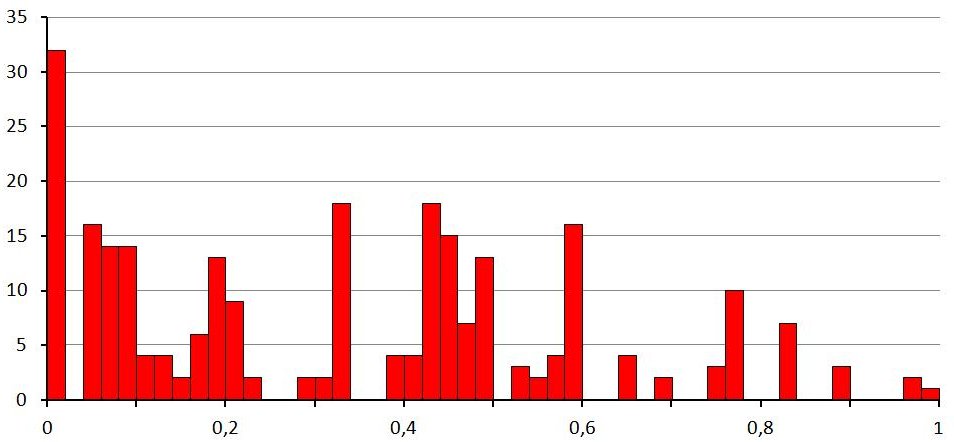}
 \put(-25, 70){\footnotesize{DOS}}
 \put(160, -10){$\langle \mathcal{H} \rangle$}
\end{overpic}
 \end{center}
 \caption{ \label{fig:Histo}
Unnormalized density of states (DOS) of a TTN parent Hamiltonian $\mathcal{H}_8$ for 
on a spin-$\frac{1}{2}$ PBC chain of 8 sites \cite{IoTree}.
Here we used the sample isometry $\Lambda: |0\rangle \to |01\rangle$, $|1\rangle \to 2^{-1/2}(|00\rangle + |11\rangle)$.
The interaction term $H$ was calculated from $\rho^{\infty}_4 (\Lambda)$, via \eqref{eq:Parentbuild},
with random positive weights $\omega_w$. Notice that the ground space is 32-fold
degenerate: its dimension is twice the lower bound given by \eqref{eq:kerbound}, the $\times 2$ factor
deriving by a hidden $\mathbb{Z}_2$ symmetry.
 }
\end{figure}

Another manifest property of our TTN parent Hamiltonian, and somehow related to the unfrustration,
is that the ground space of $\mathcal{H}$ is highly degenerate, and we can characterize it to some
extent. To show this, we will consider for simplicity the case 
of a Binary Tree and a local dimension $d\geq 3$,
and assume that the construction \eqref{eq:Parentbuild} is being developed at the thermodynamic limit
$\mu = \infty$.
We will also request, as additional hypothesis, that $\rho^{\infty}_2$ has full rank.
This assumption is typically weak, practically guaranteed in numerical settings, where stochastic noise
makes every matrix full rank.

Now, via \eqref{eq:Parentbuild} we derive a positive interaction term $H$ having support in the
kernel of $\rho^{\infty}_3$. To begin with, let us prove that $H$ is in the Kernel of
the ascending map $\Ai_{3 \to 2}$, adjoint to $\Di_{2 \to 3}$ of eq.~\eqref{eq:TNrank1}:
by definition we have
\begin{equation}
 0 = \trace\left[ H \cdot \rho^{\infty}_3 \right] =
 \trace\left[ H \cdot \Di_{2 \to 3} \left( \rho^{\infty}_2 \right) \right]
 = \trace\left[ \Ai_{3 \to 2}\left(H \right) \cdot \rho^{\infty}_2 \right].
\end{equation}
where since $\Ai_{3 \to 2}$ map is completely positive and
unital, $\Ai_{3 \to 2}\left(H \right)$ is surely a positive operator.
$\rho^{\infty}_2$ is also positive, and being full-rank, its smallest eigenvalue
$\lambda_0$ is strictly greater than zero; then
\begin{equation}
 0 = \trace\left[ \Ai_{3 \to 2}\left(H \right) \cdot \rho^{\infty}_2 \right]
 \geq \lambda_0 \,\trace\left[ \Ai_{3 \to 2}\left(H \right) \right] \geq 0.
\end{equation}
So it must be that the trace of $\Ai_{3 \to 2}\left(H \right)$ is zero, but
the only positive traceless operator is the null operator, thus $\Ai_{3 \to 2}\left(H \right) = 0$.
This very argument will let us characterize ground spaces of Hamiltonians $\mathcal{H}$
generated by this interaction term $H$.

In fact, let us consider a finite system now, with the same local dimension $d$ and
an ever number of sites $L = 2\ell$. we will define our \emph{trial} state as
a generic pure state on $\ell$ sites $|\Psi_0\rangle$, that we grow to $2\ell$ sites
by means of a single layer of isometries $\Lambda$: the same $\Lambda$ tensor
we used to build $H$. Let us sketch the Tensor Network design of this trial state as follows:
\begin{equation} \label{eq:Pluslayer}
\begin{overpic}[width = 250pt, unit=1pt]{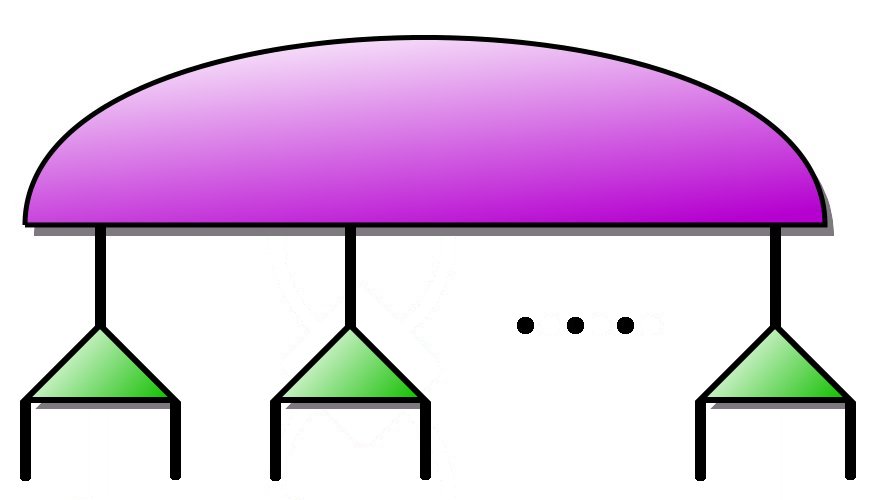}
 \put(115, 100){$| \Psi_0 \rangle$}
 \put(25, 33){$\Lambda$}
 \put(96, 33){$\Lambda$}
 \put(218, 33){$\Lambda$}
\end{overpic}
\end{equation}
It is trivial to show that this trial state is definitely a ground state for the Hamiltonian
$\mathcal{H} = \sum_{\ell_0 = 1}^{L} H_{\ell_0 + 1 \ldots \ell_0 + 3}$. Indeed, let us write
\begin{equation} \label{eq:Pluslayer2}
 \langle \mathcal{H} \rangle = \trace[H \cdot \bar{\rho}_{3}^{\downarrow}] =
 \trace[H \cdot \Di_{2 \to 3} ( \bar{\rho}_{2}^{\uparrow})] =
 \trace[\Ai_{3 \to 2}(H) \cdot \bar{\rho}_{2}^{\uparrow}] = 0,
\end{equation}
where $\bar{\rho}_{\nu}^{\uparrow}$ (resp. $\bar{\rho}_{\nu}^{\downarrow}$) is the reduced density
matrix, $\nu$-sites translationally averaged, of the trial state before (after) applying the layer of 
isometries $\Lambda$.
Equation \eqref{eq:Pluslayer2} is telling that the trial state is a ground state of the Hamiltonian,
regardless from $| \Psi_0 \rangle$. Actually every state that is written in the form \eqref{eq:Pluslayer}
is a ground state of the system, and the layer of isometries preserves orthonormality, so
we can identify (at least) a set of $d^{\ell}$ orthogonal ground states. In conclusion, the ground space of any
$\mathcal{H}$ has a wide degeneracy, namely:
\begin{equation} \label{eq:kerbound}
 \mbox{dim}\left[ \mbox{Ker}\left(\mathcal{H}\right) \right] \geq d^{L/2},
\end{equation}
with $L$ the size of the system. The same discussion can be applied to the MERA case,
and leads to the same result, although the ground states this time are built
by attaching to $|\Psi_0\rangle$ a full MERA layer, with renormalizators $\Lambda$ and
disentanglers $X$ together.

\section{Open boundary MERA} \label{sec:BERA}

In section \ref{sec:MERAintro}, we mentioned that Tree Tensor Networks are equally suitable to simulate open boundary
systems as well as periodic boundary systems. This is mainly due to the fact that there are pairs of adjacent sites
that renormalize separately for an arbitrary number of layers, thus actually
constituting an inner-boundary when approaching TD-limit (honestly, a thermodynamical TTN is geometrically
equivalent to the frontier of the Cantor set). At the same time, we stated that MERA have a natural attitude for
periodic topologies, since a single MERA layer couples every pair of neighboring sites.

It is easy to see, however, that a MERA state needs only a little adjustment to its network geometry,
to take properly into account the presence of an open boundary. This is naturally done by embedding
the boundaries, interpreted in the most general setting as a pair of additional degrees of freedom (ancillae),
in the MERA picture, while preserving the bulk network pattern, and thus the critical bulk properties.
The present idea leads to the following network design:
\begin{equation}
 \begin{overpic}[width = \textwidth, unit=1pt]{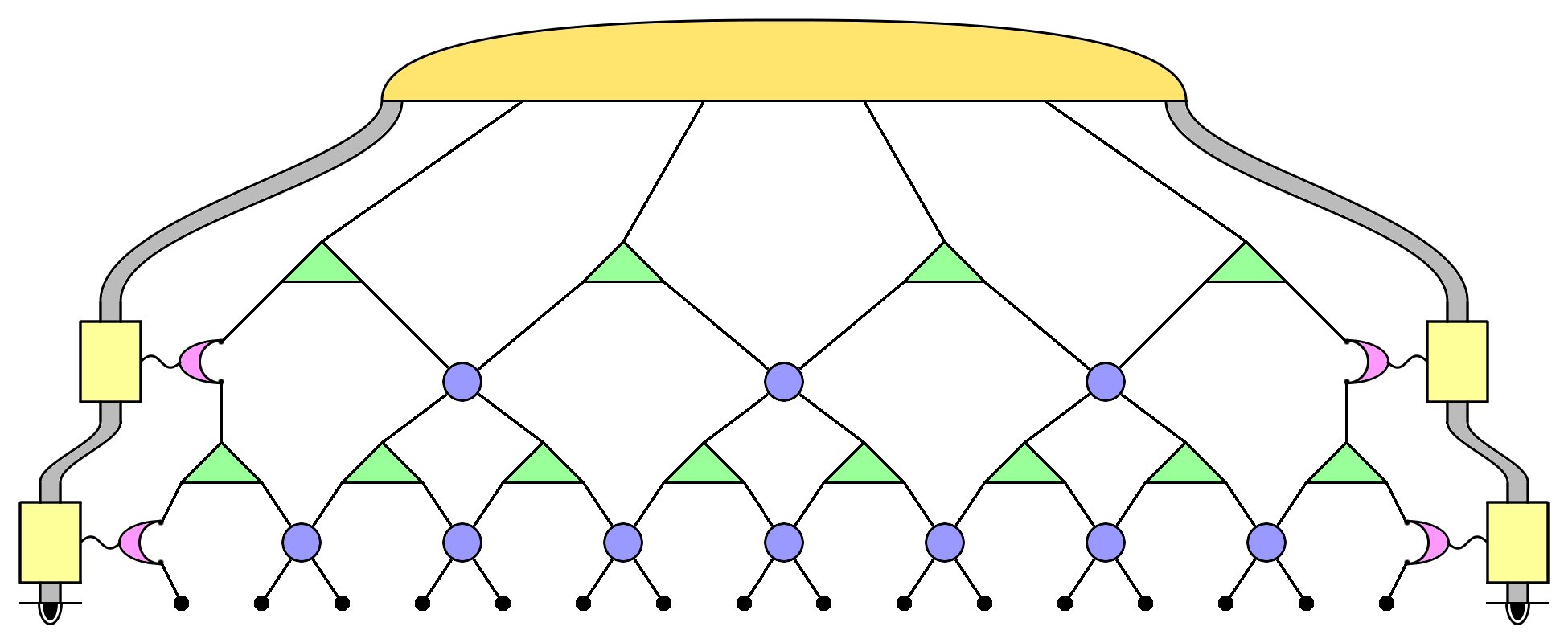}
 \put(190, 141){$\mathcal{C}$}
\end{overpic}
\end{equation}
which we refer to as (open) boundary-MERA \cite{BERA}. As you see from the diagram, during the same
algorithm-step when we apply disentanglers, we also allow the boundary ancilla to couple
locally with the system. We will add the constraint that this coupling operation is represented
by a unitary gate. Such additional requirement is necessary to preserve the causal cone relations
we argumented in section \ref{sec:causalcone}. Clearly, isometricity and unitarity conditions still hold
for renormalizers $\Lambda$ and disentanglers $X$ as before, i.e.
\begin{equation}
 \begin{overpic}[width = 350pt, unit=1pt]{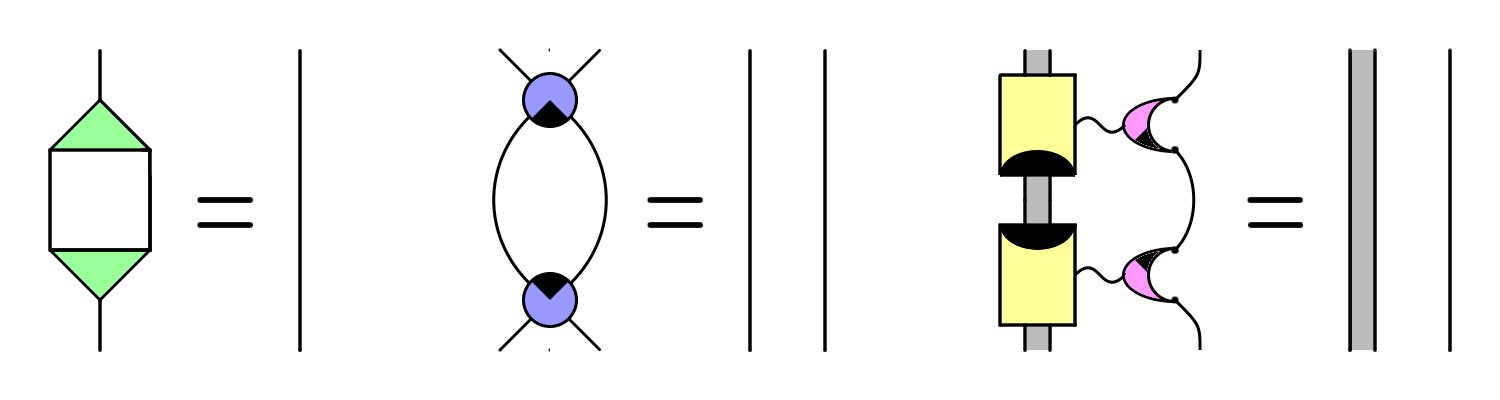}
 %\put(190, 14){$\mathcal{C}$}
\end{overpic}
\end{equation}
where the bottom tensors are complex conjugate and
up-down reversed versions of the ones standing directly above.
Preservation of the causal cone property, together with the fact that the binary MERA pattern
is identical to the PBC case in the bulk, tell us that descending map equations
\eqref{eq:TNmap6} still hold when the three involved sites
(causal cone width of 3 is still stable) are far from the boundaries, i.e.
\begin{equation} \label{eq:BERAmapbulk}
 \begin{aligned}
\rho^{[\mu]}_{2\ell, 2\ell+1, 2\ell+2} &= \mathcal{D}_{3 \to 3, L}
 \left( \rho^{[\mu-1]}_{\ell, \ell+1, \ell+2} \right)
 \qquad \mbox{and} \\
\rho^{[\mu]}_{2\ell+1, 2\ell+2, 2\ell+3} &= \mathcal{D}_{3 \to 3, R}
 \left( \rho^{[\mu-1]}_{\ell, \ell+1, \ell+2} \right)
 \end{aligned}
\end{equation}
where maps $\mathcal{D}^{\text{MERA}}_{3 \to 3}$ are those of eq.~\eqref{eq:TNmap7}.
When approaching the boundaries we must accordingly define descending quantum channels
that involve the new system-ancilla coupling element. For instance, close to the left boundary we get:
\begin{equation} \label{eq:BERAmapanci}
 \rho^{[\mu]}_{1,2,3} = \mathcal{K}_L \left( \rho^{[\mu-1]}_{A,1,2} \right)
 \qquad \mbox{and} \qquad
 \rho^{[\mu]}_{A,1,2} = \mathcal{B}_L \left( \rho^{[\mu-1]}_{A,1,2} \right),
\end{equation}
where the subscript $A$ refers to the left-ancilla degree of freedom, 
and the completely positive trace-preserving maps $\mathcal{K}_L$ and $\mathcal{B}_L$ are given by
\begin{equation} \label{eq:Beranci}
 \begin{overpic}[width = \textwidth, unit=1pt]{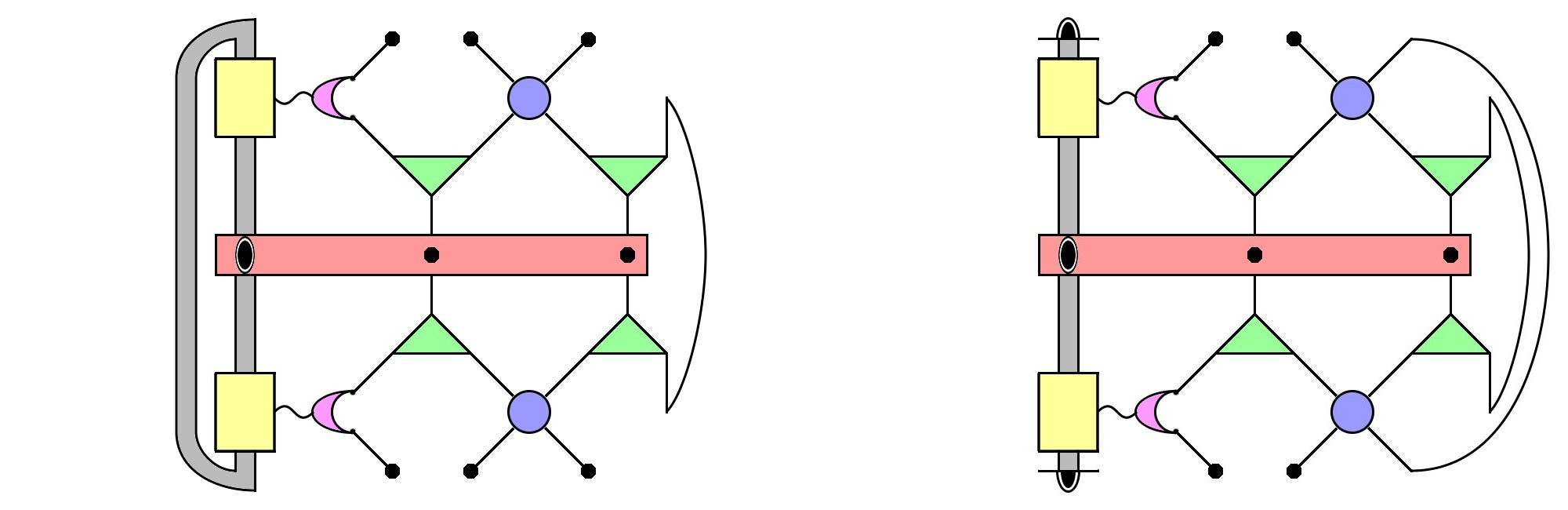}
 \put(3, 60){\large$\mathcal{K}_L = $}
 \put(220, 60){\large$\mathcal{B}_L = $}
\end{overpic}
\end{equation}
Similarly, the maps $\mathcal{K}_R$ and $\mathcal{B}_R$,
ruling the layer-recursive relations of density matrices at the right boundary,
are the left-right specular versions of \eqref{eq:Beranci}.
Notice from \eqref{eq:BERAmapanci} that once a causal cone touches a boundary,
it sticks to it along its upward propagation.

\subsection{Consistency of the thermodynamical limit}

Before speculating on how the presence of the boundary influences the system, we want to
show that as we approach the thermodynamical limit (and consider translationally-averaged
quantities), we recover the same physics of the corresponding PBC-MERA.
In particular, we will assume that the boundary-MERA is homogeneous, i.e. renormalizers $\Lambda$
are identical, and so disentanglers $X$, and even boundary-ancilla couplings; then move to $\mu \to \infty$.

We will sketch the consistency proof for size 3, which extends trivially to any size thanks to the
shrinking properties of the causal cone width. Consider the translationally averaged
3-site reduced density matrix $\bar{\rho}^{[\mu]}_3$:
\begin{equation}
 \bar{\rho}^{[\mu]}_3 = \frac{1}{2^{\mu + 2} - 2} \sum_{\ell_0 = 1}^{2^{\mu + 2} -2} \rho^{\mu}_{\ell_0, \ell_0+1, \ell_0+2}.
\end{equation}
We are actually not considering the ancillae as part of our system in this framework,
but the result would not change even if we did. We can now adopt recursive relations
\eqref{eq:BERAmapbulk} and \eqref{eq:BERAmapanci} and write:
\begin{multline}
 \bar{\rho}^{[\mu]}_3 = \frac{1}{2^{\mu + 2} - 2}
 \left[ \mathcal{K}_L \left( \rho^{[\mu-1]}_{A,1,2} \right) + \mathcal{K}_R \left( \rho^{[\mu-1]}_{L-1,L,A'} \right) \right]
 + \\ +
 \left( 1 - \frac{1}{2^{\mu + 1} - 1} \right) \cdot \Di_{3 \to 3} \left( \bar{\rho}^{[\mu-1]}_3 \right).
\end{multline}
In order to identify the thermodynamical limit of this quantity, it is important to notice that
the following expression converges to zero in trace-norm as $\mu$ grows:
\begin{equation}
 \lim_{\mu \to \infty} \left\| \bar{\rho}^{[\mu]}_3 - \Di_{3 \to 3} \left( \bar{\rho}^{[\mu-1]}_3 \right) \right\|
 \leq \lim_{\mu \to \infty} \left| \frac{4}{2^{\mu + 2} - 2} \right| = 0,
\end{equation}
where we used the fact that CPT maps are contractive, that the trace norm of a density matrix is 1,
and triangular inequality.
The previous equation guarantees that $\bar{\rho}^{[\mu]}_3$ goes in the limit to the fixed point
of $\Di_{3 \to 3}$, as
\begin{equation}
 \bar{\rho}^{\infty}_3 = \lim_{\mu \to \infty} \bar{\rho}^{[\mu]}_3 =
 \lim_{\mu \to \infty} \Di_{3 \to 3} \left( \bar{\rho}^{[\mu-1]}_3 \right) =
 \Di_{3 \to 3} \left( \bar{\rho}^{\infty}_3 \right),
\end{equation}
but the fixed point of $\Di_{3 \to 3}$ also characterized the thermodynamical limit in the periodic MERA,
and since we assume for $\Di_{3 \to 3}$ to be mixing, it has a unique fixed point, thus the two states must be the same.
This proves that PBC and OBC MERA manifest the same averaged physics in the thermodynamical limit,
which is the TD-consistency argument we requested.

\subsection{Boundary fluctuations and permeation}

Boundary conformal field theory prescribes a direct relation between critical exponents
of two-point correlation functions in the bulk, and one-point fluctuations close to
the boundary of an OBC critical system. We are now going to investigate such one-point
expectation values as a function of the distance from a boundary, say the left one,
and compare these analytical results with our previous acknowledgements involving
correlation functions in MERA, namely \eqref{eq:TNcorr5} and \eqref{eq:Polypoly}.

To do this, we consider the expectation value of a 3-site observable
applying at a distance $\ell \ll L \sim \infty$ from the, say left, boundary:
\begin{equation}
 \mathfrak{P}^{\infty}_{\ell} (\Theta) =
 \lim_{\mu \to \infty} \langle \Theta^{[\mu]}_{\ell, \ell+1, \ell+2} \rangle =
 \lim_{\mu \to \infty} \trace \left[ \Theta \cdot \rho^{[\mu]}_{\ell, \ell+1, \ell+2}  \right],
\end{equation}
which appears as a function, upon $\ell$, on how the influence of the boundary
permeates inside the  (infinite) system. We will now assume, for simplicity, that
such distance $\ell$ is a power of 2, say $\ell = 2^{\tau}$. Then, by adopting the
formalism of \eqref{eq:BERAmapbulk}, we obtain
\begin{multline} \label{eq:Leftdominance}
 \mathfrak{P}^{\infty}_{\ell = 2^{\tau}} (\Theta) =
 \lim_{\mu \to \infty} \trace \left[ \Theta \cdot
 \Di^{\tau}_{3 \to 3, L} \left( \rho^{[\mu-\tau]}_{1,2,3} \right) \right]
 = \\ =
 \lim_{\mu \to \infty} \trace \left[ \Theta \cdot \Di^{\tau}_{3 \to 3, L}
  \circ \mathcal{K}_L \left( \rho^{[\mu-\tau-1]}_{A,1,2} \right) \right] =
 \trace \left[ \Ai_{3 \to 3, L}^{\tau} \left( \Theta \right) \cdot
  \mathcal{K}_L \left( \rho^{\infty}_{A,1,2} \right) \right],
\end{multline}
where $\rho^{\infty}_{A,1,2}$ is the fixed point of $\mathcal{B}_L$, unique if the map is mixing.
The only residual dependence on $\tau$ is in the number of times the map
$\Ai_{3 \to 3, L}$ is to be applied to the operator (or, equivalently, $\Di_{3 \to 3, L}$ is to be applied
to the boundary density matrix). We can now proceed by adopting an argument similar
to \eqref{eq:TNcorr4}, i.e. let us assume that $\Theta_{\alpha,L}$ is an eigenoperator
of $\Ai_{3 \to 3, L}$, then the permeation function obeys an exact power-law behavior:
\begin{equation}
 \mathfrak{P}^{\infty}_{\ell = 2^{\tau}} (\Theta_{\alpha,L})
 = \lambda_{\alpha}^{\tau} \;\,\mathfrak{P}^{\infty}_{1} (\Theta_{\alpha,L}) =
 \ell^{\,\log_2 \lambda_{\alpha}} \;\,\mathfrak{P}^{\infty}_{1} (\Theta_{\alpha,L}),
\end{equation}
where $\lambda_{\alpha}$ is the relative eigenvalue, $|\lambda_{\alpha}| \leq 1$.
It is evident that permeation functions near the boundary are critical indeed,
as the ancillary degree of freedom strongly correlates with the system. By expanding
a generic observable $\Theta$ in the generalized eigenoperator basis of $\Ai_{3 \to 3, L}$ we can write
\begin{equation} \label{eq:Poyone}
 \bar{\mathfrak{P}}^{\infty}_{\ell}(\Theta) =
 \sum_{\alpha} \ell^{\,\zeta_{\alpha}} \cdot
 \sum_{\partial, w} \mathcal{P}^{[\Delta_{\partial}-w]}_{\lambda_{\alpha}}(\log_2 \ell),
\end{equation}
with critical exponents $\zeta_{\alpha}$ determined by the spectrum of $\Ai_{3 \to 3,L}$ via
logarithmic relation $\zeta_{\alpha} = \log_2 \lambda_{\alpha}$.

Now we would require somehow that the previous result does not hold only for $2^{\tau}$ distances,
but that it generalizes smoothly to other distance, so that our boundary-MERA ansatz would be able
to capture conformal symmetry as closest as possible. In particular, let us now consider distances
of the form $\ell = 2^{\tau} - 1$. Then, \eqref{eq:BERAmapbulk} tells us that the correct
recursive relation reads:
\begin{equation}
 \rho^{[\mu]}_{2^{\tau}-1, 2^{\tau}, 2^{\tau}+1} = \Di_{3 \to 3, R}
 \left( \rho^{[\mu-1]}_{2^{\tau-1}-1, 2^{\tau-1}, 2^{\tau-1}+1} \right).
\end{equation}
Notice that now is the other size-3 descending map 
($\Di_{3 \to 3, R}$ and not $\Di_{3 \to 3, L}$) that rules the recursion, unlike \eqref{eq:Leftdominance}.
Thus, the permeation function becomes
\begin{equation} \label{eq:Rightdominance}
 \mathfrak{P}^{\infty}_{\ell = 2^{\tau} - 1} (\Theta) =
 \trace \left[ \Ai_{3 \to 3, R}^{\tau - 1} \left( \Theta \right) \cdot
  \mathcal{K}_L \left( \rho^{\infty}_{A,1,2} \right) \right],
\end{equation}
stating that now the critical exponents are logarithms of the spectrum of $\Ai_{3 \to 3, R}$,
and its eigenoperators $\Theta_{\alpha,R}$ are one-point primary fields.

Let us introduce the translational regularity requirement.
For instance we demand that critical exponents
and primary fields do not depend on the point $\ell$ at which the permeation function is computed.
It is easy to see that, when accepting this, it immediately follows that
\begin{equation} \label{eq:splittaben}
 \begin{aligned}
 \Di_{3 \to 3, L} &= \Di_{3 \to 3, R} = \Di_{3 \to 3}, \qquad \mbox{thus}\\
 \Dsla &= \Di_{3 \to 3} \otimes \Di_{3 \to 3}.
 \end{aligned}
\end{equation}
But now, recall that by \eqref{eq:TNcorr5}, the $\Dsla$ was the map ruling the two-point
correlation functions in the bulk, two point critical exponents given by $\xi_{\alpha}$ the logarithms
of its spectrum. Thanks to \eqref{eq:splittaben} we know that two point primary fields in
the bulk correspond to the one-point ones at the boundary, and critical exponents satisfy
\begin{equation}
 \zeta_{\alpha} = \frac{1}{2} \;\xi_{\alpha}.
\end{equation}
As was pointed out by P. Calabrese \cite{CFTCalab, BERA}, this is one of the fundamental properties
prescribed by boundary conformal field theory. One-point critical exponents at the boundary $\zeta_{\alpha}$
are exactly \emph{half} of the two-point ones $\xi_{\alpha}$ in the bulk. We re-derived independently this feature
by solely exploiting geometrical features of boundary-MERA network and smoothness requirements.

\section{Hybrid MPS$\,\Leftrightarrow\,$TTN networks}

When a quantum many-body system draws near to a second-order quantum phase transition,
noncritical and critical properties begin to overlap. Often, as system parameters approach
the critical region (especially in 1D settings, where phase transitions are allowed only
at zero temperature) islands characterized by strong correlation within start to appear,
out of a noncritical long-range landscape. Then the characteristic size of these regions
themselves increases, until they reach the lenghtscale of the whole system when the
critical point is achieved.

In order to simulate efficiently this type of quantum behavior with a Tensor Network ansatz,
one would like reproduce both power-law like correlation scalings up to a tunable finite distance,
and exponential decay rates beyond. A promising candidate for this goal would be a structure which
embeds the self-similar geometry of Matrix Product States, which express noncritical character in a natural way
as we saw in section \ref{sec:MPScorr}, when observed at large distances, and resembles a hierarchical
Tensor Network, either TTN or MERA which both bear strong correlation capabilities, when observed in proximity of
the physical bondlinks. Such idea leads, almost obviously, to the design of a Hybrid
MPS$\leftrightarrow$TTN structure, picted as follows:
\begin{equation} \label{eq:HybridTTN}
 \begin{overpic}[width = 260pt, unit=1pt]{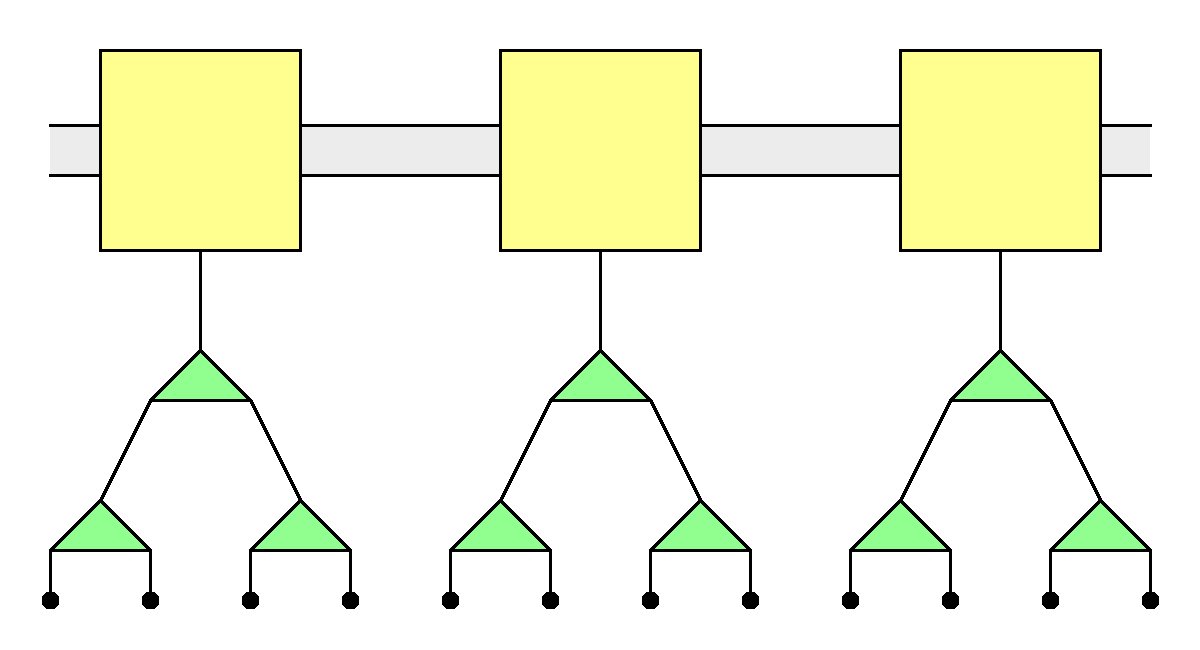}
 %\put(3, 60){$\mathcal{K}_L = $}
\end{overpic}
\end{equation}
where the system size $L$ is given by the interplay of two parameters: $n$, the
number of Matrix Product blocks, and $\mu$, the full depth of the binary
tree '\emph{curtain}' hanging from the MPS layer. So that $L = 2^{\mu}\,n$.
In order to suit as best as possible translational invariance, despite the presence
of TI-breaking tree geometry, we will consider the MPS tensors $A$ to be homogeneous,
and the isometries $\Lambda$ of the tree as well.
It is clear that, as we want for the system to reach thermodynamical limit but
keep strongly-correlated islands at finite size, we will fix the curtain height $\mu$, and
increase $n \to \infty$, so that $L$ goes to infinity too.

The algebraic ingredients in our analytical study of \eqref{eq:HybridTTN} hybrid networks
are again the translationally-averaged density matrices $\bar{\rho}^{[\mu]}_1$,
$\bar{\sigma}^{[\mu]}_{\ell}$ and $\eta^{[\mu]}_{\ell}$,
respectively defined in \eqref{eq:transpla} and \eqref{eq:sigmaeta}.
Recursive relations involving these elements will let us figure out
correlation behavior and scaling-laws.

\vspace{.5em}
\emph{\textbf{MPS as starting point - }}
In section \ref{sec:MPSTD} we learned that the fastest way to keep under control
normalization and physical quantities in thermodynamical limit MPS is to choose
the appropriate peripheral gauge, say the left one.
When $\mu = 0$, the involved density matrices
are completely defined by the matrix product properties. Namely
\begin{equation} \label{eq:Hybzero}
 \begin{aligned}
 \bar{\rho}_1^{[0]} &= \sum_{r,s = 1}^{d} ( \Phi^{+} | A_s \otimes A^{\star}_r | Q )\; |s\rangle\langle r| \\
 \bar{\sigma}_{\ell}^{[0]} &= \sum_{r_1, r_2, s_1, s_2}^{d}
 ( \Phi^{+} | \left( A_{s_1} \otimes A^{\star}_{r_1} \right) \mathbb{E}^{\ell - 1}
 \left( A_{s_1} \otimes A^{\star}_{r_1} \right)| Q )\; |s_1 s_2\rangle\langle r_1 r_2| \\
 \eta_{\ell}^{[0]} &= \bar{\rho}_1^{(0)} \otimes \bar{\rho}_1^{(0)},
 \end{aligned}
\end{equation}
where the MPS matrices $\{A_s \}$ are chosen so that they
satisfy the gauge condition $\sum_{s} A^{\dagger}_s A_s = \Id$.
and $\mathbb{E} = \sum_{s} A_s \otimes A^{\star}_s$ is the transfer matrix of the Identity operator.
Correlation vectors $| Q )$ and $( \Phi^{+} |$ are respectively the right and left
fixed points of $\mathbb{E}$, they are both positive when read as matrices, and in particular
$( \Phi^{+} | = \sum_{\alpha} (\alpha \alpha|$ corresponds to the identity itself.

\vspace{.5em}
\emph{\textbf{Adding the TTN curtain - }} we can now increase $\mu$ to a finite nonzero value.
Recursive relations between density matrices standing different tree layers are as usual given by
descending CPT-map formalism. So that
\begin{equation}
 \begin{aligned}
 \bar{\rho}_1^{[\nu]} &= \Di (\bar{\rho}_1^{[\nu-1]})\\
 \bar{\sigma}_{2\ell}^{[\nu]} &= \Dsla (\bar{\sigma}_{\ell}^{[\nu-1]})\\
 \eta_{2\ell}^{[\nu]} &= \Dsla (\eta_{\ell}^{[\nu-1]}),
 \end{aligned}
\end{equation}
for every $0 < \nu \leq \mu$.
It is clear that, while $\bar{\rho}_1^{[\mu]}$ can be always expressed in terms of \eqref{eq:Hybzero} as
$\Di^{\mu} (\bar{\rho}_1^{[0]} )$, expressions for
$\bar{\sigma}_{\ell}^{[\mu]}$ and $\eta_{\ell}^{[\mu]}$ depend whether the distance $\ell$
is smaller or larger than the island size $2^{\mu}$.

\vspace{.5em}
\emph{\textbf{Small scale regime}} ($\ell < 2^{\mu}$) - For simplicity, let us choose $\ell = 2^p$,
clearly with $p < \mu$ ($p$ integer).
Then the appropriate recursive relation read
\begin{equation} \label{eq:sfilza1}
\begin{aligned}
 \bar{\sigma}_{2^p}^{[\mu]} &= \Dsla^p (\bar{\sigma}_{1}^{[\mu-p]}),\\
 \eta_{2^p}^{[\mu]} &= \Dsla^p (\eta_{1}^{[\mu-p]}),
\end{aligned}
\end{equation}
We still have to exhibit an expression of $\bar{\sigma}_{1}^{[\nu]}$
and $\eta_{1}^{[\nu]}$, for $0 < \nu \leq \mu$. To do this in a clever way we
introduce a new density matrix $\theta^{[\nu]}$, defined as
\begin{equation} \label{eq:sfilza0}
 \theta^{[\nu]} = \frac{1}{2^p n} \sum_{\ell_0 = 1}^{2^p n} \rho_{\{\ell_0, \ell_0+1\}}^{[\nu]}
 \otimes \rho_{\{\ell_0, \ell_0+1\}}^{[\nu]}
\end{equation}
Interestingly enough, both $\bar{\sigma}_{1}^{[\nu]}$ and $\eta_{1}^{[\nu]}$
can be extracted from $\theta^{[\nu]}$ by partial trace, precisely
\begin{equation}
 \bar{\sigma}_{1}^{[\nu]} = \trace_{3,4} \left[ \theta^{[\nu]} \right]
 \qquad \mbox{and} \qquad
 \eta_{1}^{[\nu]} = \trace_{2,3}\left[ \theta^{[\nu]} \right]. 
\end{equation}
At the same time, $\theta^{[\nu]}$ satisfies the following recursion
\begin{equation} \label{eq:sfilza00}
 \theta^{[\nu]} = \mathcal{W} \left( \theta^{[\nu-1]} \right) \equiv \frac{
 \Di_R \otimes \Di_L \otimes \Di_R \otimes \Di_L
 + (S \otimes S)\, \trace_{2,4}}{2} \left[ \theta^{[\nu-1]} \right],
\end{equation}
starting at the MPS layer from $\theta^{[0]} = \bar{\sigma}_{1}^{[0]} \otimes \bar{\sigma}_{1}^{[0]}$.

\vspace{.5em}
\emph{\textbf{Large scale regime}} ($\ell > 2^{\mu}$) - Things are easier now, provided
that we choose, again for simplicity, $\ell = 2^{\mu} \tau$ ($\tau$ integer), with no finite-size effects
$\ell \ll L \sim \infty$. The correct expressions are just as follows:
\begin{align} \label{eq:nosfilza}
 \bar{\sigma}_{2^{\mu} \tau}^{(n)} =&\; \Dsla^{\mu} (\bar{\sigma}_{\tau}^{[0]}),\\
 \eta_{2^{\mu} \tau}^{(n)} =&\; \Dsla^{\mu} (\eta_{\tau}^{[0]}).
\end{align}
We formally achieved every quantity we are interested in. We are now ready to combine together
the previous results to investigate correlation properties of Hybrid Tensor Networks.

\subsection{Two-point correlation functions}

We are interested in calculating two-point correlators at fixed distance and translationally
averaged on variational states of the form \eqref{eq:HybridTTN}. Let us then recall
\begin{multline} \label{eq:avecordef2}
 \bar{\mathfrak{C}}^{[\mu]}_{\ell}(\Theta, \Theta')
  \equiv \frac{1}{L} \,\sum_{\ell_0 = 1}^{L} \langle \Theta_{[\ell_0]} \otimes \Theta'_{[\ell_0 + \ell]} \rangle
  - \langle \Theta_{[\ell_0]} \rangle \langle \Theta'_{[\ell_0 + \ell + 1]} \rangle
 = \\ =
 \trace \left[ \left( \Theta \otimes \Theta' \right) \cdot \left( 
 \bar{\sigma}_{\ell}^{[\mu]} - \eta_{\ell}^{[\mu]} \right)\right],
\end{multline}
the appropriate correlation function of $\Theta$ and $\Theta'$ at distance $\ell$.
Let us focus on the scaling laws of this quantity, in either of the two regimes we mentioned,
rather than an exact formal expression for it.

\vspace{.5em}
\emph{\textbf{Small scale regime -}} Again $\ell = 2^p$, with $p < \mu$.
By using the recursive relation \eqref{eq:sfilza00} for $\theta^{[\nu]}$ we can write
\begin{equation} \label{eq:lunghetto}
 \bar{\mathfrak{C}}^{[\mu]}_{\ell = 2^p}(\Theta, \Theta')
 = \trace \left[ \vphantom{\sum} \left( \Theta \otimes \Theta' \right) \cdot
 \Dsla^{p} \circ \left( \trace_{3,4} - \trace_{2,3} \right)
 \left[ \mathcal{W}^{\mu - p} \left( \theta^{[0]} \right) \right] \right].
\end{equation}
the residual dependence on $p$ (and $\mu$) is left only in the number of times the maps
$\Dsla$ and $\mathcal{W}$ are to be applied. This allows us to exploit the Jordan block
expansion scheme for multiple actions of the same matrix, as we did before, leading us to
\begin{equation} \label{eq:larguzzo}
 \bar{\mathfrak{C}}^{[\mu]}_{\ell = 2^p}(\Theta, \Theta') =
 \sum_{\lambda_{\Dsla}} \sum_{\lambda_{\mathcal{W}}}
 \left( \frac{\lambda_{\Dsla}}{\lambda_{\mathcal{W}}}\right)^p \cdot \mathcal{P}_{\{\lambda\}}(p)
\end{equation}
where $\lambda_{\Dsla}$ (resp. $\lambda_{\mathcal{W}}$) are eigenvalues of $\Dsla$ (of $\mathcal{W}$),
and $\mathcal{P}_{\{\lambda\}}$ are finite-degree polynomials.
Now recalling that $p$ is the logarithm of the distance $\ell$, it is clear that \eqref{eq:larguzzo}
is dominated by power-law behaving functions, apart logarithmic corrections, whose
(quasi-critical) exponents are determined by
\begin{equation}
 \chi_{\alpha} = \log_2 \left( \frac{\lambda_{\Dsla}}{\lambda_{\mathcal{W}}}\right).
\end{equation}

\vspace{.5em}
\emph{\textbf{Large scale regime -}} The system behave differently when $\ell = 2^\mu \tau$,
for $1 < \tau \ll L$. In fact at these distances eq.~\eqref{eq:avecordef2} becomes
\begin{multline}
 \bar{\mathfrak{C}}^{[\mu]}_{2^{\mu} \tau}(\Theta, \Theta') = \trace \left[
 (\Theta \otimes \Theta') \cdot \Dsla^{\mu} \left( \bar{\sigma}^{[0]}_{\tau} - \bar{\eta}^{[0]}_{\tau} \right) \right]
 = \\ = 
 f(\mu) (\Phi^{+}| \mathbb{E}_{\Id}^{\tau} - \mathbb{E}_{\Id}^{\infty} | Q).
\end{multline}
In the end we are actually calculating a $\tau$-fixed distance correlator upon the MPS layer
of the observable $\Asla^{\mu}(\Theta \otimes \Theta')$. But since we chose a homogeneous, left-gauge
MPS description in the thermodynamical limit, we can perfectly recover the corresponding result \eqref{eq:linfiz2}, i.e.
\begin{equation} \label{eq:linfizzi}
   \mathfrak{C}_{2^{\mu} \tau}(\Theta, \Theta') =
   \sum_{|\lambda_{\mathbb{E}}|<1} \lambda_{\mathbb{E}}^{\tau} \cdot \mathcal{P}_{\lambda_{\mathbb{E}}}(\tau).
\end{equation}
In fact, instead of \eqref{eq:larguzzo}, now $\tau$ is proportional to $\ell$ and we obtain
an explicit exponential decay, going to zero at infinite distance since the sum spans
only the eigenvalues $\lambda_{\mathbb{E}}$ of $\mathbb{E}_\Id$ smaller of 1 in modulus.

In conclusion, we discovered that the hybrid Tensor Network state geometry, designed in \eqref{eq:HybridTTN},
manifests a quasi-critical character. At short ranges, strong correlations identified by power-law two-point
functions, arise; their exponents characterized by the isometry element $\Lambda$ of the TTN portion.
At long ranges, the behavior is evidently noncritical, correlations vanish exponentially
and ruled by the MPS block $A$.
The number of TTN layers we adopted in the scheme effectively determines the lenghtscale $\ell_0$ of the
strong-correlation islands $\ell_0 \sim 2^{\mu}$.

\vspace{.5em}
\emph{\textbf{MPS$\,\leftrightarrow\,$MERA Hybrids - }} The previous discussions were formulated
for a binary tree curtain attached to the MPS-basis layer. But it is easy to see that
the same results can be generalized to other tree geometries and for MERA as well.
So even if we had a network structure of the form
\begin{equation} \label{eq:HybridMera}
\begin{overpic}[width = 240pt, unit=1pt]{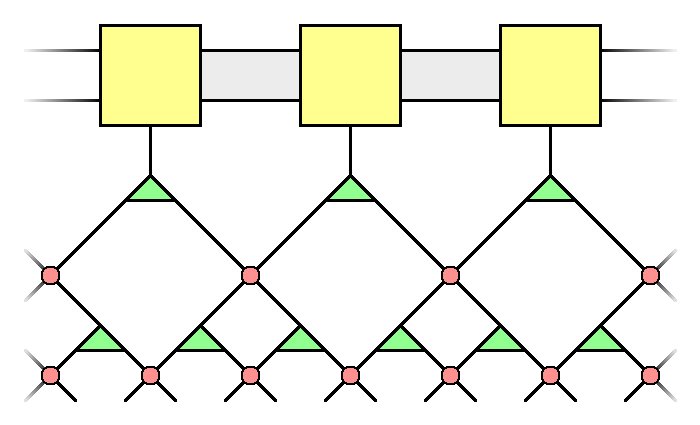}
 %\put(68, 23){${A^{\star}}^{[\ell]}$}
\end{overpic}
\end{equation}
we would similarly obtain a quasi critical regime. In fact, descending maps for MERA replace
the role of those for TTN, and the scaling laws do not sensibly change.

\section[Higher dimensions]{Higher dimensions and comparison\\between TTN/MERA and PEPS}

So far in this chapter, we discussed on how tree Tensor Networks and MERA can be used as a
suitable and efficient variational for many-body ground states of one dimensional system.
But it is clear that their power is not limited to 1D systems only, indeed generalizations
of Trees and MERA to higher dimensions is natural and intuitive.

The idea remains the same \cite{VidalTree1}: we may think to perform a real-space renormalization group of the quantum
lattice state, expressed in the language of density matrices, for \emph{any} amount of
spatial dimensions \#D. By performing this iterated process, we obtain a class of states which is
equivalently expressed as a Tensor Network, namely, a tree network in \#D dimensions.
For instance, let us consider a 2D square lattice, then the simplest 2D tree geometry we can think of
is the one that maps a 4-adjacent sites plaquette into a single renormalized site $(a)$.
The renormalization element $\Lambda$ becomes then a 5-link tensor, and read as a $D^4 \times D$
matrix is an isometry $\Lambda^{\dagger} \Lambda = \Id$.

Alternatively, again dealing with a square lattice, one could prefer to renormalize together
pairs of sites which are horizontally adjacent (so that we are actually coarse-graining only
one of the two dimensions), and at the following step those who are vertically adjacent $(b)$.
Clearly, the latter tree geometry has the advantage on the former that is easier to contract
and thus more efficient, for the same refinement parameter $D$. At the same time description capabilities
of the second choice (in terms of variational manifold) are reduced. In diagrammatic expression
these two TTN designs read respectively:
\begin{equation} \label{eq:Pyramids}
\begin{overpic}[width = 270pt, unit=1pt]{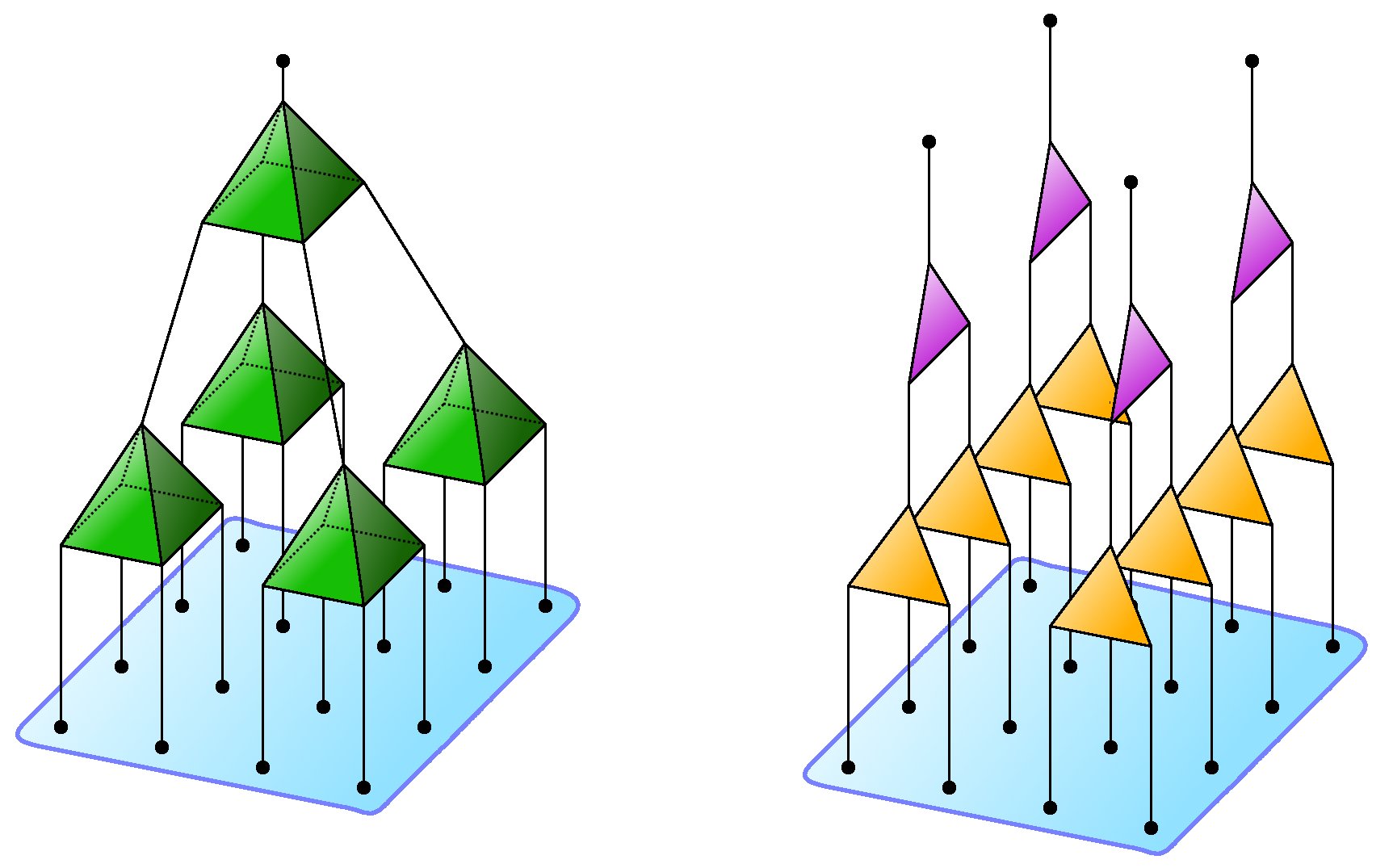}
 \put(0, 150){$(a)$}
 \put(145, 150){$(b)$}
\end{overpic}
\end{equation}

The possibilities are many. The user is encouraged to adopt for her simulation the
\#D TREE/MERA geometry that suits most symmetries (see appendix \ref{app:symchap})
and other global properties of the problem under
study; for example, a plaquette-tree geometry would embed more naturally the square group
symmetry.

Many of the algebraic properties we discussed in this chapter for 1D TTN/MERA are preserved
by their higher-dimensionality versions. The causal cone property is one of these:
as TTN remain Tensor Networks without closed loops, transforming them according to the peripheral
gauge (with respect to the hat as nucleus) is always possible without loss of generality,
no matter the geometry or dimensionality. Thanks to the causal cones, it is clear that
even in high-D, the contraction of a Tree network, or MERA, stays efficient:
the number of required elementary operations scaling logarithmically with the volume.

This is probably the most convincing argument for preferring TTN/MERA to PEPS, from 2D up.
Despite the two families of Tensor Network manifest promising description capabilities
for physically-relevant ground states, PEPS are computationally complex objects,
while hierarchical Tensor Networks keep their high efficiency. It is worth mentioning that
it was shown in \cite{EisMERAPEPS} that starting from 2D up, MERA (and thus Trees) satisfy
an entanglement area law, and therefore can be efficiently mapped, through a well-defined
formal algorithm, into PEPS.

%\vspace{.5em}
\newpage
In the end, Trees and MERA are simulation tools hard to ignore, for any physical setting: in 1D they are the answer
on how to simulate critical systems in a natural and cheap way; in two or more spatial dimensions, they constitute
an intriguing sub-class of finitely correlated states that are efficiently contractible.

\chapter*{Conclusions}

In this thesis we introduced, developed, and analyzed a large class of variational tailored
quantum many-body wavefunctions, called Tensor Network states. These states are meant to be used as
variational ansatze for interacting particles problems on a lattice; minimization algorithms
adopting these trial functions do not require any a priori knowledge on the model we are studying.
Predicting the success of Tensor Network states for a certain setting is based
upon arguments borrowed from quantum information theory: primarily entanglement.

We discussed profoundly how Tensor Network designs are somehow the variational counterpart of
some numerical renormalization group procedure. In fact, the manifold of quantum many-body states
that can be constructed via a RG-algorithm, can be equivalently identified by a tailored
analytic expression, where the variational descriptors are tied together through simple
linear algebraic relations. For these reasons Tensor Network states preserve all the faithfulness
and simulation power of numerical renormalization groups, yet the variational picture
presents several practical advantages. Two important example are:
a more immediate way to access physical information, and a more free numerical manipulability
with a consequent computational speed-up.

In particular, we showed in chapter \ref{chap:OBCMPS} that Matrix Product States arise from White's DMRG.
Since Matrix Product States correspond to finitely-correlated states in 1D, they are allowed to manifest
only the correct amount of entanglement of 1D non-critical ground states, and at the same time
to be extremely efficient for computation. This explains why DMRG (and MPS-based algorithms alike)
are so successful in one-dimensional systems.

In chapter \ref{chap:PBCMPS} we presented a generalization of MPS formalism that extends
to Periodic boundary systems, with the care of keeping the right amount of allowed correlation.
This led us to the definition of a homogeneous PBC-MPS representation, which was instrumental
in the definition of thermodynamical limit for Matrix Product States. Once we defined an infinite
MPS, we could investigate the correlations behavior of these states, and showed that under
the assumption of keeping a finite refinement parameter, the target state is distinctly non-critical.

Matrix Product States are considered the fundamental template that led to the more general definition
of Tensor Network states. Various classes of Tensor Networks manifest properties depending strictly
on their geometry. Still, it is possible to determine general features common to every TN-state;
those we sketched in chapter \ref{chap:TN}.

Finally, chapter \ref{chap:TTNMERA} contained most of the analytical advancements I contributed.
There we discussed about Tree Tensor Networks, and MERA, the latter geometry presented as an enhancement
of the former, but still sharing most of its scaling properties.
We explained how these TN-variational states arise from a real-space renormalization group
technique, an origin which leads unavoidably to scale invariant physics for these states.
In fact, we showed that by adopting a CPT map formalism, we can well-define the thermodynamical
limit for TTN/MERA states, and this state manifest critical behavior, identified by
power-law decay rates of two-point correlations.

\section*{Acknowledgements}

I want to thank my colleagues D.~Rossini, M.~Rizzi and S.~Montangero for their numerical support,
S.~Peotta for his stimulating ideas,
G.~Santoro for his trust,
R.~Fazio for his protection,
and V.~Giovannetti for his unfaltering helpfulness.

\appendix

\chapter{Completely Positive Trace preserving maps} \label{app:cptchap}

Here we list some features of Completely Positive Trace preserving (CPT) maps, as they are common and useful tools
in quantum information theory, and thus thoroughly studied \cite{Terhal}. They represent quantum channels,
having the property that they map density matrices
into density matrices, and often describe the time evolution of quantum states
when both coherent and incoherent sources couple with the system.
In Tensor Network settings, CPT maps are primarily used
as inverse transformation of some numerical renormalization procedure, typically applied to a density matrix,
and thus work as propagators for renormalized density matrices within the network structure.

\section{Definition}

A homomorphism between matrix spaces $\mathcal{M}: \mathbb{C}^{n \times n} \to \mathbb{C}^{m \times m}$
is said to be a CPT map if satisfies the following requirements
\begin{enumerate}
 \item $\mathcal{M}$ is positive: $\forall\, A \geq 0, A \in \mathbb{C}^{n \times n} \Rightarrow \mathcal{M}(A) \geq 0$
 \item $\mathcal{M}$ is completely positive, meaning that its identity extension on any larger space
 is still positive. Let us define $\mathcal{I}_k \otimes \mathcal{M}:
 \mathbb{C}^{k \times k} \otimes \mathbb{C}^{n \times n} \to \mathbb{C}^{k \times k} \otimes \mathbb{C}^{m \times m}$
 defined on separable matrices $[\mathcal{I}_k \otimes \mathcal{M}] (B \otimes A) = B \otimes \mathcal{M}(A)$.
 Then the complete positivity reads:
 $\forall\, C \geq 0,C \in \mathbb{C}^{k \times k} \otimes \mathbb{C}^{n \times n}
  \Rightarrow [\mathcal{I}_k \otimes \mathcal{M}](C) \geq 0$, for every extension $k$.
 \item $\mathcal{M}$ preserves the trace: $\trace[\mathcal{M}(A)] = \trace[A]$.
\end{enumerate}
Of course, a matrix $A$ is positive if its spectrum lies in $\mathbb{R}^{+}$, or also
$\langle \psi | A |\psi \rangle \geq 0$ $\forall |\psi\rangle$ if we are in a Hilbert space.
Choi's theorem \cite{Choi}
states that a mapping between matrices $\mathcal{M}$ is completely positive \emph{iff} it exists a set
of Kraus operators $\{V_s\}_s$ for $\mathcal{M}$, namely a set of matrices $V_s \in \mathbb{C}^{m \times n}$
which satisfy:
\begin{equation} \label{eq:Kraus}
 \sum_s V_s^{\dagger} V_s = \Id_{n \times n}
 \qquad \mbox{and} \qquad \sum_s V_s A V_s^{\dagger} = \mathcal{M}(A) \quad \forall\,A.
\end{equation}
For later reference, it is also convenient to define also the map adjoint of $\mathcal{M}$ with respect
to the trace scalar product between matrices $(A|B) = \trace[A^{\dagger} B]$; by definition
$\mathcal{M}^{\text{adj}}: \mathbb{C}^{m \times m} \to \mathbb{C}^{n \times n}$ so that
\begin{equation}
 (A|\mathcal{M}^{\text{adj}}(B)) = (\mathcal{M}( A)|B) =
 \trace[ (\mathcal{M} (A))^{\dagger} B] =
 \sum_s \trace[V_s A^{\dagger} V_s^{\dagger} B].
\end{equation}
Then, by using cyclicity of the trace, you immediately see that $\mathcal{M}^{\text{adj}}$ must read:
\begin{equation}
 \mathcal{M}^{\text{adj}}(B) =
 \sum_s V^{\dagger}_s B V_s \quad \forall\,B.
\end{equation}
Which tells us that $\mathcal{M}^{\text{adj}}$ is again completely positive, and \emph{unital},
i.e. it maps the identity operator onto itself, thanks to \eqref{eq:Kraus}, but not necessarily
trace preserving.

\section{Spectral properties} \label{app:CPTspec}

When CPT maps $\mathcal{M}$ are endomorphisms ($n = m$), they can be expressed as square matrices, and expanded
in a basis of eigenoperators and generalized eigenoperators as usual.
However, spectrum and eigenmatrices of a CPT map are bound to undergo certain properties:
\begin{itemize}
 \item \emph{Every eigenoperator $O_{\lambda}$ with eigenvalue $\lambda \neq 1$, must have null trace}.
This is clear from the fact that
$\trace[O_{\lambda}] = \trace[\mathcal{M}(O_{\lambda})] = \trace[ \lambda O_{\lambda}] =
\lambda \,\trace[O_{\lambda}]$, and since $\lambda \neq 1$ its only solution is $\trace[O_{\lambda}] = 0$.
Also, it follows that such a $O_{\lambda}$ can not be positive, because the only positive
traceless matrix is the null matrix. The traceless requirement trivially extends to generalized
eigenoperators as well.

 \item \emph{The spectrum of $\mathcal{M}$ must be symmetric with respect to the real axis}.
Indeed, $\mathcal{M}(A^{\dagger}) = \sum_s V_s A^{\dagger} V_s^{\dagger} =
\sum_s ( V_s A V_s^{\dagger})^{\dagger} = (\mathcal{M}(A))^{\dagger}$.
Then, if $O_{\lambda}$ is eigenoperator with eigenvalue $\lambda$, we have
$\mathcal{M}(O_{\lambda}^{\dagger}) = (\mathcal{M}(O_{\lambda}))^{\dagger} =
(\lambda O_{\lambda})^{\dagger} = \lambda^{\star} O_{\lambda}^{\dagger}$,
so also $\lambda^{\star}$ is in the spectrum. In particular, a hermitian eigenoperator
has necessarily a real eigenvalue. This also tells us that the spectrum of
$\mathcal{M}$ and $\mathcal{M}^{\text{adj}}$ coincide.

 \item \emph{1 always belongs to the spectrum of $\mathcal{M}$}. By absurd if $\mathcal{M}$
had no eigenvalue 1, then its entire basis of generalized eigenoperators would be traceless,
but that would necessarily be an incomplete set because the identity operator $\Id$ can not be generated.
Another way to see this is that $\Id$ is always eigenoperator of $\mathcal{M}^{\text{adj}}$,
and 1 its eigenvalue; but spectra of $\mathcal{M}^{\text{adj}}$ and $\mathcal{M}$ are equivalent,
so 1 is also eigenvalue for $\mathcal{M}$. 

\item \emph{The generalized eigenspace of $\lambda = 1$ coincides with the strict eigenspace}
In fact if we assume that there is a Jordan block of dimension 2 or greater,
we can define the generalized eigenoperator
$O'_{1}$ for which $\mathcal{M}^{\text{adj}}(O'_{1}) = O'_{1} + \Id$. but the equation
$\alpha = \trace[O'_{1}] = \trace[O'_{1}] + \trace[\Id] = \alpha + n$ has no solution.

\item \emph{The spectral radius is 1}, or equivalently, CPT maps are contractive.
To prove this, we are going to show that if $\mathcal{M}^{\text{adj}}$ had an eigenvalue $\lambda$
greater in modulus than 1, it could not be a positive map. Indeed let $O_\lambda$ be the related
eigenoperator, and consider $O_{+} = O_\lambda + O_{\lambda}^{\dagger}$,
which is hermitian, but traceless and thus not positive, meaning that it exists a (normalized) vector $|\psi\rangle$
for which $\langle \psi | O_+ |\psi\rangle = w < 0$. Now, the operator $(\Id + \varepsilon\, O_{+})$
is definitely positive for $0 < \varepsilon < \| O_{+} \|$.
Therefore ${[\mathcal{M}^{\text{adj}}]}^q(\Id + \varepsilon\, O_{\lambda})
= \Id + \varepsilon ( \lambda^{q} O_{\lambda} + \lambda^{\star\,q} O^{\dagger}_{\lambda}) = \Gamma_q$
should be positive for every $q$, but
\begin{equation} \label{eq:aquat}
 \langle \psi | \Gamma_q | \psi \rangle = 1 + \varepsilon |\lambda|^{q}
 \left( w \cos(\phi q) + z \sin(\phi q) \right),
\end{equation}
where $\phi = \arg(\lambda)$.
This expression either oscillates with exponentially increasing amplitude, with period
$2\pi / \phi$, or it is monotonically, and exponentially, decreasing if $\phi = 0$, since $w < 0$.
Either way, sooner or later we encounter some integer $q$ for which \eqref{eq:aquat} is negative,
telling us $\mathcal{M}^{\text{adj}}$ mapped a positive operator into a non-positive one,
which is the desired absurd.
\end{itemize}

This final remark on contractivity of CPT maps is particularly important, it tells us that
within the space of matrices, there is a proper subspace which is attractive, i.e.
that every operator is driven towards it by multiple applications of the map $\mathcal{M}$,
its distance from such set exponentially decreasing. Also we know that the attraction subspace is
generated by generalized eigenoperators of eigenvalues $\lambda$ of modulus $|\lambda|=1$.

Obviously, the case that gathers greatest interest from this point of view is when this
subspace is one-dimensional, so that the map contracts the projective space into a point. We will see
that this additional request is also related to spectral properties of $\mathcal{M}$.

\section{The Mixing requirement} \label{app:mixin}

A CPT map $\mathcal{M}$ is said to be \emph{mixing} when infinite applications of
$\mathcal{M}^q \equiv \mathcal{M} \circ \ldots \circ \mathcal{M}$
contract the whole space of matrices (modulus the trace) into a unique point $\Lambda$, i.e.
\begin{equation} \label{eq:mixing}
 \lim_{q \to \infty} \mathcal{M}^q (A) = \Lambda \; \trace[A] \qquad \forall\,A.
\end{equation}
It is clear that when this condition holds $\Lambda$ is obviously a fixed point of the map, since
\begin{equation}
 \mathcal{M}(\Lambda) = \mathcal{M} \circ \lim_{q \to \infty} \mathcal{M}^q (\Id/n) =
 \lim_{q \to \infty} \mathcal{M}^{q+1} (\Id/n) = \Lambda.
\end{equation}
which immediately tells us that $\Lambda$ is positive and $\trace[\Lambda] = 1$,
so $\Lambda$ is a density matrix. The mixing condition
has consequences on the spectral periphery of $\mathcal{M}$,
which are also related to well-known physical properties \cite{mixing}:
\begin{itemize}
 \item \emph{\textbf{Ergodicity} - 
 The eigenvalue 1 of $\mathcal{M}$ is simple}, meaning that the related eigenspace is one dimensional,
 i.e. $\mathcal{M}$ has a unique fixed point.
 When this statement fails, it is clear that it exists a whole manifold (at least one dimensional)
 of operators $O_{\mathbb{M}}$,
 with $\trace[O_{\mathbb{M}}]=1$, which are all fixed points of the map.
 Thus eq.~\eqref{eq:mixing} breaks down
 since $\mathcal{M}^{\infty}(O_{\mathbb{M}}) = O_{\mathbb{M}}$ and $O_{\mathbb{M}}$ not unique.

 \item \emph{\textbf{Relaxation} - 
 $\mathcal{M}$ has no modulus $|\lambda_{\alpha}| = 1$ eigenvalue $\lambda_{\alpha}$, except for 1 itself.}
 This tells us that every component of a given observable $O$, expanded in the generalized eigenbasis,
 other than $\Lambda$, decays exponentially with a rate governed
 by the second greatest modulus eigenvalue $\lambda_2$: $|\lambda_{\alpha > 2}| \leq |\lambda_{2}| < 1$.
 By contradiction, if an eigenvalue $\bar{\lambda} = e^{i\varphi}$ existed, one can construct a whole
 set of operators that \emph{rotate} infinitely around the fixed point $\Lambda$, like
 \begin{equation}
  \mathcal{M}^{q}(\Lambda + \alpha O_{\bar{\lambda}}) = \Lambda + \alpha \,e^{iq\varphi}\, O_{\bar{\lambda}}
 \end{equation}
 and the sequence in $q$ has no limit.
\end{itemize}

When the map $\mathcal{M}$ is mixing, expanding the action of several, but finite, applications of $\mathcal{M}$
in the generalized eigenbasis $O_{\alpha, \partial, w}$ 
($\alpha$ - eigenvalue index, $\partial$ - Jordan block index, $w$ - position within the block)
is particularly useful, and it reads
\begin{equation} \label{eq:APolyexpand}
 \mathcal{M}^{q}(A) = \Lambda \trace[A] + \sum_{\alpha = 2} \lambda_{\alpha}^q
 \sum_{\partial, w} \mathcal{P}^{[\Delta_{\partial}-w]}_{\lambda_{\alpha}}(q)
 \; O_{\alpha, \partial, w},
\end{equation}
with $\mathcal{P}^{[x]}_{\lambda_{\alpha}}(q)$ is a polynomial function of degree $x$, with coefficients
depending on $A$ and $\lambda_{\alpha}$; obviously all the sum terms vanish at $q \to \infty$.

\chapter{Symmetries in Tensor Networks} \label{app:symchap}

Exploiting symmetries in numerical simulations of quantum problems is one of the best and most
valuable techniques we can adopt to drastically improve the efficiency of computation, with
no actual loss of accuracy. When the Hamiltonian of the many-body problem is invariant under a group
of unitary transformations, then an immediate characterization of the ground states, under the action
of the group itself, emerges. Even in the rare cases when a discrete symmetry is spontaneously broken, it is
always possible, thanks to the wave-like formulation of quantum mechanics, to identify a symmetry-invariant
ground state.

In variational contexts, where the efficiency of every protocol is extremely sensitive to the number
of effective parameters, the total amount of such descriptors is heavily reduced every time a symmetry
constraint is embedded in the framework. Of course, the capability of upholding a symmetry in a given variational
ansatz is not guaranteed a priori: the manifold of states belonging to the variational family must
be able to recognize, distinguish, and capture the physics of the symmetry group as a whole.

Tensor Networks, as we discussed thoroughly, are variational counterparts of numerical renormalization groups.
Since RG-processes naturally manifest a notion of locality preservation, TN-states seem most suitable to
reproduce symmetries which are global, but act locally as an uncorrelated product of on-site transformations.
The strategies to implant symmetries within Matrix Product States had been known from the DMRG era
\cite{SchollDMRG, Singh1}; generalizations to other Tensor Network geometries were studied in depth by
S.~Singh \emph{et al.} \cite{Singh2, Singh3}. It was shown that the overall effect of inserting a
symmetry constraint in a TN-ansatz is that one can easily isolate residual variational degrees of freedom
out of structure factors and selection rules. In practice, every tensor decomposes, into a variational
and a structural part (or \emph{fragment}), which are again tied together by network linking relations.

\section[Pointwise symmetries and representations]{Pointwise symmetries and\\representations} \label{sec:irrep}

We will now define the class of symmetry transformation groups we will take into account
through the following sections, and then review some basic, but useful, results of representation theory.
Let $\mathcal{G}$ be our compact symmetry group; it can be a finite group as well as a Lie group:
typical candidates are $\mathbb{Z}_n$, O$(n)$, SO$(n)$, U$(n)$, SU$(n)$ but also dihedral groups are common.
We identify $\{U_g\}_{g \in \mathcal{G}}$ as its unitary
representation on the one-site $d$-dimensioned Hilbert space:
$U_{g_1 g_2} = U_{g_1} \cdot U_{g_2}$
Then the overall transformations are given by tensor products of on-site unitaries
\begin{equation} \label{eq:pointwise}
 U^{\otimes L}_g = \bigotimes_{\ell = 1}^{L} U^{[\ell]}_{g},
\end{equation}
where $L$ is the total number of sites, and $U^{[\ell]}_{g}$
does not depend on the site $\ell$ on which it acts, as we are requiring homogeneity of the local representation.
In literature, these transformation are often referred to either as \emph{local} symmetries, as $U^{\otimes L}_g$
is a product of local terms, or as \emph{global} symmetries, as $U^{\otimes L}_g$ has global support.
To avoid misunderstandings, we prefer to call transformations of the form \eqref{eq:pointwise},
\emph{pointwise symmetries}, as they are an overall operation acting on every site singularly.
Pointwise symmetries are extremely relevant from a physical point of view: they can uphold
most of the extensive constraints of a physical framework, like total particle conservation, parity, or spin conservation.
Trivially, the $U^{\otimes L}_g$ form a group and are an actual representation of $\mathcal{G}$,
although it is hardly an irreducible representation, even when the original $U_{g}$ is irreducible.

Before moving further, let us briefly recall some basic principles of matricial group representation theory;
in particular, we summon the natural expansion in irreducible components and subspaces.
Precisely, let $\mathbb{V}$ be any given vector space, on complex field $\mathbb{C}$,
and $\{W_g\}_{g \in \mathcal{G}}$ a unitary representation of $\mathcal{G}$ on $\mathbb{V}$.
Then $\mathbb{V}$ decomposes naturally as the Cartesian sum of irreducible subspaces $\mathbb{V}^{[c]}$:
\begin{equation} \label{eq:irrepdeco}
 \mathbb{V} \approx \bigoplus_{c} \left[ \bigoplus_{\partial = 1}^{\bar{\partial}_c} \mathbb{V}^{[c]} \right]
 \approx \bigoplus_{c} \left( \mathbb{D}^{[c]} \otimes \mathbb{V}^{[c]} \right),
\end{equation}
where the $\mathbb{V}^{[c]}$ are the smallest invariant subspaces of $\mathbb{V}$ under the action of
$U_{\mathcal{G}}$. In \eqref{eq:irrepdeco}, $c$ is a scalar integer labeling which of the various
irreducible representations (irreps) of $\mathcal{G}$ is related to the given subspace $\mathbb{V}^{[c]}$,
which, in turn, determines the dimension $\bar{m}_c$ of $\mathbb{V}^{[c]}$.
Index $c$ is commonly called \emph{charge}, or \emph{sector}; for comfort, we will always label with $c = 0$ the trivial
representation $V^{[0]}_g = 1$, of dimension $\bar{m}_0 = 1$,  which always exists, regardless of $\mathcal{G}$.
Moreover, in a generic representation, the same $c$-charged irrep can appear an arbitrary
number of times $\bar{\partial}_c$; this (integer positive) $\bar{\partial}_c$
number actually defines the \emph{degeneracy} of sector $c$ within representation $W_{\mathcal{G}}$.

In conclusion we can build a complete basis for $\mathbb{V}$, in accordance to expansion \eqref{eq:irrepdeco},
listed as $|c,\partial_c,m_c\rangle$, where $\partial_c \in \{1 .. \bar{\partial}_c \}$ labels
degeneracy-space vectors, while $m_c \in \{1 .. \bar{m}_c \}$ labels irrep-space vectors.
Then the action of $W_{g}$ is respectful of this decomposition:
\begin{equation}
 \begin{aligned}
 W_{g} |\varphi \rangle &=
 W_{g} \left( \sum_{c} \sum_{\partial_c = 1}^{\bar{\partial}_c} \sum_{m_c = 1}^{\bar{m}_c}
        \varphi_{c,\partial_c, m_c} |c,\partial_c,m_c\rangle \right)
 \\ &=
 \sum_{c} \sum_{\partial_c = 1}^{\bar{\partial}_c} \sum_{m'_c = 1}^{\bar{m}_c}
 \left( \sum_{m_c = 1}^{\bar{m}_c} \left[V^{[c]}_g \right]_{m'_c}^{m_c} \;\varphi_{c,\partial_c, m_c}  \right)
 |c,\partial_c,m'_c\rangle,
\end{aligned}
\end{equation}
or more simply, by exploiting the direct sum formalism of \eqref{eq:irrepdeco} we can write
\begin{equation} \label{eq:irrepdeco2}
 W_{g} = \bigoplus_{c} \left( \Id_{\bar{\partial}_c \times \bar{\partial}_c} \otimes V^{[c]}_g \right),
\end{equation}
where $\Id$ is the identity operator, and the unitary matrix $V^{[c]}_g$ is the $c$-charge irrep  of
the group element $g$.
An important remark is the following: undoubtedly, the set
of allowed $c$ values in expansion \eqref{eq:irrepdeco} and \eqref{eq:irrepdeco2} is a peculiar property
of the representation $W$ we chose, and similarly also the degeneracy numbers $\bar{\partial}_c$.
Instead, the irrep dimensions $\bar{m}_c$ depend only the group $\mathcal{G}$ itself
(after irreps and charges $c$ have been associated once and for all), not specifically on $W$.
In particular, let us recall that if $\mathcal{G}$ is an \emph{abelian group}, then all its irreps have
always dimension one, i.e. $\bar{m}_c = 1$ regardless from $c$.

\section{Network geometry preservation}

We will now select an arbitrary network geometry,
and focus on the manifold of Tensor Network states $\{ |\Psi_{\text{TN}}\rangle \}$
which can be generated through that geometry. A preliminary argument that must
be introduced, is that $\{ |\Psi_{\text{TN}}\rangle \}$ is closed under the action
of any pointwise symmetry group:
\begin{equation}
 | \Psi' \rangle = U_{g}^{\otimes L} |\Psi_{\text{TN}}\rangle \quad
 \Longrightarrow \quad | \Psi' \rangle \in \{ |\Psi_{\text{TN}}\rangle \},
\end{equation}
telling us that the target state $| \Psi' \rangle$ admits an exact tensor
network representation (with the same graph design).
This follows trivially from the fact that the operation $U_{g}^{\otimes L}$ is
a tensor product of local and invertible transformations. Therefore the target state $| \Psi' \rangle$
must yield the same entanglement properties of the original state $|\Psi_{\text{TN}}\rangle$.
Another way to see the equivalence, from an algebraic viewpoint,
is that every of the local $U_g$ transformations can be adsorbed into the nearest linked tensor,
thus effectively recovering the original network structure (and unaltered bondlink dimensions $D$).

\section{Zero charge Tensor Network states} \label{app:zerocharge}

The first, and simplest, step to undertake
if we want match Tensor Network ansatze and symmetries is to provide a characterization
of TN-states which are invariant under the pointwise symmetry group,
i.e. those $|\Psi^{[0]}_{\text{TN}}\rangle$ for which it holds
\begin{equation}
 |\Psi^{[0]}_{\text{TN}}\rangle = U_{g}^{\otimes L} |\Psi^{[0]}_{\text{TN}}\rangle \qquad \forall\,g \in \mathcal{G}.
\end{equation}
It is clear that, in order for the unitary symmetry group to have no effect on them,
these $|\Psi^{[0]}_{\text{TN}}\rangle$ must strictly belong to the ($\bar{\partial}_{0}$-dimensioned)
sector $c = 0$ of $U_{g}$. Therefore, we will call states $|\Psi^{[0]}_{\text{TN}}\rangle$
belonging to this sub-manifold, zero charge Tensor Network states.

Clearly, if all tensors of $|\Psi^{[0]}_{\text{TN}}\rangle$
are invariant under the action of $U_g$ (applied at the physical links solely), the state $|\Psi^{[0]}_{\text{TN}}\rangle$
is invariant. This condition is sufficient, but not necessary. Indeed the application of $U_g$
could map the tensors into a \emph{gauge-transformed version} of the original ones, and the state
would be unaltered anyway. In practice, tensors $T$ of $|\Psi^{[0]}_{\text{TN}}\rangle$ must be such that,
for every $g \in \mathcal{G}$, applying $U_{g}^{\otimes L}$ is equivalent to performing a gauge transformation.
For obvious reasons the selected gauges must form a group
(intended as a subgroup of the whole gauge group), representing $\mathcal{G}$.
But since gauge transformations of Tensor Networks are made of local, independent isomorphisms
upon doubly-connected links, every link is implicitly carrying a group representation separately.

In conclusion, we may associate to every non-physical link $\alpha$ a  unitary representation
$W_{(\alpha)}$ of $\mathcal{G}$ and a direction, and tensors $T$ must be symmetric,
i.e. invariant under the action of $\mathcal{G}$ on \emph{all} of their indices
(with their respective representation and direction):
\begin{equation} \label{eq:TNsymtransform}
 T^{\{q\},\{p\}}_{\{s\}} \! = \!\!\!
 \sum_{\{r\},\{v\},\{w\}} \!
 \left( \prod_{\alpha}^{\bar{\alpha}_{\text{phys}}} \left[U_g\right]_{s_{\alpha}}^{r_{\alpha}} \right)
 \left( \prod_{\alpha}^{\bar{\alpha}_{\text{in}}} \left[W^{\dagger}_{(\alpha),g}\right]_{q_{\alpha}}^{v_{\alpha}} \right)
 \left( \prod_{\alpha}^{\bar{\alpha}_{\text{out}}} \left[W_{(\alpha),g}\right]_{p_{\alpha}}^{w_{\alpha}} \right)
 T^{\{v\},\{w\}}_{\{r\}}
\end{equation}
for every $g \in \mathcal{G}$,
where tensor $T$ is connected to $\bar{\alpha}_{\text{phys}}$ physical links,
$\bar{\alpha}_{\text{in}}$ incoming virtual links, and $\bar{\alpha}_{\text{out}}$ outcoming virtual links.
The representation $U$ is the only one fixed by the physics of the problem,
while the $W_{(\alpha)}$ are arbitrary, chosen by the user. We are strongly encouraged
to model the $W_{(\alpha)}$ on the problem we are investigating, especially if we are implementing a simulation algorithm.
It is instructive to rewrite equation \eqref{eq:TNsymtransform} as follows:
\begin{equation} \label{eq:Symgauge}
\begin{overpic}[width = \textwidth, unit=1pt]{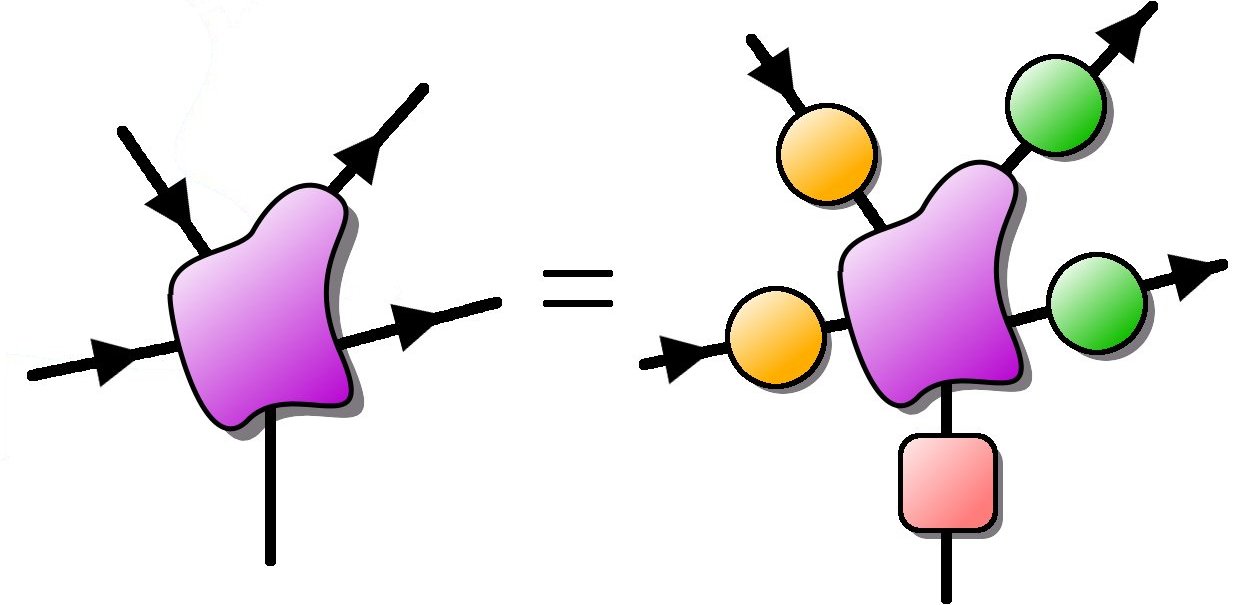}
 \put(77, 83){$T$}
 \put(286, 91){$T$}
 \put(230, 80){\footnotesize{$W^{\dagger}_{(1)g}$}}
 \put(246, 138){\footnotesize{$W^{\dagger}_{(2)g}$}}
 \put(318, 153){\footnotesize{$W_{(3)g}$}}
 \put(330, 90){\footnotesize{$W_{(4)g}$}}
 \put(291, 35){\footnotesize{$U_{g}$}}
\end{overpic}
\end{equation}
The diagrammatic formulation of Tensor Network formalism will help us to understand
why equation \eqref{eq:TNsymtransform} is the standard requirement for a zero charge TN-state.
Precisely, consider the following example:
\begin{equation} \label{eq:Symequiv}
\begin{overpic}[width = \textwidth, unit=1pt]{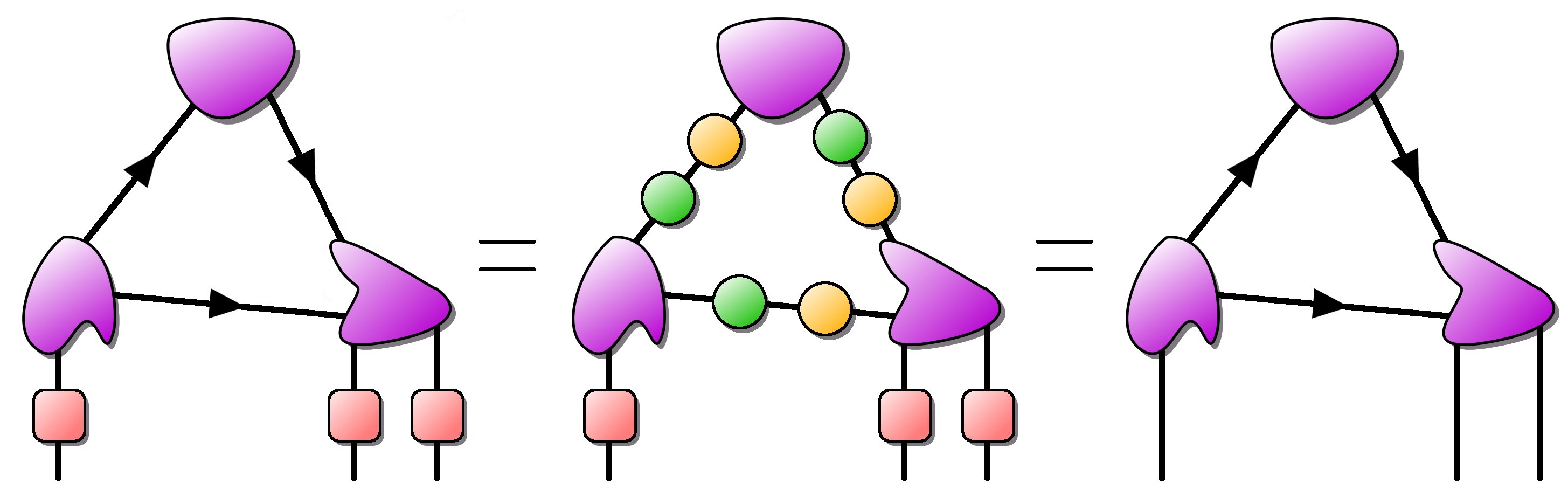}
% \put(79, 93){$T$}
\end{overpic}
\end{equation}
where a three-sites state is involved, on which we are applying the pointwise transformation
$U^{\otimes 3}_g$ for a given $g \in \mathcal{G}$. The first equality in \eqref{eq:Symequiv}
is just creation of pairs operator-antioperator, which is freely allowed since it is a gauge
transformation (as discussed in section \ref{sec:TNgauge}). By choice, we are picking exactly the unitary
operators $W_{(\alpha)g}$ (green circle) and $W_{(\alpha)g}^{-1} = W_{(\alpha)g}^{\dagger}$ (orange circle), with the
selected representation for each virtual link $\alpha$. Now we apply the tensor symmetricity requirement of
eq. \eqref{eq:Symgauge}, which automatically gives the second equality. Then the state is
invariant under the action of the whole pointwise symmetry group $U^{\otimes 3}_{\mathcal{G}}$.
We showed an example with a given network geometry, but it is obvious that this argument
applies equivalently to every Tensor Network state whose tensors satisfy \eqref{eq:TNsymtransform}, \eqref{eq:Symgauge}.

A peculiar ingredient in this framework
is selecting the virtual representations $W_{(\alpha)}$ of the symmetry group. It is true, as we stated,
that for any choice of those representations the resulting TN-state would be zero charged;
but at the same time, the description capabilities of the symmetric Tensor Network ansatz
may depend on such choice, and typically will.

\section{Symmetric tensors fragmentation} \label{sec:fragment}

The prescription \eqref{eq:TNsymtransform} of using symmetric tensors in order to generate invariant
network states, can be interpreted \cite{grouptheory} as a generalization of Schur's lemma to a linear operand (tensor) with
an an arbitrary number of indices. Precisely, consider the case investigated by Schur, where an operator $O$
commutes with a unitary representation $W$ of a compact symmetry group $\mathcal{G}$; the $O$
can have nontrivial support only on the degeneracy space, i.e.
\begin{equation} \label{eq:schur}
 [O, W_{\mathcal{G}}] = 0 \quad \Longrightarrow \quad O = \bigoplus_{c} \left( O^{[c]} \otimes
 \Id_{\bar{m}_c \times \bar{m}_c} \right),
\end{equation}
using the same irrep subspace decomposition of \eqref{eq:irrepdeco}. It is easy to check that operators
in this form are the only ones commuting with every $W_g$ of equation \eqref{eq:irrepdeco2}.
Then, as it was discussed in refs.~\cite{Singh2, grouptheory} a similar argument applies to every
symmetric tensor, whichever its amount of indices might be. Ultimately, symmetric tensors
decompose in such a way that the degrees of freedom which are not fixed by symmetry
(variational DOF) are isolated, and separated from the symmetry constraints (structural DOF).
This decomposition takes place at every node of the network structure, substantially \emph{fragmenting}
a single tensor $T$ into a fully-variational tensor $R$, and one (or more) structural-tensor $S$;
thus actually splitting the original network into a pair of connected superimposed graphs.

We will now explain how this symmetric tensor fragmentation scheme is performed in practice,
by considering tensors with a limited correlation number (up to four attached links). We will
proceed step-by-step starting from the simplest cases, and sketching diagrams whenever possible for clarity and comfort.

\vspace{.5em}
\emph{\textbf{Bondlink fragmentation -}} Before considering fragmentation of tensors, it is useful to
understand how symmetry relations decompose network bondlinks themselves. Indeed, we mentioned
that to every link $\alpha$ we associated a representation $W_{(\alpha)}$ of $\mathcal{G}$,
as well as a direction which helps us to discriminate between the application of the direct unitary transformation
$W_{(\alpha)g}$ and its inverse $W_{(\alpha)g}^{\dagger}$ (according to this picture, we
may think open physical links as \emph{outgoing} links). Then, every value $j$ of the link $\alpha$,
intended as a virtual state $|j)_{\alpha}$ decomposes according to the irrep subspace expansion:
$|j)_{\alpha} \rightarrow |c, \partial_c, m_c)_{\alpha}$. Therefore, the network link literally splits into:
\begin{equation} \label{eq:Fraglink}
\begin{overpic}[width = 270pt, unit=1pt]{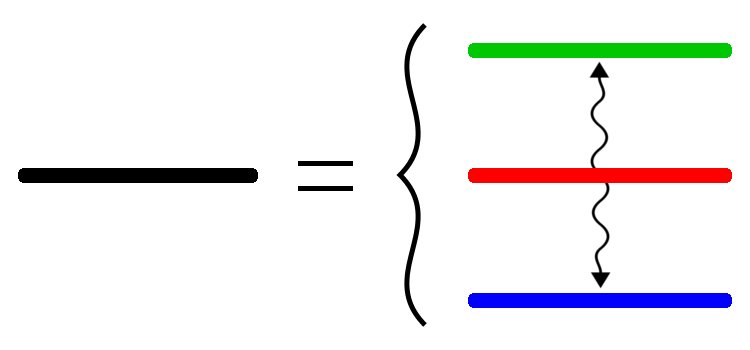}
 \put(36, 73){link $\alpha$}
 \put(-4, 60){$j$}
 \put(271, 60){$c$}
 \put(271, 106){$\partial_c$}
 \put(271, 15){$m_c$}
\end{overpic}
\end{equation}
where the red fragment-link carries the charge index $c$, the green one holds the degeneracy index
$\partial_c$, and the blue one keeps track of the irrep basis vector label $m_c$.
As we stated, the number  $\bar{m}_c$ of allowed values for the index $m_c$ depends on the
value of $c$, in other words the blue link has a \emph{dimension} depending on the value of the
red link. Similarly, the degeneracy number $\bar{\partial}^{(\alpha)}_c$ depends on both $c$ and
the representation chosen on link $\alpha$.
This is the purpose of the wavy arrows in \eqref{eq:Fraglink}, to remember that the value of
the red link influences the dimension of the blue and green links tied to it. The total bondlink
dimension becomes then
\begin{equation}
 D_{\alpha} = \sum_{c \;\in\; \text{irreps}(\mathcal{G})} \bar{m}_c \times \bar{\partial}_c^{(\alpha)},
\end{equation}
where we can sum over all irreducible representations of $\mathcal{G}$. with $c$ being the charge.
To control $D_{\alpha}$ and keep it finite, even when there are infinite independent irreps for $\mathcal{G}$, it is sufficient
to set the $\bar{\partial}_c^{(\alpha)}$ so that only a finite number of them are nonzero.

To make a practical example, let us assume that $\mathcal{G}$ is SO(3), then $\bar{m}_c = 2c + 1$
with $c \in \mathbb{N}$. Assume we are performing a renormalization
process, keeping track of pointwise spin rotation symmetry; and three spin-1 sites were already renormalized
into the given link $\alpha$. Then, by spin-sum rules, we might want to describe for $W_{(\alpha), \mathcal{G}}$:
one singlet, three triplets, two quintuplets, and one
7-plet, so that $\bar{\partial}_0 = 1$, $\bar{\partial}_1 = 3$, $\bar{\partial}_3 = 2$ and
$\bar{\partial}_4 = 1$ ($\bar{\partial}_c = 0$ for $c \geq 5$). 
Clearly, if we take these degeneracies, we are keeping all the state information, and not actually
renormalizing anything: indeed the total bondlink dimension $D$ is 27, equal to $d^3$.

As we stated, if the symmetry group $\mathcal{G}$ we are considering is abelian, $\bar{m}_c$ is always 1,
regardless from $c$ or the link $\alpha$; so the blue link-fragment allows only one value, and therefore
is futile. Indeed in abelian symmetry frameworks, typically only the red and green sub-links appear,
as they are the only ones needed.

\vspace{.5em}
\emph{\textbf{One-link tensors -}} a tensor having only one index behaves like a vector.
This means that it is symmetric only if $\mathcal{G}$ acts trivially on it. This implies that its
support must be restricted to the $c = 0$ sector, and $j = \partial_0$ labels completely the
space $|0, \partial_0, 1)$, which is fully degenerate

\vspace{.5em}
\emph{\textbf{Two-links tensors -}} This is the case considered in Schur's lemma.
Equation \eqref{eq:schur} can be read in the following terms: a two-link symmetric tensor $T$, written in the basis
$|j\rangle\langle k|\rightarrow|c,\partial_c, m_c\rangle\langle q, \partial'_q, m'_q|$
must preserve both charge and irrep label $m$.
Therefore $T$ naturally decomposes as
\begin{equation} \label{eq:twofrag}
 T_{k}^{j} =
 (R^c_{q})^{\partial_c}_{\partial'_q} \times (S^c_{q})^{m_c}_{m'_q}
 \qquad \mbox{with} \qquad
 (S^c_{q})^{m_c}_{m'_q} = \delta_{c,q}\; \delta_{m,m'}
\end{equation}
where $R$ is the variational fragment of $T$, and $S$, the structural one, is equal to the identity $\Id$.
Let us sketch fragmentation diagrams for one-link and two-link tensors:
\begin{equation} \label{eq:Frag12}
\begin{overpic}[width = \textwidth, unit=1pt]{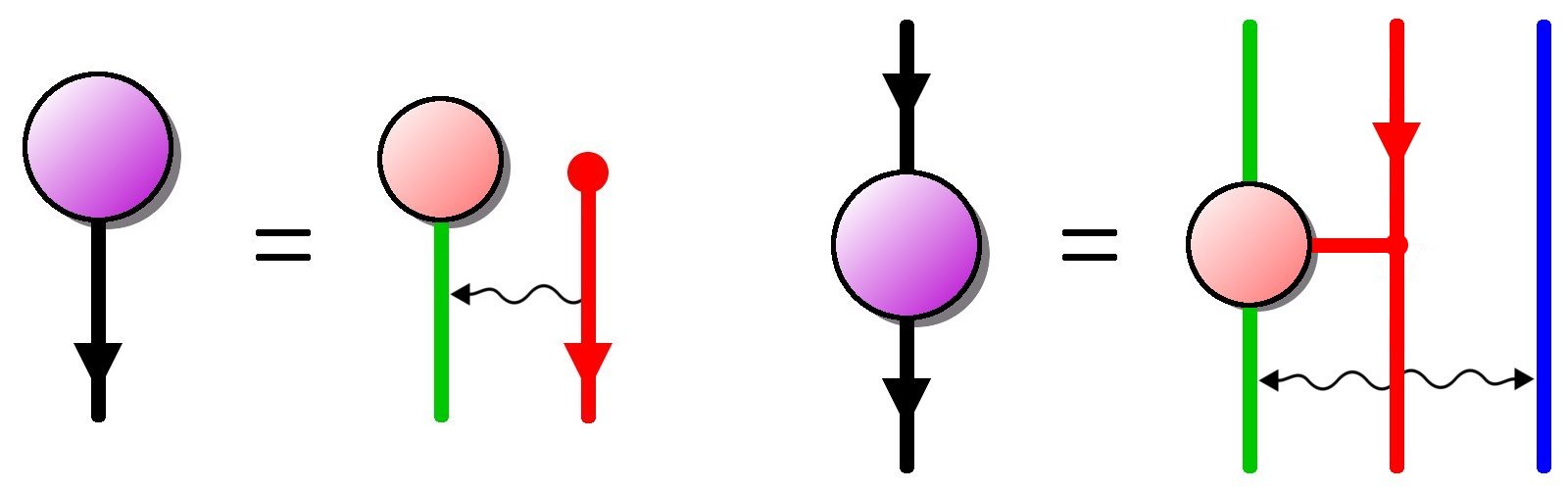}
 \put(0, 110){a)}
 \put(200, 110){b)}
 \put(21, 81){$T$}
 \put(105, 78){$R$}
 \put(140, 89){$\delta_{c,0}$}
 \put(222, 57){$T$}
 \put(306, 57){$R$}
 \put(353, 57){\footnotesize $\delta^{[3]}$}
\end{overpic}
\end{equation}
Notice that in the case of one-link symmetric tensor, we disregarded the irrep vector sub-link (blue one) since the only
relevant charge is $c = 0$ which is the trivial irrep, and thus 1-dimensioned.

The peculiar double-delta form for $S$ in \eqref{eq:twofrag} actually depends on the fact that we
are considering one ingoing link and one outgoing, which is the most common setting.
Other choices of directing links
(say two incoming or two outcoming links) lead to different structure tensors:
$(S^c_{q})^{m_c}_{m'_q} = \delta_{c,q}\; Q_{m}^{m'}$.
The best way to recover these setups is by contracting a 3-link
symmetric tensor (we will discuss it shortly) with a 1-link one.

The two-leg symmetric tensor case is the first where we encounter an actual \emph{reduction of variational parameters},
by employing symmetries, while preserving the original total bondlink dimensions $D_{\alpha}$, and
thus entanglement features as well. Indeed consider respectively $T$ and $R$ of (\ref{eq:Frag12}.b):
\begin{equation}
 T \longleftarrow D_{\alpha} D_{\beta} \;\mbox{ descriptors}, \qquad
 R \longleftarrow \bar{D}_{\alpha, \beta}^{\text{II}} \equiv \sum_c
 \bar{\partial}^{[\alpha]}_c \bar{\partial}^{[\beta]}_c \;\mbox{ descriptors}.
\end{equation}
Then, since $D_{\alpha} = \sum_{c} \bar{\partial}^{[\alpha]}_c \bar{m}_c$, we typically end up with an
effective amount of parameters $\bar{D}_{\alpha, \beta}^{\text{II}} \ll D_{\alpha} D_{\beta} \sim D^2$
way smaller than the original one,
especially when the \emph{active} sectors ($c$ values for which $\bar{\partial}^{[\alpha]}_c > 0$) are many.

\begin{figure}
 \begin{center}
\begin{overpic}[width = 360pt, unit=1pt]{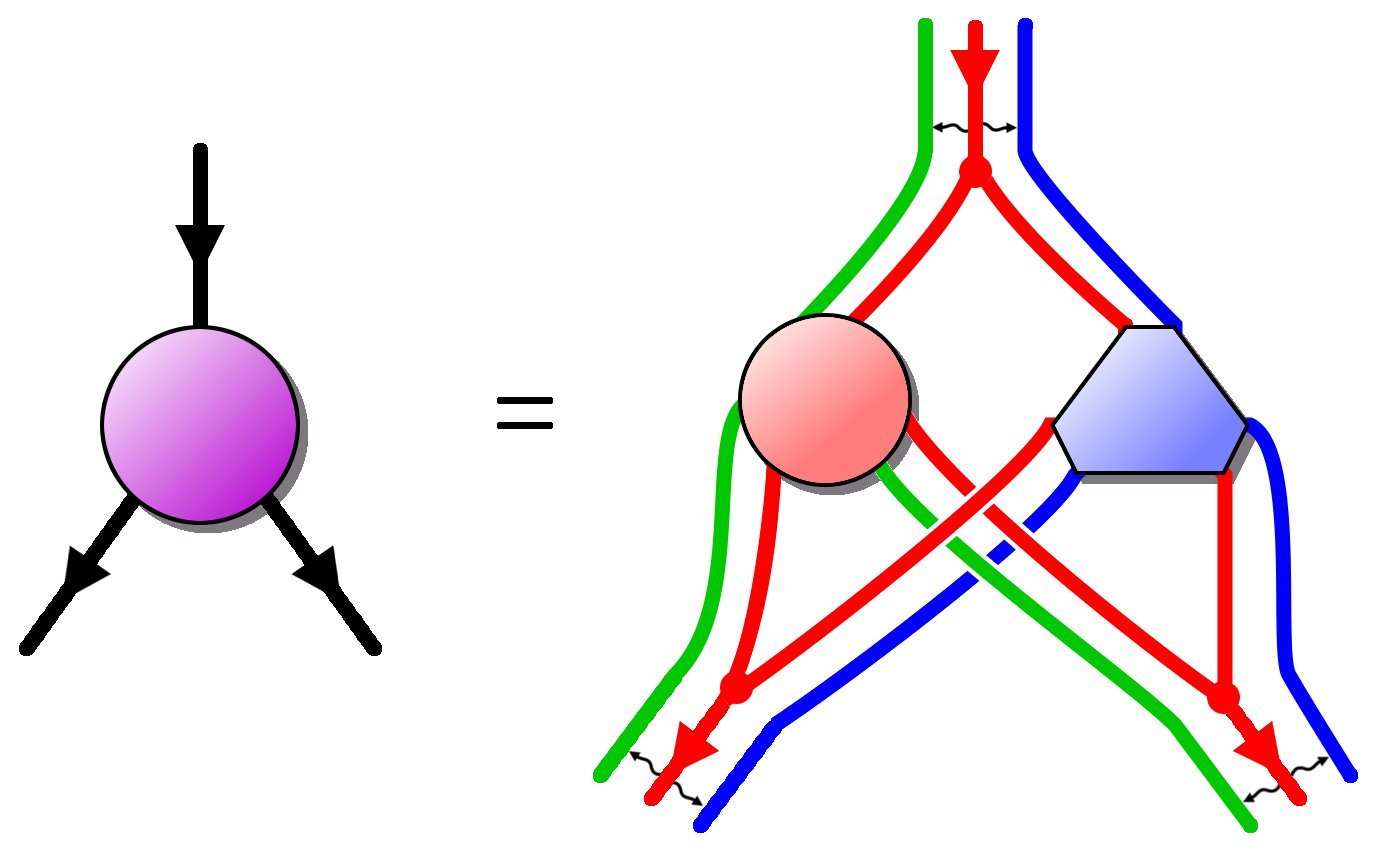}
 \put(48, 106){$T$}
 \put(211, 113){$R$}
 \put(297, 112){$S$}
 \put(222, 170){\footnotesize $\partial_c$}
 \put(282, 170){\footnotesize $m_c$}
 \put(254, 159){\footnotesize $c$}
 \put(206, 62){\footnotesize $a$}
 \put(171, 62){\footnotesize $\partial'_{a}$}
 \put(216, 31){\footnotesize $m'_{a}$}
 \put(286, 31){\footnotesize $\partial''_{b}$}
 \put(309, 56){\footnotesize $b$}
 \put(342, 56){\footnotesize $m''_b$}
 \end{overpic}
\end{center}
\caption{  \label{fig:Frag3}
 Fragmentation diagram for a three-link symmetric tensor $T$. The structure tensor $S$ is made of
 Clebsh-Gordan coefficients for irreps of the group $\mathcal{G}$. We are assuming $\mathcal{G}$ to be
 multiplicity free; otherwise, a third tensor fragment appears.
}
\end{figure}
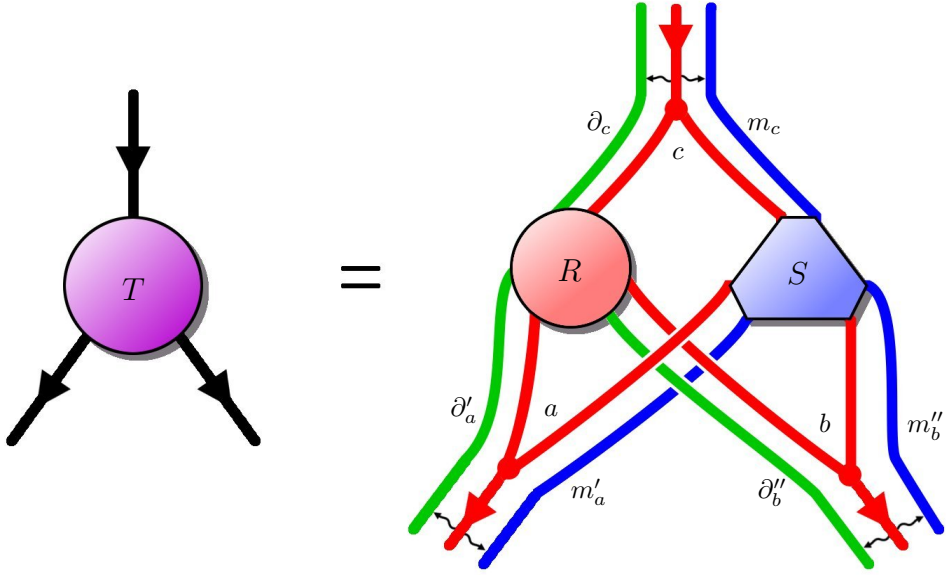

\vspace{.5em}
\emph{\textbf{Three-links tensors -}} The tensor product of two irreps, with charges $a$ and $b$ respectively,
is still a representation of $\mathcal{G}$; and thus can be decomposed in a direct sum of irreps according to
\eqref{eq:irrepdeco} and \eqref{eq:irrepdeco2}:
\begin{equation} \label{eq:mfedco}
 \mathbb{V}^{[a]} \otimes \mathbb{V}^{[b]} \approx \bigoplus_c \left[
 \bigoplus_{\partial_c = 1}^{\mu_{a, b}^c} \mathbb{V}^{[c]} \right]
\end{equation}
where $\mu_{a, b}^c$ is the number of copies of $\mathbb{V}^{[c]}$ appearing in the
tensor product representation.

Here it is comfortable to assume that the group $\mathcal{G}$ is
multiplicity free, i.e. that $\mu_{a, b}^c$ can be either zero or 1, no matter the sectors.
This is quite a typical case for physically relevant symmetries: SO(3) and SU(2) which take into
account rotational invariance, as well as every abelian group that keeps track of particle number
and parity conservation, are multiplicity free symmetries (but not SU(3), for example).
In this framework, the Wigner-Eckart theorem provides a remarkable fragmentation law for a three-leg
tensor $T$:
\begin{equation} \label{eq:3frag}
 T^{\,i}_{j,k} = ( R^{\,c}_{a,b} )^{\partial_c}_{\partial'_a, \partial''_b}
 \times ( S^{\,c}_{a,b} )^{m_c}_{m'_a, m''_b},
\end{equation}
where the structural tensor fragment $S$ contains the Clebsh-Gordan coefficients for irreps
of $\mathcal{G}$, which are well defined thanks to \eqref{eq:mfedco} and the multiplicity freedom requirement:
\begin{equation}
  ( S^{\,c}_{a,b} )^{m_c}_{m'_a, m''_b} = \langle c, m_c | a, m'_a; b, m''_b \rangle.
\end{equation}
An analogous decomposition with different structural tensors $S$ holds for other direction configuration
of the links connected to $T$. Figure \ref{fig:Frag3} shows the diagram for \eqref{eq:3frag}, the case we considered.

We want to check, in this three-link tensor scenario, the effective reduction of variational descriptors
caused by enforcing symmetry relations. So, let us compare the amount of parameters:
\begin{equation}
 T \longleftarrow D_{\alpha} D_{\beta} D_{\gamma} \sim D^3 \,, \quad
 R \longleftarrow \bar{D}^{\text{III}} \equiv
 \sum_a \sum_b \sum_{c \in \{a \oplus b\}}\bar{\partial}^{[\alpha]}_a\, \bar{\partial}^{[\beta]}_b\,
 \bar{\partial}^{[\gamma]}_c,
\end{equation}
where the innermost sum spans the sole sectors $c$ that are achievable by fusing together
charges $a$ and $b$, via \eqref{eq:mfedco}.
It is clear that $\bar{D}^{\text{III}}$ is very small compared to $D^3$; it is more likely to scale
like $D^2$ rather than $D^3$, especially if the symmetry $\mathcal{G}$ is abelian.

\vspace{.5em}
\emph{\textbf{Four-links tensors -}} The tensor product of three irreps
$\mathbb{V}^{[a]}$, $\mathbb{V}^{[b]}$ and $\mathbb{V}^{[c]}$ may contain several copies
of the same irrep $\mathbb{V}^{[q]}$, even when the symmetry group $\mathcal{G}$ is multiplicity free.
However, we can workaround this issue by fusing together two charges, say $a$ and $b$, beforehand,
operation which is well-defined through Clebsh-Gordan sum rules.

\begin{figure}
 \begin{center}
\begin{overpic}[width = \textwidth, unit=1pt]{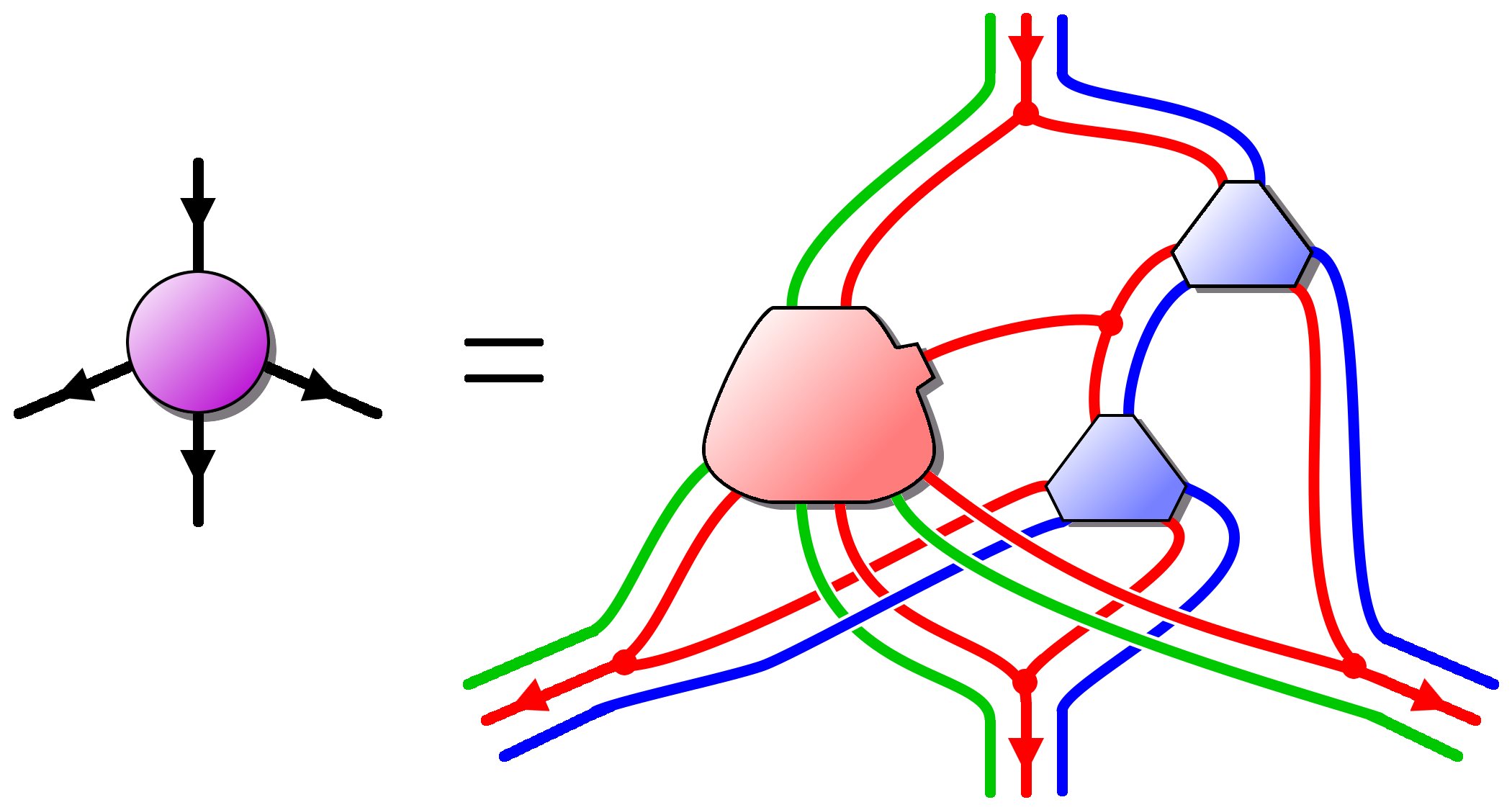}
 \put(48, 116){$T$}
 \put(206, 100){$R$}
 \put(284, 83){$S_0$}
 \put(316, 143){$S_0$}
 \put(177, 47){\footnotesize $a$}
 \put(146, 55){\footnotesize $\partial'_a$}
 \put(162, 19){\footnotesize $m'_a$}
 \put(258, 42){\footnotesize $b$}
 \put(241, 17){\footnotesize $\partial''_b$}
 \put(279, 17){\footnotesize $m''_b$}
 \put(333, 46){\footnotesize $c$}
 \put(330, 17){\footnotesize $\partial'''_c$}
 \put(358, 52){\footnotesize $m'''_c$}
 \put(270, 130){\footnotesize $e$}
 \put(298, 112){\footnotesize $m^{+}_e$}
 \put(263, 167){\footnotesize $q$}
 \put(241, 187){\footnotesize $\partial_q$}
 \put(280, 191){\footnotesize $m_q$}
\end{overpic}
 \end{center}
\caption{  \label{fig:Frag4}
 Fragmentation diagram for a four-legs symmetric tensor $T$. The charges $a$ and $b$ fuse into a set
 of intermediate charges $e$, by means of Clebsh-Gordan tensor $S_0$.
 Index $e$ also enters in the variational fragment $R$, to keep track of degeneracies arising from
 the product of three irreps.
}
\end{figure}
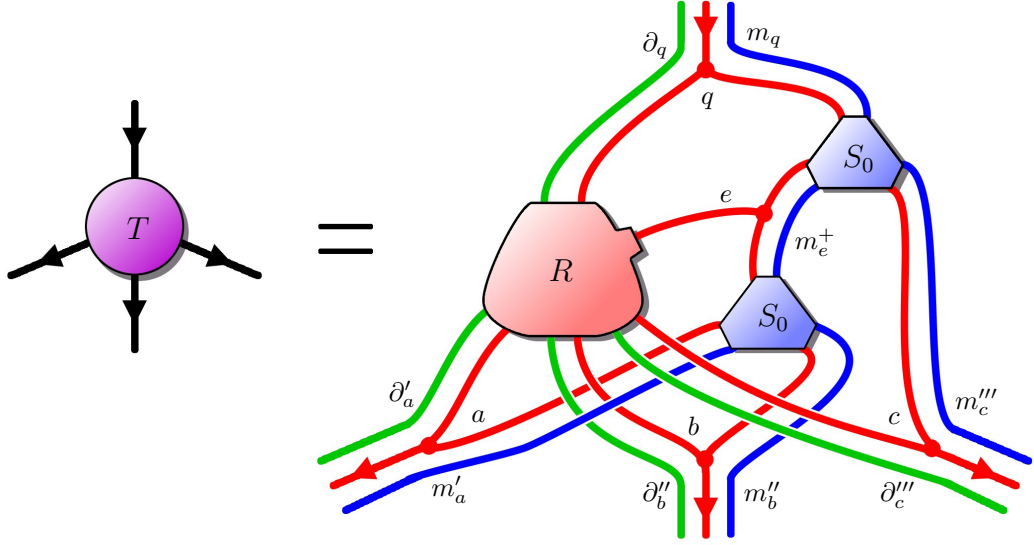

Let us focus on a directing configuration having one ingoing link, and three outgoing ones.
Let then $e = a \oplus b$, where $\oplus$ stands for charge fusion, and clearly $q = e \oplus c$.
We have to take into account all
the allowed intermediate charges, i.e. those $e$ for which
\begin{equation}
 \mu^{q}_{e,c} \times \mu^{e}_{a,b} \neq 0.
\end{equation}
Adopting the intermediate charge scheme is sufficient to label and address separately the different copies
of the same irrep $\mathbb{V}^{[q]}$. Then the symmetric tensor $T$ fragments as follows:
\begin{equation}
 T^{\,i}_{j,k,l} = \sum_e \left[ ( R^{\,q,e}_{a,b,c} )^{\partial_q}_{\partial'_a, \partial''_b, \partial'''_c}
 \times ( S^{\,q,e}_{a,b,c} )^{m_q}_{m'_a, m''_b, m'''_c} \right],
\end{equation}
but, at the same time, we know that the structural fragment is obtained by two consecutive
fusions, so in the end
\begin{equation}
 ( S^{\,q,e}_{a,b,c} )^{m_c}_{m'_a, m''_b, m'''_c} =
 \sum_{m^{+}_{e}} \; \langle e, m^{+}_e| a, m'_a, b; m''_b \rangle \;
 \langle q, m_q| e, m^{+}_e; c, m'''_c \rangle,
\end{equation}
i.e. $S$ further decomposes into a pair of sub-fragmented structure tensors $S_0$, coinciding with the Clebsh-Gordan
tensor we used for the three-links case, and we must contract over the $m^{+}_e$ levels related to the $e$-charge irrep.
The fragmentation scheme we just introduced is picted in figure \ref{fig:Frag4}.

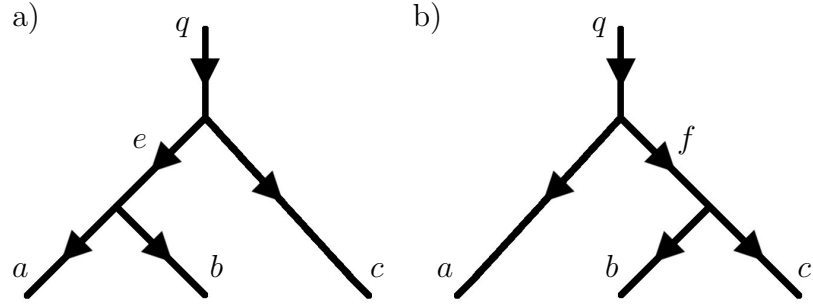
\begin{figure}
 \begin{center}
 \begin{overpic}[width = 300pt, unit=1pt]{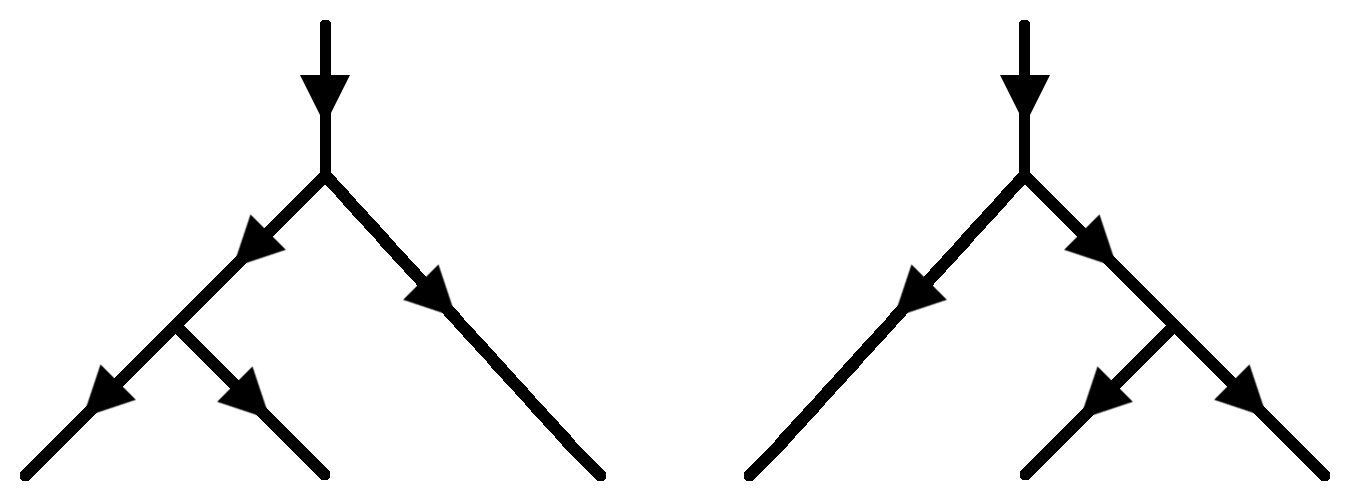}
 \put(0, 107){a)}
 \put(150, 107){b)}
 \put(61, 105){$q$}
 \put(217, 105){$q$} % +156
 \put(0, 12){$a$}
 \put(159, 12){$a$}
 \put(74, 12){$b$}
 \put(222, 12){$b$}
 \put(134, 12){$c$}
 \put(294, 12){$c$}
 \put(45, 61){$e$}
 \put(249, 61){$f$}
 \end{overpic}
 \end{center}
\caption{  \label{fig:branch} Two schemes to group together three irreps, of charge $a$, $b$, and $c$,
 by adopting the intermediate fusion charge framework, either $e$ or $f$ depending on which irrep is
 the last to fuse. The intermediate charge takes into account arising degeneracies in the irrep $q$.
}
\end{figure}

Of course, we could alternatively decompose $T$ by fusing together $b$ and $c$ at a first step,
and then fuse $a$ with $f = b \oplus c$, as in figure \ref{fig:branch}.b.
In this case, we have different variational $\bar{R}$ and structural $\bar{S}$ tensor fragments, according to:
\begin{equation}
 T^{\,i}_{j,k,l} = \sum_f \left\{ ( \bar{R}^{\,q,f}_{a,b,c} )^{\partial_q}_{\partial'_a, \partial''_b, \partial'''_c} \times
   \sum_{m^{-}_f}   \left[ ( S^{\,q}_{0\,a,f} )^{m_q}_{m'_a, m^{-}_f} 
 \times ( S^{\,f}_{0\,b,c} )^{m^{-}_f}_{m''_b, m'''_c} \right] \right\},
\end{equation}
which is the right-left specular of the diagram \ref{fig:Frag4}.

The two intermediate fusion schemes we showed are in strict connection. Indeed, if we consider the partial fusion
basis $|e, m^{+}_e \rangle$ and $|f, m^{-}_f \rangle$ they are related \cite{Singh2} by the 6-index tensor $F$,
(e.g. for the group $\mathcal{G} =$ SU(2), $F$ coincides with Wigner's 6-j symbols), so that:
\begin{equation}
 \bar{S}^{q,f}_{a,b,c} = \sum_e ( F^{a,b,c}_d )^{e}_f \times S^{q,e}_{a,b,c}.
\end{equation}
At the same time, since the global tensor $T$ is uniquely defined and does not depend on the
choice of intermediate fusion basis, the variational fragments $R$ and $\bar{R}$ must transform accordingly:
\begin{equation}
 \bar{R}^{q,e}_{a,b,c} = \sum_e ( F^{a,b,c}_d )^{\star\,e}_f \times R^{q,e}_{a,b,c}.
\end{equation}
Theoretically, we could also study the scenario where $a$ and $c$ fuse together
into an intermediate charge $h$, which fuses with $b$ afterwards. But in order to investigate
this framework, we should exchange the order of links, say $a$ and $b$, before the first fusion takes place.
As we discussed in section \ref{sec:linkexchange}, when swapping Tensor Network indices is needed,
the inner nature of the degrees of freedom we are describing (i.e. whether they are spins, fermions, bosons or
even anyons) manifests as an exchange-statistic of links themselves. Therefore, the resulting fragmentation picture
will depend on the statistic obeyed by the particles we are describing .

Generalization of the equations we encountered in this section to other tensor topologies can be done by
hand following the same intermediate charge rules.
This concludes our discussion regarding Tensor Network fragmentation due to symmetry relations.

\section{Finite charge Tensor Network states}

The fragmentation schemes we just analyzed for different tensor topologies, derived from
the argument that all the tensors in the network ought to be symmetric to make the global state
$|\Psi_{\text{TN}}\rangle$ invariant under the application of $U^{\otimes L}_{\mathcal{G}}$.
Although this is an intriguing context, it is quite limited for practical purposes of studying physical settings.
Indeed, in most variational problems, we are actually interested in working with an arbitrary,
fixed, finite symmetry charge $q$.

For instance, assume our model Hamiltonian $\mathcal{H}$ manifests a pointwise U(1) invariance,
i.e. undergoes a particle conservation law: this means that global states, expanded in irreps subspaces as
$|c, \partial_c, m_c\rangle$, belonging to different sectors $c$ are not coupled by $\mathcal{H}$. Then
we might wish to achieve the ground state of $\mathcal{H}$ \emph{restricted} to a given charge, i.e. number of particles;
this is definitely a physical question. Analogously, we could be interested in describing the lowest
energy state of a SU(2)-invariant $\mathcal{H}$, on a spin-$\frac{1}{2}$ lattice, with total spin, say, $\frac{7}{2}$.

Achieving a \emph{charge-selective ground state} is no trivial task from a variational point of view.
It can not be performed blindly, by starting from the correct sector and hoping that the symmetry-invariant
dynamics will prevent other sectors to be explored. This approach typically fails, as numerical errors will
inevitably introduce other-sectors fluctuations, which will ultimately break the symmetry, and push the algorithm
towards the \emph{absolute} ground state.
The right way to deal with this charge-selective quantum problem is by forcing the variational wavefunctions
to belong to the right sector. Here is were Tensor Networks succeed: we will be able to recover
fragmentation rules we learned in the previous section, and apply them even for finite charge $c$ TN-states,
with $c$ chosen by the user.

Therefore, let $|\Psi_{\text{TN}}\rangle$ be our Tensor Network state, for a given graph geometry, bondlink
dimensions $D$, and consequent entanglement bounds. We will require that $|\Psi_{\text{TN}}^{[q]}\rangle$, when expanded
in irrep-subspace basis $|c, \partial_c, m_c\rangle$ of the pointwise symmetry group $U^{\otimes L}_{\mathcal{G}}$,
has nonzero components only for a given charge $q$:
\begin{equation} \label{eq:chargeselect}
 |\Psi_{\text{TN}}^{[q]}\rangle = \sum_{c} \sum_{\partial_c}^{\bar{\partial}_c}
 \sum_{m_c}^{\bar{m}_c} \left( \delta_{c,q} \; \mathcal{T}^{[q]}_{\partial_{q}, m_{q} } \right)\;
 |c, \partial_c, m_c \rangle.
\end{equation}
Let us recall that $\partial_c$ are degeneracy indices, while $m_c$ are irrep basis indices. So when
a symmetry transformation occurs, it interferes with the $m_c$ but leaves the $\partial_c$ unaltered,
separately for every sector $c$. Then, our fixed-charge TN-state transforms as
\begin{equation} \label{eq:chargeselect2}
 U_g^{\otimes L} |\Psi_{\text{TN}}^{[q]}\rangle =
 \sum_{\partial_q}^{\bar{\partial}_q} \sum_{m_q}^{\bar{m}_q}
 \left(  \sum_{m'_q}^{\bar{m}_q} ( V^{[q]}_g )_{m_q, m'_q} \; \mathcal{T}^{[q]}_{\partial_{q}, m'_{q} } \right)
 |q, \partial_q, m_q \rangle
\end{equation}
with $V^{[q]}_g$ being the irrep unitary matrix, charge $q$, group element $g$.

Now, we will give a prescription on the Tensor Network itself that will allow \emph{only} states
in the form $|\Psi_{\text{TN}}^{[q]}\rangle$, i.e. \eqref{eq:chargeselect}, to be variationally generated.

\begin{figure}
 \begin{center}
 \begin{overpic}[width = 260pt, unit=1pt]{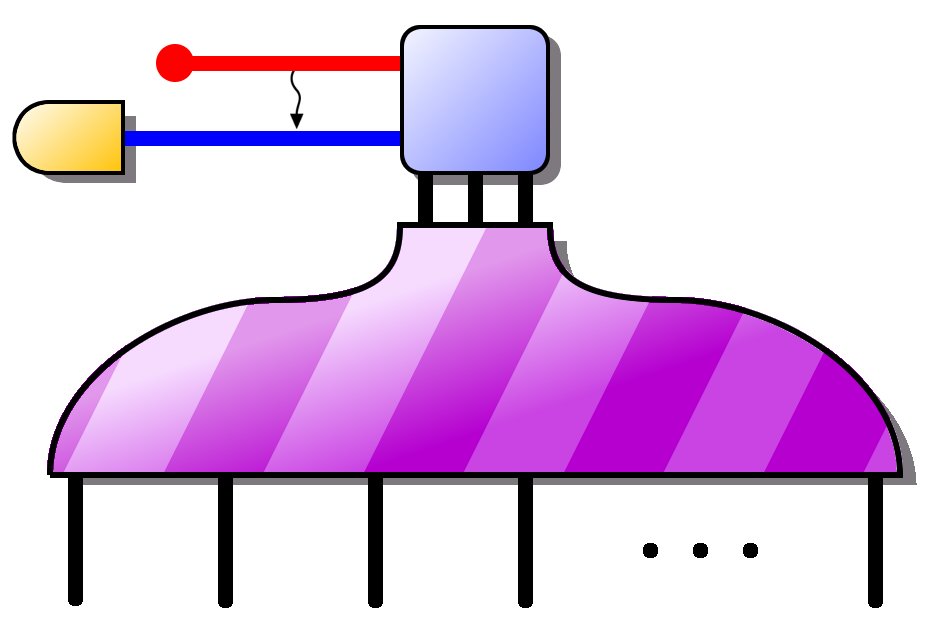}
 \put(16, 133){$C$}
 \put(26, 159){$\delta_{c,q}$}
 \put(80, 163){\footnotesize $c$}
 \put(80, 127){\footnotesize $m_c$}
 \put(118, 145){source}
 \put(92, 65){Tensor Network}
 \put(74, 65){\Large $\downarrow$}
 \put(182, 65){\Large $\downarrow$}
 \put(8, 13){\footnotesize $s_1$}
 \put(50, 13){\footnotesize $s_2$}
 \put(252, 13){\footnotesize $s_L$}
 \end{overpic}
 \end{center}
\caption{ \label{fig:chgfix}
 Diagram of the finite charge $q$ Tensor Network prescription. The network is partially contracted into the violet
 dashed tensor, only the source node and the (structured) charge selector node are highlighted.
}
\end{figure}
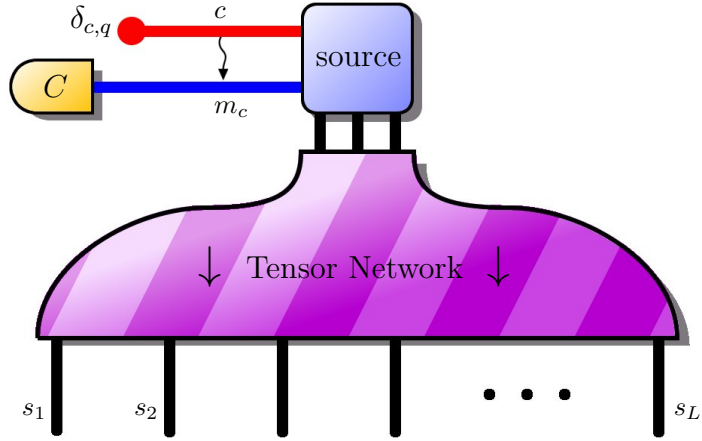

\begin{enumerate}
 \item Choose a single tensor in the network, a node in the graph. We will refer to this tensor as \emph{source node}.
 \item Direct the graph, i.e. associate a direction to every network link. Also, we require that every
    tensor in the network has \emph{at least one incoming link}, except for the source node, which instead must have
    no incoming links. Such a directing scheme always exists, as long as the graph is fully connected. Moreover,
    if there are no closed loops, this scheme is unique.
    Physical links are meant as outgoing.

\begin{figure}
 \begin{center}
 \begin{overpic}[width = 320pt, unit=1pt]{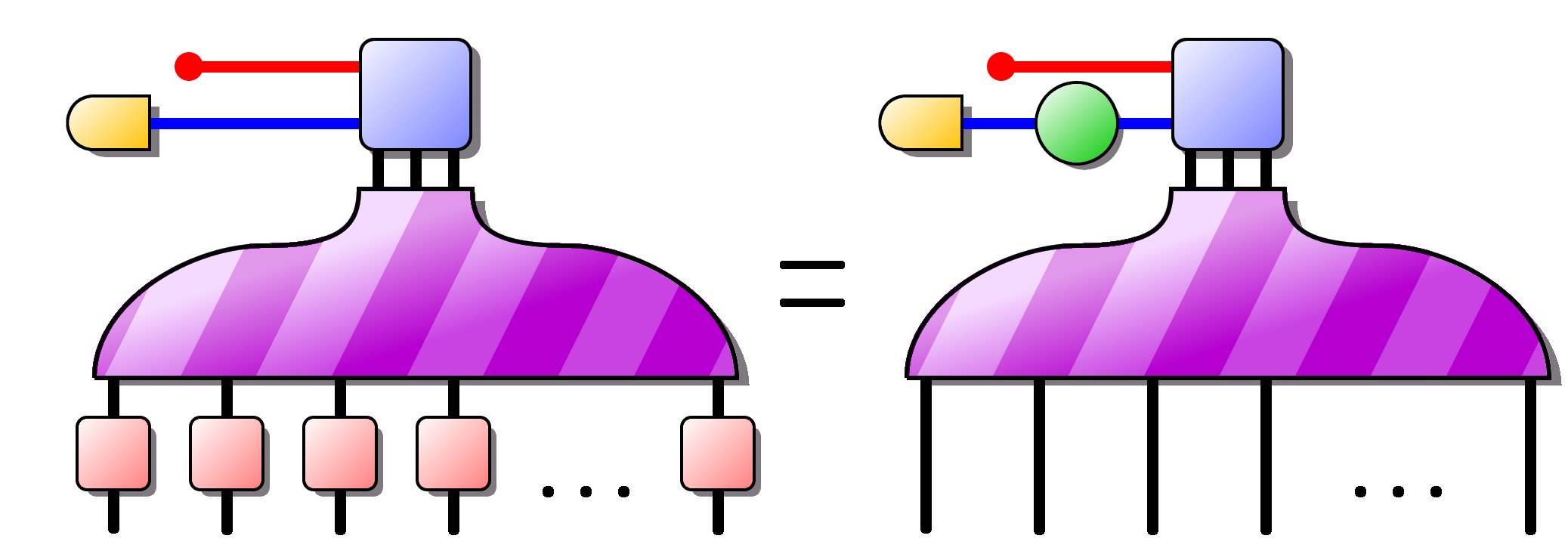}
 \put(16, 133){$C$}
 \end{overpic}
 \end{center}
\caption{ \label{fig:chgfix2}
 The pointwise symmetry transformation $U^{\otimes L}_{g}$ (pink boxes), when applied to $|\Psi_{\text{TN}}^{[q]}\rangle$,
 literally 'jumps over' the symmetric part of the network, which is formed by the original graph, including the source node.
 On the added link, it becomes $V^{[q]}_g$ (green circle), as stated by equation \eqref{eq:cfpass}.
}
\end{figure}

 \item Associate a representation $W_{(\alpha)}$ of $\mathcal{G}$ to every link $\alpha$.
 \item Add a single one-leg tensor to the network (\emph{selector node}), which connects to the source node.
    It is easy to see that TN-entanglement bounds are unaltered by this graph geometry adjustment. The link we just
    added is directed from the selector node to the source node. The representation of $\mathcal{G}$ associated
    to the new link is the $q$-charge irrep $V^{[q]}_{\mathcal{G}}$.
 \item Symmetrize every tensor in the network, except for the selector node (but including the source tensor),
    according to directions and representations $W_{(\alpha)}$ of the links touched. This
    means that every tensor $T$ decomposes into its variational $R$ and structural $S$ fragments, as we
    discussed in section \ref{sec:fragment}.
 \item Fix the charge selector tensor to be $T_{c, m_c} = \delta_{c,q}\,C_{m_q}$, $C$ arbitrary.
\end{enumerate}
This concludes the prescription. The Tensor Network state then reads as in figure \ref{fig:chgfix}.
The charge selector tensor can not be symmetric (unless $C_{m_q} = 0$, or $q = 0$),
as its only connected link has a nontrivial representation $V^{[q]}$. And in fact,
$|\Psi_{\text{TN}}^{[q]}\rangle$ will not be $U^{\otimes L}_{\mathcal{G}}$ invariant since not
every tensor is symmetric (actually, all but one are).

Then, consider $\mathcal{Z}$, the contraction of all the tensors in the final network except for the selector node.
Clearly $\mathcal{Z}$ is symmetric, as contraction of symmetric tensors; i.e. it holds
\begin{equation}
 \mathcal{Z}^{m_q}_{s_1 \ldots s_L} = \sum_{m'_q}^{\bar{m}_q} \sum_{\{r\}}^d
 ( V^{[q]\,\star}_{g} )_{m'_q, m_q} \left( \prod_{\ell} ( U_{g} )_{s_{\ell}, r_{\ell}} \right)
 \mathcal{Z}^{m'_q}_{r_1 \ldots r_L} \qquad \forall g \in \mathcal{G}, 
\end{equation}
or, in matricial form, $U^{\otimes L}_{g} \cdot \mathcal{Z} \cdot V^{[q]\,\dagger}_{g} = \mathcal{Z}$.
In conclusion, when we apply the pointwise symmetry group to the Tensor Network state, we obtain
\begin{multline} \label{eq:cfpass}
 U^{\otimes L}_{g} |\Psi_{\text{TN}}^{[q]}\rangle =
 U^{\otimes L}_{g} \sum_{s_1 \ldots s_L}^d \sum_{m_q}^{\bar{m}_q}
 \left( \mathcal{Z}^{m_q}_{s_1 \ldots s_L} C_{m_q} \right) |s_1 \ldots s_L\rangle
 = \\ =
 \sum_{s_1 \ldots s_L}^d \sum_{m_q}^{\bar{m}_q}
 \mathcal{Z}^{m_q}_{s_1 \ldots s_L}
 \left( \sum_{m'_q}^{\bar{m}_q} ( V^{[q]}_g )_{m_q, m'_q} \;C_{m'_q} \right) |s_1 \ldots s_L\rangle,
\end{multline}
which is formally equivalent to \eqref{eq:chargeselect2}, thus proving that our prescription is sound.

The reason why we needed to direct the graph so that it had a single source of directions
(the source node) is to spread the information about $q$ to the whole network. If this is not
the case, then one can identify regions of the network insensitive to $q$ thus actually behaving
like symmetry-invariant zones: e.g. a party of sites which are always empty.
Although the resulting state would still be a $q$-charge Tensor Network, it would be far more trivial.

With the construction we just introduced, we are finally able to understand and exploit the strict
relationship that ties symmetries and Tensor Networks through representation theory. In these sections
we developed selection rules and manipulation techniques to embed symmetries into
Tensor Network variational ansatze, allowing us to address charge-specific problems, and to
meet a drastic speed-up in computational time.

\section{Example: Symmetries in MPS}

We would like to conclude this appendix chapter by applying the symmetry arguments and techniques
upon a most common template in the family of Tensor Network, namely on Matrix Product States, the
variational counterpart of DMRG algorithms. Methods for dealing with symmetries within the
Density Matrix Renormalization Group framework were already known before the acknowledgement
of Tensor Network states \cite{SchollDMRG}, still, the in-depth understanding of both MPS representations
and symmetric Tensor Network states, allows us to build a formulation for finite charge MPS
which is compact, elegant, and efficient.

Here we will work with open boundary conditions MPS, as the no-closed loop geometry encounters
less accidents, and show the MPS-fragmentation scheme respectively for an abelian symmetry group $\mathcal{G}$,
and then for a non-abelian one. Generalization to PBC is not trivial but possible nevertheless.

\subsection{MPS with pointwise U(1)}

The abelian symmetry U(1) has infinite one-dimensional ($\bar{m}_c = 1$, $\forall\,c$)
non-equivalent representations labeled by integer numbers
$c \in \mathbb{Z}$. U(1) is used to take care of particle conservation, when the Hamiltonian
$\mathcal{H}$ has only terms that preserve particle number; indeed its fusion rule $\oplus$ corresponds
to the simple sum of two integer numbers, i.e. $c \oplus c' \equiv c + c'$.

A natural way to choose the source node (defined in the previous section), in order to characterize MPS
states with an arbitrary particle number $q$, is to choose one of the edge blocks, say the one at right boundary.
MPS tensors have three connection links, so we can use three-leg fragmentation rule \eqref{eq:3frag} to
split a block into structural and variational part. Also, recall that since $\bar{m}_c = 1$,
we have no need for blue (irrep vector) links. Ultimately, the resulting fragmented-MPS reads:
\begin{equation} \label{eq:Mpssym}
\begin{overpic}[width = \textwidth, unit=1pt]{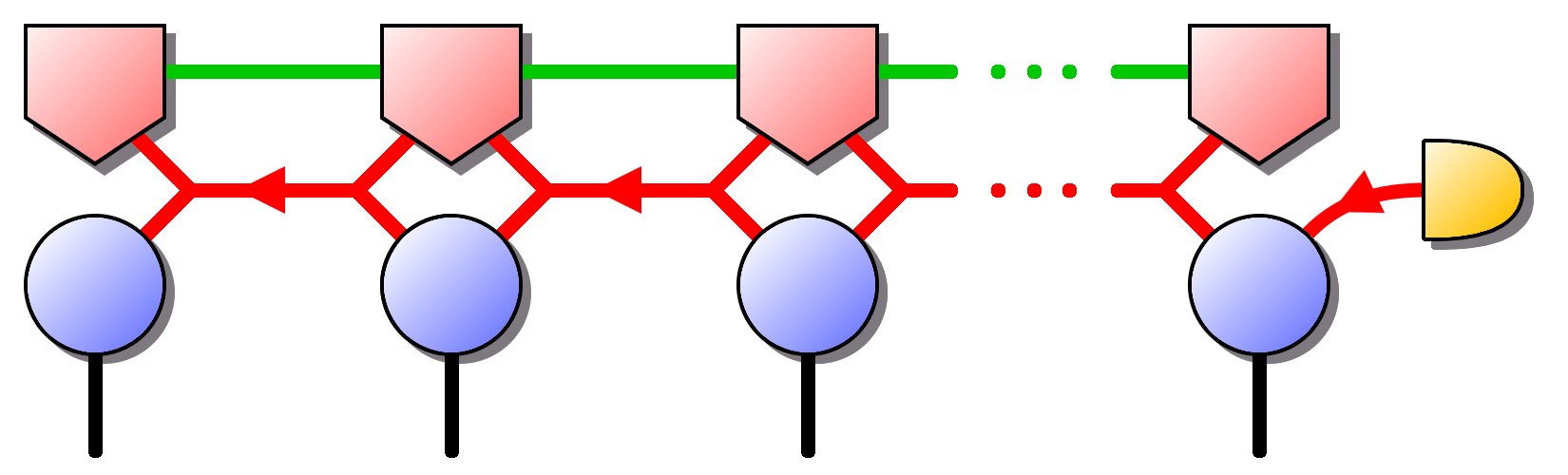}
 \put(17, 94){$R^{[1]}$}
 \put(19, 43){$S$}
 \put(64, 58){\footnotesize $c_1$}
 \put(64, 106){\footnotesize $\partial_{c_1}^{[1]}$}
 \put(106, 94){$R^{[2]}$}
 \put(108, 43){$S$}
 \put(153, 58){\footnotesize $c_2$} 
 \put(153, 106){\footnotesize $\partial_{c_2}^{[2]}$}
 \put(195, 94){$R^{[3]}$}
 \put(197, 43){$S$}
 \put(307, 94){$R^{[L]}$}
 \put(309, 43){$S$}
 \put(335, 79){\footnotesize $c_L$} 
 \put(360, 67){$Q$}
 \put(29, 6){\footnotesize $s_1$}
 \put(118, 6){\footnotesize $s_2$}
 \put(207, 6){\footnotesize $s_3$}
 \put(319, 6){\footnotesize $s_L$}
\end{overpic}
\end{equation}
where the structural tensors are homogeneously defined $S^{s_j}_{c_{j-1}, c_j} = \delta_{c_j, c_{j-1} + s_j}$.
The yellow tensor $Q$ is the charge selector node, properly connected to the source node $A^{[L]}$;
its purpose is to select the global sector: $Q_{c_L} = \delta_{c_L,q}$ with a total charge $q$ chosen
by the user. The tensor fragments $R^{[j]}$ are completely variational, and we can freely manipulate their parameters,
for instance, to lower the total energy, without constraints: we will always remain
forcefully in the correct $q$ sector due to the presence of $S$ fragments.
%The total bondlink dimension $D_j$ (on physical bond $j$)) we are using,
%and consequent partition entanglement $\log D_{j}$, is given by
%$D_j = \sum_{c} \bar{\partial}_c^{[j]}$, where $\bar{\partial}_{c_j}^{[j]}$ is the $j$-th green link
%dimension, influenced by $c_j$, i.e. $\partial_{c_j}^{[j]} \in \{ 1 \,..\,\bar{\partial}_{c_j}^{[j]}\}$.

An intriguing feature of this abelian symmetric-MPS is that it is always operationally possible to gauge-transform
it into the left (or right) gauge, while preserving the fragmentation scheme \eqref{eq:Mpssym}.
The basic idea is to perform a singular value decomposition of $R^{[j]}$ tensors
separately for every $c_{j}$:
\begin{equation}
 ( R^{[j]}_{c_j} )^{\partial_{c_j}^{[j]}}_{c_{j-1}, \partial_{c_{j-1}}^{[j-1]}} =
 \sum_{\beta_{c_j}} = ( U^{[j]}_{c_j} )^{\beta_{c_j}}_{c_{j-1}, \partial_{c_{j-1}}^{[j-1]}}
 \;\lambda_{\beta_{c_j}}\;
 ( V^{[j]\,\dagger}_{c_j} )^{\partial_{c_j}^{[j]}}_{\beta_{c_j}}.
\end{equation}
This is equivalent to performing an SVD of a block diagonal matrix, by actually singular value decomposing
every diagonal block separately. This is not only formally meaningful, but also cheaper in terms of computational time.
In the end we can recover all the engineering we developed in section \ref{sec:minimizOBC},
which strongly exploited left and right gauges, and further enhance its computational power by
embedding symmetries.

\subsection{MPS with pointwise SU(2)}

The Heisenberg model $\vec{\sigma}_i \cdot \vec{\sigma}_j$ is the archetype of an SU(2)-invariant lattice Hamiltonian.
Being a continuous symmetry, SU(2) can not be spontaneously broken, so its ground state has to be a total spin 0
(provided it is possible by fusion rules). Nevertheless, we might be interested to describe either this ground state,
or maybe the lowest energy level at fixed total spin $q$. 

SU(2) irrep charges $c$ are typically labeled
by integer and half-integer positive numbers, i.e. $c \in \mathbb{N}/2$, and
corresponding irrep (blue link) dimension $\bar{m}_c = 2c + 1$. The MPS fragmentation scheme \cite{Diasym} then reads:
\begin{equation} \label{eq:Mpssym2}
\begin{overpic}[width = \textwidth, unit=1pt]{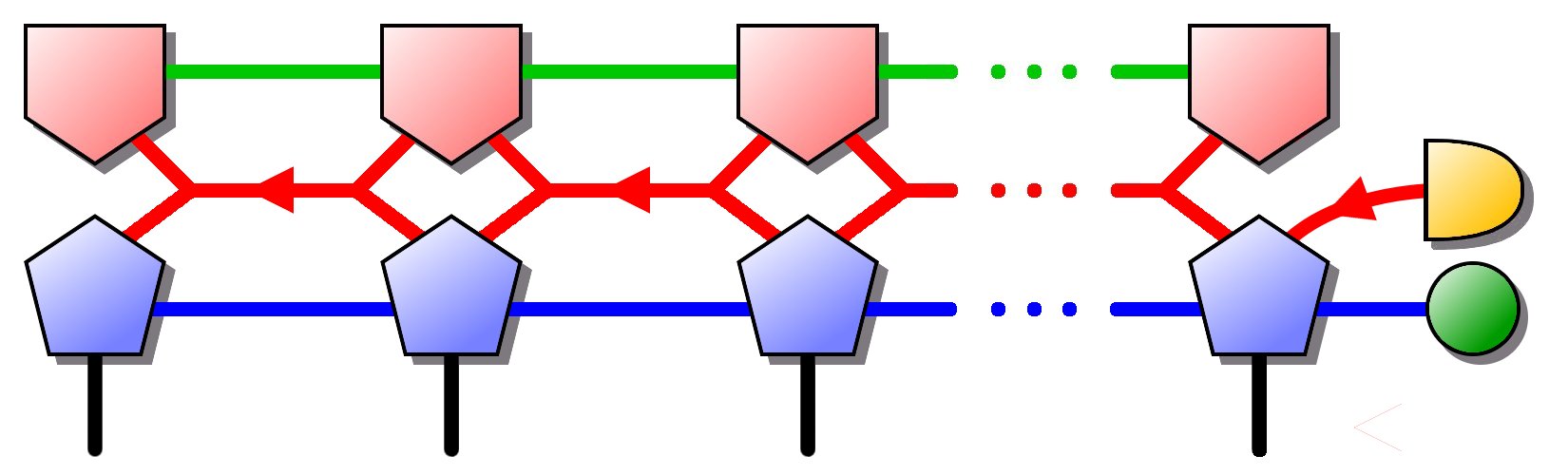}
 \put(17, 94){$R^{[1]}$}
 \put(19, 42){$S$}
 \put(64, 58){\footnotesize $c_1$}
 \put(64, 26){\footnotesize $m^{[1]}_{c_1}$}
 \put(64, 106){\footnotesize $\partial_{c_1}^{[1]}$}
 \put(106, 94){$R^{[2]}$}
 \put(108, 42){$S$}
 \put(153, 58){\footnotesize $c_2$} 
 \put(153, 26){\footnotesize $m^{[2]}_{c_2}$}
 \put(153, 106){\footnotesize $\partial_{c_2}^{[2]}$}
 \put(195, 94){$R^{[3]}$}
 \put(197, 42){$S$}
 \put(307, 94){$R^{[L]}$}
 \put(309, 42){$S$}
 \put(335, 79){\footnotesize $c_L$} 
 \put(360, 67){$Q$}
 \put(363, 37){$Y$}
 \put(29, 6){\footnotesize $s_1$}
 \put(118, 6){\footnotesize $s_2$}
 \put(207, 6){\footnotesize $s_3$}
 \put(319, 6){\footnotesize $s_L$}
\end{overpic}
\end{equation}
where structure fragments $S$ are Clebsh-Gordan coefficients.
The tensor $Y$ can be any random tensor; no algorithm based on a SU(2) invariant benchmark can determine
or variate $Y$ because it is the only non SU(2) invariant component of the MPS network.

\end{document}